\documentclass[12pt,a4paper]{article}

\usepackage{cite}
\usepackage{amsmath}
\usepackage{graphicx}
\usepackage{geometry}
\usepackage{color}

\newcommand{\half}{\frac{1}{2}}
\newcommand{\ep}{\epsilon}
\newcommand{\al}{\alpha_{s}}
\newcommand{\ale}{\alpha_{e}}
\newcommand{\qs}{q^{2}}

\makeatletter
    
    \@addtoreset{equation}{section}
\makeatother

\begin{document}
\setlength{\baselineskip}{18pt}

\begin{titlepage}

\begin{flushright}
HRI-P-14-03-001\\
KOBE-TH-15-06
\end{flushright}
\vspace{1.0cm}
\begin{center}
{\Large\bf Dipole Splitting Algorithm \\[1ex]
-- A practical algorithm to use the dipole subtraction procedure --
} 
\end{center}
\vspace{20mm}

\begin{center}
{\large
K. Hasegawa
}\footnote{{\it E-mail address}\,: kouhei@phys.sci.kobe-u.ac.jp\,.}
\end{center}
\vspace{1cm}
\centerline{{\it
Department of Physics, Kobe University,
Kobe 657-8501, Japan.}}

%
%
\vspace{2.0cm}
\centerline{\large\bf Abstract}
\vspace{0.5cm}
The Catani--Seymour dipole subtraction is a general and 
powerful procedure to calculate the QCD 
next-to-leading order corrections for collider 
observables. We clearly define a practical algorithm 
to use the dipole subtraction. The algorithm is called 
the Dipole splitting algorithm (DSA). The DSA is 
applied to an arbitrary process by following well 
defined steps. The subtraction terms created by the 
DSA can be summarized in a compact form by tables. 
We present a template for the summary tables. One 
advantage of the DSA is to allow a straightforward 
algorithm to prove the consistency relation of all 
the subtraction terms. The proof algorithm is 
presented in the subsequent article. We demonstrate 
the DSA in two collider processes, 
$pp \to \mu^{-}\mu^{+}$ and $2\,jets$. 
Further as a confirmation of the DSA it is shown 
that the analytical results obtained by the DSA at 
the Drell-Yan process exactly agree with the 
well known results obtained by the traditional method.

\end{titlepage}

\tableofcontents
\newpage
\section{Introduction}
The operating CERN Large Hadron Collider (LHC) 
discovered a new boson whose mass is around 125\,GeV
in the year 2012.
The new boson is identified as the Higgs boson in the 
standard model. In order to specify the property of the 
field and the interactions precisely we need more 
results from the LHC experiments and the various 
examinations of the results must be carried by 
comparing them to the theoretical predictions. In the 
present article, we study the theoretical 
prediction for an arbitrary process at a hadron collider 
like LHC. The calculation for the prediction
consists of two ingredients, the parton distribution 
function (PDF) and the partonic cross section. 
The PDF is a process-independent quantity and 
is determined as a numerical function from the 
experimental data. 
Recent reviews on the PDF can be found in Refs. 
\cite{DeRoeck:2011na,Perez:2012um,Blumlein:2012bf,Forte:2013wc}.
The partonic cross section is calculated by 
the perturbative expansion of the strong coupling 
constants of quantum chromodynamics\,(QCD).
The prediction, which includes only 
the leading order (LO), has a large dependence on
the renormalization scale $\mu_{R}$ 
in the strong coupling constants $\al(\mu_{R})$,
and the factorization scale $\mu_{F}$ 
in the PDF $f(x,\mu_{F})$. 
The large dependence on $\mu_{R}$ and $\mu_{F}$ 
leads to the large uncertainty of the prediction. 
The QCD next-to-leading order (NLO) corrections reduce 
the $\mu_{R}$ and $\mu_{F}$ dependence and makes 
the prediction more precise.
At next-to-next-to-leading order (NNLO),
the $\mu_{R}$ and $\mu_{F}$ dependence is more reduced. 
Actually, 
the cutting edge of the calculation technique is now 
at NNLO. The general environment for the NLO calculation
is still under development and has much room for
improvement and sophistication. In the present paper, 
we deal with the QCD NLO corrections only.
\vspace{1.5mm}

There are two main prediction schemes:
the {\em matrix element prediction} and 
the {\em showered prediction}, as mentioned in 
Ref.\cite{Frederix:2009yq}. 
The matrix element prediction is as follows.
The partonic cross section includes the matrix element 
that represents the transition amplitude from the initial 
partons to the final partons.  
A jet algorithm defines the jet observables.
Typical jet algorithms are constructed in Refs.
\cite{Catani:1993hr,Ellis:1993tq,Dokshitzer:1997in,
Wobisch:1998wt,Cacciari:2008gp}. 
A jet algorithm is directly 
applied to the partons in the final state of a matrix 
element and the distributions of the jet observables are 
compared to the experimental results.
The showered prediction is as follows. The final-state partons 
of a matrix element are showered by a 
shower algorithm. Then a jet algorithm is applied to 
the partons after showering and finally the distributions 
of the jet observables are compared to the experimental 
results. The hadronization effect may also be included 
for better simulation. The merit of the matrix element 
prediction is that it is less involved and simpler than 
the showered prediction. The merit of the 
showered prediction is that it simulates the phenomena 
happening after the collision better than the matrix 
element prediction. In the present article, we deal with QCD 
NLO corrections only in the matrix element prediction 
scheme for simplicity. 
\vspace{1.5mm}

The tradition of the calculation of
QCD NLO corrections at hadron 
colliders probably started with the Drell-Yan process,
$pp/p\bar{p} \to \mu^{-}\mu^{+}$,
in the pioneering works of Refs.
\cite{Altarelli:1978id,Altarelli:1979ub,Abad:1978nr,
Abad:1978ke,Humpert:1979qk,Humpert:1979hb,Humpert:1980uv,
KubarAndre:1978uy,Harada:1979bj}; 
this process might be the simplest of the hadron collider 
processes. In the pioneering works, the analytical 
expression of the NLO corrections is obtained by using 
the dimensional regularization in $d$-dimensions
throughout. All the ultraviolet (UV) and
infrared, or, more precisely, soft and collinear, 
divergences are regulated as the poles of 
$1/\ep$ and $1/\ep^{2}$. 
The NLO corrections to the cross section are generally 
written as
\begin{equation}
\sigma_{\mbox{{\tiny NLO}}} 
= \sigma_{\mbox{{\tiny R}}} 
+ \sigma_{\mbox{{\tiny V}}} 
+ \sigma_{\mbox{{\tiny C}}}\,,
\label{mast}
\end{equation}
where the symbols, 
$\sigma_{\mbox{{\tiny R}}}, \sigma_{\mbox{{\tiny V}}}$, 
and 
$\sigma_{\mbox{{\tiny C}}}$
represent the real emission correction, the virtual 1-loop 
correction, and the collinear subtraction term, 
respectively. The 1-loop matrix element in the virtual 
correction includes the UV divergences that are 
subtracted by the renormalization program. 
Each of the three terms,
$\sigma_{\mbox{{\tiny R}}}, \sigma_{\mbox{{\tiny V}}}$,
and 
$\sigma_{\mbox{{\tiny C}}}$
includes the soft/collinear divergences.
Complete cancellation of the soft and collinear divergences 
from the three terms can be realized 
only after analytical integration of the $d$-dimensional
phase space (PS). Then we obtain the finite 
physical cross section. For more complicated multiparton 
processes, this method mainly encounters the two
difficulties:
the evaluation of the 1-loop matrix element and 
the analytical integration of the $d$-dimensional PS.
This is because the matrix elements for the multiparton leg 
processes become complex and long expressions.
The difficulty with the evaluation of the 1-loop 
matrix element has been solved by the technical 
developments. Reviews on the recent developments 
can be found in Refs.
\cite{Bern:2007dw,Ellis:2011cr,Ossola:2013jea}. 
\vspace{1.5mm}

During the 1990s, the methods were
invented to overcome the difficulty 
with the analytical integration of the $d$-dimensional PS: 
the phase-space slicing method 
\cite{Giele:1991vf,Giele:1993dj,Keller:1998tf,Harris:2001sx}
and the subtraction method 
\cite{Catani:1996vz,Catani:2002hc,Frixione:1995ms}.
Among these, the Catani--Seymour dipole subtraction 
procedure 
\cite{Catani:1996vz,Catani:2002hc} 
in the subtraction method is quite successful and has 
been widely used. In the dipole subtraction procedure, 
subtraction terms are introduced and the NLO 
corrections are reconstructed as
\begin{equation}
\sigma_{\mbox{{\tiny NLO}}} 
= (\sigma_{\mbox{{\tiny R}}} 
- \sigma_{\mbox{{\tiny D}}})
+ (\sigma_{\mbox{{\tiny V}}} 
+ \sigma_{\mbox{{\tiny I}}}) 
+ \sigma_{\mbox{{\tiny P}}} 
+ \sigma_{\mbox{{\tiny K}}}\,,
\label{dipmast}
\end{equation}
where the symbols
$\sigma_{\mbox{{\tiny D}}}, \sigma_{\mbox{{\tiny I}}}$,
$\sigma_{\mbox{{\tiny P}}}$,
and 
$\sigma_{\mbox{{\tiny K}}}$
represent the the dipole\,(D), I, P, and K terms, 
respectively. The terms 
$\sigma_{\mbox{{\tiny D}}}$ 
and 
$\sigma_{\mbox{{\tiny I}}}$ 
subtract all the soft/collinear divergences from
the terms
$\sigma_{\mbox{{\tiny R}}}$ 
and 
$\sigma_{\mbox{{\tiny V}}}$ 
respectively, 
at the integrand level 
on each phase-space point.
Then the PS integration of the subtracted cross 
sections, 
$(\sigma_{\mbox{{\tiny R}}} 
- \sigma_{\mbox{{\tiny D}}})$ 
and 
$(\sigma_{\mbox{{\tiny V}}} 
+ \sigma_{\mbox{{\tiny I}}})$,
can be executed in 4 dimensions to be finite. 
The $\sigma_{\mbox{{\tiny P}}}$ 
and
$\sigma_{\mbox{{\tiny K}}}$ 
are separately finite under the 4 dimensional 
integration. In this way, in the dipole subtraction, all 
the PS integrations are done in 4 dimensions and the 
procedure can avoid the difficulty of $d$-dimensional 
analytical integration. Thus, the procedure makes possible 
the calculation of the NLO corrections for multiparton 
processes. The dipole subtraction has already been
used for many processes at LHC. The complete list of
achievements is too long to show here, 
so we select only the most impressive ones:
processes with massless quarks
\cite{
Nagy:2001fj,
DelDuca:2003uz,
Figy:2003nv,
Nagy:2003tz,
Oleari:2003tc,
Dittmaier:2007th,
Binoth:2008kt,
Binoth:2009rv,
Binoth:2009wk,
Harlander:2010cz,
Greiner:2011mp,
Badger:2012pf,
Denner:2012dz,
Altenkamp:2012sx,
Cullen:2012eh,
Greiner:2013gca,
Campanario:2013fsa,
Gehrmann:2013bga,
Badger:2013yda
};
processes with massive quarks
\cite{
Beenakker:2001rj,
Harris:2002md,
Beenakker:2002nc,
Dittmaier:2007wz,
Ellis:2009zw,
Bredenstein:2009aj,
KeithEllis:2009bu,
Bevilacqua:2009zn,
Bevilacqua:2010ve,
Greiner:2012im,
GoncalvesNetto:2012nt,
Bevilacqua:2012em,
Denner:2012yc,
Bevilacqua:2013taa,
vanDeurzen:2013xla
};
and processes done by projects,
the MCFM
\cite{
Campbell:2002tg,
Campbell:2004ch,
Campbell:2005bb,
Campbell:2006xx,
Campbell:2007ev,
Campbell:2009ss,
Melnikov:2010iu,
Melia:2010bm,
Badger:2010mg,
Melnikov:2011ta,
Campbell:2012ft
},
the VBFNLO
\cite{
Bozzi:2007ur,
Figy:2007kv,
Hankele:2007sb,
Campanario:2008yg,
Arnold:2008rz,
Jager:2009xx,
Campanario:2010hp,
Campanario:2011ud,
Campanario:2013qba
},
and the BlackHat
\cite{
Berger:2009zg,
Berger:2010vm,
Berger:2010zx,
Bern:2011ep,
Bern:2013gka,
Bern:2013bha}.
Of these, the highest achievements
up to now may be those in the following processes:
$pp \to$ 5jets \cite{Badger:2013yda}, 
$pp \to W+$ 5jets \cite{Bern:2013gka},
$pp \to t \bar{t} b\bar{b}$ \cite{Bredenstein:2009aj},
and
$pp \to t \bar{t}$ + 2jets \cite{Bevilacqua:2010ve}.
The large number of successful calculations proves 
the generality and the strong power of the dipole 
subtraction. The price that we must pay to employ the 
procedure is the construction of the subtraction terms, 
the D, I, P, and K terms. For multiparton leg processes,
the total number of all the subtraction terms 
sometimes exceeds 
one hundred and the expression of each subtraction 
term is not so simple. Construction by hand requires 
too much working time and we usually make some mistakes.
Thus, automation as a computer program is really 
desirable. Fortunately, the construction 
of all the subtraction terms is given in a general 
algorithm and hence automation as a computer program 
is possible. Several computer packages have
already been made 
\cite{Gleisberg:2007md,Seymour:2008mu,Hasegawa:2008ae,
Hasegawa:2009tx,Hasegawa:2010dz,Frederix:2008hu,
Frederix:2010cj,Czakon:2009ss}.
Among these, the publicly available packages are those
in Refs.
\cite{Seymour:2008mu,Hasegawa:2009tx,Hasegawa:2010dz,
Frederix:2008hu,Czakon:2009ss}.
\vspace{1.5mm}

Now that the dipole subtraction has been applied for 
so many processes, we can see some drawbacks. Among 
them, we would like to point out the following 
difficulties.
The person who has obtained the results for 
the NLO corrections by dipole subtraction
sometimes has difficulty confirming the
results, because many subtraction terms 
are involved and a large amount of
calculations are executed as numerical
evaluation for the Monte Carlo integration.
For the other person who does not do the calculations
him- or herself, confirmation of the results shown
is more difficult.
All the soft and collinear divergences
from the D and I terms must exactly 
cancel the divergences
from the real correction 
$\sigma_{\mbox{{\tiny R}}}$ 
and the virtual correction 
$\sigma_{\mbox{{\tiny V}}}$, 
respectively. 
If we use any wrong collection or any wrong expression for 
the D and I terms, the cancellation is spoiled.
A successful cancellation provides one check on the 
singular parts of the D and I terms. The P and K terms are 
separately finite by 
themselves and a check by the cancellation 
is impossible.
In this sense, the uses of the P and K terms are the place 
where we can easily make mistakes.
In many articles, it is not 
clear which and how many subtraction terms are used. 
Of course, this explanation would be too long and
it is unreasonable to expect that 
explicit expressions of all subtraction terms are given 
in the articles. However, it is possible that minimal information 
specifying the subtraction terms is shown. At least the 
total number of subtraction terms used should be 
clearly mentioned. 
In many works, the dipole subtraction is applied by using 
automated programs. In this case, similar 
criticisms should be made about the implementation, 
especially the algorithm to create the subtraction terms. 
In the packages in Refs.
\cite{Frederix:2008hu} and \cite{Czakon:2009ss},
the creation algorithms are not so strictly documented,
and, on the execution of the packages information on
the terms created is not printed out. 
We cannot know which and how many D, I, and P/K terms are 
created under a given input process. 
In the package in Ref.
\cite{Hasegawa:2009tx}, 
the creation algorithm is presented in the article.
In the output of the Mathematica program information 
on all the subtraction terms created and the total number
of the terms are printed out separately. 
The output codes of the D, I, 
and P/K terms are saved  separately in the corresponding 
folders and can be easily identified. In all three packages, 
no algorithm to check the consistency of all the 
subtraction terms created is provided. 
\vspace{1.5mm}

In order to solve one part of the difficulties
and the criticisms,
we need a clear 
definition of a practical algorithm to use the dipole 
subtraction. Although the general algorithm is given in the 
original articles
\cite{Catani:1996vz,Catani:2002hc},
we need a more practical algorithm that can be directly 
used step by step. Also, to automatize the 
procedure into a computer program, we need such a practical 
algorithm to be applied to an arbitrary process. 
In consequence,
we desire an algorithm that provides
clear definitions of the following items:
\begin{enumerate}
\item Input, output, creation order, and
all formulae in the document,
\item Necessary information to specify each subtraction term,
\item Summary table of all subtraction terms created,
\item Associated proof algorithm.
\end{enumerate}
In the last entry in the wishlist,
an associated proof algorithm 
means the following.
When the NLO cross sections
in Eqs.\,(\ref{mast}) and (\ref{dipmast}) 
are equated,
we obtain the relation of the cancellation as
\begin{equation}
\sigma_{\mbox{{\tiny subt}}} 
= \sigma_{\mbox{{\tiny D}}} 
+ \sigma_{\mbox{{\tiny C}}}
- \sigma_{\mbox{{\tiny I}}} 
- \sigma_{\mbox{{\tiny P}}} - 
\sigma_{\mbox{{\tiny K}}}=0\,. 
\label{subtmast}
\end{equation}
We call this relation the consistency relation
of the subtraction terms.
The consistency relation is an identity
that means that the relation stands
in an arbitrary process.
For the factorization scale $\mu_{F}$
in Eq.\,(1.3),
apart from the PDFs, only the two terms
$\sigma_{\mbox{{\tiny C}}}$
and 
$\sigma_{\mbox{{\tiny P}}}$
have the same $\mu_{F}$ dependence,
which cancel each other.
An associated proof algorithm is 
an algorithm to prove 
the consistency relation
in Eq.\,(\ref{subtmast}) 
for an arbitrary process.
\vspace{1.5mm}

The purpose of this article is to present a practical 
algorithm that satisfies all the requirements 
in the above wishlist. We actually present such an 
algorithm, called the dipole splitting algorithm 
(DSA). In the DSA, the input is all real emission processes
that contribute to a collider process like $pp \to n\,jets$. 
Each real emission process, such as like
$u\bar{u} \to u\bar{u}g$, 
creates the output of the D, I, P/K terms, 
and all of them
have the same initial states as the input real 
process. 
In this sense, the subtraction terms are sorted by the 
initial-state partons. All the subtraction 
terms are also sorted by the kinds of the splittings that
each subtraction term possesses as a part. The sorting by
splittings is equivalent with the sorting by the reduced 
Born process that each subtraction term possesses
as a remaining part when the splittings part is removed. 
In order to specify each subtraction term uniquely we 
introduce a bijection mapping, called {\em field mapping}. 
Each field mapping is made for each subtraction term.
The field mapping exactly specifies the connection between
the legs of the Born process reduced from an input real 
process and the fixed legs of the same reduced Born process. 
By using the field mapping, we can specify each term in a 
compact form without confusion. 
Thanks to this well defined compact form,
we can summarize the subtraction terms in tables.
We will present a template for the summary tables.
We intend that the person who does not create
the subtraction terms him- or herself
can specify and write down them
only by reading the tables in a document
without any direct communication with
the author of the tables.
For the last entry in 
the wishlist, we have constructed a straightforward algorithm to 
prove the consistency relation of the subtraction terms
created by the DSA.
We present the proof algorithm (PRA)
in the following paper
\cite{Hasegawa:2014nna}.
\vspace{1.5mm}

We mention here the relation between the DSA and the 
algorithm implemented in the AutoDipole package 
\cite{Hasegawa:2009tx}. 
The creation algorithm of the D and I terms in the DSA is
essentially the same as the algorithm in AutoDipole. 
In this sense, the DSA originates in the algorithm of
AutoDipole. In the DSA, the concrete expressions are more 
clearly defined by using the field mapping
and the necessary information to specify each subtraction 
is given in a compact form. 
The creation algorithm for the P and K terms in the DSA is 
different from the algorithm in AutoDipole.
In the DSA, the initial states of all the created subtraction 
terms under one input real process are the same as the initial 
state of the input real process. In the 
algorithm of AutoDipole, the initial states of the P and 
K terms with the nondiagonal splittings are different from
the initial state of the input real process. 
The difference in the creation of the P and K terms
makes possible the construction of the proof algorithm
associated with the DSA.
\vspace{1.5mm}

The present article is organized as follows.
The DSA is defined in Sec.\,\ref{sec_2}.
The formulae for the subtraction terms 
are collected
in Appendix \ref{ap_A}.
The DSA is demonstrated
in the collider processes, 
$pp \to \mu^{-}\mu^{+}$ and $2 \, jets$,
in Sec.\,\ref{sec_3} and \ref{sec_4}
respectively.
The summary tables
for the dijet process are shown
in Appendix \ref{ap_B}.
We give a confirmation that
the analytical results obtained by the DSA 
in the Drell-Yan process coincide 
with the results by the traditional method
in Sec.\,\ref{sec_5}.
We give a summary in
Sec.\,\ref{sec_6}.

\clearpage





\section{Dipole splitting algorithm \label{sec_2}}
In this section, the DSA is defined.
In Sec.\,\ref{s2_1}, all the steps of the DSA and 
the master formulae are presented. 
In Sec.\,\ref{s2_2}, \ref{s2_3}, and \ref{s2_4}, 
the creation algorithms of the D, I, and P/K terms 
are explained, respectively. 
The advantages of the DSA are clarified 
in Sec.\,\ref{s2_5}.
The formulae for all the subtraction terms 
are collected in Appendix\,\ref{ap_A}.

\subsection{Definition and Step\,1: List of $\mbox{R}_{i}$ \label{s2_1}}
The DSA consists of the following steps\,:
\begin{align}
\mbox{{\bf Step 1.}}  & \ \ \mbox{List of real emission processes} 
\ \{\mbox{R}_{i}\}\,, \nonumber\\[5pt] 
\mbox{{\bf 2.}} & \ \ \mbox{D}\,(\mbox{R}_{i})\,, \nonumber\\[5pt] 
\mbox{{\bf 3.}} & \ \ \mbox{I}\,(\mbox{R}_{i})\,, \nonumber\\[5pt] 
\mbox{{\bf 4.}} & \ \ \mbox{P}\,(\mbox{R}_{i}) 
\ \mbox{and} \ \mbox{K}\,(\mbox{R}_{i})\,, \nonumber\\[5pt] 
\mbox{{\bf 5.}} & \ \ \sigma_{\mbox{{\tiny NLO}}}= \sum_{i} 
\sigma\,(\mbox{R}_{i})\,.
\label{dsastep}
\end{align}
We assume that we wish to make a prediction 
for the observables in a collider process. 
Once a collider process is 
selected, the contributing partonic real emission processes are 
specified. {\bfseries Step 1} is to specify all the real 
processes denoted as $\mbox{R}_{i}$ and to write down the list.
The number of real processes is written as $n_{\mbox{{\tiny real}}}$
and the number of fields in the final states is denoted as $(n+1)$. 
For example, the collider process $pp \to 2 jets$ has the real 
emission processes as
\begin{align}
\mbox{R}_{1} &= u\bar{u} \to u\bar{u}g\,, \nonumber\\
\mbox{R}_{2} &= uu \to uug\,,   \nonumber\\
& \ \ \vdots \nonumber\\
\mbox{R}_{n_{\mbox{{\tiny real}}}} &= gg \to ggg\,.
\label{listex}
\end{align}
{\bfseries Steps 2}, {\bfseries 3}, and {\bfseries 4}, are 
explained in Sec.\,\,\ref{s2_2}, \ref{s2_3}, and 
\ref{s2_4}, respectively. The last step of the DSA, 
{\bfseries Step 5,} is to obtain the NLO corrections as
\begin{equation}
\sigma_{\mbox{{\tiny NLO}}} = \sum_{i=1}^{n_{\mbox{{\tiny real}}}} 
\sigma(\mbox{R}_{i})\,, \label{masternlo}
\end{equation}
where each cross section $\sigma(\mbox{R}_{i})$ is defined as
\begin{align}
\sigma(\mbox{R}_{i}) 
&= \int dx_{1} \int dx_{2} \ 
f_{\mbox{{\tiny F}}(x_{a})}(x_{1}) 
f_{\mbox{{\tiny F}}(x_{b})}(x_{2}) \ \times  \nonumber \\
& \ \ \biggl[
\bigl(\hat{\sigma}_{\mbox{{\tiny R}}}(\mbox{R}_{i}) - 
\hat{\sigma}_{\mbox{{\tiny D}}}(\mbox{R}_{i}) \bigr) + 
\bigl(\hat{\sigma}_{\mbox{{\tiny V}}}(\mbox{B}1(\mbox{R}_{i})) + 
\hat{\sigma}_{\mbox{{\tiny I}}}(\mbox{R}_{i}) \bigr) + 
\hat{\sigma}_{\mbox{{\tiny P}}}(\mbox{R}_{i}) + 
\hat{\sigma}_{\mbox{{\tiny K}}}(\mbox{R}_{i}) \biggr]\,,  
\label{master}
\end{align}
where $f_{\mbox{{\tiny F}}(x_{a/b})}(x_{1/2})$ is the PDF and 
the subscript $\mbox{F}(x_{a/b})$ represents the field 
species of the initial-state parton 
of the leg a/b, which is again defined in the 
next section. The symbols $\hat{\sigma}(\mbox{R}_{i})$ with 
subscripts R, D, V, I, P, and K represent the partonic 
cross sections of the real correction, the dipole (D) term, the 
virtual correction, the I term, the P term, and the K term, 
respectively. Each partonic cross section is defined as
\begin{align}
\hat{\sigma}_{\mbox{{\tiny R}}}(\mbox{R}_{i}) 
&= \frac{1}{S_{\mbox{{\tiny R}}_{i}}} \ \Phi(\mbox{R}_{i})_{4} 
\cdot |\mbox{M}(\mbox{R}_{i})|_{4}^{2}\,,
 \label{hatr} \\
\hat{\sigma}_{\mbox{{\tiny D}}}(\mbox{R}_{i}) 
&= \frac{1}{S_{\mbox{{\tiny R}}_{i}}} \ \Phi(\mbox{R}_{i})_{4} 
\cdot \frac{1}{n_{s}(a) n_{s}(b)} \cdot \mbox{D}(\mbox{R}_{i})\,,
\label{hatd} \\
\hat{\sigma}_{\mbox{{\tiny V}}}(\mbox{B}1) 
&= \frac{1}{S_{\mbox{{\tiny B1}} }} \ \Phi( \mbox{B}1  )_{d} 
\cdot |\mbox{M}_{\mbox{{\tiny virt}}}(\mbox{B}1)|_{d}^{2}\,,  
\label{hatv} \\
\hat{\sigma}_{\mbox{{\tiny I}}}(\mbox{R}_{i}) 
&= \frac{1}{S_{\mbox{{\tiny B1}} }} \ \Phi(\mbox{B}1)_{d} 
\cdot \mbox{I}(\mbox{R}_{i})\,, 
\label{hati} \\
\hat{\sigma}_{\mbox{{\tiny P}}}(\mbox{R}_{i}) 
&= \int_{0}^{1}dx \sum_{\mbox{{\tiny B}}_{j}} 
\frac{1}{S_{\mbox{{\tiny B}}_{j} }}
\Phi_{a}(\mbox{R}_{i}:\mbox{B}_{j},x)_{4} \cdot 
\mbox{P}(\mbox{R}_{i},x_{a}:\mbox{B}_{j},xp_{a}) \ + \ 
(a \leftrightarrow b)\,, 
 \label{hatp} \\
\hat{\sigma}_{\mbox{{\tiny K}}}(\mbox{R}_{i}) 
&= \int_{0}^{1}dx \sum_{\mbox{{\tiny B}}_{j}} 
\frac{1}{S_{\mbox{{\tiny B}}_{j} }}
\Phi_{a}(\mbox{R}_{i}:\mbox{B}_{j},x)_{4} \cdot 
\mbox{K}(\mbox{R}_{i},x_{a}:\mbox{B}_{j},xp_{a})  \ + \ 
(a \leftrightarrow b)\,. 
\label{hatk}
\end{align} 
In Eqs.\,(\ref{hatr}) and (\ref{hatd}), the symbol
$S_{\mbox{{\tiny R}}_{i}}$
is the symmetric factor of the real process $\mbox{R}_{i}$
and the symbol
$\Phi(\mbox{R}_{i})_{4}$ is the 4-dimensional 
($n$+1)-body PS including the flux factor as
\begin{equation}
\Phi(\mbox{R}_{i})_{4} = \frac{1}{{\cal F}(p_{a},p_{b})} 
\prod_{i=1}^{n+1} \int \frac{d^{3}p_{i}}{(2\pi)^{3}} 
\frac{1}{2E_{i}} \cdot (2\pi)^{4} \delta^{(4)} 
\Bigl(p_{a}+p_{b}- \sum_{i=1}^{n+1} p_{i} \Bigr)\,.
\end{equation}
The energy is denoted as 
$E_{I}$ for $I=a,b,1,...,n+1$.
The flux factor is written as
${\cal F}(p_{a},p_{b})=4E_{a}E_{b}$\,.
$|\mbox{M}(\mbox{R}_{i})|_{4}^{2}$ is the square of 
the matrix element of the real emission process 
$\mbox{R}_{i}$ after 
the average over spin and color in 4 dimensions. 
The $n_{s}(a/b)$ 
represents the spin degree of freedom of the leg a/b 
in $\mbox{R}_{i}$ in 4 dimensions. 
In Eqs.\,(\ref{hatv}) and 
(\ref{hati}), the process $\mbox{B}1$ is an abbreviation of
$\mbox{B}1(\mbox{R}_{i})$, which is made by removing one gluon 
from the final states of $\mbox{R}_{i}$, denoted as 
$\mbox{B}1(\mbox{R}_{i})=\mbox{R}_{i} - g$.
$S_{\mbox{{\tiny B1}} }$ is the symmetric factor of
process $\mbox{B}1$, and the symbol $\Phi( \mbox{B}1 )_{d}$
represents the $d$-dimensional $n$-body PS with the flux factor as
\begin{equation}
\Phi(\mbox{B}1)_{d} = \frac{1}{{\cal F}(p_{a},p_{b})} 
\prod_{i=1}^{n} \int \frac{d^{d-1}p_{i}}{(2\pi)^{d-1}} 
\frac{1}{2E_{i}} \cdot (2\pi)^{d} \delta^{(d)} 
\Bigl(p_{a}+p_{b}- \sum_{i=1}^{n} p_{i} \Bigr)\,.
\label{b1ps}
\end{equation}
The symbol $|\mbox{M}_{\mbox{{\tiny virt}}}(\mbox{B}1)|_{d}^{2}$ is 
the abbreviation
of the quantity $( \mbox{M}_{\mbox{{\tiny LO}}}(\mbox{B}1)$ 
$\mbox{M}_{\mbox{{\tiny 1-loop}}}(\mbox{B}1)^{*}$ + 
$\mbox{M}_{\mbox{{\tiny LO}}}(\mbox{B}1)^{*}$ 
$\mbox{M}_{\mbox{{\tiny 1-loop}}}(\mbox{B}1))$
after the average over spin and color in $d$ dimensions, 
where 
$\mbox{M}_{\mbox{{\tiny LO}}}(\mbox{B}1)$ and
$\mbox{M}_{\mbox{{\tiny 1-loop}}}(\mbox{B}1)$
are the matrix elements of the LO and the 1-loop correction 
of process $\mbox{B}1$, respectively. 
In Eqs.\,(\ref{hatp}) and (\ref{hatk}), $\mbox{B}_{j}$ is 
an abbreviation of $\mbox{B}_{j}(\mbox{R}_{i})$ and is 
a Born process reduced from $\mbox{R}_{i}$, 
which is precisely defined in Sec.\,\ref{s2_2}. 
$S_{\mbox{{\tiny B}}_{j} }$ is the symmetric factor of
$\mbox{B}_{j}$ and
$\Phi_{a}(\mbox{R}_{i}:\mbox{B}_{j},x)_{4}$ 
is the 4-dimensional $n$-body PS with the scaled momentum 
$xp_{a}$ and the flux factor 
${\cal F}(xp_{a},p_{b})=4\,x E_{a}E_{b}$ as 
\begin{equation}
\Phi_{a}(\mbox{R}_{i}:\mbox{B}_{j},x)_{4} = 
\frac{1}{{\cal F}(xp_{a},p_{b})} \prod_{i=1}^{n}
\int \frac{d^{3}p_{i}}{(2\pi)^{3}} \frac{1}{2E_{i}}
\cdot (2\pi)^{4} \delta^{(4)} 
\Bigl( xp_{a}+p_{b}- \sum_{i=1}^{n} p_{i} \Bigr).
\label{pkps}
\end{equation}
In Eqs.\,(\ref{hatd}), (\ref{hati}), (\ref{hatp}), and 
(\ref{hatk}),
concrete expressions of the subtraction terms
$\mbox{D}(\mbox{R}_{i})$, $\mbox{I}(\mbox{R}_{i})$, and
$\mbox{P/K}(\mbox{R}_{i}:\mbox{B}_{j},xp_{a})$ are presented in 
Sec.\,\,\ref{s2_2}, \ref{s2_3}, and \ref{s2_4},
respectively.
The jet functions
$F_J^{(n/n+1)}(p_{1},...,p_{n/n+1})$
must be multiplied to the partonic 
cross sections
in Eqs.\,(\ref{hatr})--(\ref{hatk}).
For the real correction in Eq.\,(\ref{hatr}),
the jet function with $(n+1)$ fields 
is multiplied as
\begin{equation}
\hat{\sigma}_{\mbox{{\tiny R}}}(\mbox{R}_{i}) 
= \frac{1}{S_{\mbox{{\tiny R}}_{i}}} \ \Phi(\mbox{R}_{i})_{d} 
\cdot |\mbox{M}(\mbox{R}_{i})|_{d}^{2}
\cdot F_J^{(n+1)}(p_{1},...,p_{n+1})
\,.
\end{equation}
For the cross sections 
in Eqs.\,(\ref{hatv})--(\ref{hatk}),
the jet function with $n$ fields 
$F_J^{(n)}(p_{1},...,p_{n})$
is multiplied.
For the dipole term in Eq.\,(\ref{hatd}),
the jet function $F_J^{(n)}$
is multiplied and the $n$ momenta
of the arguments are identified 
with the $n$ reduced momenta
$(\mbox{P}(y_{1}),...,\mbox{P}(y_{n}))$.
The details of the use of the jet functions
in the dipole subtraction 
are explained in Ref. \cite{Catani:1996vz}.
For compact notation, we do not show
the jet functions explicitly hereafter.

The PS integration in Eq.\,(\ref{master}) is finite in 4 dimensions 
and we see here the finite parts separately. The real correction
$\hat{\sigma}_{\mbox{{\tiny R}}}(\mbox{R}_{i})$ has soft and
collinear divergences,
which are subtracted by the dipole terms 
$\hat{\sigma}_{\mbox{{\tiny D}}}(\mbox{R}_{i})$.
The subtracted cross section 
$(\hat{\sigma}_{\mbox{{\tiny R}}}(\mbox{R}_{i}) - 
\hat{\sigma}_{\mbox{{\tiny D}}}(\mbox{R}_{i}))$
is finite in 4 dimensions as
\begin{equation}
\hat{\sigma}_{\mbox{{\tiny R}}}(\mbox{R}_{i}) - 
\hat{\sigma}_{\mbox{{\tiny D}}}(\mbox{R}_{i}) =
\frac{1}{S_{\mbox{{\tiny R}}_{i}}} \Phi(\mbox{R}_{i})_{4} 
\cdot \Bigl[ \
|\mbox{M}_{\mbox{{\tiny real}}}(\mbox{R}_{i})|_{4}^{2} \ - \ 
\frac{1}{n_{s}(a) n_{s}(b)} \mbox{D}(\mbox{R}_{i}) \ 
\Bigr] \ < \ \infty \,.
\end{equation}
The virtual correction
$\hat{\sigma}_{\mbox{{\tiny V}}}(\mbox{B}1)$
includes the poles of the soft and collinear divergences 
$1/\ep$ and $1/\ep^{2}$ after 
the subtraction of the UV 
divergences $1/\ep_{{\tiny \mbox{UV}}}$ by 
the renormalization program. The I term
$\hat{\sigma}_{\mbox{{\tiny I}}}(\mbox{R}_{i})$ 
cancels all the soft and collinear poles 
in $d$ dimensions as
\begin{equation}
\hat{\sigma}_{\mbox{{\tiny V}}}(\mbox{B}1) + 
\hat{\sigma}_{\mbox{{\tiny I}}}(\mbox{R}_{i}) = 
\frac{1}{S_{\mbox{{\tiny B1}} }} \ 
\Phi( \mbox{B}1  )_{d} \cdot 
\Bigl[ \ |\mbox{M}_{\mbox{{\tiny virt}}}(\mbox{B}1)|_{d}^{2} \ + \ 
\mbox{I}(\mbox{R}_{i}) \ \Bigr].
\label{viddim}
\end{equation}
After the cancellation of the poles, we can integrate 
the PS in 4 dimensions as
\begin{equation}
\hat{\sigma}_{\mbox{{\tiny V}}}(\mbox{B}1) + 
\hat{\sigma}_{\mbox{{\tiny I}}}(\mbox{R}_{i}) = 
\frac{1}{S_{\mbox{{\tiny B1}} }} \ 
\Phi( \mbox{B}1  )_{4} \cdot 
\Bigl[ \ |\mbox{M}_{\mbox{{\tiny virt}}}(\mbox{B}1)|^{2} 
\ + \ \mbox{I}(\mbox{R}_{i})\ \Bigr]_{4}   
\ < \ \infty \,.
\end{equation}
The cross sections of the P and K terms, 
$\hat{\sigma}_{\mbox{{\tiny P}}}(\mbox{R}_{i})$ and 
$\hat{\sigma}_{\mbox{{\tiny K}}}(\mbox{R}_{i})$, 
are themselves finite separately in 4 dimensions.
In this way, all the PS integrations in 
Eq.\,(\ref{master}) are executed in 4 dimensions.

To complete the master formulae, we add the LO 
contribution as 
\begin{align}
\sigma_{\mbox{{\tiny LO}}} 
&= \sum_{i=1}^{n_{\mbox{{\tiny LO}}}} 
\sigma(\mbox{L}_{i}), 
\label{masterlo}\\
\sigma(\mbox{L}_{i}) 
&= \int dx_{1} \int dx_{2} \, f_{F(x_{a})}(x_{1})\, 
f_{F(x_{b})}(x_{2}) \ 
\hat{\sigma}(\mbox{L}_{i}),  
\label{mastlocr}\\
\hat{\sigma} (\mbox{L}_{i}) 
&= \frac{1}{S_{\mbox{{\tiny L}}_{i}}} \ 
\Phi(\mbox{L}_{i})_{4} \cdot |\mbox{M} 
(\mbox{L}_{i})|_{4}^{2},
 \label{mastlopa}
\end{align}
where the subpartonic LO processes that contribute 
to the selected collider process are denoted as 
$\mbox{L}_{i}$ and the number of LO processes 
is denoted as $n_{\mbox{{\tiny LO}}}$.
$\Phi(\mbox{L}_{i})_{4}$ is the 4-dimensional
$n$-body PS including the flux factor.
Then the prediction at NLO accuracy is written as
\begin{equation}
\sigma_{\mbox{{\tiny prediction}}} = 
\sigma_{\mbox{{\tiny LO}}} + \sigma_{\mbox{{\tiny NLO}}}, 
\end{equation}
where $\sigma_{\mbox{{\tiny LO}}}$ and 
$\sigma_{\mbox{{\tiny NLO}}}$ are defined in 
Eqs.\,(\ref{masterlo}) and (\ref{masternlo}),
respectively.
%
%
%
%
%
\subsection{Step\,2: D term creation  \label{s2_2}}
In this section, \textquoteleft{\bfseries Step 2}. 
D($\mbox{R}_{i}$)\textquoteright \ is explained. The input 
and output of the step are written as
\begin{description}
\setlength{\itemsep}{-1mm}
\item[\ \ \ \ Input:] \ \ \ $\mbox{R}_{i}$\,,
\item[\ \ \ \ Output:] \ D($\mbox{R}_{i}$)\,.
\end{description}
The input process is each real emission process 
$\mbox{R}_{i}$ 
among all real processes,
$\mbox{R}_{1}$,...,$\mbox{R}_{n_{\mbox{{\tiny real}}}}$, 
which are specified in
{\bfseries Step 1}.
The input defines set \{x\} with 
the field species 
$\mbox{F}(x_{\mbox{{\tiny I}}})$
and the momenta $p_{_{\mbox{{\tiny I}}}}$ 
for the indices $I=a,b,1,..., n+1$, as
\begin{align}
\{x\} &= \{x_{a},x_{b};x_{1}, ..., x_{n+1} \}, 
\label{xdef} \\ 
\mbox{F}(\{x\}) 
&= \{\mbox{F}(x_{a}),\mbox{F}(x_{b});\mbox{F}(x_{1}), 
..., \mbox{F}(x_{n+1}) \}\,, \\
\mbox{Momenta}: 
& \ \ \ \     \{p_{a},p_{b} \ ;p_{1},...,p_{n+1} \}\,. 
\label{orimom}
\end{align}
The output D($\mbox{R}_{i}$) is the summation of all 
the created dipole terms. 
The creation is repeated over all 
the input processes $\mbox{R}_{i}$ with 
$i=1,...,n_{\mbox{{\tiny real}}}$,
and the outputs are the corresponding dipole terms
D($\mbox{R}_{i}$) with 
$i=1,...,n_{\mbox{{\tiny real}}}$\,.

In the original article by Catani and Seymour 
\cite{Catani:1996vz}, each dipole term is 
specified by three legs $(I,J,K)$, where 
we call the pair of legs $(I,J)$
and the combined leg $\widetilde{IJ}$,
the emitter pair and the emitter,
respectively.
Leg $K$ is called the spectator.
All possible combinations
of $(I,J,K)$ are chosen without duplicate from all 
($n$+3) legs of $\mbox{R}_{i}$. In the DSA, the 
creation algorithm of the dipole terms is 
divided into substeps as follows:
\begin{enumerate}
\setlength{\itemsep}{0mm}
\item Choose all the possible emitter pairs 
$(x_{I},x_{J})$ from set 
\{x\} in the order of the splittings
from (1) to (7)
in Fig.\,\ref{fig_Dterm} in Appendix \ref{ap_A_1}.
\item Choose all the possible spectators $x_{K}$ 
from \{x\} for each choice of the pair $(x_{I},x_{J})$.
\item Make one field mapping for each combination
$(x_{I},x_{J},x_{K})$.
\item Write down the concrete expressions of all 
the dipole terms.
\end{enumerate}
Substeps 1 and 2 are explained in Sec.\,\ref{sec221}.
Substep 3 is explained in Sec.\,\ref{sec222},
and substep 4 is in Sec.\,\ref{sec223}.
Some concrete examples are shown 
in Sec.\,\ref{sec224}.
Finally we give a summary in Sec.\,\ref{sec225}.
The formulae for the dipole terms are collected 
in Appendix\,\ref{ap_A_1}.

\subsubsection{Creation order \label{sec221}}
In the DSA, the creation of the dipole terms is 
sorted by the kind of the splittings of 
the emitter pairs $(x_{I},x_{J})$.
The sorting order is shown in Fig. \ref{fig_Dterm}.
The created 
dipoles are grouped into categories, 
{\tt Dipoles\,1,2,3,4}, and subcategories, 
(1)--(7), as follows\,:
\begin{verbatim}
 Dipole 1 (1),(2),(3),(4),
 Dipole 2 (5),
 Dipole 3 (6),
 Dipole 4 (7).
\end{verbatim}
{\tt Dipole\,1} includes the splitting of
a gluon emission, which is sometimes called
diagonal splitting.
{\tt Dipole\,2,3,4} have further subcategories for
the quark flavors in the five-massless-flavor scheme 
as follows\,:
\begin{description}
\setlength{\itemsep}{-1mm}
\item {\tt Dipole 2} : $u, c, d, s, b,$
\item {\tt Dipole 3} : $u, c, d, s, b, 
\bar{u}, \bar{c}, \bar{d}, \bar{s}, \bar{b}$,
\item {\tt Dipole 4} : $u, c, d, s, b, 
\bar{u}, \bar{c}, \bar{d}, \bar{s}, \bar{b}$,
\end{description}
where, e.g., {\tt Dipole2}$u$ means that 
the species of fields
$x_{i}$ and $x_{j}$ are the up- and anti-up-quarks
as $(\mbox{F}(x_{i}),\mbox{F}(x_{j}))=(u,\bar{u})$,
and {\tt Dipole3}$\bar{u}$ means
$(\mbox{F}(x_{a}),\mbox{F}(x_{i}))=(\bar{u},\bar{u})$.
The spectator $x_{K}$ can be a quark or gluon
in either the final or the initial state.
If the spectator is in the final/initial 
state, we denote the case as subcategory\,--1/2. 
Then the category of the dipole terms are 
further divided into subcategories as
\begin{align}
& {\tt Dipole 1 \ (1)-1/2, \ (2)-1/2, \ (3)-1/2, \ (4)-1/2}\,,
\nonumber\\
& {\tt Dipole 2 \ (5)-1/2}\,, \nonumber\\
& {\tt Dipole 3 \ (6)-1/2}\,, \nonumber\\
& {\tt Dipole 4 \ (7)-1/2}\,.  
\label{creord}
\end{align}
The summation of the dipole terms that belong to 
the same category is written, respectively, as
\begin{align*}
& \mbox{D} (\mbox{R}_{i},\,{\tt dip1},(1)\,)\,, \ 
  \mbox{D} (\mbox{R}_{i},\,{\tt dip1},(2)\,)\,, \ 
  \mbox{D} (\mbox{R}_{i},\,{\tt dip1},(3)\,)\,, \ 
  \mbox{D} (\mbox{R}_{i},\,{\tt dip1},(4)\,)\,, \\
&\mbox{D}  (\mbox{R}_{i},\,{\tt dip2}\,)\,, \\
&\mbox{D}  (\mbox{R}_{i},\,{\tt dip3}\,)\,, \\
&\mbox{D}  (\mbox{R}_{i},\,{\tt dip4}\,)\,,
\end{align*}
where the subcategories for the quark flavors and 
for the spectator in the final or initial states are all 
summed. The dipole terms 
belonging to {\tt Dipole\,1} are summed as 
\begin{equation}
\mbox{D} (\mbox{R}_{i},\,{\tt dip1})=
\sum_{j=1}^{4} \,\mbox{D}(\mbox{R}_{i},\,{\tt dip1},\,(j)\,)\,,
\end{equation}
and the summation of all the dipole terms as output 
is written as
\begin{equation}
\mbox{D} (\mbox{R}_{i})=
\sum_{j=1}^{4} \,\mbox{D}(\mbox{R}_{i},\,{\tt dip}j)\,.
\end{equation}
The concrete expression for each dipole is given in
the original article \cite{Catani:1996vz} in the form
\begin{equation}
\mbox{D} (\mbox{R}_{i},\,{\tt dip}j)_{IJ,K} =
 -\frac{1}{s_{{\scriptscriptstyle IJ}}} 
\frac{1}{x_{{\scriptscriptstyle IJK}}}
\langle \mbox{B}j \ | 
\frac{\mbox{T}_{{\scriptscriptstyle IJ}} \cdot 
\mbox{T}_{{\scriptscriptstyle K}}}{\mbox{T}_{{\scriptscriptstyle IJ}}^{2}} 
\mbox{V}_{{\scriptscriptstyle IJ,K}}
| \ \mbox{B}j \rangle\,, 
\label{dipform}
\end{equation}
where $s_{IJ}$ is defined as
$s_{IJ}=2p_{I} \cdot p_{J}$, and $x_{IJK}$ 
is specified in 
Sec.\,\ref{sec223}.
$\mbox{B}j$ is a Born 
process that is reduced from the input process 
$\mbox{R}_{i}$ by 
removing the splitting part. Then the dipole terms 
$\mbox{D}(\mbox{R}_{i},{\tt dip}j)$
in the category {\tt Dipole}\,$j$ have the reduced 
Born process $\mbox{B}j$, which is made from
$\mbox{R}_{i}$ with the following rules\,:
\begin{align}
&{\tt Dipole 1} \ \ : \mbox{B}1=\mbox{R}_{i} -g_{f}, 
\nonumber\\
&{\tt Dipole 2}u: \mbox{B}2u=\mbox{R}_{i} - u_{f} - \bar{u}_{f} +g_{f}, 
\nonumber\\
&{\tt Dipole 3}u: \mbox{B}3u=\mbox{R}_{i} - u_{f} - u_{i} + g_{i}, 
\nonumber\\
&{\tt Dipole 4}u: \mbox{B}4u=\mbox{R}_{i} - u_{f} - g_{i} + \bar{u}_{i},
\label{makerb}
\end{align}
where the symbols $g_{f/i}$ and $u_{f/i}$ represent
a gluon and an up-quark in the final/initial state.
The operation $\pm g_{f}$ means to add/remove
a gluon to/from the final state. The other operations 
such as $\pm u_{f}$, are similarly defined.
For $\mbox{B}2u,\mbox{B}3u$, and $\mbox{B}4u$, 
other subcategories with other quark flavors also 
exist. The symbol
$(\mbox{T}_{IJ} \cdot \mbox{T}_{K})/\mbox{T}_{IJ}^{2}$
represents the operators of the color factor insertions 
and $\mbox{V}_{IJ,K}$ is the dipole splitting function
with the helicity correlation. 
The actions of the color and helicity operators on the amplitude 
of the reduced Born process are clearly defined by 
using the {\em field mapping}, explained in the next section.

\subsubsection{Field mapping \label{sec222}}
Each dipole term includes the square of a reduced Born amplitude
shown in Eq.\,(\ref{dipform}).
The original ($n$+3)-legs of the input process $\mbox{R}_{i}$ 
are connected to the ($n$+2)-legs of the reduced Born process.
In order to specify the connection clearly in
a compact form we introduce a bijection mapping for 
each dipole, called the {\em field mapping}.
For each combination choice $(x_{I},x_{J},x_{K})$, 
we can make a new set $\{\tilde{x}\}$ by the unification 
of the elements 
$(x_{I},x_{J}) \to x_{\widetilde{IJ}}$
and the replacement
$x_{K} \to x_{\widetilde{K}}$.
To explain the definition of set $\{\tilde{x}\}$
precisely,
we separate the dipole terms into four categories,
$(IJ,K)$
$=(ij,k),(ij,a),(ai,k)$, and $(ai,b)$,
where the indices $i,j,k$ represent the legs in the final
state and the indices $a$ and $b$ represent the legs
in the initial state.
The four categories are called
the final-final, final-initial,
initial-final, and initial-initial
dipole terms, respectively.
The relations between the four categories and
the categories defined in Sec.\,\ref{sec221} 
are shown as
\begin{align}
 \mbox{Final--final}:
(ij,k) \  \supset \ & {\tt Dipole 1} \ (1)\mbox{-}1, \ (2)\mbox{-}1,  
\nonumber\\
 & {\tt Dipole 2} \ (5)\mbox{-}1,  \nonumber\\
 \mbox{Final--initial}:
(ij,a) \  \supset \ & {\tt Dipole 1} \ (1)\mbox{-}2, \ (2)\mbox{-}2,  
\nonumber\\
 &{\tt Dipole 2} \ (5)\mbox{-}2,  \nonumber\\
 \mbox{Initial--final}:
(ai,k) \  \supset \ & {\tt Dipole 1} \ (3)\mbox{-}1, \ (4)\mbox{-}1,  
\nonumber\\
 &{\tt Dipole 3} \ (6)\mbox{-}1,  \nonumber\\
 &{\tt Dipole 4} \ (7)\mbox{-}1,  \nonumber\\
 \mbox{Initial--initial}:
(ai,b) \  \supset \  &{\tt Dipole 1} \ (3)\mbox{-}2, \ (4)\mbox{-}2,  
\nonumber\\
 &{\tt Dipole 3} \ (6)\mbox{-}2,  \nonumber\\
 &{\tt Dipole 4} \ (7)\mbox{-}2\,.
\end{align}
\begin{itemize}
\item {\bf Final--final} : $(ij,k)$ \\
In this category, set $\{\tilde{x}\}$ is defined as
\begin{align}
\{\tilde{x}\} 
&=\{ \tilde{x}_{a},\tilde{x}_{b};\tilde{x}_{1}, ..., 
\tilde{x}_{n-2}, \tilde{x}_{n-1},\tilde{x}_{n} \} 
\nonumber\\
&= \{x_{a},x_{b};x_{1}, ..., x_{n+1}, \ 
x_{\widetilde{ij}}, \ x_{\widetilde{k}} \}, 
\label{ffxt} \\
\mbox{F}(\{\tilde{x}\}) 
&= \{\mbox{F}(\tilde{x}_{a}),\mbox{F}(\tilde{x}_{b}); 
\mbox{F}(\tilde{x}_{1}), ..., \mbox{F}(\tilde{x}_{n-2}), 
\mbox{F}(\tilde{x}_{n-1}),\mbox{F}(\tilde{x}_{n}) \}  
\nonumber\\
&= \{\mbox{F}(x_{a}),\mbox{F}(x_{b}) ;\mbox{F}(x_{1}), 
..., \mbox{F}(x_{n+1}), \ \mbox{F}(x_{\widetilde{ij}}),
\ \mbox{F}(x_{\widetilde{ k}}) \}, \label{fffs}  \\
\mbox{P}(\{\tilde{x}\}) 
&= \{\mbox{P}(\tilde{x}_{a}),\mbox{P}(\tilde{x}_{b}); 
\mbox{P}(\tilde{x}_{1}), 
...,  \mbox{P}(\tilde{x}_{n-2}), \mbox{P}(\tilde{x}_{n-1}),
\mbox{P}(\tilde{x}_{n}) \}  \nonumber\\
&= \{\mbox{P}(x_{a}),\mbox{P}(x_{b}); \mbox{P}(x_{1}), 
..., \mbox{P}(x_{n+1}), \ \mbox{P}(x_{\widetilde{ij}}),
\ \mbox{P}(x_{\widetilde{k}}) \},  \label{ffmo}
\end{align}
where the symbols $\mbox{F}(\tilde{x}_{\alpha})$ and 
$\mbox{P}(\tilde{x}_{\alpha})$
represent the field species and the momenta of the 
elements 
$\tilde{x}_{\alpha}$ with the indices
$\alpha=a,b,1,...,n$. 
The field species are determined as
\begin{align}
\mbox{F}(\tilde{x}_{n-1}) 
&=\mbox{F}(x_{\widetilde{ij}}), 
\label{ffijF} \\
\mbox{F}(\tilde{x}_{n})
&=\mbox{F}(x_{\widetilde{k}}) =\mbox{F}(x_{k}), 
\label{ffkF} \\
\mbox{F}(\tilde{x}_{\alpha})
&=\mbox{F}(x_{L}) 
\ \ \mbox{for} \ \ \alpha=a,b,1,..., (n-2)\,.
\label{ffotF} 
\end{align}
In Eq.\,(\ref{ffijF}), the field specifies of the element
$x_{\widetilde{ij}}$,
$\mbox{F}(x_{\widetilde{ij}})$,
is defined as the field species of the root of the 
splitting $x_{\widetilde{ij}} \to x_{i} + x_{j}$, 
where the legs $x_{i}$ and $x_{j}$ are external legs
in the final state, 
and 
$x_{\widetilde{ij}}$ is the internal line that is
attached to the gray circle at the center of
Fig.\,\ref{fig_Dterm}.
The relation in Eq.\,(\ref{ffkF}) means that
the field species of the spectator 
$x_{\widetilde{k}}$ is identical to the species of the 
original leg $x_{k}$ in set $\{x \}$.
The relation in Eq.\,(\ref{ffotF}) 
means that
the field species of the other elements 
$\tilde{x}_{\alpha}$, with
$\alpha=a,b,1,...,n-2$,
are the same as the element $x_{L}$ in 
$\mbox{F}(\tilde{x}_{\alpha})=\mbox{F}(x_{L})$,
where the element $x_{L}$ is specified as
$\tilde{x}_{\alpha}=x_{L}$, in Eq.\,(\ref{ffxt})
for the indices $L=a,b,1,...,n+1$, 
skipping the indices $i,j,k$.
The momenta of the elements
in set $\{\tilde{x}\}$ are determined as
\begin{align}
\mbox{P}(\tilde{x}_{n-1}) 
&=\mbox{P}(x_{\widetilde{ij}}) =\tilde{p}_{ij}, \\
\mbox{P}(\tilde{x}_{n})
&=\mbox{P}(x_{\widetilde{k}}) =\tilde{p}_{k}, \\
\mbox{P}(\tilde{x}_{\alpha})
&=\mbox{P}(x_{L}) = p_{L},
\ \ \mbox{for} \ \ \alpha=a,b,1,..., (n-2),
\end{align}
where the reduced momenta $\tilde{p}_{ij}$ and 
$\tilde{p}_{k}$ are defined in 
Eqs.\,(\ref{rmffem}) and (\ref{rmffsp}) 
in Appendix \ref{ap_A_1}.
Similarly to the relation in
Eq.\,(\ref{ffotF}), the momenta of the
other elements $\tilde{x}_{\alpha}$,
with $\alpha=a,b,1,...,n-2$,
are the same as the corresponding 
original legs $x_{L}$.
It is noted that the order of the elements in set 
$\{\tilde{x}\}$ in Eq.\,(\ref{ffxt}) is not strict
and other orders are also possible as long as 
the elements in the initial and final states
are not mixed. What is necessary is that 
the field species and the momenta 
of all the elements in $\{\tilde{x}\}$
are properly determined.
To demonstrate the other possibilities
we here take one example with 
$n=3$. Set $\{x\}$ is defined as 
\begin{align}
\{x\} &= \{x_{a},x_{b};x_{1},x_{2},x_{3},x_{4} \}. 
\\ \mbox{Momenta}: 
& \ \ \ \     \{p_{a},p_{b} \ ;p_{1},p_{2},p_{3},p_{4} \} 
\\
\mbox{F}(\{x\}) 
&= \{\mbox{F}(x_{a}),\mbox{F}(x_{b});\mbox{F}(x_{1}), 
\mbox{F}(x_{2}), \mbox{F}(x_{3}), \mbox{F}(x_{4}) \}. 
\end{align}
According to the relations in
Eqs.\,(\ref{ffxt}),\,(\ref{fffs}), 
and (\ref{ffmo}),
set $\{\tilde{x}\}$ is made on the choice 
$(I,J,K)=(1,2,3)$, for instance, as
\begin{align}
\{\tilde{x}\} 
&=\{ \tilde{x}_{a},\tilde{x}_{b};\tilde{x}_{1},\tilde{x}_{2},
\tilde{x}_{3} \} 
\nonumber\\
&= \{x_{a},x_{b};x_{4}, x_{\widetilde{12}}, x_{\widetilde{3}} \}, 
 \\
\mbox{F}(\{\tilde{x}\}) 
&= \{\mbox{F}(\tilde{x}_{a}),\mbox{F}(\tilde{x}_{b}); 
\mbox{F}(\tilde{x}_{1}), 
\mbox{F}(\tilde{x}_{2}), 
\mbox{F}(\tilde{x}_{3}) \}  
\nonumber\\
&= \{\mbox{F}(x_{a}),\mbox{F}(x_{b}) ;\mbox{F}(x_{4}), 
\mbox{F}(x_{\widetilde{12}}),
\mbox{F}(x_{\widetilde{3}}) \},   \\
\mbox{P}(\{\tilde{x}\}) 
&= \{\mbox{P}(\tilde{x}_{a}),\mbox{P}(\tilde{x}_{b}); 
\mbox{P}(\tilde{x}_{1}), 
\mbox{P}(\tilde{x}_{2}), 
\mbox{P}(\tilde{x}_{3})
\}  \nonumber\\
&= \{\mbox{P}(x_{a}),\mbox{P}(x_{b}); \mbox{P}(x_{4}), 
\mbox{P}(x_{\widetilde{12}}),
\mbox{P}(x_{\widetilde{3}}) \}.  
\end{align}
The following choice of set $\{\tilde{x}\}$ is also
possible:
\begin{align}
\{\tilde{x}\} 
&=\{ \tilde{x}_{a},\tilde{x}_{b};\tilde{x}_{1},\tilde{x}_{2},
\tilde{x}_{3} \} 
\nonumber\\
&= \{x_{a},x_{b};x_{\widetilde{12}}, x_{\widetilde{3}},x_{4} \}, 
 \\
\mbox{F}(\{\tilde{x}\}) 
&= \{\mbox{F}(\tilde{x}_{a}),\mbox{F}(\tilde{x}_{b}); 
\mbox{F}(\tilde{x}_{1}), 
\mbox{F}(\tilde{x}_{2}), 
\mbox{F}(\tilde{x}_{3}) \}  
\nonumber\\
&= \{\mbox{F}(x_{a}),\mbox{F}(x_{b}) ; 
\mbox{F}(x_{\widetilde{12}}),
\mbox{F}(x_{\widetilde{3}}), \mbox{F}(x_{4}) \},   \\
\mbox{P}(\{\tilde{x}\}) 
&= \{\mbox{P}(\tilde{x}_{a}),\mbox{P}(\tilde{x}_{b}); 
\mbox{P}(\tilde{x}_{1}), 
\mbox{P}(\tilde{x}_{2}), 
\mbox{P}(\tilde{x}_{3})
\}  \nonumber\\
&= \{\mbox{P}(x_{a}),\mbox{P}(x_{b}); 
\mbox{P}(x_{\widetilde{12}}),
\mbox{P}(x_{\widetilde{3}}), \mbox{P}(x_{4})\}.  
\end{align}
The field species and the momenta in this case
are similarly determined. The note about the 
freedom of the order of the elements 
in $\{\tilde{x}\}$ is also valid for
cases $(ij,a),(ai,k)$, and $(ai,b)$.

\item {\bf Final--Initial} : $(ij,a)$\\
For the category of the final-initial dipoles
$(ij,a)$, set $\{\tilde{x}\}$ is made as
\begin{align}
\{\tilde{x}\} 
&=\{ \tilde{x}_{a},\tilde{x}_{b};\tilde{x}_{1}, 
..., \tilde{x}_{n-1}, \tilde{x}_{n} \} \nonumber \\
&= \{x_{\widetilde{a}},x_{b};x_{1}, ..., 
x_{n+1},x_{\widetilde{ij}} \}, \label{fixtil} \\
\mbox{F}(\{\tilde{x}\}) 
&= 
\{\mbox{F}(x_{\widetilde{a}}),\mbox{F}(x_{b}); 
\mbox{F}(x_{1}), ..., \mbox{F}(x_{n+1}),
\mbox{F}(x_{\widetilde{ij}}) \}.  \\
\mbox{P}(\{\tilde{x}\}) 
&= 
\{\mbox{P}(x_{\widetilde{a}}),\mbox{P}(x_{b}); 
\mbox{P}(x_{1}), ..., \mbox{P}(x_{n+1}),
\mbox{P}(x_{\widetilde{ij}}) \}.
\end{align}
The field species are defined in a similar way 
to the previous case in 
Eqs.\,(\ref{ffijF}), (\ref{ffkF}), and (\ref{ffotF}) as
\begin{align}
\mbox{F}(\tilde{x}_{n}) 
&=\mbox{F}(x_{\widetilde{ij}}), 
\label{fiijF} \\
\mbox{F}(\tilde{x}_{a})
&=\mbox{F}(x_{\widetilde{a}}) =\mbox{F}(x_{a}), 
\label{fiaF} \\
\mbox{F}(\tilde{x}_{\alpha})
&=\mbox{F}(x_{L}) 
\ \ \mbox{for} \ \ \alpha=b,1,..., (n-1)\,,
\label{fiotF} 
\end{align}
where the $\mbox{F}(x_{\widetilde{ij}})$ is the same as
in Eq.\,(\ref{ffijF}).
For the relation in Eq.\,(\ref{fiotF})
the elements $\tilde{x}_{\alpha}$,
with $\alpha=b,1,...,n-1$,
are identified with the elements $x_{L}$ 
as $\tilde{x}_{\alpha}=x_{L}$ in Eq.\,(\ref{fixtil})
for $L=b,1,...,n+1$,
skipping the indices $i$ and $j$.
The momenta are also similarly defined as
\begin{align}
\mbox{P}(\tilde{x}_{n}) 
&=\mbox{P}(x_{\widetilde{ij}}) =\tilde{p}_{ij}, \\
\mbox{P}(\tilde{x}_{a})
&=\mbox{P}(x_{\widetilde{a}}) =\tilde{p}_{a}, \\
\mbox{P}(\tilde{x}_{\alpha})
&=\mbox{P}(x_{L}) = p_{L}
\ \ \mbox{for} \ \ \alpha=b,1,..., (n-1),
\end{align}
where the reduced momenta $\tilde{p}_{ij}$ and 
$\tilde{p}_{a}$ are defined in Eqs.\,(\ref{rmfiem}) 
and (\ref{rmfisp}).

\item {\bf Initial--final} : $(ai,k)$\\
For the initial-final dipoles $(ai,k)$,  
set $\{\tilde{x}\}$ is made as
\begin{align}
\{\tilde{x}\} 
&=\{ \tilde{x}_{a},\tilde{x}_{b};\tilde{x}_{1}, 
..., \tilde{x}_{n-1}, \tilde{x}_{n} \} \nonumber \\
&= \{x_{\widetilde{ai}},x_{b};x_{1}, ..., 
x_{n+1},x_{\widetilde{k}} \}, \\
\mbox{P}(\{\tilde{x}\}) 
&= \{\mbox{P}(x_{\widetilde{ai}}),\mbox{P}(x_{b}); 
\mbox{P}(x_{1}), ..., \mbox{P}(x_{n+1}),
\mbox{P}(x_{\widetilde{k}}) \},  \\
\mbox{F}(\{\tilde{x}\}) 
&= 
\{\mbox{F}(x_{\widetilde{ai}}),\mbox{F}(x_{b}); 
\mbox{F}(x_{1}), ..., \mbox{F}(x_{n+1}),
\mbox{F}(x_{\widetilde{k}}) \}.  
\end{align}
The field species are defined as
\begin{align}
\mbox{F}(\tilde{x}_{a}) 
&=\mbox{F}(x_{\widetilde{ai}}), 
\label{ifaiF} \\
\mbox{F}(\tilde{x}_{n})
&=\mbox{F}(x_{\widetilde{k}}) =\mbox{F}(x_{k}), 
\label{ifkF} \\
\mbox{F}(\tilde{x}_{\alpha})
&=\mbox{F}(x_{L}). 
\ \ \mbox{for} \ \ \alpha=b,1, ..., (n-1)\,.
\label{ifotF} 
\end{align}
For the relation in Eq.\,(\ref{ifaiF}),
the field species of the element $x_{\widetilde{ai}}$, 
$\mbox{F}(x_{\widetilde{ai}})$, 
is defined as the field species of the root of the 
splitting
$x_{a}  \to x_{\widetilde{ai}} + x_{i}$,
where the legs $x_{a}$ and $x_{i}$ are
external legs, and $x_{\widetilde{ai}}$ is the internal 
line that is attached to the gray circle at the center 
of Fig.\,\ref{fig_Dterm}. 
Similarly to the previous cases,
the field species of the spectator 
and the other elements are the same as 
those of the original legs.
The momenta are defined as
\begin{align}
\mbox{P}(\tilde{x}_{a}) 
&=\mbox{P}(x_{\widetilde{ai}}) =\tilde{p}_{ai}, \\
\mbox{P}(\tilde{x}_{n})
&=\mbox{P}(x_{\widetilde{k}}) =\tilde{p}_{k}, \\
\mbox{P}(\tilde{x}_{\alpha})
&=\mbox{P}(x_{L}) = p_{L},
\ \ \mbox{for} \ \ \alpha=b,1,..., (n-1),
\end{align}
where the reduced momenta $\tilde{p}_{ai}$ and 
$\tilde{p}_{k}$ are defined in
Eqs.\,(\ref{rmifem}) and (\ref{rmifsp}). 

\item {\bf Initial--Initial} : $(ai,b)$\\
For the initial-initial dipoles $(ai,b)$,
set $\{\tilde{x}\}$ is made as
\begin{align}
\{\tilde{x}\} 
&=\{ \tilde{x}_{a},\tilde{x}_{b};\tilde{x}_{1}, ... 
..., \tilde{x}_{n} \} \nonumber\\
&= \{x_{\widetilde{ai}}, x_{\widetilde{b}};x_{1}, 
..., x_{n+1} \}, \\
\mbox{P}(\{\tilde{x}\}) 
&= \{\mbox{P}(x_{\widetilde{ai}}),\mbox{P}(x_{\widetilde{b}}); 
\mbox{P}(x_{1}), ..., \mbox{P}(x_{n+1}) \},  \\
\mbox{F}(\{\tilde{x}\})
&= \{\mbox{F}(x_{\widetilde{ai}}),\mbox{F}(x_{\widetilde{b}});
\mbox{F}(x_{1}), ..., \mbox{F}(x_{n+1}) \}.
\end{align}
The field species are defined as
\begin{align}
\mbox{F}(\tilde{x}_{a}) 
&=\mbox{F}(x_{\widetilde{ai}}), 
\label{iiaiF} \\
\mbox{F}(\tilde{x}_{b})
&=\mbox{F}(x_{\widetilde{b}}) =\mbox{F}(x_{b}), 
\label{iibF} \\
\mbox{F}(\tilde{x}_{\alpha})
&=\mbox{F}(x_{L}) 
\ \ \mbox{for} \ \ \alpha=1, ..., n.
\label{iiotF} 
\end{align}
In Eq.\,(\ref{iiaiF}), the definition of 
$\mbox{F}(x_{\widetilde{ai}})$
is the same as the previous case of $(ai,k)$. 
$\tilde{x}_{a}$ represent
the other elements with $\alpha=1, ...,n$.
The momenta are defined as
\begin{align}
\mbox{P}(\tilde{x}_{a}) 
&=\mbox{P}(x_{\widetilde{ai}}) =\tilde{p}_{ai}, \\
\mbox{P}(\tilde{x}_{b})
&=\mbox{P}(x_{\widetilde{b}}) =p_{b}, \label{iispem}\\
\mbox{P}(\tilde{x}_{\alpha}) 
&=\mbox{P}(x_{L}) = \tilde{k}_{L} 
\ \ \mbox{for} \ \ \alpha=1, ..., n,
\label{iiotm}
\end{align}
where the reduced momenta $\tilde{p}_{ai}$ and 
$\tilde{k}_{L}$ are defined in 
Eqs.\,(\ref{rmiiem}) and (\ref{rmiisp}). 
It is noted that in this case the momentum of the spectator
is not changed as shown in Eq.\,(\ref{iispem})
and the momenta of all the other elements 
$\tilde{x}_{\alpha}$ 
with $\alpha=1, ...,n$ are changed 
into $\tilde{k}_{L}$ with $L=1,2, ...,(n+1)$,
skipping the index $i$, as shown in
Eq.\,(\ref{iiotm}).
\end{itemize}

\begin{figure}[t]
\begin{center}
\includegraphics[width=6.5cm]{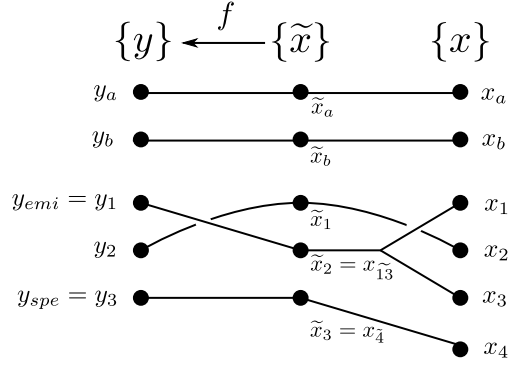}
\end{center} 
\caption{ 
The number of elements of set $\{x\}$ 
is assumed to be $(n+3)=6$.
The dipole term that is specified by 
the combination
$(IJ,K)=(x_{1}x_{3},x_{4})$ is 
selected.
\label{fig_s2_2}}
\end{figure}

\begin{figure}[t]
\begin{center}
\includegraphics[width=6.5cm]{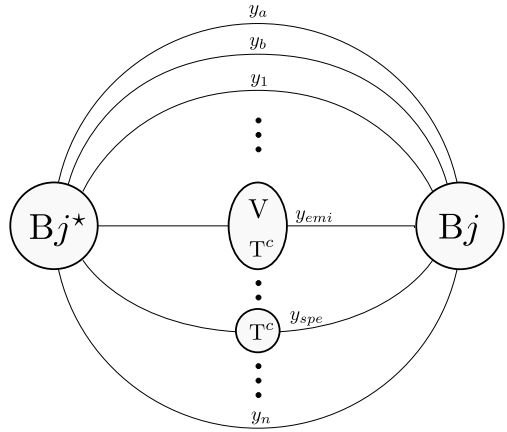}
\end{center} 
\caption{ The structure of the square of 
the reduced Born process
with the helicity and color correlation, 
$\langle \mbox{B}j | 
\mbox{T}_{y_{emi}} \cdot 
\mbox{T}_{y_{spe}}
\ \mbox{V}^{y_{emi}}
 | \ \mbox{B}j \rangle$,
is shown.
The inner product of two color operators
is denoted as
$\mbox{T} \cdot \mbox{T}=
\sum_{c} \mbox{T}^{c}\,\mbox{T}^{c}$.
\label{fig_s2_3}}
\end{figure}

Next we deal with the definition of 
set $\{y\}$, which is made
from a reduced Born process.
Under the input $\mbox{R}_{i}$ 
the reduced Born process $\mbox{B}j$ that 
belongs to {\tt Dipole}\,$j$ is made 
by the rules shown in Eq.\,(\ref{makerb}).
One Born process $\mbox{B}j$ determines set 
$\{y\}$ with the field species and the momenta as
\begin{align}
\{ y \} 
&= \{y_{a},y_{b};y_{1},  ..., y_{n} \}, \\
\mbox{F}(\{ y \}) 
&= \{\mbox{F}(y_{a}),\mbox{F}(y_{b}); 
\mbox{F}(y_{1}), ..., \mbox{F}(y_{n}) \}, 
\label{yfield}\\
\mbox{P}(\{ y \})  
&= \{\mbox{P}(y_{a}),\mbox{P}(y_{b}); 
\mbox{P}(y_{1}), ..., \mbox{P}(y_{n}) \}.
\end{align}
The number of elements of the set $\{y\}$ is
($n$+2), which is the same as the set 
$\{\tilde{x}\}$. 
We can always find a bijection 
(one-to-one correspondence) mapping from set
$\{\tilde{x}\}$ to set $\{y\}$ as
\begin{equation}
y_{\beta}=f(\tilde{x}_{\alpha})\,,
\end{equation}
with the indices $\alpha,\beta=a,b,1, ...,n$,
which satisfies two conditions\,:
\begin{enumerate}
\item[-] $\mbox{F}(y_{\beta})=\mbox{F}(\tilde{x}_{\alpha})$\,, 
\item[-] The argument $\tilde{x}_{\alpha}$ and the image
$y_{\beta}$
are both in either the final or the initial state.
\end{enumerate}
We call the mapping, $f$, the field mapping.
The two conditions mean that the mapping, 
$f$, connects the 
elements whose species are identical,
and it
does not mix the elements in the initial and 
final states.
The inverse mapping is denoted as 
$\tilde{x}_{\alpha}=f^{-1}(y_{\beta})$.
After the construction of a mapping, the element 
$\tilde{x}_{\alpha}$
is identified with the image $y_{\beta}$. 
Using the inverse mapping, the identification of the 
elements is generally written as
\begin{equation}
\mbox{B}j : (y_{a},y_{b};y_{1}, ..., y_{n})
=(f^{-1}(y_{a}),f^{-1}(y_{b});f^{-1}(y_{1}), 
..., f^{-1}(y_{n})),
\end{equation}
where the elements are sorted by the order of elements of 
set $\{y\}$. The momenta of set $\{y\}$ are defined as
\begin{equation}
\mbox{P}(y_{\beta}) = \mbox{P}(f^{-1}(y_{\beta})) = \mbox{P}(\tilde{x}_{\alpha}).
\label{pydef}
\end{equation}
For convenience, we introduce the notations
$y_{emi}$ and $y_{spe}$, defined as
$y_{emi}=f(x_{\widetilde{{\scriptscriptstyle IJ}}})$,
and
$y_{spe}=f(x_{\widetilde{{\scriptscriptstyle K}}})$.
One example of the field mapping is shown in Fig.\ref{fig_s2_2},
where the number of final states
of an input process $\mbox{R}_{i}$ is 
assumed to be $(n+1)=4$, 
and the dipole term with the combination
$(IJ,K)=(x_{1}x_{3},x_{4})$ is selected.
Using the field mapping,
the concrete form of each dipole term can be 
clearly expressed as
\begin{align}
\mbox{D}({\tt dip}j)_{IJ,K}
= -\frac{1}{s_{{\scriptscriptstyle IJ}}} 
\frac{1}{x_{{\scriptscriptstyle IJK}}}
\frac{1}{\mbox{T}_{\mbox{{\tiny F}}(x_{\widetilde{{\scriptscriptstyle IJ}}})}^{2}} 
\langle \mbox{B}j =\{ y_{a},y_{b} ; y_{1}, ..., y_{n} \} 
\ | 
\mbox{T}_{f(x_{\widetilde{{\scriptscriptstyle IJ}}} )} \cdot 
\mbox{T}_{f(x_{\widetilde{{\scriptscriptstyle K}}} )}
\ \mbox{V}_{{\scriptscriptstyle IJ,K}}^{f(x_{\widetilde{{\scriptscriptstyle IJ}}})}
| \ \mbox{B}j \rangle. \nonumber\\
\end{align}
Using the notation $y_{emi}$ and $y_{spe}$,
the concrete form of each dipole term is slightly 
simplified as
\begin{align}
\mbox{D}({\tt dip}j)_{IJ,K}= -\frac{1}{s_{{\scriptscriptstyle IJ}}} 
\frac{1}{x_{{\scriptscriptstyle IJK}}}
\frac{1}{\mbox{T}_{\mbox{{\tiny F}}(y_{emi})}^{2}} 
\langle \mbox{B}j =\{ y \} \ | 
\mbox{T}_{y_{emi}} \cdot 
\mbox{T}_{y_{spe}}
\ \mbox{V}_{{\scriptscriptstyle IJ,K}}^{y_{emi}}
 | \ \mbox{B}j \rangle\,.  
\label{dipmas}
\end{align}
The subscript indices $I,J$, and $K$ in the quantities 
$s_{{\scriptscriptstyle IJ}}$, 
$x_{{\scriptscriptstyle IJ,K}}$, and 
$\mbox{V}_{{\scriptscriptstyle IJ,K}}^{y_{emi}}$
refer to the elements of set $\{x\}$
and the momenta
$\{p_{a},p_{b},p_{1}, ...,p_{n+1} \}$ 
in Eq.\,(\ref{orimom}).
The indices $y_{emi}$ and $y_{spe}$
in the operators
$\mbox{T}_{y_{emi}}$,
$\mbox{T}_{y_{spe}}$,
and $\mbox{V}_{{\scriptscriptstyle IJ,K}}^{y_{emi}}$
refer to the elements of set $\{y\}$.
The Casimir operator
$\mbox{T}_{\mbox{{\tiny F}}(y_{emi})}^{2}$
is defined as the constants 
$\mbox{C}_{\mbox{{\tiny F}}}=4/3$ 
in the case $\mbox{F}(y_{emi})=$quark, and 
$\mbox{C}_{\mbox{{\tiny A}}}=3$
in the case $\mbox{F}(y_{emi})=$gluon.
The momenta input into the reduced Born amplitude 
$\mbox{B}j$ are the momenta
$\mbox{P}(y_{\beta})$ defined in 
Eq.\,(\ref{pydef}). The legs of the Born process,
on which the color and 
helicity operators act, are clearly specified on 
the basis of set $\{y\}$. 
The action of the color and
helicity operators in the square of 
the reduced Born process
is illustrated in Fig.\,\ref{fig_s2_3}.
In the DSA, the field 
species $\mbox{F}(\{y\})$ of the reduced Born process
$\mbox{B}j$ are fixed in one category {\tt Dipole}$j$;
namely, an identical set of field species,
$\mbox{F}(\{ y \}) = \{\mbox{F}(y_{a}),\mbox{F}(y_{b}); 
\mbox{F}(y_{1}), ..., \mbox{F}(y_{n}) \}$, 
is used for all the dipole terms that belong to the 
category {\tt Dipole}$j$.
The momenta of $\mbox{B}j$, 
$\mbox{P}(\{ y \}) = \{\mbox{P}(y_{a}),
\mbox{P}(y_{b}); \mbox{P}(y_{1}), 
..., \mbox{P}(y_{n}) \}$, are generally different 
functions of the original momenta
$\{p_{a},p_{b},p_{1}, ...,p_{n+1} \}$ 
associated with set $\{x\}$ depending on 
the choice $(IJ,K)$.

\subsubsection{Concrete formulae \label{sec223}}
In this section, we specify the full expressions 
of the dipole terms.
The expressions are separated into cases in which
the field species of the emitter,
$\mbox{F}(y_{emi})$, is a quark or a gluon.
In the case in which $\mbox{F}(y_{emi})=$quark, the expression
in Eq.\,(\ref{dipmas}) is written as
\begin{align}
\mbox{D}({\tt dip}j)_{IJ,K}= -\frac{1}{s_{{\scriptscriptstyle IJ}}} 
\frac{1}{x_{{\scriptscriptstyle IJK}}}
\frac{1}{\mbox{C}_{\mbox{{\tiny F}}}} 
\mbox{V}_{{\scriptscriptstyle IJ,K}} \
\langle \mbox{B}j  \ | 
\mbox{T}_{y_{emi}} \cdot 
\mbox{T}_{y_{spe}}
 | \ \mbox{B}j \rangle,
\label{dipnohel}
\end{align}
where the helicity correlation 
of the dipole splitting function
$\mbox{V}_{{\scriptscriptstyle IJ,K}}^{y_{emi}}$ 
disappears and 
the function is fully factorized to the reduced Born 
amplitude. In the case in which $\mbox{F}(y_{emi})=$gluon, 
the expression is 
written as
\begin{align}
\mbox{D}({\tt dip}j)_{IJ,K}= -\frac{1}{s_{{\scriptscriptstyle IJ}}} 
\frac{1}{x_{{\scriptscriptstyle IJK}}}
\frac{1}{\mbox{C}_{\mbox{{\tiny A}}}} 
\langle \mbox{B}j | 
\mbox{T}_{y_{emi}} \cdot 
\mbox{T}_{y_{spe}}
\ \mbox{V}_{{\scriptscriptstyle IJ,K}}^{y_{emi}}
 | \ \mbox{B}j \rangle.
\end{align}
The formulae for the dipole terms in all categories 
in Eq.\,(\ref{creord}) are collected in Appendix\,\ref{ap_A_1}.
Here we select two examples.

The first example is in the category {\tt Dipole\,1}\,(1)-1.
In this category, the dipole terms are given 
in Eq.\,(\ref{dp111}) as
\begin{align}
\mbox{D}({\tt dip}1,\,(1)\mbox{-}1)_{ij,k}
= -\frac{1}{s_{{\scriptscriptstyle ij}}} 
\frac{1}{\mbox{C}_{\mbox{{\tiny F}}}} 
\mbox{V}_{{\scriptscriptstyle ij,k}} \
\langle \mbox{B}1  \ | 
\mbox{T}_{y_{emi}} \cdot 
\mbox{T}_{y_{spe}}
 | \ \mbox{B}1 \rangle,
\end{align}
where the dipole splitting function is given in 
Eq.\,(\ref{ds111}) as
\begin{equation}
V_{ij,k} = 8 \pi \al \,\mbox{C}_{\mbox{{\tiny F}}}
\biggl[ \frac{2}{1 - z_{i} (1 - y_{ij,k})} - 1 - z_{i} \biggr].
\end{equation}
The reduced momenta are given  
in Eqs.\,(\ref{rmffem})
and (\ref{rmffsp}).
The scalars
$y_{ij,k}$ and $z_{i}$
appearing in the above formulae are 
defined 
in Eqs.\,(\ref{ffyijk})
and (\ref{ffzi}).

The second example is the category 
{\tt Dipole\,3} (6)-2, 
which is given 
in Eq.\,(\ref{dp362}) as
\begin{align}
\mbox{D}(
{\tt dip}3,\,(6)\mbox{-}2)_{ai,b} 
= -\frac{1}{s_{{\scriptscriptstyle ai}}} 
\frac{1}{x_{{\scriptscriptstyle i,ab}}}
\frac{1}{\mbox{C}_{\mbox{{\tiny A}}}} 
\langle \mbox{B}3 | 
\mbox{T}_{y_{emi}} \cdot 
\mbox{T}_{y_{spe}}
\ \mbox{V}_{{\scriptscriptstyle ai,k}}^{y_{emi}}
 | \mbox{B}3 \rangle,
\end{align}
where the dipole splitting function is given 
in Eq.\,(\ref{ds362}) as
\begin{align}
V_{ai,b}^{y_{emi}, \ \mu \nu} 
&= 8 \pi \al C_F \biggl[
 -g^{\mu \nu}x_{i,ab}  \nonumber \\
& + \frac{1-x_{i,ab}}{x_{i,ab}}
\frac{2p_a \cdot p_b}{p_i \cdot p_a \ p_i \cdot p_b} 
\left( p_i^{\mu} - \frac{p_ip_a}{p_bp_a} p_b^{\mu} \right)
\left( p_i^{\nu} - \frac{p_ip_a}{p_bp_a} p_b^{\nu} \right)
\biggr].
\end{align}
The reduced momenta $\tilde{p}_{ai}$ and 
$\tilde{k}_{L}$ are defined 
in Eqs.\,(\ref{rmiiem})
and (\ref{rmiisp}).
The scalar $x_{i,ab}$ is defined in 
Eq.\,(\ref{iixiab}).
Sometimes it is convenient that the gluon polarization vector
in the matrix element is taken in the basis of the helicity 
eigenstate as the circular polarization vector.
Formulae for the dipole terms 
with the helicity correlation in the helicity basis 
are constructed in Refs. \cite{Weinzierl:1998ki}
and \cite{Hasegawa:2009tx}.

\subsubsection{Examples \label{sec224}}
In order to demonstrate the creation algorithm,
we take the dijet process 
$pp \to 2\,jets$, and select one input real process as
\begin{equation}
\mbox{Input}: \ \mbox{R}_{1} = u\bar{u} \to u\bar{u}g\,.
\label{dipexin}
\end{equation}
The input defines set $\{ x \}$ with the field 
species and the momenta as
\begin{align}
\{x\} &= \{x_{a},x_{b};x_{1},x_{2},x_{3}  \}\,,
\label{dipex1x}\\
\mbox{F}(\{x\}) &= \{ u,\bar{u}; u,\bar{u}, g \}\,,
\label{dipex1f} \\
\mbox{Momenta}: & \ \ \ \ \ \{ p_{a},p_{b};p_{1},p_{2},p_{3} \}\,. 
\label{dipex1m}
\end{align}
Following the creation order in Eq.\,(\ref{creord}) 
and in Fig.\,\ref{fig_Dterm},
we create all the dipole terms from the input 
$\mbox{R}_{1}$ by choosing the 
three legs $(x_{I}x_{J},x_{K})$ as
\begin{align}
 {\tt Dipole1} \ \      (1)&-1: \ \ (13,2),(23,1),                \nonumber\\[-4pt]
                           &-2: \ \ (13,a),(13,b),(23,a),(23,b),  \nonumber\\[-4pt]
                        (3)&-1: \ \ (a3,1),(a3,2),(b3,1),(b3,2),  \nonumber\\[-4pt]
                           &-2: \ \ (a3,b),(b3,a),               \nonumber \\[-4pt]
 {\tt Dipole2}u \       (5)&-1: \ \ (12,3),                       \nonumber\\[-4pt]
                           &-2: \ \ (12,a),(12,b),               \nonumber \\[-4pt]
 {\tt Dipole3}u \       (6)&-1: \ \ (a1,2),(a1,3),               \nonumber \\[-4pt]
                           &-2: \ \ (a1,b),                      \nonumber \\[-4pt]
 {\tt Dipole3}\bar{u} \ (6)&-1: \ \ (b2,1),(b2,3),              \nonumber  \\[-4pt]
                           &-2: \ \ (b2,a)\,,
\label{dipex1}
\end{align}
where the notation $(IJ,K)$ is an abbreviation of $(x_{I}x_{J},x_{K})$.
We select four dipole terms and write down the concrete expressions.
\begin{itemize}
%
%
\item Example 1: \ {\tt Dipole1} (1)-1: (13,2)  \\
Set $\{\tilde{x}\}$ is defined with 
the field species and
the momenta as
\begin{align}
\{\tilde{x}\} 
&=\{ \tilde{x}_{a},\tilde{x}_{b};\tilde{x}_{1}, \tilde{x}_{2} \} 
\nonumber\\
&= \{x_{a},x_{b};x_{\widetilde{13}},x_{\widetilde{2}}  \}\,, \\
\mbox{F}(\{\tilde{x}\}) 
&= \{\mbox{F}(\tilde{x}_{a}),\mbox{F}(\tilde{x}_{b}); 
\mbox{F}(\tilde{x}_{1}), 
\mbox{F}(\tilde{x}_{2}) \}  
\nonumber\\
&= \{\mbox{F}(x_{a}),\mbox{F}(x_{b}) ; 
\mbox{F}(x_{\widetilde{13}}),
\mbox{F}(x_{\widetilde{2}}) \}   
\nonumber\\ 
&= \{ u,\bar{u};u,\bar{u} \}\,,  
\label{dipex132} \\
\mbox{P}(\{\tilde{x}\}) 
&= \{\mbox{P}(\tilde{x}_{a}),\mbox{P}(\tilde{x}_{b}); 
\mbox{P}(\tilde{x}_{1}), 
\mbox{P}(\tilde{x}_{2}) \}  \nonumber\\
&= \{\mbox{P}(x_{a}),\mbox{P}(x_{b}); 
\mbox{P}(x_{\widetilde{13}}),
\mbox{P}(x_{\widetilde{2}})\} 
\nonumber\\
&= \{ p_{a},p_{b};\tilde{p}_{13},\tilde{p}_{2} \},  
\end{align}
where the reduced momenta $\tilde{p}_{13}$ and 
$\tilde{p}_{2}$ are defined in 
Eqs.\,(\ref{rmffem}) and  (\ref{rmffsp}).
Next we fix the reduced Born process as
$\mbox{B}1=u\bar{u} \to u\bar{u}$,
which defines set $\{y\}$ with the field species 
and the momenta as
\begin{align}
\{ y \} 
&\equiv \{y_{a},y_{b};y_{1}, y_{2} \}\,, \label{ex1y} \\
\mbox{F}(\{ y \}) 
&= \{ u,\bar{u};u,\bar{u} \}\,,  \label{ex1yF} \\
\mbox{P}(\{ y \}) 
&= \{\mbox{P}(y_{a}),\mbox{P}(y_{b}); \mbox{P}(y_{1}), 
\mbox{P}(y_{2}) \}\,.  \label{ex1yP}
\end{align}
A field mapping $y_{\beta}=f(\tilde{x}_{\alpha})$ 
is found as
\begin{align}
f(\tilde{x}_{a}) &= y_{a}, \nonumber\\
f(\tilde{x}_{b}) &= y_{b}, \nonumber\\
f(\tilde{x}_{1}) &= y_{1}, \nonumber\\
f(\tilde{x}_{2}) &= y_{2}.
\end{align}
The field mapping is unique in this case.
The field mapping is interpreted as the identification
between the elements in sets, $\{\tilde{x}\}$ and 
$\{y\}$ as
\begin{align}
(y_{a},y_{b};y_{1}, y_{2}) =(\tilde{x}_{a},\tilde{x}_{b};
\tilde{x}_{1}, \tilde{x}_{2})
=(x_{a},x_{b};x_{\widetilde{13}},x_{\widetilde{2}} ).
\label{ex1fm}
\end{align}
Since the field species of set $\{y\}$ are fixed in the 
category {\tt Dipole\,1}, the expression 
in Eq.\,(\ref{ex1fm}) can 
be abbreviated without confusion as
\begin{equation}
(a,b \ ;\widetilde{13},\widetilde{2}).
\end{equation}
The momenta $\mbox{P}(\{ y \})$ are determined as
\begin{align}
\mbox{P}(\{ y \}) 
&= \{\mbox{P}(y_{a}),\mbox{P}(y_{b}); \mbox{P}(y_{1}), 
\mbox{P}(y_{2}) \} \nonumber\\
&= \{\mbox{P}(x_{a}),\mbox{P}(x_{b}); 
\mbox{P}(x_{\widetilde{13}}), 
\mbox{P}(x_{\widetilde{2}}) \} \nonumber\\
&= \{ p_{a},p_{b};\tilde{p}_{13},\tilde{p}_{2} \}. \label{ex1rm}
\end{align}
Then the dipole term is given in 
Eq.\,(\ref{dp111}) and 
is written down as 
\begin{align}
\mbox{D}( {\tt dip}1\mbox{-}(1)\mbox{-}1)_{13,2} 
&= -\frac{1}{s_{13}}
\frac{1}{\mbox{T}_{\mbox{{\tiny F}}{ (x_{\widetilde{13}} )}}^{2}} 
\ \mbox{V}_{13,2} \ 
\langle \mbox{B}1 =\{ y_{a},y_{b} ; y_{1}, y_{2} \} \ | 
\mbox{T}_{f{ (x_{\widetilde{13}} )}} \cdot 
\mbox{T}_{f{ (x_{\widetilde{2}} )}}
| \ \mbox{B}1 \rangle   \nonumber\\
&= -\frac{1}{s_{13}}
\frac{1}{\mbox{C}_{\mbox{{\tiny F}}}} 
\ \mbox{V}_{13,2} \ 
\langle \mbox{B}1 =\{ y \} \ | 
\mbox{T}_{y_{1}} \cdot \mbox{T}_{y_{2}} | \ \mbox{B}1 \rangle.
\label{ex1dip}
\end{align}
where the dipole splitting function $\mbox{V}_{13,2}$ 
is written in Eq.\,(\ref{ds111}) 
and the reduced momenta are determined in Eq.\,(\ref{ex1rm}).
The legs on which two 
color insertion operators act are clearly specified 
by referring to the elements $y_{1}$ and $y_{2}$ 
of set $\{y\}$. 
It is again noted that, in the same category, {\tt Dipole}$j$, 
in the present case, {\tt Dipole}$1$, the same field species, 
$\mbox{F}(\{ y \})$, is used. 
Under the agreements in the DSA,
we can abbreviate 
the expression in Eq.\,(\ref{ex1dip}) as
\begin{equation}
\mbox{D}_{13,2} = -\frac{1}{s_{13}} 
\frac{1}{\mbox{C}_{\mbox{{\tiny F}}}} 
\ \mbox{V}_{13,2} \ \langle 1,2 \rangle,
\label{dipcom1}
\end{equation}
with the definition 
$\langle 1,2 \rangle =\langle \mbox{B}1 \,| 
\,\mbox{T}_{y_{1}} \cdot \mbox{T}_{y_{2}} | \ 
\mbox{B}1 \rangle$.
While the indices $1,2$, and $3$ in
the quantities
$s_{13}$ and $\mbox{V}_{13,2}$
refer to the momenta of the original legs of 
$\mbox{R}_{1}$, 
$\{ p_{a},p_{b};p_{1},p_{2},p_{3} \}$,
the arguments \textquoteleft 1,2\textquoteright \ inside 
the brackets $\langle \ \rangle$ 
refer to the elements of set $\{ y \}$.
%
%
\item Example 2: \ {\tt Dipole1} (1)-1: (23,1) \\
Set $\{\tilde{x}\}$ is defined as
\begin{align}
\{\tilde{x}\} 
&=\{ \tilde{x}_{a},\tilde{x}_{b};\tilde{x}_{1}, 
\tilde{x}_{2} \} \nonumber\\
&= \{x_{a},x_{b};x_{\widetilde{23}},
x_{\widetilde{1}} \}\,, \\
\mbox{F}(\{\tilde{x}\}) &= \{ u,\bar{u};
\bar{u},u \}\,,  \\
\mbox{P}(\{\tilde{x}\}) &= \{ p_{a},p_{b};
\tilde{p}_{23},\tilde{p}_{1} \}\,.  
\end{align}
The definition of set $\{ y \}$ is the same as in
Eqs.\,(\ref{ex1y}), (\ref{ex1yF}), and (\ref{ex1yP}).
The field mapping is found as
\begin{align}
f(\{\tilde{x}\})=f(x_{a},x_{b};x_{\widetilde{23}},
x_{\widetilde{1}}) = (y_{a},y_{b};y_{2},y_{1})\,. 
\end{align}
The mapping determines the identification of the elements as
\begin{align}
(y_{a},y_{b};y_{1},y_{2})=(x_{a},x_{b};x_{\widetilde{1}},
x_{\widetilde{23}})\,,  
\end{align}
which can be abbreviated as
$(a,b;\widetilde{1},\widetilde{23})$.
The momenta of set $\{ y \}$ are determined as
\begin{align}
\mbox{P}(\{ y \}) 
&= \{\mbox{P}(y_{a}),\mbox{P}(y_{b}); \mbox{P}(y_{1}), 
\mbox{P}(y_{2}) \} \nonumber\\
&= \{\mbox{P}(x_{a}),\mbox{P}(x_{b}); 
\mbox{P}(x_{\widetilde{1}}), 
\mbox{P}(x_{\widetilde{23}}) \} \nonumber\\
&= \{ p_{a},p_{b};\tilde{p}_{1}, \tilde{p}_{23}\}.  
\end{align}
The dipole term is written as
\begin{equation}
\mbox{D}_{23,1} = -\frac{1}{s_{23}} 
\frac{1}{\mbox{C}_{\mbox{{\tiny F}}}} \ \mbox{V}_{23,1} \ 
\langle 2,1 \rangle.
\end{equation}
%
%
\item Example 3: \ {\tt Dipole2}$u$ (5)-2: (12,a) \\
Set $\{\tilde{x}\}$ is defined 
with the field species and the momenta as
\begin{align}
\{\tilde{x}\} 
&=\{ \tilde{x}_{a},\tilde{x}_{b};
\tilde{x}_{1}, \tilde{x}_{2} \} \nonumber\\
&= \{x_{\widetilde{a}}, x_{b};
x_{\widetilde{12}}, x_{3} \}\,, \\
\mbox{F}(\{\tilde{x}\}) 
&= \{ u,\bar{u};g,g \}\,,  \\
\mbox{P}(\{\tilde{x}\}) 
&= \{ \tilde{p}_{a},p_{b};
\tilde{p}_{12},p_{3} \}\,.  
\end{align}
The reduced Born process is fixed as
$\mbox{B}2=u \bar{u} \to gg$, which  
determines set $\{ y \}$ with the 
field species and the momenta as
\begin{align}
\{ y \} 
&= \{y_{a},y_{b};y_{1}, y_{2} \}, \\
\mbox{F}(\{ y \}) 
&= \{ u,\bar{u};g,g \}\,, \\
\mbox{P}(\{ y \}) 
&= \{\mbox{P}(y_{a}),\mbox{P}(y_{b}); 
\mbox{P}(y_{1}), \mbox{P}(y_{2}) \}\,.
\end{align}
In this case,
the field mapping has two possibilities 
due to the two identical fields in the final state, 
two gluons, as
\begin{align}
f(\{\tilde{x}\})=f(x_{\tilde{a}},x_{b}\,;x_{\widetilde{12}},
x_{3}) = (y_{a},y_{b}\,;y_{1},y_{2}) \ 
\mbox{or} \ (y_{a},y_{b}\,;y_{2},y_{1}). \label{ex3fm}
\end{align}
Both field mappings are equally allowed. To write down
the concrete expression, either of the two possibilities
must be chosen. Here we choose the first case in 
Eq.\,(\ref{ex3fm}). The identification of the elements 
is written as 
\begin{equation}
(y_{a},y_{b};y_{1}, y_{2})=(x_{\widetilde{a}}, x_{b};
x_{\widetilde{12}},x_{3})\,,
\end{equation}
which is abbreviated as $(\widetilde{a}, b\,;
\widetilde{12},3)$.
The momenta is determined as
\begin{align}
\mbox{P}(\{ y \}) 
&= \{\mbox{P}(y_{a}),\mbox{P}(y_{b}); \mbox{P}(y_{1}), 
\mbox{P}(y_{2}) \} 
= \{ \tilde{p}_{a},p_{b};\tilde{p}_{12},p_{3} \}.  
\end{align}
Referring to the formula in
Eq.\,(\ref{dp252}),
the dipole term is written as
\begin{align}
\mbox{D}({\tt dip}2u,\,(5)\mbox{-}2)_{12,a} 
&= -\frac{1}{s_{12}} \frac{1}{x_{12,a}}
\frac{1}{\mbox{T}_{\mbox{{\tiny F}}{(x_{\widetilde{12}})}}^{2}} \ 
\langle \mbox{B}2u =\{ y \} \ | 
\mbox{T}_{f{ (x_{\widetilde{12}} )}} \cdot 
\mbox{T}_{f{ (x_{\widetilde{a}} )}} \
\mbox{V}_{{\scriptscriptstyle 12,a}}^{f{ (x_{\widetilde{12}} )}}
| \ \mbox{B}2u \rangle   \nonumber \\
&= -\frac{1}{s_{12}} \frac{1}{x_{12,a}}
\frac{1}{\mbox{C}_{\mbox{{\tiny A}}}} \ 
\langle \mbox{B}2u \ 
| \mbox{T}_{y_{1}} \cdot \mbox{T}_{y_{a}} \ 
\mbox{V}_{{\scriptscriptstyle 12,a}}^{y_{1}}| \ 
\mbox{B}2u \rangle.
\end{align}
Keeping in mind the reduced Born process
$\mbox{B}2u$, the expression is abbreviated
as
\begin{equation}
\mbox{D}_{12,a} =- \frac{1}{s_{12}} \frac{1}{x_{12,a}}  
\frac{1}{\mbox{C}_{\mbox{{\tiny A}}}} \ \langle 1,a, 
\mbox{V}_{{\scriptscriptstyle 12,a}}^{1} \rangle.
\end{equation}
%
%
\item Example 4: \ {\tt Dipole3}$u$ (6)-2: (a1,b) \\
Set $\{\tilde{x}\}$ is defined as
\begin{align}
\{\tilde{x}\} 
&=\{ \tilde{x}_{a},\tilde{x}_{b};\tilde{x}_{1}, 
\tilde{x}_{2} \} 
\nonumber\\
&= \{x_{\widetilde{a1}},x_{\widetilde{b}};x_{2},x_{3} 
\}, \\
\mbox{F}(\{\tilde{x}\}) 
&= \{ g,\bar{u}; \bar{u},g \}, \\
\mbox{P}(\{\tilde{x}\}) 
&= \{\mbox{P}(\tilde{x}_{a}),\mbox{P}(\tilde{x}_{b}); 
\mbox{P}(\tilde{x}_{1}), \mbox{P}(\tilde{x}_{2} ) \}  
\nonumber\\
& =\{ \tilde{p}_{a1},p_{b};\tilde{k}_{2}, \tilde{k}_{3} \}.  
\end{align}
Next we fix the reduced Born process as 
\begin{equation}
\mbox{B}3u= g \bar{u} \to \bar{u} g, 
\end{equation}
which determines set $\{ y \}$ as
\begin{align}
\{ y \} &= \{y_{a},y_{b};y_{1}, y_{2} \}, \\
\mbox{F}(\{ y \}) 
&= \{ g,\bar{u}; \bar{u},g \}, 
\label{dipexa1b} \\
\mbox{P}(\{ y \}) 
&= \{\mbox{P}(y_{a}),\mbox{P}(y_{b}); \mbox{P}(y_{1}), 
\mbox{P}(y_{2}) \}.
\end{align}
The field mapping is uniquely found as
\begin{equation}
f(\{\tilde{x}\})=f(x_{\widetilde{a1}},x_{\widetilde{b}};
x_{2},x_{3})=(y_{a},y_{b};y_{1},y_{2}),
\end{equation}
which is interpreted as the identification as 
\begin{equation}
(y_{a},y_{b};y_{1}, y_{2})=
(x_{\widetilde{a1}},x_{\widetilde{b}};x_{2},x_{3})\,.
\end{equation}
The expression is abbreviated as 
$(\widetilde{a1},\widetilde{b} \ ;2,3)$.
The momenta are also determined as
\begin{align}
\mbox{P}(\{ y \}) 
&= \{\mbox{P}(x_{\widetilde{a1}}),\mbox{P}(x_{\widetilde{b}}); 
\mbox{P}(x_{2}), \mbox{P}(x_{3}) \}\,, 
\nonumber\\
&= \{ \tilde{p}_{a1},p_{b};\tilde{k}_{2}, \tilde{k}_{3} \}\,.
\end{align}
The dipole term is written as
\begin{align}
\mbox{D}({\tt dip3}u,\,(6)\mbox{-}2)_{a1,b} 
&= -\frac{1}{s_{a1}} \frac{1}{x_{1,ab}}
\frac{1}{\mbox{T}_{\mbox{{\tiny F}}{ (x_{\widetilde{a1}} )}}^{2}} \ 
\langle \mbox{B}3u | 
\mbox{T}_{f{ (x_{\widetilde{a1}} )}} \cdot 
\mbox{T}_{f{ (x_{\widetilde{b}} )}} \ 
\mbox{V}_{{\scriptscriptstyle a1,b}}^{f{ (x_{\widetilde{a1}} )}}
| \ \mbox{B}3u \rangle   \nonumber \\
&= -\frac{1}{s_{a1}} \frac{1}{x_{1,ab}}
\frac{1}{\mbox{C}_{\mbox{{\tiny A}}}} \ 
\langle \mbox{B}3u | 
\mbox{T}_{y_{a}} 
\cdot \mbox{T}_{y_{b}} \ 
\mbox{V}_{{\scriptscriptstyle a1,b}}^{y_{a}} | 
\ \mbox{B}3u \rangle,  
\end{align}
which is abbreviated as 
\begin{equation}
\mbox{D}_{a1,b} =- \frac{1}{s_{a1}} \frac{1}{x_{1,ab}}  
\frac{1}{\mbox{C}_{\mbox{{\tiny A}}}} \ \langle a,b, 
\mbox{V}_{{\scriptscriptstyle a1,b}}^{a} \rangle.
\end{equation}

For this dipole term, we
have another possibility to fix
the reduced Born process as
\begin{equation}
\mbox{B}3u'= \bar{u} g \to \bar{u} g, 
\end{equation}
which defines set $\{ y \}$ as
\begin{align}
\{ y \} &= \{y_{a},y_{b};y_{1}, y_{2} \}, \\
\mbox{F}(\{ y \}) 
&= \{ \bar{u},g; \bar{u},g \},  \\
\mbox{P}(\{ y \}) 
&= \{\mbox{P}(y_{a}),\mbox{P}(y_{b}); \mbox{P}(y_{1}), 
\mbox{P}(y_{2}) \}.  
\end{align}
In this case the field mapping is found as
\begin{equation}
f(\{\tilde{x}\})=f(x_{\widetilde{a1}},x_{\widetilde{b}};
x_{2},x_{3})=(y_{b},y_{a};y_{1},y_{2}), 
\end{equation}
which is interpreted as
\begin{equation}
(y_{a},y_{b};y_{1}, y_{2})=
(x_{\widetilde{b}},x_{\widetilde{a1}};x_{2},x_{3})\,.
\end{equation}
This is abbreviated as $(\widetilde{b},\widetilde{a1} \ ;2,3)$.
The momenta are determined as
\begin{align}
\mbox{P}(\{ y \}) 
&= \{\mbox{P}(x_{\widetilde{b}}),\mbox{P}(x_{\widetilde{a1}}); 
\mbox{P}(x_{2}), \mbox{P}(x_{3}) \}\,, 
\nonumber\\
&= \{ p_{b}, \tilde{p}_{a1};\tilde{k}_{2}, \tilde{k}_{3} \}\,.
\end{align}
The dipole term is written as
\begin{align}
\mbox{D}_{a1,b} &= -\frac{1}{s_{a1}} \frac{1}{x_{1,ab}}
\frac{1}{\mbox{T}_{\mbox{{\tiny F}}{ (x_{\widetilde{a1}} )}}^{2}} \ 
\langle \mbox{B}3u' \,| \mbox{T}_{f{ (x_{\widetilde{a1}} )}} \cdot 
\mbox{T}_{f{ (x_{\widetilde{b}} )}} \ 
\mbox{V}_{{\scriptscriptstyle a1,b}}^{f{ (x_{\widetilde{a1}} )}}
| \ \mbox{B}3u' \rangle   \nonumber \\
&= -\frac{1}{s_{a1}} \frac{1}{x_{1,ab}}
\frac{1}{\mbox{C}_{\mbox{{\tiny A}}}} \ \langle \mbox{B}3u'\,| 
\mbox{T}_{y_{b}} 
\cdot \mbox{T}_{y_{a}} \ 
\mbox{V}_{{\scriptscriptstyle a1,b}}^{y_{b}} | \ 
\mbox{B}3u' \rangle.
\end{align}
which is abbreviated, keeping $\mbox{B}3u'$ in mind, as
\begin{equation}
\mbox{D}_{a1,b} =- \frac{1}{s_{a1}} \frac{1}{x_{1,ab}} 
\frac{1}{\mbox{C}_{\mbox{{\tiny A}}}} \ \langle b,a, 
\mbox{V}_{{\scriptscriptstyle a1,b}}^{b} \rangle.
\label{dr1last}
\end{equation}
In the first case of B3$u$,
the element $x_{\widetilde{a1}}$
is identified with the element $y_{a}$ and, 
in the second case of $\mbox{B}3u'$,
$x_{\widetilde{a1}}$ is 
identified with $y_{b}$. 
The first case appears to be a simpler 
expression in the sense that the leg $a$ of the input 
process $\mbox{R}_{1}$ is connected to the leg $a$ of
the reduced Born process $\mbox{B}3u$.
For this reason, the first case may be more favored than the
second.
\end{itemize}

\subsubsection{Summary \label{sec225}}
The hadronic cross section of a real correction
subtracted by the dipole terms is written as
\begin{align}
\sigma_{\mbox{{\tiny R}}}(\mbox{R}_{i}) - 
\sigma_{\mbox{{\tiny D}}}(\mbox{R}_{i})
= \int dx_{1} \int dx_{2} \ f_{\mbox{{\tiny F}}(x_{a})}(x_{1}) 
f_{\mbox{{\tiny F}}(x_{b})}(x_{2}) \
\bigl(\hat{\sigma}_{\mbox{{\tiny R}}}(\mbox{R}_{i}) - 
\hat{\sigma}_{\mbox{{\tiny D}}}(\mbox{R}_{i}) \bigr), \label{dipmas2}
\end{align}
which is part of Eq.\,(\ref{master}). 
The partonic cross section is written as 
\begin{equation}
\hat{\sigma}_{\mbox{{\tiny R}}}(\mbox{R}_{i}) - 
\hat{\sigma}_{\mbox{{\tiny D}}}(\mbox{R}_{i}) 
=\frac{1}{S_{\mbox{{\tiny R}}_{i}}} 
\Phi(\mbox{R}_{i})_{4} \cdot \Bigl[ \ 
|\mbox{M}(\mbox{R}_{i})|_{4}^{2} \ - 
\ \frac{1}{n_{s}(a) n_{s}(b)} \mbox{D}(\mbox{R}_{i}) 
\ \Bigr]\,. \label{dippar}
\end{equation}
It is sufficient that the real correction
$|\mbox{M}(\mbox{R}_{i})|_{4}^{2}$ and 
the dipole term $\mbox{D}(\mbox{R}_{i})$
are obtained in 4 dimensions. The real correction
$|\mbox{M}(\mbox{R}_{i})|_{4}^{2}$ is summed and 
averaged over both the spin and the color. 
The dipole term $\mbox{D}(\mbox{R}_{i})$
is the summation
of all the dipole terms under the input $\mbox{R}_{i}$
and is separated into subcategories as
\begin{equation}
{\tt D} (\mbox{R}_{i})=\sum_{j=1}^{4}\, 
{\tt D} (\mbox{R}_{i},\,{\tt dip} j\,).
\end{equation}
In each category ${\tt Dipole}j$,
the reduced Born process $\mbox{B}j$ is fixed, 
which defines set $\{ y\}$ with the field species as
\begin{align}
\mbox{B}j \ \to \   
\left\{
\begin{array}{l}
\ \ \ \ \{ y \} = \{y_{a},y_{b};y_{1}, ..., y_{n} \}\,, \\
\mbox{F}(\{ y \}) = \{\mbox{F}(y_{a}),\mbox{F}(y_{b}); 
\mbox{F}(y_{1}), ..., \mbox{F}(y_{n}) \}\,.
\end{array}
\right.
\end{align}
Once the reduced Born process and set $\{ y\}$
are fixed, 
the necessary information to specify each dipole is
the three elements of set 
$\{ x\}$ and the field mapping as
\begin{align}
&1. \ (x_{I}x_{J},x_{K})\,,  \\
&2. \ (y_{a},y_{b};y_{1}, ..., y_{n})=
(f^{-1}(y_{a}),f^{-1}(y_{b});f^{-1}(y_{1}), ..., 
f^{-1}(y_{n})) \nonumber \\
& \hspace{35mm}= ( x_{a},x_{b};x_{1}, ...,
x_{\widetilde{{\scriptscriptstyle IJ}}}, ..., 
x_{\widetilde{{\scriptscriptstyle K}}},
..., x_{n+1} ). 
\end{align}
The form to specify the information can be
abbreviated as
\begin{align}
&1. \ (IJ,K)\,,  \nonumber \\
&2. \ (a,b;1, ...,
\widetilde{IJ}, ..., 
\widetilde{ K},
..., n+1)\,. 
\label{dinfo}
\end{align}
Using the notation,
$y_{emi}=f(x_{\widetilde{{\scriptscriptstyle IJ}}})$
and 
$y_{spe}=f(x_{\widetilde{{\scriptscriptstyle K}}})$,
each dipole term is simply written down as
\begin{equation}
\mbox{D}(\mbox{R}_{i}\,,{\tt dip}j\ )_{IJ,K} =
 -\frac{1}{s_{{\scriptscriptstyle IJ}}} 
\frac{1}{x_{{\scriptscriptstyle IJK}}}
\frac{1}{\mbox{T}_{\mbox{{\tiny F}}(y_{emi})}^{2}} 
\langle \mbox{B}j \ | 
\mbox{T}_{y_{emi}} \cdot 
\mbox{T}_{y_{spe}}
\ \mbox{V}_{{\scriptscriptstyle IJ,K}}^{y_{emi}}
 | \ \mbox{B}j \rangle,  
\end{equation}
which is abbreviated as
\begin{equation}
\mbox{D}({\tt dip}j )_{IJ,K} =
 -\frac{1}{s_{{\scriptscriptstyle IJ}}} 
\frac{1}{x_{{\scriptscriptstyle IJK}}}
\frac{1}{\mbox{T}_{y_{emi}}^{2}} 
\langle  
y_{emi} \cdot y_{spe},
 \mbox{V}_{{\scriptscriptstyle IJ,K}}^{y_{emi}}
 \rangle.
\end{equation}
The square of the reduced Born amplitude with color
and spin correlations,  
$\langle \mbox{B}j \ | \mbox{T} \cdot \mbox{T} \ \mbox{V}
| \ \mbox{B}j \rangle $, is summed and averaged over 
the color degree of freedom. It is also summed over the spin
configurations, but not averaged.
Instead, the dipole terms are divided by 
the spin average factor of the input real process,
as shown in Eq.\,(\ref{dippar}).
It is also noted that the symmetric factor
by which the dipole terms are divided 
is not the symmetric factor
of the reduced Born processes $S_{\mbox{{\tiny B}}_{j}}$,
but the symmetric factor of the input
process $S_{\mbox{{\tiny R}}_{i}}$,
as shown in Eq.\,(\ref{dippar}).
%
%
%
%
%
\subsection{Step\,3: I term creation  \label{s2_3}}
In this section, 
\textquoteleft{\bfseries Step 3}. 
I\,($\mbox{R}_{i}$)\textquoteright \ is explained.
The input and output of this step are:
\begin{description}
\setlength{\itemsep}{-1mm}
\item[\ \ \ \ Input:] \ \ \ $\mbox{B}1\,(\mbox{R}_{i})$\,,
\item[\ \ \ \ Output:] \ I\,($\mbox{R}_{i}$)\,.
\end{description}
$\mbox{B}1(\mbox{R}_{i})$ is the reduced
Born process of the category {\tt Dipole\,1}, 
which is made from the input process $\mbox{R}_{i}$ 
on the rule
$\mbox{B}1=\mbox{R}_{i} -g_{f}$, as shown 
in Eq.\,(\ref{makerb}).

The creation algorithm given in
the original article \cite{Catani:1996vz} 
is to choose all combinations of two elements
$(y_{I},y_{K})$ without duplicate
from the set 
$\{ y\}=\{y_{a},y_{b}; 
y_{1}, ..., y_{n} \}$
of the process $\mbox{B}1\,(\mbox{R}_{i})$,
where the elements to be chosen
are quark or gluon
as $\mbox{F}(y_{I/K})=$ quark or gluon.
In the DSA, the creation algorithm of the I terms
is divided into substeps as follows\,:
\begin{enumerate}
\setlength{\itemsep}{0mm}
\item Choose all possible elements $y_{I}$
from set $\{ y\}$
in the order in Fig.\,\ref{fig_Iterm} in 
Appendix \ref{ap_A_2}. 
\item Choose all the possible elements $y_{K}$
from set $\{ y\}$ for each choice of the 
element $y_{I}$.
\item Write down the concrete expressions of all 
the I terms.
\end{enumerate}
Substeps 1 and 2 are explained in 
Sec.\,\ref{sec231}.
Substep 3 is explained in 
Sec.\,\ref{sec232} and \ref{sec233}.
Some examples are shown in 
Sec.\,\ref{sec234}.
Finally, we give a summary in 
Sec.\,\ref{sec235}.
The formulae for the I terms are collected 
in Appendix \ref{ap_A_2}.
\subsubsection{Creation order \label{sec231}}
In the DSA, the order to choose the first element
$y_{I}$ is determined as
\begin{align}
{\tt (1)} \ f_{fin}, \ \ \ {\tt (2)} \ g_{fin}, \ \ \  
{\tt (3)} \ f_{ini}, \ \ \ {\tt (4)} \ g_{ini},
\end{align}
where the $f_{fin/ini}$ represents a quark 
in the final/initial state and the 
$g_{fin/ini}$ represents a gluon 
in the final/initial state, as
shown in Fig.\,\ref{fig_Iterm}.
Each choice of $y_{I}$ is followed by
the choice of the second element
$y_{K}$. 
The choice of $y_{K}$ in the
final state is first and the choice
in the initial state is second,
which are denoted as subcategories
$1$ and $2$ respectively. Then the creation 
order is written as
\begin{align}
{\tt (1)-1/2, \ \ \ (2)-1/2, \ \ \ (3)-1/2, \ \ \ (4)-1/2.} 
\label{icror} 
\end{align}
Each pair $(y_{I},y_{K})$ specifies
each I term, which is denoted as $\mbox{I}\,(\mbox{R}_{i})_{I,\,K}$.
The summation of all the created I terms is the output
$\mbox{I}\,(\mbox{R}_{i})$, which is written as
\begin{equation}
\mbox{I}\,(\mbox{R}_{i})=
\sum_{(I,K)} \mbox{I}\,(\mbox{R}_{i})_{I,\,K}\,.
\end{equation}
When all ($n$+2) legs of $\mbox{B}1$, and, equally,
all ($n$+2) elements of set $\{y\}$,
are quark or gluon, the indices $I$ and $K$
run over $I,K=a,b,1,...,n$,
with the condition $I \not= K$,
and the total number of I terms is $(n+2)(n+1)$.

\subsubsection{Concrete formulae \label{sec232}}
The concrete expression of each I term 
is given in the universal form as
\begin{align}
\mbox{I}\,_{I,K}
= - A_{d} \,
\frac{1}{\mbox{T}_{\mbox{{\tiny F}}(
y_{{\scriptscriptstyle I}})}^{2}} 
\,
{\cal V}_{\mbox{{\tiny F}}(
y_{{\scriptscriptstyle I}})}
\, s_{{\scriptscriptstyle IK}}^{-\ep} \cdot
\langle \mbox{B}1=\{ y_{a},y_{b} ; y_{1}, ..., y_{n} \} 
\ | 
\mbox{T}_{y_{{\scriptscriptstyle I}}} \cdot 
\mbox{T}_{y_{{\scriptscriptstyle K}}}
| \ \mbox{B}1 \rangle,
\label{iexp}
\end{align}
where the common factor $A_{d}$ is defined as 
\begin{equation}
A_{d}=\frac{\al}{2\pi} \frac{(4\pi \mu^{2})^{\ep}}{\Gamma(1-\ep)},
\end{equation}
with the free parameter $\mu$ introduced in the dimensional
regularization with $d=4-2\ep$. 
The definition of the Casimir operator 
$\mbox{T}_{\mbox{{\tiny F}}(y_{{\scriptscriptstyle I}})}^{2}$
is the same as in Eq.\,(\ref{dipmas}).
The universal singular function 
${\cal V}_{\mbox{{\tiny F}}(y_{{\scriptscriptstyle I}})}$
is defined as
\begin{align}
{\cal V}_{f} &= {\cal V}_{fg} (\ep)  \\
{\cal V}_{g} &= \half {\cal V}_{gg} (\ep) + 
N_{f}{\cal V}_{f\bar{f}} (\ep), 
\end{align}
where the singular functions 
${\cal V}_{fg} (\ep), {\cal V}_{gg} (\ep)$,
and ${\cal V}_{f\bar{f}} (\ep)$ are written in 
Eqs.\,(\ref{isfg}),\,(\ref{isgg}), 
and (\ref{isff}), respectively.
The symbol $N_{f}$ represents the number of massless quark 
flavors. 
The momenta of the reduced Born process 
$\mbox{B}1$ are written as
\begin{equation}
\mbox{P}(\{ y \})=
\{\mbox{P}(y_{a}),\mbox{P}(y_{b}); \mbox{P}(y_{1}), ..., 
\mbox{P}(y_{n}) \},
\end{equation}
which are defined as the momenta of the $n$-body PS.
The scalar $s_{{\scriptscriptstyle IK}}$ is denoted as
$s_{{\scriptscriptstyle IK}}=2\mbox{P}(y_{I}) \cdot \mbox{P}(y_{k})$.
The partonic cross section of the I term is written in
Eq.\,(\ref{hati}) as
\begin{equation}
\hat{\sigma}_{\mbox{{\tiny I}}}(\mbox{R}_{i}) 
= \frac{1}{S_{\mbox{{\tiny B1}} }} \ \Phi(\mbox{B}1)_{d} 
\cdot \mbox{I}\,(\mbox{R}_{i})\,,
\end{equation}
where the $n$-body PS integration is defined in Eq.\,(\ref{b1ps}) as
\begin{equation}
\Phi(\mbox{B}1)_{d} = \frac{1}{{\cal F}(\mbox{P}(y_{a}),\mbox{P}(y_{b}))} 
\prod_{i=1}^{n} \int \frac{d^{d-1}\mbox{P}(y_{i}) }{(2\pi)^{d-1}} 
\frac{1}{2E_{i}} \cdot (2\pi)^{d} \delta^{(d)} 
\Bigl(\mbox{P}(y_{a})+\mbox{P}(y_{b})- \sum_{i=1}^{n} \mbox{P}(y_{i}) 
\Bigr)\,.
\label{b1psd}
\end{equation}
The square of the color-correlated Born amplitude, 
$\langle \mbox{B}1 | 
\mbox{T}_{y_{{\scriptscriptstyle I}}} \cdot 
\mbox{T}_{y_{{\scriptscriptstyle K}}}
| \ \mbox{B}1 \rangle$,
is obtained in $d$ dimensions,
and summed and averaged over the spin and color
including the spin average factor,
which is in contrast to the case of
the dipole terms in Eq.\,(\ref{hatd}).
Again, in contrast to 
\textquoteleft{\bfseries Step 2}.
D($\mbox{R}_{i}$)\textquoteright,
in {\bfseries Step 3}
only one process, $\mbox{B}1$, 
and the only one set, $\{ y\}$, appear.
Then we can drop 
the specification of $\mbox{B}1$
in Eq.\,(\ref{iexp}) as
\begin{align}
\mbox{I}_{I,\,K}
= - A_{d}\,
\frac{1}{\mbox{T}_{\mbox{{\tiny F}}(I)}^{2}} 
\,{\cal V}_{\mbox{{\tiny F}}(I)}
\,[ \,I , K
]\,,
\end{align}
where we introduce the following notation for convenience\,:
\begin{equation}
[ \,I , K]=s_{{\scriptscriptstyle IK}}^{-\ep} 
\,\langle \mbox{B}1 | 
\mbox{T}_{y_{{\scriptscriptstyle I}}} \cdot 
\mbox{T}_{y_{{\scriptscriptstyle K}}}
| \ \mbox{B}1 \rangle
\,.
\label{sqbk}
\end{equation}

\subsubsection{Complete set \label{sec233}}
Each I term $\mbox{I}_{I,\,K}$ includes the square
of a reduced Born amplitude with color correlations
as $\langle \mbox{T}_{I} \cdot 
\mbox{T}_{K} \rangle$.
The summation of I terms $\mbox{I}=
\sum_{(I,K)} \mbox{I}_{I,\,K}$
includes the square of a Born amplitude
with all combinations of the pairs
$(I,K)$. We call the set that
consists of the elements
$\langle \mbox{T}_{I} \cdot 
\mbox{T}_{K} \rangle$
with all the combinations of $(I,K)$,
\textquoteleft
the complete set of the square
of the color-correlated Born 
amplitude $\mbox{B}1$\textquoteright.
The name is sometimes abbreviated as
\textquoteleft
the complete set of $\mbox{B}1$\textquoteright,
which is explicitly written down as
\begin{align}
\{ \langle \mbox{B}1 | 
\mbox{T}_{I} \cdot 
\mbox{T}_{K}
| \ \mbox{B}1 \rangle \}_{{\footnotesize \mbox{comp}}}
&=\{ 
\langle a,b \rangle,
\langle a,1 \rangle,
\langle a,2 \rangle,
...,
\langle a,n \rangle, 
\nonumber\\
&\hspace{8mm}\langle b,a \rangle,
\langle b,1 \rangle,
\langle b,2 \rangle,
...,
\langle b,n \rangle, 
\nonumber\\
&\hspace{8mm}\langle 1, a \rangle, 
\langle 1, b \rangle,
\langle 1, 2 \rangle, 
...,
\langle 1, n \rangle,
\nonumber \\
&\hspace{8mm}\langle 2, a \rangle, 
\langle 2, b \rangle,
\langle 2, 1 \rangle, 
...,
\langle 2, n \rangle,
\nonumber \\
&\hspace{8mm}..., 
\nonumber \\
&\hspace{8mm}\langle n, a \rangle, 
\langle n, b \rangle,
\langle n, 1 \rangle, 
...,
\langle n, n-1 \rangle
\}\,,
\end{align}
where all the legs of the reduced Born process 
$\mbox{B}1$ are assumed to be quark or gluon.
The number of elements is $(n+2)(n+1)$,
which is the same as the number of I terms.
The complete set of $\mbox{B}1$ is always included in 
the dipole terms in the category {\tt Dipole\,1}, 
because the dipole terms
$\mbox{D} (\mbox{R}_{i},{\tt dip1})$
include the square of the reduced Born amplitude  
$\mbox{B}1$ 
with all combinations of the pair
$(y_{emi},y_{spe})$
shown in Eq.\,(\ref{dipmas}) as
$\langle \mbox{B}1 \ | 
\mbox{T}_{y_{emi}} \cdot 
\mbox{T}_{y_{spe}}
| \ \mbox{B}1 \rangle$.
Once we have obtained analytical
or numerical expressions for
the complete set of $\mbox{B}1$
as a function of the arbitrary 
input momenta, 
$\{\mbox{P}(y_{a}),\mbox{P}(y_{b}); \mbox{P}(y_{1}), ..., 
\mbox{P}(y_{n}) \}$,
for the calculation of the dipole terms,
the expressions can be used again 
for the calculation of the I terms as well.
Such a reuse of expressions 
can save a certain amount of work
in constructing the subtraction terms.

\subsubsection{Examples \label{sec234}}
We show some examples in the same process
in Eq.\,(\ref{dipexin}),
$\mbox{R}_{1} = u\bar{u} \to u\bar{u}g$.
The input for {\bfseries Step 3} is 
\begin{equation}
\mbox{B}1(\mbox{R}_{1}) = u\bar{u} \to u\bar{u},
\label{iex1}
\end{equation}
which determines set $\{ y\}$ with the field species in 
Eqs.\,(\ref{ex1y}),\,(\ref{ex1yF}), and (\ref{ex1yP}) 
as
\begin{align}
\{ y \} 
&= \{y_{a},y_{b};y_{1}, y_{2} \}\,,  \\
\mbox{F}(\{ y \}) 
&= \{ u,\bar{u};u,\bar{u} \}\,,   \\
\mbox{P}(\{ y \}) 
&= \{\mbox{P}(y_{a}),\mbox{P}(y_{b}); \mbox{P}(y_{1}), 
\mbox{P}(y_{2}) \}\,. 
\end{align}
Following the creation order in Eq.\,(\ref{icror}), 
the I terms are created as
\begin{align*}
(1)&-1: \ \ (1,2),(2,1),            \\[-4pt]
   &-2: \ \ (1,a),(1,b),(2,a),(2,b), \\[-4pt]
(3)&-1: \ \ (a,1),(a,2),(b,1),(b,2),  \\[-4pt]
   &-2: \ \ (a,b),(b,a).             \\[-4pt]
\end{align*}
The concrete expression for the I term, 
$\mbox{I}_{1,\,2}$, for instance,
is written as
\begin{align}
\mbox{I}_{1,\,2}
&= - A_{d} \cdot
\frac{1}{\mbox{T}_{\mbox{{\tiny F}}(
y_{1})}^{2}} \cdot
{\cal V}_{\mbox{{\tiny F}}(
y_{1})}
\cdot s_{12}^{-\ep} \cdot
\langle \mbox{B}1=\{ y_{a},y_{b} ; y_{1}, ..., y_{n} \} 
\ | 
\mbox{T}_{y_{1}} \cdot 
\mbox{T}_{y_{2}}
| \ \mbox{B}1 \rangle,
\nonumber \\
&= - A_{d} \cdot
\frac{1}{\mbox{C}_{\mbox{{\tiny F}}}} \cdot
{\cal V}_{f}
\cdot s_{12}^{-\ep} \cdot
\langle \mbox{B}1 \ | 
\mbox{T}_{y_{1}} \cdot 
\mbox{T}_{y_{2}}
| \ \mbox{B}1 \rangle,
\end{align}
which is abbreviated as
\begin{align}
\mbox{I}_{1,\,2}
= - A_{d} \,
\frac{{\cal V}_{f}}{\mbox{C}_{\mbox{{\tiny F}}}} 
\,[1,2]\,.
\label{icom1}
\end{align}
The input momenta into the Born amplitude $\mbox{B}1$,
$\{\mbox{P}(y_{a}),\mbox{P}(y_{b}); \mbox{P}(y_{1}), 
\mbox{P}(y_{2}) \}$, are the momenta in the
2-body phase space in Eq.\,(\ref{b1psd}).
The output $\mbox{I} (\mbox{R}_{1})$ is simply 
written as
\begin{align}
\mbox{I} (\mbox{R}_{1})
&=\sum_{(I,K)} \mbox{I}\,(\mbox{R}_{1})_{I,\,K}
\nonumber\\
&=- A_{d} \,\frac{{\cal V}_{f}}{\mbox{C}_{\mbox{{\tiny F}}}} 
\cdot \bigl( \
[1,2]+[2,1]+[1,a]+[1,b]+
[2,a]+[2,b] 
\nonumber\\
&\hspace{24mm}+[a,1]+[a,2]+
[b,1]+[b,2]+[a,b]+[b,a]      
\bigr)\,.
\end{align}
$\mbox{I} (\mbox{R}_{1})$ is calculated
in terms of the complete set of 
$\mbox{B}1=u\bar{u} \to u\bar{u}$,
$\{ \langle \mbox{B}1 | 
\mbox{T}_{I} \cdot 
\mbox{T}_{K}
| \ \mbox{B}1 \rangle \}_{{\footnotesize \mbox{comp}}}$.
The number of elements of the complete set is twelve,
which exactly corresponds to the twelve,
$\langle \mbox{B}1 \ | 
\mbox{T}_{y_{emi}} \cdot 
\mbox{T}_{y_{spe}}
| \ \mbox{B}1 \rangle$,
included in the dipole terms in category
{\tt Dipole\,1} in Eq.\,(\ref{dipex1}).

\subsubsection{Summary \label{sec235}}
The contributions of the virtual correction
and the I term to the hadronic cross section
are written as
\begin{align}
\sigma_{\mbox{{\tiny V}}}(\mbox{B}1) + 
\sigma_{\mbox{{\tiny I}}}(\mbox{R}_{i}) 
&= \int dx_{1} \int dx_{2} \ 
f_{\mbox{{\tiny F}}(x_{a})}(x_{1}) \,
f_{\mbox{{\tiny F}}(x_{b})}(x_{2}) \ 
\bigl(\hat{\sigma}_{\mbox{{\tiny V}}}(\mbox{B}1) + 
\hat{\sigma}_{\mbox{{\tiny I}}}(\mbox{R}_{i}) \bigr)\,,
\end{align}
which is part of Eq.\,(\ref{master}).
The partonic cross section is written as
\begin{equation}
\hat{\sigma}_{\mbox{{\tiny V}}}(\mbox{B}1) + 
\hat{\sigma}_{\mbox{{\tiny I}}}(\mbox{R}_{i}) = 
\frac{1}{S_{\mbox{{\tiny B1}} }} \ 
\Phi( \mbox{B}1  )_{d} \cdot 
\Bigl[ \ |\mbox{M}_{\mbox{{\tiny virt}}}(\mbox{B}1)|_{d}^{2}
\ + \ \mbox{I}(\mbox{R}_{i}) \ \Bigr],
\end{equation}
where the virtual correction, 
$|\mbox{M}_{\mbox{{\tiny virt}}}(\mbox{B}1)|_{d}^{2}$,
is obtained in $d$ dimensions, and summed
and averaged over the spin and color.
The I term, $\mbox{I}\,(\mbox{R}_{i})$,
is also obtained in $d$ dimensions,
and summed and averaged over the spin and color
including the spin average factor.
The output $\mbox{I} \,(\mbox{R}_{i})$
is the summation of all the created I terms
as
\begin{equation}
\mbox{I} (\mbox{R}_{i})=
\sum_{(I,K)} \mbox{I}\,(\mbox{R}_{i})_{I,\,K}\,.
\end{equation}
Once we determine the reduced Born process 
$\mbox{B}1$ and the associated set $\{y\}$,
each I term 
$\mbox{I}\,(\mbox{R}_{i})_{I,\,K}$
is specified by the information of the pair
\begin{equation}
(I,K)\,.
\label{iinfo}
\end{equation}
The concrete expression of each I term
is written in the universal form as
\begin{align}
\mbox{I}(\mbox{R}_{i})_{I,\,K}
= - A_{d}\,
\frac{1}{\mbox{T}_{\mbox{{\tiny F}}(I)}^{2}} 
\,{\cal V}_{\mbox{{\tiny F}}(I)}
\,[ \,I , K
]\,,
\end{align}
with the notation in  Eq.\,(\ref{sqbk}).
The universal singular functions are defined 
in Eqs.\,(\ref{usf1}) and (\ref{usf2}).
The virtual correction has soft and collinear 
singularities in the form $1/\ep^{2}$ and $1/\ep$, 
which are subtracted by the same poles with
opposite signs in the I term.
After the cancellation of the poles,
PS integration is carried out,
in the 4 dimensions to 
be finite, as follows\,:
\begin{equation}
\hat{\sigma}_{\mbox{{\tiny V}}}(\mbox{B}1) + 
\hat{\sigma}_{\mbox{{\tiny I}}}(\mbox{R}_{i}) = 
\frac{1}{S_{\mbox{{\tiny B1}} }} \ 
\Phi( \mbox{B}1  )_{4} \cdot 
\Bigl[ \ |\mbox{M}_{\mbox{{\tiny virt}}}(\mbox{B}1)|^{2} 
\ + \ \mbox{I}\,(\mbox{R}_{i}) \ \Bigr]_{4}\,.
\end{equation}
%
%
%
%
%
\subsection{Step\,4: P and K terms creation  \label{s2_4}}
In this section, \textquoteleft{\bfseries Step 4}. 
P($\mbox{R}_{i}$) and K($\mbox{R}_{i}$)\textquoteright \ is explained.
The input and output are written as
\begin{description}
\setlength{\itemsep}{-1mm}
\item[\ \ \ \ Input:] \ \ \ $\mbox{R}_{i}$ and 
$\mbox{B}j\,(\mbox{R}_{i})$\,,
\item[\ \ \ \ Output:] \ P($\mbox{R}_{i}$) and K($\mbox{R}_{i}$)\,.
\end{description}
The symbol $\mbox{B}j(\mbox{R}_{i})$ represents the 
reduced Born process of the category {\tt Dipole}\,$j$, 
which is made from the input process $\mbox{R}_{i}$ 
by the rules in Eq.\,(\ref{makerb}).
The creation algorithm in the DSA
is divided into the following substeps\,:
\begin{enumerate}
\setlength{\itemsep}{0mm}
\item Take set $\{x\}$ of the process $\mbox{R}_{i}$
and choose all possible pairs $(x_{a},x_{i})$
of the splittings in the order in Fig.\,\ref{fig_PKterm}
in Appendix \ref{ap_A_3}.
\item For each pair $(x_{a},x_{i})$,
take set $\{ y\}$ of the corresponding
Born process $\mbox{B}j$,
which is determined in \textquoteleft{\bfseries Step 2}. 
D($\mbox{R}_{i}$)\textquoteright.
\item If $\mbox{F}(x_{\widetilde{ai}})=\mbox{F}(y_{a})$,
create the pairs $(y_{a},y_{K})$ with $K=0,1,...,n,b$. \\
If $\mbox{F}(x_{\widetilde{ai}})=\mbox{F}(y_{b})$,
create the pairs $(y_{b},y_{K})$ with $K=0,1,...,n,a$.
\item Write down concrete expressions for all 
the P and K terms.
\end{enumerate}
Substeps 1, 2, and 3 are explained in 
Sec.\,\ref{sec241}.
Substep 4 is explained in 
Sec.\,\ref{sec242} and \ref{sec243}.
Some examples are shown in
Sec.\,\ref{sec244}.
Finally, we give a summary in
Sec.\,\ref{sec245}.
The formulae for the P and K terms are collected 
in Appendix \ref{ap_A_3}.

\subsubsection{Creation order \label{sec241}}
We take set $\{ x\}$ of the process $\mbox{R}_{i}$
in Eq.\,(\ref{xdef}). The creation order is divided
into cases with leg-$a\,(x_{a})$ and 
leg-$b\,(x_{b})$.
We start with the leg-$a$ case.
We choose the possible pairs $(x_{a},x_{i})$ 
from set $\{ x\}$
in the order of the splittings
(3),\,(4),\,(6) and (7) shown in
Fig.\,\ref{fig_PKterm}.
The possible splittings are the same as 
those chosen with leg-$a$
in \textquoteleft{\bfseries Step 2}. 
D($\mbox{R}_{i}$)\textquoteright. 
For each choice
of one pair $(x_{a},x_{i})$, we can always find 
the corresponding reduced Born process $\mbox{B}j$
and set $\{ y\}$, which have already been
fixed in {\bfseries Step 2}.
Then, for each choice of one pair $(x_{a},x_{i})$, we take the 
corresponding set $\{ y\}$ and check which of the two 
relations
$\mbox{F}(x_{\widetilde{ai}})=\mbox{F}(y_{a})$,
$\mbox{F}(x_{\widetilde{ai}})=\mbox{F}(y_{b})$
stands. Here $\mbox{F}(x_{\widetilde{ai}})$
represents the field species of the element 
$x_{\widetilde{ai}}$, defined in Eqs.\,(\ref{ifaiF})
and (\ref{iiaiF}). $\mbox{F}(y_{a/b})$ 
are defined in Eq.\,(\ref{yfield}).
Then we create the P and K terms in the following way\,:
\begin{equation}
\mbox{If} \ \ \mbox{F}(x_{\widetilde{ai}})=\mbox{F}(y_{a}) \ \
\rightarrow \ \mbox{Create pairs}:(y_{a},y_{K}) \ 
\mbox{for} \ K=0,1,2,...,n,b\,,
\end{equation}
where each pair creates each P and K terms as
\begin{align}
\left\{
 \begin{array}{l}
 \mbox{P}(\mbox{R}_{i},x_{a}:\mbox{B}j,y_{a},y_{K})  
\ \ \mbox{for} \ \ K=1,2,...,n,b\,, \\
 \mbox{K}(\mbox{R}_{i},x_{a}:\mbox{B}j,y_{a},y_{K}) 
\ \ \mbox{for} \ \ K=0,1,2,...,n,b\,,
 \end{array}
\right.
\label{pkdef}
\end{align}
or
\begin{equation}
\mbox{If} \ \ \mbox{F}(x_{\widetilde{ai}})=\mbox{F}(y_{b}) \ \
\rightarrow \ \mbox{Create pairs}:(y_{b},y_{K}) \ 
\mbox{for} \ K=0,1,2,...,n,a\,,
\end{equation}
where each pair creates the P and K terms as
\begin{align}
\left\{
 \begin{array}{l}
 \mbox{P}(\mbox{R}_{i},x_{a}:\mbox{B}j,y_{b},y_{K})  
\ \ \mbox{for} \ \ K=1,2,...,n,a\,, \\
 \mbox{K}(\mbox{R}_{i},x_{a}:\mbox{B}j,y_{b},y_{K}) 
\ \ \mbox{for} \ \ K=0,1,2,...,n,a\,.
 \end{array}
\right.
\end{align}
Among the elements $y_{K}$ with $K=1,2,...,n,a,$ 
and $b$,
only the colored fields are taken, namely,
$\mbox{F}(y_{K})=$ quark or gluon.
For convenience, we categorize the 
P and K terms by the type of second
leg $y_{K}$.
The P and K terms with $y_{K}=y_{0}$,
$y_{k}$ for $k=1,...,n$ and $y_{a/b}$,
are categorized with the labels
-0, -1, and -2, respectively.
It is noted that,
when the final state of $\mbox{R}_{i}$
includes identical fields,
only one pair of $(x_{a},x_{i})$ is taken
and the other pairs must be discarded.
For example, we take the process 
$\mbox{R}_{i}=u\bar{u} \to ggg$
and the set 
$\{ x\}=\{x_{a},x_{b}; x_{1},x_{2},x_{3} \}$. 
From the input, we can find three pairs of
splitting (3) as
\begin{equation}
(x_{a},x_{1}), \ (x_{a},x_{2}), \ \mbox{and} \ \ (x_{a},x_{3}).
\end{equation}
Among these three pairs, we can take only one pair, 
for instance, $(x_{a},x_{1})$, and must
discard the other two pairs 
$(x_{a},x_{2})$ and $(x_{a},x_{3})$.
The discard rule is in contrast to the 
{\bfseries Step 2},
where all three pairs must be taken
for the creation of the dipole terms.
The creation order with leg-$b$ is completely
analogous to the case with leg-$a$.

\subsubsection{Concrete formula for P term \label{sec242}}
The concrete formula for the P term with
leg-$a \, (x_{a})$ in the case of
$\mbox{F}(x_{\widetilde{ai}})=\mbox{F}(y_{a})$
is written
in the universal form as
\begin{align}
\mbox{P}(\mbox{R}_{i},x_{a}:\mbox{B}j,y_{a},y_{K}) 
&=\frac{\al}{2\pi} \,
\frac{1}{\mbox{T}_{\mbox{{\tiny F}}(y_{a})}^{2}} \,
P^{
\mbox{{\tiny F}}(x_{a})\mbox{{\tiny F}}(y_{a})
}(x) \,
\ln \frac{\mu_{F}^{2}}{x\,s_{x_{a} y_{{\scriptscriptstyle K}}}}
\ \cdot \nonumber \\
&\hspace{10mm}\langle \mbox{B}j=\{ y_{a},y_{b} ; y_{1}, ..., y_{n} \} 
\ | 
\mbox{T}_{y_{{\scriptscriptstyle a}}} \cdot 
\mbox{T}_{y_{{\scriptscriptstyle K}}}
| \ \mbox{B}j \rangle,
\label{pmast}
\end{align}
where the definition of the Casimir operator 
$\mbox{T}_{\mbox{{\tiny F}}(y_{a})}^{2}$
is the same as in Eq.\,(\ref{dipmas}) and
the symbol $P^{ab}(x)$ represents the {\em four-dimensional}
Altarelli-Parisi splitting function shown in 
Eqs.\,(\ref{alpff}), (\ref{alpgg}), (\ref{alpfg}), 
and (\ref{alpgf}).
The Lorentz scalar $s_{x_{a} y_{{\scriptscriptstyle K}}}$
is defined as 
$s_{x_{a} y_{{\scriptscriptstyle K}}}=2 \, p_{a} \cdot \mbox{P}(y_{K})$
with $p_{a}$ in Eq.\,(\ref{orimom}).
The square of the reduced Born amplitude with color
correlation, $\langle \mbox{B}j| 
\mbox{T}_{y_{{\scriptscriptstyle a}}} \cdot 
\mbox{T}_{y_{{\scriptscriptstyle K}}}
| \ \mbox{B}j \rangle$,
is obtained in 4 dimensions, which 
is the same function of the momenta, $\mbox{P}(\{y\})$,
in the dipole term in Eq.\,(\ref{dipnohel}),
except for the spin average factor.
In the P term,
the squared amplitude
$\langle \mbox{T} \cdot 
\mbox{T} \rangle$
is summed and averaged over
the spin and color including the spin 
average factor.
The input momenta into the Born amplitude are written as
\begin{equation}
\mbox{P}(\{ y \})=
\{\mbox{P}(y_{a}),\mbox{P}(y_{b}); \mbox{P}(y_{1}), ..., 
\mbox{P}(y_{n}) \},
\end{equation}
which are defined in the contribution to the cross section
as 
\begin{align}
\hat{\sigma}_{\mbox{{\tiny P}}}(
\mbox{R}_{i},x_{a}:\mbox{B}j,y_{a},y_{K}
) 
&= \int_{0}^{1}dx \,
\frac{1}{S_{\mbox{{\tiny B}}_{j} }}
\,\Phi_{a}(\mbox{P}(y_{a}),\mbox{P}(y_{b}) \to \mbox{P}(y_{1}), 
..., \mbox{P}(y_{n}))_{4} \ \cdot 
\nonumber \\
&\hspace{20mm} \mbox{P}(\mbox{R}_{i},x_{a}:\mbox{B}j,y_{a},y_{K}).
\end{align}
The $n$-body phase space including the flux factor is defined as
\begin{align}
\Phi_{a}(\mbox{P}(y_{a}),\mbox{P}(y_{b}) \to \mbox{P}(y_{1}), 
..., \mbox{P}(y_{n}))_{4}
&= \frac{1}{{\cal F}(\mbox{P}(y_{a}),\mbox{P}(y_{b}))} 
\nonumber \\
\prod_{i=1}^{n}
\int \frac{d^{3}\mbox{P}(y_{i})}{(2\pi)^{3}} \frac{1}{2E_{y_{i}}}
\cdot& (2\pi)^{4} \delta^{(4)} 
\Bigl( \mbox{P}(y_{a})+\mbox{P}(y_{b}) - \sum_{i=1}^{n} 
\mbox{P}(y_{i}) \Bigr)\,,
\label{pkaps}
\end{align}
with the initial momenta
$(\mbox{P}(y_{a}),\mbox{P}(y_{b}))=(xp_{a},p_{b})$
and the energy $E_{y_{i}}=\mbox{P}(y_{i})^{\mu=0}$.
The phase space is the same as in Eq.\,(\ref{pkps})
with the identification $p_{i}=\mbox{P}(y_{i})$
for $i=1,2,...$ and $n$.
The expression for the P term in the case 
$\mbox{F}(x_{\widetilde{ai}})=\mbox{F}(y_{b})$
is similarly written as
\begin{align}
\mbox{P}(\mbox{R}_{i},x_{a}:\mbox{B}j,y_{b},y_{K}) 
&=\frac{\al}{2\pi} \,
\frac{1}{\mbox{T}_{\mbox{{\tiny F}}(y_{b})}^{2}} \,
P^{
\mbox{{\tiny F}}(x_{a})\mbox{{\tiny F}}(y_{b})
}(x) \,
\ln \frac{\mu_{F}^{2}}{x\,s_{x_{a} y_{{\scriptscriptstyle K}}}}
\ \cdot \nonumber \\
&\hspace{10mm}\langle \mbox{B}j=\{ y_{a},y_{b} ; y_{1}, ..., y_{n} \} 
\ | 
\mbox{T}_{y_{{\scriptscriptstyle b}}} \cdot 
\mbox{T}_{y_{{\scriptscriptstyle K}}}
| \ \mbox{B}j \rangle\,.
\end{align}
The phase space in this case, 
$\Phi_{a}(\mbox{P}(y_{a}),\mbox{P}(y_{b}) \to \mbox{P}(y_{1}), 
..., \mbox{P}(y_{n}))_{4}$,
is the same expression as in Eq.\,(\ref{pkaps})
and the initial momenta are defined as
$(\mbox{P}(y_{a}),\mbox{P}(y_{b}))=(p_{b},xp_{a})$.
The concrete formulae for the P term
with leg-$b$, 
$\mbox{P}(\mbox{R}_{i},x_{b}:\mbox{B}j,y_{b/a},y_{K})$,  
and the phase space,
$\Phi_{b}(\mbox{P}(y_{a}),\mbox{P}(y_{b}) \to \mbox{P}(y_{1}), 
..., \mbox{P}(y_{n}))_{4}$\,,
are completely analogous to the case 
with leg-$a$.

We define the output P($\mbox{R}_{i}$)
as the set that consists of all the created P terms,
\begin{equation}
\mbox{P}(\mbox{R}_{i})=\{ \,
\mbox{P}(\mbox{R}_{i},x_{a}), \,
\mbox{P}(\mbox{R}_{i},x_{b}) \, \},
\label{outp}
\end{equation}
where the subset $\mbox{P}(\mbox{R}_{i},x_{a})$
is defined as
\begin{equation}
\mbox{P}(\mbox{R}_{i},x_{a})=\{ \,
\mbox{P}(\mbox{R}_{i},x_{a}:\mbox{B}1), \,
\mbox{P}(\mbox{R}_{i},x_{a}:\mbox{B}3), \,
\mbox{P}(\mbox{R}_{i},x_{a}:\mbox{B}4) \}\,.
\end{equation}
Each element 
$\mbox{P}(\mbox{R}_{i},x_{a}:\mbox{B}j)$
is the summation over the P terms
including $\mbox{B}j$ as
\begin{equation}
\mbox{P}(\mbox{R}_{i},x_{a}:\mbox{B}j) =
\sum_{k=1}^{n} 
\mbox{P}(\mbox{R}_{i},x_{a}:\mbox{B}j,y_{a/b},y_{k})
+ \mbox{P}(\mbox{R}_{i},x_{a}:\mbox{B}j,y_{a/b},y_{b/a})\,.
\end{equation}
In the case with leg-$b$, the set 
$\mbox{P}(\mbox{R}_{i},x_{b})$
and
the summation
$\mbox{P}(\mbox{R}_{i},x_{b}:\mbox{B}j)$
are similarly defined.

\subsubsection{Concrete formula for K term \label{sec243}}
The concrete formula for the K term with 
leg-$a \, (x_{a})$, in the case of
$\mbox{F}(x_{\widetilde{ai}})=\mbox{F}(y_{a})$,
is separated into the three categories, -0, -1, and -2,
introduced above.
The formulae are written as
\begin{align}
\mbox{K}(\mbox{R}_{i},x_{a}:\mbox{B}j,y_{a},y_{0}) 
&=\frac{\al}{2\pi} \,
\bar{K}^{
\mbox{{\tiny F}}(x_{a})\mbox{{\tiny F}}(y_{a})
}(x) \ \cdot \langle \mbox{B}j \,|  \mbox{B}j \rangle, \\
\mbox{K}(\mbox{R}_{i},x_{a}:\mbox{B}1,y_{a},y_{k}) 
&=\frac{\al}{2\pi} \,
\frac{\gamma_{\mbox{{\tiny F}}(y_{k})}}
{\mbox{T}_{\mbox{{\tiny F}}(y_{k})}^{2}} \, h(x)
\ \cdot \langle \mbox{B}1 | 
\mbox{T}_{y_{{\scriptscriptstyle a}}} \cdot 
\mbox{T}_{y_{{\scriptscriptstyle k}}}
| \ \mbox{B}1 \rangle, 
\label{ktif} \\
\mbox{K}(\mbox{R}_{i},x_{a}:\mbox{B}j,y_{a},y_{b}) 
&=-\frac{\al}{2\pi} \,
\frac{1}{\mbox{T}_{\mbox{{\tiny F}}(y_{a})}^{2}} \, 
\widetilde{K}^{
\mbox{{\tiny F}}(x_{a})\mbox{{\tiny F}}(y_{a})
}(x) \ \cdot
\langle \mbox{B}j | 
\mbox{T}_{y_{{\scriptscriptstyle a}}} \cdot 
\mbox{T}_{y_{{\scriptscriptstyle b}}}
| \ \mbox{B}j \rangle\,.
\end{align}
The symbol
$\langle \mbox{B}j \,|  \mbox{B}j \rangle$
is the abbreviation of 
$\langle \mbox{B}j=\{ y_{a},y_{b} ; y_{1}, ..., y_{n} \} 
\ | \ \mbox{B}j \rangle$,
which is the usual squared amplitudes in the LO process.
The symbol
$\langle \mbox{B}j | 
\mbox{T}_{y_{{\scriptscriptstyle a}}} \cdot 
\mbox{T}_{y_{{\scriptscriptstyle K}}}
| \ \mbox{B}j \rangle$
is the abbreviation of
$\langle \mbox{B}j=\{ y_{a},y_{b} ; y_{1}, ..., y_{n} \} 
\ | 
\mbox{T}_{y_{{\scriptscriptstyle a}}} \cdot 
\mbox{T}_{y_{{\scriptscriptstyle K}}}
| \ \mbox{B}j \rangle$,
which is the same quantity as in 
Eq.\,(\ref{pmast}).
The functions of the argument $x$,
$\bar{K}^{
\mbox{{\tiny F}}(x_{a})\mbox{{\tiny F}}(y_{a})
}(x)$,
$h(x)$, and
$\widetilde{K}^{
\mbox{{\tiny F}}(x_{a})\mbox{{\tiny F}}(y_{a})
}(x)$, and
the symbol $\gamma_{\mbox{{\tiny F}}(y_{k})}$
are defined in Appendix \ref{ap_A_3}.
It is noted that the K terms with pair $(y_{a},y_{k})$
with $k=1,2,...,$ and $n$ exist only 
in the case of diagonal splittings, 
namely, the case including process B1
shown in Eq.\,(\ref{ktif}).

In the same way as the P term, the input momenta into the Born amplitude,
$\mbox{P}(\{ y \})$, are given in the contribution to the 
cross section as 
\begin{align}
\hat{\sigma}_{\mbox{{\tiny K}}}(
\mbox{R}_{i},x_{a}:\mbox{B}j,y_{a},y_{K}
) 
&= \int_{0}^{1}dx \,
\frac{1}{S_{\mbox{{\tiny B}}_{j} }}
\,\Phi_{a}(\mbox{P}(y_{a}),\mbox{P}(y_{b}) \to \mbox{P}(y_{1}), 
..., \mbox{P}(y_{n}))_{4} \ \cdot 
\nonumber \\
&\hspace{20mm} \mbox{K}(\mbox{R}_{i},x_{a}:\mbox{B}j,y_{a},y_{K}),
\end{align}
where the $n$-body PS, $\Phi_{a}$, is the same 
as in Eq.\,(\ref{pkaps}).
The formulae in the case of
$\mbox{F}(x_{\widetilde{ai}})=\mbox{F}(y_{b})$
are similarly given as
\begin{align}
\mbox{K}(\mbox{R}_{i},x_{a}:\mbox{B}j,y_{b},y_{0}) 
&=\frac{\al}{2\pi} \,
\bar{K}^{
\mbox{{\tiny F}}(x_{a})\mbox{{\tiny F}}(y_{b})
}(x) \ \cdot \langle \mbox{B}j \,|  \mbox{B}j \rangle, \\
\mbox{K}(\mbox{R}_{i},x_{a}:\mbox{B}1,y_{b},y_{k}) 
&=\frac{\al}{2\pi} \,
\frac{\gamma_{\mbox{{\tiny F}}(y_{k})}}
{\mbox{T}_{\mbox{{\tiny F}}(y_{k})}^{2}} \, h(x)
\ \cdot \langle \mbox{B}1 | 
\mbox{T}_{y_{{\scriptscriptstyle b}}} \cdot 
\mbox{T}_{y_{{\scriptscriptstyle k}}}
| \ \mbox{B}1 \rangle, \\
\mbox{K}(\mbox{R}_{i},x_{a}:\mbox{B}j,y_{b},y_{a}) 
&=-\frac{\al}{2\pi} \,
\frac{1}{\mbox{T}_{\mbox{{\tiny F}}(y_{b})}^{2}} \, 
\widetilde{K}^{
\mbox{{\tiny F}}(x_{a})\mbox{{\tiny F}}(y_{b})
}(x) \ \cdot
\langle \mbox{B}j | 
\mbox{T}_{y_{{\scriptscriptstyle b}}} \cdot 
\mbox{T}_{y_{{\scriptscriptstyle a}}}
| \ \mbox{B}j \rangle.
\end{align}
The formulae for the K term
with leg-$b$, 
$\mbox{K}(\mbox{R}_{i},x_{b}:\mbox{B}j,y_{b/a},y_{K})$,  
and the phase space
$\Phi_{b}$
are completely analogous to the leg-$a$ case.

Again, similar to the case of the P term
we define the output K($\mbox{R}_{i}$)
as the set that consists of all the created K terms,
\begin{equation}
\mbox{K}(\mbox{R}_{i})=\{ \,
\mbox{K}(\mbox{R}_{i},x_{a}), \,
\mbox{K}(\mbox{R}_{i},x_{b}) \, \},
\label{outk}
\end{equation}
where the subset $\mbox{K}(\mbox{R}_{i},x_{a})$
is defined as
\begin{equation}
\mbox{K}(\mbox{R}_{i},x_{a})=\{ \,
\mbox{K}(\mbox{R}_{i},x_{a}:\mbox{B}1), \,
\mbox{K}(\mbox{R}_{i},x_{a}:\mbox{B}3), \,
\mbox{K}(\mbox{R}_{i},x_{a}:\mbox{B}4) \}\,,
\end{equation}
where the elements
$\mbox{K}(\mbox{R}_{i},x_{a}:\mbox{B}j)$
are the summation of the K terms with
the process B$j$ as
\begin{align}
\mbox{K}(\mbox{R}_{i},x_{a}:\mbox{B}j) &
=
\sum_{K=0}^{n} 
\mbox{K}(\mbox{R}_{i},x_{a}:\mbox{B}j,y_{a/b},y_{K}) 
 \ \ + \mbox{K}(\mbox{R}_{i},x_{a}:\mbox{B}j,y_{a/b},y_{b/a})\,.
\end{align}
The set 
$\mbox{K}(\mbox{R}_{i},x_{b})$ and
the summation
$\mbox{K}(\mbox{R}_{i},x_{b}:\mbox{B}j)$
are similarly defined.

\subsubsection{Examples \label{sec244}}
To demonstrate the creation of 
the P and K terms,
we take the same input process used for 
the dipole term creation in Eq.\,(\ref{dipexin}) as
\begin{equation}
\mbox{Input}: \ \mbox{R}_{1} = u\bar{u} \to u\bar{u}g\,,
\label{pkr1}
\end{equation}
which defines set $\{ x \}$ with 
the field species and
the momenta 
in Eqs.\,(\ref{dipex1x}),\,(\ref{dipex1f}), and
(\ref{dipex1m}) as
\begin{align}
\{x\} &= \{x_{a},x_{b};x_{1},x_{2},x_{3}  \}\,, 
\\
\mbox{F}(\{x\}) &= \{ u,\bar{u}; u,\bar{u}, g \}\,,
\\
\mbox{Momenta}: & \ \ \ \ \{ p_{a},p_{b};p_{1},p_{2},p_{3} \}\,.
\end{align}
The possible reduced Born processes $\mbox{B}j$ 
and the associated set $\{ y\}=\{y_{a},y_{b};y_{1},y_{2}\}$
are specified in {\bfseries Step 2} as
\begin{align}
\mbox{B}1  &: \, \mbox{F}(\{ y\})= \{u,\bar{u}; u,\bar{u} \}\,,
 \\
\mbox{B}3u &: \, \mbox{F}(\{ y\})= \{g,\bar{u}; \bar{u},g \}\,,
 \\
\mbox{B}3\bar{u} &: \, \mbox{F}(\{ y\})= \{u,g; u,g \}\,,
\end{align}
where B1 and B3$u$ are explicitly 
defined in Eqs.\,(\ref{dipex132}) and (\ref{dipexa1b}),
respectively. 
Then we start the creation of the P and K terms with
leg-$a\, (x_{a})$ as
\begin{description}
\item ${\tt Dipole1}$ (3): $\mbox{B}1=
\{u,\bar{u}; u,\bar{u} \}$ \\
$(x_{a},x_{3}) \to \mbox{F}(x_{\widetilde{a3}})=u=\mbox{F}(y_{a})
\ \to \ 1.(y_{a},y_{0}), 2.(y_{a},y_{1}), 
3.(y_{a},y_{2}), 4.(y_{a},y_{b})$\,,
\item ${\tt Dipole3u}$ (6): $\mbox{B}3u=
\{g,\bar{u}; \bar{u},g\}$ \\
$(x_{a},x_{1}) \to \mbox{F}(x_{\widetilde{a1}})=g=\mbox{F}(y_{a})
\ \to \ 5.(y_{a},y_{0}), 6.(y_{a},y_{1}), 
7.(y_{a},y_{2}), 8.(y_{a},y_{b})$\,.
\end{description}
Next we proceed to the creation with leg-$b$ as
\begin{description}
\item ${\tt Dipole1}$ (3): $\mbox{B}1=
\{u,\bar{u}; u,\bar{u} \}$ \\
$(x_{b},x_{3}) \to \mbox{F}(x_{\widetilde{b3}})=\bar{u}=\mbox{F}(y_{b})
\ \to \ 9.(y_{b},y_{0}), 10.(y_{b},y_{1}), 
11.(y_{b},y_{2}), 12.(y_{b},y_{a})$\,,
\item ${\tt Dipole3\bar{u}}$ (6): $\mbox{B}3\bar{u}=
\{ u,g; u,g \}$ \\
$(x_{b},x_{2}) \to \mbox{F}(x_{\widetilde{b2}})=g=\mbox{F}(y_{b})
\ \to \ 13.(y_{b},y_{0}), 14.(y_{b},y_{1}), 
15.(y_{b},y_{2}), 16.(y_{b},y_{a})$\,.
\end{description}
Sixteen pairs are created; each pair corresponds to
one P and one K term. Two kinds of exceptions
have already been noted above.
The first exception is the pairs of type
$(y_{a/b},y_{0})$, in the present example, 
1,\,5,\,9,\, and 13,
which produce only a K term. 
The second exception is the following.
As noted, the K terms with the pair $(y_{a/b},y_{k})$
with $k=1,2,...,$ and $n$ exist only 
for diagonal splittings. Then the K terms with the pairs
with nondiagonal splittings,
pairs, 6,\,7,\,14,\,and 15, do not exist.
We select the three pairs, 
$1.(y_{a},y_{0})$, $2.(y_{a},y_{1})$ and $16.(y_{b},y_{a})$,
for instance,
and show their concrete expressions.
\begin{itemize}
\item Example: $(x_{a},x_{3})$, $\mbox{B}1=\{u,\bar{u}; u,\bar{u} \}$, 
$1.(y_{a},y_{0})$ 
\begin{equation}
\mbox{K}(\mbox{R}_{1},x_{a}:\mbox{B}1,y_{a},y_{0}) 
=\frac{\al}{2\pi} \,
\bar{K}^{uu}(x) \ \cdot \langle \mbox{B}1 \,|  \mbox{B}1 \rangle \,,
\end{equation}
where the concrete expression of $\bar{K}^{uu}(x)=\bar{K}^{ff}(x)$,
is given in Eq.\,(\ref{apK13b})
\footnote{
The P and K terms with the Born process B1
include the \textquoteleft +\textquoteright-distribution like
$(1/(1-x))_{+}$.
This is defined in Ref.\cite{Catani:1996vz}
and also in Eq.\,(5.16) in the present paper.
When we use the definition directly for the Monte Carlo
integration, the multiplied matrix element 
$\langle \mbox{B}1 \,|  \mbox{B}1 \rangle$
must be evaluated twice at points $x$ and $x=1$.
An excellent technique to avoid this
might be available, where the matrix element
is evaluated only once at $x=1$.
The technique is briefly explained in Sec.4
of Ref.\cite{Frederix:2010cj}. 
We suggest that interested readers
consult that reference.
}.
\item Example: $(x_{a},x_{3})$, $\mbox{B}1=\{u,\bar{u}; u,\bar{u} \}$, 
$2.(y_{a},y_{1})$ 
\begin{align}
\mbox{P}(\mbox{R}_{1},x_{a}:\mbox{B}1,y_{a},y_{1}) 
&=\frac{\al}{2\pi} \,
\frac{1}{\mbox{C}_{\mbox{{\tiny F}}}} \,
P^{uu}(x) \,
\ln \frac{\mu_{F}^{2}}{x\,s_{x_{a} y_{1}}}
\ \cdot \langle \mbox{B}1 \ | 
\mbox{T}_{y_{{\scriptscriptstyle a}}} \cdot 
\mbox{T}_{y_{{\scriptscriptstyle 1}}}
| \ \mbox{B}1 \rangle, \\
\mbox{K}(\mbox{R}_{1},x_{a}:\mbox{B}1,y_{a},y_{1}) 
&=\frac{\al}{2\pi} \,
\frac{\gamma_{u}}
{\mbox{C}_{\mbox{{\tiny F}}}} \, h(x)
\ \cdot \langle \mbox{B}1 | 
\mbox{T}_{y_{{\scriptscriptstyle a}}} \cdot 
\mbox{T}_{y_{1}}
| \ \mbox{B}1 \rangle\,,
\end{align}
where the quantities $P^{uu}(x)=P^{ff}(x)$, 
$\gamma_{u}=\gamma_{f}$,
and $h(x)$ are defined in Eqs.\,(\ref{alpff}),\,(\ref{gamf}), 
and (\ref{hxd}), respectively.
The Lorentz scalar $s_{x_{a} y_{1}}$ 
is defined as 
$s_{x_{a} y_{1}}=2 p_{a} \cdot \mbox{P}(y_{1})$.
\item Example: $(x_{b},x_{2})$, $\mbox{B}3\bar{u}=\{u,g; u,g \}$, 
$16.(y_{b},y_{a})$ 
\begin{align}
\mbox{P}(\mbox{R}_{1},x_{b}:\mbox{B}3\bar{u},y_{b},y_{a}) 
&=\frac{\al}{2\pi} \,
\frac{1}{\mbox{C}_{\mbox{{\tiny A}}}} \,
P^{\bar{u}g}(x) \,
\ln \frac{\mu_{F}^{2}}{x\,s_{x_{b} y_{a}}}
\ \cdot \langle \mbox{B}3\bar{u} \ | 
\mbox{T}_{y_{b}} \cdot 
\mbox{T}_{y_{a}}
| \ \mbox{B}3\bar{u} \rangle, 
\label{pcom1}\\
\mbox{K}(\mbox{R}_{1},x_{b}:\mbox{B}3\bar{u},y_{b},y_{a}) 
&=-\frac{\al}{2\pi} \,
\frac{1}{\mbox{C}_{\mbox{{\tiny A}}}} \, 
\widetilde{K}^{\bar{u}g}(x) \ \cdot
\langle \mbox{B}3\bar{u} | 
\mbox{T}_{y_{{\scriptscriptstyle b}}} \cdot 
\mbox{T}_{y_{{\scriptscriptstyle a}}}
| \ \mbox{B}3\bar{u} \rangle.
\label{kcom1}
\end{align}
where the functions
$P^{\bar{u}g}(x)=P^{fg}(x)$ and 
$\widetilde{K}^{\bar{u}g}(x)=\widetilde{K}^{fg}(x)$
are given in Eqs.\,(\ref{alpfg}) and (\ref{apKtil36}).
The Lorentz scalar $s_{x_{b} y_{a}}$ is defined as
$s_{x_{b} y_{a}}=2 p_{b} \cdot \mbox{P}(y_{a})=2 p_{b} \cdot p_{a}$.
\end{itemize}

\subsubsection{Summary \label{sec245}}
The contributions of the P and K terms to the hadronic
cross section are written as 
\begin{align}
\sigma_{\mbox{{\tiny P}}}(\mbox{R}_{i}) +
\sigma_{\mbox{{\tiny K}}}(\mbox{R}_{i})
&= \int dx_{1} \int dx_{2} \ 
f_{\mbox{{\tiny F}}(x_{a})}(x_{1}) 
f_{\mbox{{\tiny F}}(x_{b})}(x_{2}) \ \biggl(
\hat{\sigma}_{\mbox{{\tiny P}}}(\mbox{R}_{i}) + 
\hat{\sigma}_{\mbox{{\tiny K}}}(\mbox{R}_{i}) 
\biggr)\,.
\end{align}
The partonic cross sections are written as 
\begin{align}
\hat{\sigma}_{\mbox{{\tiny P}}}(\mbox{R}_{i}) 
+
\hat{\sigma}_{\mbox{{\tiny K}}}(\mbox{R}_{i}) 
&= \int_{0}^{1}dx \sum_{\mbox{{\tiny B}}_{j}} 
\frac{1}{S_{\mbox{{\tiny B}}_{j} }}
\,\Phi_{a}(\mbox{R}_{i}:\mbox{B}_{j},x)_{4} \ \cdot \nonumber \\
&\biggl(\mbox{P}(\mbox{R}_{i},x_{a}:\mbox{B}_{j}) \ + \
\mbox{K}(\mbox{R}_{i},x_{a}:\mbox{B}_{j})
\biggr) \ + \
(a \leftrightarrow b)\,,
\label{hatpk2}
\end{align} 
where the PS, $\Phi_{a}$, is defined in Eq.\,(\ref{pkaps}).
The PS integrations of the P and K terms
are separately finite in 4 dimensions.
The outputs, $\mbox{P}(\mbox{R}_{i})$
and $\mbox{K}(\mbox{R}_{i})$, are the sets
defined in Eqs.\,(\ref{outp}) and (\ref{outk}),
respectively.
Once we select an input process $\mbox{R}_{i}$,
a leg-$a$ or -$b$, and 
a reduced Born process, $\mbox{B}j$,
each P and K term,
$\mbox{P/K}(\mbox{R}_{i},x_{a/b}:\mbox{B}j,y_{a/b},y_{K})$,
is specified by information on the pair 
$(y_{a/b},y_{K})$,
which is abbreviated as
\begin{equation}
(a/b,K)\,.
\label{pkinfo}
\end{equation}
Concrete formulae for the P and K terms
are collected in Appendix \ref{ap_A_3}.
%
%
%
%
%
\subsection{Advantages of the DSA \label{s2_5}}
In the present section, the advantages of the DSA 
are clarified. 
For this purpose, we first point out the special 
features of the DSA presented in
Sec.\,\ref{s2_1},\ \ref{s2_2},\ \ref{s2_3},\, and 
\ref{s2_4}. The master formula of the DSA is
shown in Eq.\,(\ref{master}) as
\begin{align}
\sigma(\mbox{R}_{i}) 
&= \int dx_{1} \int dx_{2} \ 
f_{\mbox{{\tiny F}}(x_{a})}(x_{1}) 
f_{\mbox{{\tiny F}}(x_{b})}(x_{2}) \ \times  \nonumber \\
& \ \ \biggl[
\bigl(\hat{\sigma}_{\mbox{{\tiny R}}}(\mbox{R}_{i}) - 
\hat{\sigma}_{\mbox{{\tiny D}}}(\mbox{R}_{i}) \bigr) + 
\bigl(\hat{\sigma}_{\mbox{{\tiny V}}}(\mbox{B}1(\mbox{R}_{i})) + 
\hat{\sigma}_{\mbox{{\tiny I}}}(\mbox{R}_{i}) \bigr) + 
\hat{\sigma}_{\mbox{{\tiny P}}}(\mbox{R}_{i}) + 
\hat{\sigma}_{\mbox{{\tiny K}}}(\mbox{R}_{i}) \biggr]\,.
\end{align}
This formula shows that the real correction
$\hat{\sigma}_{\mbox{{\tiny R}}}(\mbox{R}_{i})$,
the virtual correction
$\hat{\sigma}_{\mbox{{\tiny V}}}(\mbox{B}1(\mbox{R}_{i}))$,
and all the subtraction terms
$\hat{\sigma}_{\mbox{{\tiny D}}}(\mbox{R}_{i}),
\hat{\sigma}_{\mbox{{\tiny I}}}(\mbox{R}_{i}),
\hat{\sigma}_{\mbox{{\tiny P}}}(\mbox{R}_{i}),$ 
and
$\hat{\sigma}_{\mbox{{\tiny K}}}(\mbox{R}_{i})$,
which are created from one input process
$\mbox{R}_{i}$\,,
have the same initial parton states, 
$\mbox{F}(x_{a})$ and $\mbox{F}(x_{b})$.
They are all multiplied by the same PDFs, 
$f_{\mbox{{\tiny F}}(x_{a})}(x_{1}) 
\, f_{\mbox{{\tiny F}}(x_{b})}(x_{2})$.
In other words, the subtraction terms
are sorted by the initial-state partons.
This is the first feature. 
The second feature is that the subtraction 
terms are also sorted by the reduced Born processes.
As defined in the previous sections,
the creation order of
the D, I, P, and K terms is sorted
by the kind of the splittings
and the reduced Born processes B$j$
with $j=1,2,3,$ and $4$, where 
the processes B2, B3, and B4, may have 
subcategories for the quark flavors.
The third feature is that 
we introduce sets $\{x\}$,\,$\{\tilde{x}\}$, and $\{y\}$
and the field mapping $y=f(\tilde{x})$.
Using the sets and the mapping, each subtraction
term is specified in a well defined compact form.

The three features of the DSA mentioned above
lead to the following three advantages of the DSA:
\begin{enumerate}
\item Consistency proof of the subtraction terms,
\item Easy construction of the codes for the Monte Carlo 
integration,
\item Compact form of the subtraction terms in
the summary tables.
\end{enumerate}
We start by explaining the first advantage.
By the construction of the dipole subtraction procedure
the summation of all the introduced subtraction terms must 
vanish as in Eq.\,(\ref{subtmast}),
\begin{equation}
\sum_{\mbox{{\tiny R}}_{i}} \hat{\sigma}_{\mbox{{\tiny subt}}}(\mbox{R}_{i})
=
\sum_{\mbox{{\tiny R}}_{i}} 
\bigl[
\hat{\sigma}_{\mbox{{\tiny D}}}(\mbox{R}_{i}) + 
\hat{\sigma}_{\mbox{{\tiny C}}}(\mbox{R}_{i})
- \hat{\sigma}_{\mbox{{\tiny I}}}(\mbox{R}_{i}) - 
\hat{\sigma}_{\mbox{{\tiny P}}}(\mbox{R}_{i}) 
- \hat{\sigma}_{\mbox{{\tiny K}}}(\mbox{R}_{i})
\bigr]
=0\,,
\label{subtmast2}
\end{equation}
which we call the consistency relation of
the subtraction terms.
The first advantage is that a straightforward proof
of the consistency relation
in Eq.\,(\ref{subtmast2}) is possible.
The cancellation among the subtraction terms
can be realized among subtraction terms
with the same initial parton states
and the same reduced Born processes.
According to the first two features of the DSA,
we can systematically identify the categories of 
the subtraction terms that cancel each other. 
Then a systematic proof of the consistency relation
becomes possible.
We have succeeded in constructing
a straightforward proof algorithm (PRA),
which is presented in the following article
\cite{Hasegawa:2014nna}.
The second advantage is the following.
In order to construct the computer codes
for the Monte Carlo integration, 
we must collect subtraction terms
with the same initial parton 
states to be multiplied by the same PDFs.
Such collection is realized in the DSA
thanks to the first feature that 
the created subtraction terms are sorted
by the initial parton states.
The third advantage is the following.
According to the third feature, we can specify all the
subtraction terms in a compact form.
For example, the compact forms for
the D, I, P, and K terms are shown in
Eqs.\,(\ref{dipcom1}),\,(\ref{icom1}),\,(\ref{pcom1})\,,
and (\ref{kcom1}), respectively.
Furthermore, on the fixed reduced Born processes,
the minimal information to specify
the subtraction terms is defined 
for the D, I, and P/K terms 
in Eqs.\,(\ref{dinfo}),\,(\ref{iinfo}),\,
and (\ref{pkinfo}), respectively.
With agreement on the form of expression,
everyone can understand the
summary tables of all the created
subtraction terms written in 
a template form.
Summary tables for 
the Drell--Yan and the dijet processes 
are explicitly shown in Sec.\,\ref{sec_3} 
and Appendix \ref{ap_B},
respectively.

Finally, we compare the DSA against the 
algorithm implemented in the AutoDipole
package, because the comparison makes the
advantages of the DSA clearer. The creation 
algorithm of the D and I terms 
in AutoDipole is essentially
the same as the DSA. The creation algorithm
of the P and K terms is different from the DSA.
In order to demonstrate the difference,
we use the same example process as
in Eq.\,(\ref{pkr1}),
$\mbox{R}_{1} = u\bar{u} \to u\bar{u}g$.
The creation algorithm of the P and K terms
in AutoDipole takes only the reduced Born
process, 
$\mbox{B}1=\mbox{R}_{1}-g_{f}=u\bar{u} \to u\bar{u}$,
as the input. Then the P and K terms are
created by adding to process B1 
the possible splittings in
Fig.\,\ref{fig_PKterm}.
Splitting (3) can be added to $y_{a}$
of B1 and the elements $y_{K}$ are chosen.
The choice creates the P and K terms as
$\mbox{P/K}(\mbox{R}_{1},x_{a}:\mbox{B}1,y_{a},y_{K})$,
which are the same as the DSA.
As the next choice, splitting (7) can be 
added and $y_{K}$ are chosen.
The choice creates the P and K terms,
which are written in the notation defined 
in the DSA in Eq.\,(\ref{pkdef}) as
\begin{equation}
\mbox{P/K}(\mbox{R}_{i}=ug\to u\bar{u}u,x_{a}:
\mbox{B}4u,y_{a},y_{K}).
\end{equation}
As the notation of the DSA shows,
these P and K terms are created 
from the input $\mbox{R}_{i}=ug\to u\bar{u}u$, 
when splitting (7) is applied.
In this way, the creation places
of the P and K terms
with the nondiagonal
splittings (6) and (7) are different between 
the AutoDipole algorithm and the DSA. 
The advantage of the AutoDipole algorithm 
is that all the P and K terms include
only one kind of reduced Born process B1.
In this sense the collected expressions
of the P and K terms, which are created from
the input $\mbox{R}_{i}$,
are simpler than the case of the DSA. 
The disadvantage of the AutoDipole algorithm
is that 
the P and K terms with the
nondiagonal splittings, created by the AutoDipole
algorithm, have different initial parton states
from the other subtraction terms.
The involvement of different initial states
spoils the first feature of the DSA and hence
causes the loss of the the first and second advantages 
of the DSA.
Namely, in the AutoDipole algorithm, the proof of the 
consistency relation becomes more complex, and 
re-collection of the P and K terms 
with nondiagonal splittings
for the Monte Carlo integration 
is required as extra work for the users.
The third advantage of the DSA,
the expressions and the summary tables in a compact form,
also holds for the AutoDipole
algorithm, because the subtraction terms
are also sorted by the reduced Born processes
in the AutoDipole algorithm.








\newpage
\section{Drell--Yan\,: $pp \to \mu^{+}\mu^{-} + X$ \label{sec_3}}
In the present section we apply the DSA to the Drell--Yan process.
The five steps in Eq.\,(\ref{dsastep}) are
executed in Sec.\,\ref{s3_1}, \ref{s3_2}, \ref{s3_3}, 
\ref{s3_4}, and \ref{s3_5}, respectively.
%
%
%
\subsection{List of $\mbox{R}_{i}$ \label{s3_1}}
In {\bfseries Step 1} we make a list of the contributing
real emission processes $\{\mbox{R}_{i}\}$ as follows\,:
\begin{align}
\mbox{R}_{1} &= u\bar{u} \to \mu^{-}\mu^{+}g\,, \nonumber\\
\mbox{R}_{2} &= ug \to \mu^{-}\mu^{+}u\,,  \nonumber\\
\mbox{R}_{3} &= \bar{u}g \to \mu^{-}\mu^{+}\bar{u}\,.
\label{listdy}
\end{align}
There are three independent processes, which sets
$n_{\mbox{{\tiny real}}}=3$\,.
The number of final states is three, which sets
$(n+1)=3$.
In order to exhaust
all independent partonic processes
in the Drell--Yan event, 
it is sufficient to
take into account one quark flavor, $u$,
for instance. 
%
%
%
\subsection{D term \label{s3_2}}
In {\bfseries Step 2}, we create the dipole terms 
D($\mbox{R}_{i}$) from the inputs
$\{\mbox{R}_{1},\mbox{R}_{2},\mbox{R}_{3}\}$
in Eq.\,(\ref{listdy}).

\subsubsection*{D($\mbox{R}_{1}$) creation}
The input process $\mbox{R}_{1} = u\bar{u} \to \mu^{-}\mu^{+}g$
determines set $\{ x \}$ with the field 
species and the momenta as
\begin{align}
\{x\} &= \{x_{a},x_{b};x_{1},x_{2},x_{3}  \}\,,
\label{dy1dipx}\\
\mbox{F}(\{x\}) &= \{ u,\bar{u}\,; \mu^{-}, \mu^{+}, g \}\,,
\label{dy1dipf} \\
\mbox{Momenta}&: \ \, \{ p_{a},p_{b}\,;p_{1},p_{2},p_{3} \}\,. 
\label{dy1dipm}
\end{align}
We create the dipole terms in the order shown
in Fig.\,1 as
\begin{equation}
{\tt Dipole\,1} \, (3)\,\mbox{-}2: \ \ 1.(a3,b),\,2.(b3,a)\,.
\end{equation}
Only two dipole terms are created.
The reduced Born process of the category
{\tt Dipole\,1} is fixed as
$\mbox{B}1(\mbox{R}_{1})=u\bar{u} \to \mu^{-}\mu^{+}$\,,
which determines set $\{y\}$ with the field species 
and the momenta as
\begin{align}
\{ y \} 
&= \{y_{a},y_{b};y_{1}, y_{2} \}\,, 
\label{dy1y} \\
\mbox{F}(\{ y \}) 
&= \{ u,\bar{u}\,; \mu^{-}, \mu^{+}\}\,,  
\label{dy1yF} \\
\mbox{P}(\{ y \}) 
&= \{\mbox{P}(y_{a}),\mbox{P}(y_{b})\,; \mbox{P}(y_{1}), 
\mbox{P}(y_{2}) \}\,.  
\label{dy1yP}
\end{align}
Then we specify the field mapping for 
each dipole term and write down the
concrete expression.
\begin{description}
\item[ 1.\,(a3,b)]
Set $\{\tilde{x}\}$ is defined with 
the field species and
the momenta as
\begin{align}
\{\tilde{x}\} 
&=\{ \tilde{x}_{a},\tilde{x}_{b}\,;\tilde{x}_{1}, \tilde{x}_{2} \} 
\nonumber\\
&= \{x_{\widetilde{a3}},x_{\widetilde{b}} \,
;x_{1},x_{2} \}\,, \\
\mbox{F}(\{\tilde{x}\}) 
&= \{ u,\bar{u} \,; \mu^{-}, \mu^{+}\}\,,  \\
\mbox{P}(\{\tilde{x}\}) 
& =\{ \tilde{p}_{a3},p_{b};\tilde{k}_{1}, \tilde{k}_{2} \}.
\end{align}
where the reduced momenta, $\tilde{p}_{a3}$ and 
$\tilde{k}_{1/2}$, are defined 
in Eqs.\,(\ref{rmiiem}) and (\ref{rmiisp}). 
Then we construct the field mapping as
\begin{equation}
f(\{\tilde{x}\})=f(x_{\widetilde{a3}},x_{\widetilde{b}}\,;
x_{1},x_{2})
=(y_{a},y_{b}\,;y_{1},y_{2}),
\end{equation}
which is interpreted as the identification 
of the elements as 
\begin{equation}
(y_{a},y_{b};y_{1}, y_{2})=
(x_{\widetilde{a3}},x_{\widetilde{b}};x_{1},x_{2})\,.
\end{equation}
The expression is abbreviated as 
$(\widetilde{a3},\widetilde{b} \ ;1,2)$.
The field mapping determines
the momenta as
\begin{align}
\mbox{P}(\{ y \}) 
&= \{\mbox{P}(x_{\widetilde{a3}}),\mbox{P}(x_{\widetilde{b}})\,; 
\mbox{P}(x_{1}), \mbox{P}(x_{2}) \}\,, 
\nonumber\\
&= \{ \tilde{p}_{a3},p_{b}\,;\tilde{k}_{1}, \tilde{k}_{2} \}\,.
\end{align}
The dipole term is written in Eq.\,(\ref{dp132}) as
\begin{align}
\mbox{D}( {\tt dip}1,\,(3)\mbox{-}2)_{a3,b} 
= -\frac{1}{s_{a3}} 
\frac{1}{x_{3,ab}}
\frac{1}{\mbox{C}_{\mbox{{\tiny F}}}} 
\mbox{V}_{a3,b} \
\langle \mbox{B}1  \ | 
\mbox{T}_{y_{a}} \cdot \mbox{T}_{y_{b}}
 | \ \mbox{B}1 \rangle\,, 
\label{dyr1dip1}
\end{align}
where
the dipole splitting function, $\mbox{V}_{a3,b}$,
and the Lorentz scalar, $x_{3,ab}$, are defined in 
Eqs.\,(\ref{ds132}) and (\ref{iixiab}).

\item[2.\,(b3,a)]
Set $\{\tilde{x}\}$ is defined as
\begin{align}
\{\tilde{x}\} 
&= \{x_{\widetilde{a}},x_{\widetilde{b3}}, \,
;x_{1},x_{2} \}\,, \\
\mbox{F}(\{\tilde{x}\}) 
&= \{ u,\bar{u} \,;\mu^{-}, \mu^{+} \}\,,  \\
\mbox{P}(\{\tilde{x}\}) 
& =\{ p_{a},\tilde{p}_{b3}\,;\tilde{k}_{1}, \tilde{k}_{2} \}\,.
\end{align}
Set $\{y \}$ is fixed in Eq.\,(\ref{dy1y})
and the field mapping is found as 
\begin{equation}
(y_{a},y_{b};y_{1}, y_{2})=
(x_{\widetilde{a}},x_{\widetilde{b3}};x_{1},x_{2})\,,
\end{equation}
which is abbreviated as 
$(\widetilde{a},\widetilde{b3} \ ;1,2)$.
The momenta are determined as
\begin{equation}
\mbox{P}(\{ y \}) = 
\{ p_{a}, \tilde{p}_{b3}\,;\tilde{k}_{1}, \tilde{k}_{2} \}\,.
\end{equation}
The dipole term is written as
\begin{equation}
\mbox{D}( {\tt dip}1,\,(3)\mbox{-}2)_{b3,a} 
= -\frac{1}{s_{b3}} 
\frac{1}{x_{3,ba}}
\frac{1}{\mbox{C}_{\mbox{{\tiny F}}}} 
\mbox{V}_{b3,a} \
\langle \mbox{B}1  \ | 
\mbox{T}_{y_{b}} \cdot \mbox{T}_{y_{a}}
 | \ \mbox{B}1 \rangle\,.
\label{dyr1dip2}
\end{equation}
\end{description}
The output D($\mbox{R}_{1}$) is written as
\begin{align}
\mbox{D} (\mbox{R}_{1})
&=\mbox{D} (\mbox{R}_{1},\,{\tt dip1}) \nonumber\\
&=\mbox{D} (\mbox{R}_{1},\,{\tt dip1},\,(3)\mbox{-}2)_{a3,b} 
+
\mbox{D} (\mbox{R}_{1},\, {\tt dip1},\,(3)\mbox{-}2)_{b3,a}\,.
\label{dy1do}
\end{align}

\subsubsection*{D($\mbox{R}_{2}$) creation}
The input process $\mbox{R}_{2} = ug \to \mu^{-}\mu^{+}u$
determines set $\{ x \}$ as
\begin{align}
\{x\} &= \{x_{a},x_{b};x_{1},x_{2},x_{3}  \}\,,
\label{dy2dipx}\\
\mbox{F}(\{x\}) &= \{ u,g\,; \mu^{-}, \mu^{+}, u \}\,,
\label{dy2dipf} \\
\mbox{Momenta}: & \ \ \ \ \ \{ p_{a},p_{b}\,;p_{1},p_{2},p_{3} \}\,. 
\label{dy2dipm}
\end{align}
We create the dipole term as
\begin{equation}
{\tt Dipole4} \, (7)\,\mbox{-}2: \ \ 1.\,(b3,a)\,.
\label{dy2dc}
\end{equation}
The dipole term has
a reduced Born process of category
{\tt Dipole\,4} as
$\mbox{B}4u(\mbox{R}_{2})=u\bar{u} \to \mu^{-}\mu^{+}$\,,
which determines set $\{y\}$ as
\begin{align}
\{ y \} 
&= \{y_{a},y_{b};y_{1}, y_{2} \}\,, 
\label{dy2y} \\
\mbox{F}(\{ y \}) 
&= \{ u,\bar{u}\,; \mu^{-}, \mu^{+}\}\,.  
\label{dy2yF} 
\end{align}
Although the dipole term
${\tt Dipole\,3} \ (6)\mbox{-}2: (a3,b)$ 
is possible as a selection of the splitting, 
the reduced Born process 
$\mbox{B}3u(\mbox{R}_{2})=gg \to \mu^{-}\mu^{+}$
does not exist at LO, and neither does
the dipole term.
Next we write down the
concrete expression for the dipole term
1.\,(b3,a) in Eq.\,(\ref{dy2dc}). 
Set $\{\tilde{x}\}$ is defined as
\begin{align}
\{\tilde{x}\} 
&= \{x_{\widetilde{a}},x_{\widetilde{b3}}, \,
;x_{1},x_{2} \}\,, \\
\mbox{F}(\{\tilde{x}\}) 
&= \{ u,\bar{u} \,;\mu^{-}, \mu^{+} \}\,,  \\
\mbox{P}(\{\tilde{x}\}) 
& =\{ p_{a},\tilde{p}_{b3}\,;\tilde{k}_{1}, \tilde{k}_{2} \}\,.
\end{align}
The field mapping to set $\{y \}$ is found in 
Eq.\,(\ref{dy2y}) as
\begin{equation}
(y_{a},y_{b};y_{1}, y_{2})=
(x_{\widetilde{a}},x_{\widetilde{b3}};x_{1},x_{2})\,,
\end{equation}
which is abbreviated as 
$(\widetilde{a},\widetilde{b3} \ ;1,2)$.
The momenta are determined as
\begin{equation}
\mbox{P}(\{ y \}) = 
\{ p_{a}, \tilde{p}_{b3}\,;\tilde{k}_{1}, \tilde{k}_{2} \}\,.
\end{equation}
The dipole term is written in Eq.\,(\ref{dp472}) as
\begin{equation}
\mbox{D}( {\tt dip}4,\,(7)\mbox{-}2)_{b3,a} 
= -\frac{1}{s_{b3}} 
\frac{1}{x_{3,ba}}
\frac{1}{\mbox{C}_{\mbox{{\tiny F}}}} 
\mbox{V}_{b3,a} \
\langle \mbox{B}4u  \ | 
\mbox{T}_{y_{b}} \cdot \mbox{T}_{y_{a}}
 | \ \mbox{B}4u \rangle\,.
\label{dy2dip1}
\end{equation}
The output D($\mbox{R}_{2}$) is written as
\begin{align}
\mbox{D} (\mbox{R}_{2})
&=\mbox{D} (\mbox{R}_{2},\,{\tt dip4}) \nonumber\\
&=\mbox{D} (\mbox{R}_{2},\,{\tt dip4},\,
(7)\mbox{-}2)_{b3,a} \,.
\label{dy2do}
\end{align}
The dipole terms D($\mbox{R}_{3}$)
are created in a similar way
to D($\mbox{R}_{2}$).

\subsubsection*{Summary of  creation}
The created dipole terms from the inputs
$\{\mbox{R}_{1},\mbox{R}_{2},\mbox{R}_{3}\}$
are summarized in Table\,\ref{s3_tab1}.

\begin{table}[t]
  \centering
\begin{align}
\mbox{D}\,&(\mbox{R}_{1}=u\bar{u} \to \mu^{-}\mu^{+}g): 
\ \ S_{\mbox{{\tiny R}}_{1}}=1
\nonumber\\
&
\begin{array}{|c|c|c|c|c|} \hline
{\tt Dip}\,j
& \mbox{B}j 
& \mbox{Splitting} 
& (x_{I}x_{J},x_{K}) 
& (y_{a},y_{b}:y_{1},y_{2}) 
\\[4pt] \hline
& & & & \\[-12pt]
{\tt Dip\,1}  
& \mbox{B}1=u\bar{u} \to \mu^{-}\mu^{+}
& (3)-2   
& 1.\,(a3,b)
& (\widetilde{a3},\widetilde{b} \ ;1,2) 
\\[4pt]  
&  
&   
& 2.\,(b3,a)
& (\widetilde{a},\widetilde{b3} \ ;1,2) 
\\[4pt]  \hline
\end{array} \nonumber
\end{align}
\begin{align}
\mbox{D}\,&(\mbox{R}_{2}=ug \to \mu^{-}\mu^{+}u): 
\ \ S_{\mbox{{\tiny R}}_{2}}=1 \nonumber\\
&
\begin{array}{|c|c|c|c|c|} \hline
{\tt Dip}\,j
& \mbox{B}j 
& \mbox{Splitting} 
& (x_{I}x_{J},x_{K}) 
& (y_{a},y_{b}:y_{1},y_{2}) 
\\[4pt] \hline
& & & & \\[-12pt]
{\tt Dip\,4}u  
& \mbox{B}4u=u\bar{u} \to \mu^{-}\mu^{+} 
& (7)u-2   
& 1.\,(b3,a)
& (\widetilde{a},\widetilde{b3} \ ;1,2)
\\[4pt]  \hline
\end{array} \nonumber
\end{align}
\begin{align}
\mbox{D}\,&(\mbox{R}_{3}=\bar{u}g \to \mu^{-}\mu^{+}\bar{u}): 
\ \ S_{\mbox{{\tiny R}}_{3}}=1 \nonumber\\
&
\begin{array}{|c|c|c|c|c|} \hline
{\tt Dip}\,j
& \mbox{B}j 
& \mbox{Splitting} 
& (x_{I}x_{J},x_{K}) 
& (y_{a},y_{b}:y_{1},y_{2}) 
\\[4pt] \hline
& & & & \\[-12pt]
{\tt Dip\,4}\bar{u} 
& \mbox{B}4\bar{u}=\bar{u}u \to \mu^{-}\mu^{+}
& (7)\bar{u}-2   
& 1.\,(b3,a)
& (\widetilde{a},\widetilde{b3} \ ;1,2)
\\[4pt]  \hline
\end{array} \nonumber
\end{align}
\caption{\small
Summary table of dipole term creation.
\label{s3_tab1}}
\end{table}
%
%
%
\subsection{I term \label{s3_3}}
In {\bfseries Step 3}, we create the I term
I\,($\mbox{R}_{i}$) from the input B1\,($\mbox{R}_{i}$).
Among the real emission processes
$\{\mbox{R}_{1},\mbox{R}_{2},\mbox{R}_{3}\}$,
only the $\mbox{R}_{1}$ has the reduced Born 
process B1 as
\begin{equation}
\mbox{B}1(\mbox{R}_{1}) = u\bar{u} \to \mu^{-}\mu^{+}.
\end{equation}
Set $\{ y \}$ is fixed in 
Eqs.\,(\ref{dy1y}) and (\ref{dy1yF}) 
as
\begin{align}
\{ y \} 
&= \{y_{a},y_{b};y_{1}, y_{2} \}\,, \\
\mbox{F}(\{ y \}) 
&= \{ u,\bar{u}\,; \mu^{-}, \mu^{+}\}\,.
\end{align}
Following the order shown in Eq.\,(\ref{icror}),
the I terms are created as
\begin{equation}
(3)\mbox{-}2: \ \ 1.(a,b),\,2.(b,a)\,.
\end{equation}
The concrete expressions are shown in Eq.\,(\ref{iform})
as
\begin{align}
1. \ \ \mbox{I}\,(\mbox{R}_{1})_{a,b}
&=-A_{d} \,
\frac{{\cal V}_{f}}{\mbox{C}_{\mbox{{\tiny F}}}} 
\, s_{ab}^{-\ep} \,
\langle \mbox{B}1 \ | 
\mbox{T}_{y_{a}} \cdot 
\mbox{T}_{y_{b}}
| \ \mbox{B}1 \rangle, \\
2. \ \ \mbox{I}\,(\mbox{R}_{1})_{b,a}
&=- A_{d} \,
\frac{{\cal V}_{f}}{\mbox{C}_{\mbox{{\tiny F}}}} 
\, s_{ba}^{-\ep} \,
\langle \mbox{B}1 \ | 
\mbox{T}_{y_{b}} \cdot 
\mbox{T}_{y_{a}}
| \ \mbox{B}1 \rangle.
\end{align}
The output I\,($\mbox{R}_{1}$) is written as
\begin{align}
\mbox{I}\,(\mbox{R}_{1})
&=\mbox{I}\,(\mbox{R}_{1})_{a,b}+\mbox{I}
\,(\mbox{R}_{1})_{b,a} 
\nonumber\\
&=-A_{d} \,
\frac{{\cal V}_{f}}{\mbox{C}_{\mbox{{\tiny F}}}} 
\, 
\bigl( \,
[ a,b ] + [ b,a ]
\, \bigr)\,,
\label{dy1io}
\end{align}
with the notation in Eq.\,(\ref{sqbk}).
The I terms created are summarized in Table \ref{s3_tab2}.
\begin{table}[t]
  \centering
\begin{align}
\mbox{I}\,&(\mbox{R}_{1}): \ 
\mbox{B}1=u\bar{u} \to \mu^{-}\mu^{+},
\ \ S_{\mbox{{\tiny B}}_{1}}=1
\nonumber\\
&
\begin{array}{|c|c|c|} \hline
 \mbox{Leg-}\,y_{I} 
& \mbox{F}(y_{I})
& (y_{I},y_{K}) 
\\[2pt] \hline
(3)-2   
& u
& 1.\,(a,b)
\\[4pt]  
& \bar{u}
& 2.\,(b,a)
\\[4pt]  \hline
\end{array} \nonumber
\end{align}
\caption{\small
Summary table of I term creation
\label{s3_tab2}}
\end{table}
%
%
%
\subsection{P and K terms \label{s3_4}}
In {\bfseries Step 4}, 
we create the P and K terms
P($\mbox{R}_{i}$) and K($\mbox{R}_{i}$)
from the inputs $\mbox{R}_{i}$ and 
$\mbox{B}_{j}(\mbox{R}_{i})$.
\subsubsection*{P/K($\mbox{R}_{1}$) creation}
The process $\mbox{R}_{1}$ defines set 
$\{ x \}$ in Eq.\,(\ref{dy1dipx}) as
\begin{align}
\{x\} &= \{x_{a},x_{b};x_{1},x_{2},x_{3}  \}\,,
\\
\mbox{F}(\{x\}) &= \{ u,\bar{u}\,; \mu^{-}, \mu^{+}, g \}\,,
\\
\mbox{Momenta}&: \ \  \{ p_{a},p_{b}\,;p_{1},p_{2},p_{3} \}\,. 
\end{align}
The only possible reduced Born process is 
B1($\mbox{R}_{1}$) as
\begin{equation}
\mbox{B}1(\mbox{R}_{1}) = u\bar{u} \to \mu^{-}\mu^{+}.
\end{equation}
Set $\{ y \}$
is fixed in Eqs.\,(\ref{dy1y}) and (\ref{dy1yF}) as
\begin{align}
\{ y \} 
&= \{y_{a},y_{b};y_{1}, y_{2} \}\,, \\
\mbox{F}(\{ y \}) 
&= \{ u,\bar{u}\,; \mu^{-}, \mu^{+}\}\,.
\end{align}
The P and K terms are created in the order 
shown in Fig.\,\ref{fig_PKterm} as
\begin{align}
&\mbox{leg-}\,a:{\tt Dipole\,1} \ (3): \mbox{B}1=
\{u,\bar{u}; \mu^{-}, \mu^{+}\} \nonumber\\
&\hspace{20mm}
(x_{a},x_{3}) \to \mbox{F}(x_{\widetilde{a3}})=u=
\mbox{F}(y_{a})
\ \to \ 1.(y_{a},y_{0}), \,2.(y_{a},y_{b})\,, \nonumber\\
&\mbox{leg-}\,b:{\tt Dipole\,1} \ (3): \mbox{B}1=
\{u,\bar{u}; \mu^{-}, \mu^{+} \} \nonumber\\
&\hspace{20mm}(x_{b},x_{3}) \to
\mbox{F}(x_{\widetilde{b3}})=\bar{u}=\mbox{F}(y_{b})
\ \to \ 3.(y_{b},y_{0}), \,4.(y_{b},y_{a})\,.
\end{align}
The concrete expressions for 
the P and K terms with leg-\,$a$
are shown in 
Eqs.\,(\ref{pform}), (\ref{k0form}), and (\ref{kbform})  as
\begin{align}
&1.(y_{a},y_{0}) \ \ \
\mbox{K}(\mbox{R}_{1},x_{a}:\mbox{B}1,y_{a},y_{0}) 
=\frac{\al}{2\pi} \,
\bar{K}^{uu}(x) \ 
\langle \mbox{B}1 \,|  \mbox{B}1 \rangle \,, \\
&2.(y_{a},y_{b}) \ \ \
\mbox{P}(\mbox{R}_{1},x_{a}:\mbox{B}1,y_{a},y_{b}) 
=\frac{\al}{2\pi} \,
\frac{1}{\mbox{C}_{\mbox{{\tiny F}}}} \,
P^{uu}(x) \,
\ln \frac{\mu_{F}^{2}}{x\,s_{x_{a} y_{b}}}
\ \langle \mbox{B}1 \ | 
\mbox{T}_{y_{{\scriptscriptstyle a}}} \cdot 
\mbox{T}_{y_{{\scriptscriptstyle b}}}
| \ \mbox{B}1 \rangle, \\
&\hspace{20mm}
\mbox{K}(\mbox{R}_{1},x_{a}:\mbox{B}1,y_{a},y_{b}) 
=
-\frac{\al}{2\pi} \,
\frac{1}{\mbox{C}_{\mbox{{\tiny F}}}} \, 
\widetilde{K}^{uu}(x) \
\langle \mbox{B}1 | 
\mbox{T}_{y_{{\scriptscriptstyle a}}} \cdot 
\mbox{T}_{y_{{\scriptscriptstyle b}}}
| \ \mbox{B}1 \rangle\,,
\end{align}
and with leg-\,$b$ as
\begin{align}
&3.(y_{b},y_{0}) \ \ \
\mbox{K}(\mbox{R}_{1},x_{b}:\mbox{B}1,y_{b},y_{0}) 
=\frac{\al}{2\pi} \,
\bar{K}^{\bar{u}\bar{u}}(x) \ 
\langle \mbox{B}1 \,|  \mbox{B}1 \rangle \,, 
\\
&4.(y_{b},y_{a}) \ \ \ 
\mbox{P}(\mbox{R}_{1},x_{b}:\mbox{B}1,y_{b},y_{a}) 
=\frac{\al}{2\pi} \,
\frac{1}{\mbox{C}_{\mbox{{\tiny F}}}} \,
P^{\bar{u}\bar{u}}(x) \,
\ln \frac{\mu_{F}^{2}}{x\,s_{x_{b} y_{a}}}
\ \langle \mbox{B}1 \ | 
\mbox{T}_{y_{{\scriptscriptstyle b}}} \cdot 
\mbox{T}_{y_{{\scriptscriptstyle a}}}
| \ \mbox{B}1 \rangle, \\
&\hspace{20mm}
\mbox{K}(\mbox{R}_{1},x_{b}:\mbox{B}1,y_{b},y_{a}) 
=-\frac{\al}{2\pi} \,
\frac{1}{\mbox{C}_{\mbox{{\tiny F}}}} \, 
\widetilde{K}^{\bar{u}\bar{u}}(x) \ 
\langle \mbox{B}1 | 
\mbox{T}_{y_{{\scriptscriptstyle b}}} \cdot 
\mbox{T}_{y_{{\scriptscriptstyle a}}}
| \ \mbox{B}1 \rangle.
\end{align}
The output for the P term is written 
in Eq.\,(\ref{outp}) as
\begin{equation}
\mbox{P}(\mbox{R}_{1})=\{ \,
\mbox{P}(\mbox{R}_{1},x_{a}), \,
\mbox{P}(\mbox{R}_{1},x_{b}) \, \},
\label{dyr1pout}
\end{equation}
where the elements $\mbox{P}(\mbox{R}_{1},x_{a/b})$
are written as
\begin{align}
\mbox{P}(\mbox{R}_{1},x_{a/b})
&=\mbox{P}(\mbox{R}_{1},x_{a/b}:\mbox{B}1) \nonumber\\
&=\mbox{P}(\mbox{R}_{1},x_{a/b}:\mbox{B}1,y_{a/b},y_{b/a})\,.
\label{dy1po}
\end{align}
The output for the K term is written 
in Eq.\,(\ref{outk}) as
\begin{equation}
\mbox{K}(\mbox{K}_{1})=\{ \,
\mbox{K}(\mbox{K}_{1},x_{a}), \,
\mbox{K}(\mbox{K}_{1},x_{b}) \, \},
\label{dyr1kout}
\end{equation}
where the elements $\mbox{K}(\mbox{R}_{1},x_{a/b})$
are written as
\begin{align}
\mbox{K}(\mbox{R}_{1},x_{a/b})
&=\mbox{K}(\mbox{R}_{1},x_{a/b}:\mbox{B}1) \nonumber\\
&=\mbox{K}(\mbox{R}_{1},x_{a/b}:\mbox{B}1,y_{a/b},y_{0}) 
+\mbox{K}(\mbox{R}_{1},x_{a/b}:\mbox{B}1,y_{a/b},y_{b/a})\,.
\label{dy1ko}
\end{align}
\subsubsection*{P/K($\mbox{R}_{2}$) creation}
The process $\mbox{R}_{2}$ defines set 
$\{ x \}$ in Eq.\,(\ref{dy2dipx}).
Only one reduced Born process,
$\mbox{B}4u(\mbox{R}_{2})=u\bar{u} \to \mu^{-}\mu^{+}$\,,
exists, which fixes set $\{y\}$ in  Eq.\,(\ref{dy2y}).
The P and K terms are created as
\begin{align}
&\mbox{leg-}b:{\tt Dipole\,4}u \ (7): \mbox{B}4u=
\{u,\bar{u}; \mu^{-}, \mu^{+} \} \nonumber\\
&\hspace{20mm}(x_{b},x_{3}) \to 
\mbox{F}(x_{\widetilde{b3}})=\bar{u}=\mbox{F}(y_{b})
\ \to \ 1.(y_{b},y_{0}), \,2.(y_{b},y_{a})\,.
\end{align}
The concrete expressions are written down as
\begin{align}
&1.(y_{b},y_{0}) \ \ \
\mbox{K}(\mbox{R}_{2},x_{b}:\mbox{B}4u,y_{b},y_{0}) 
=\frac{\al}{2\pi} \,
\bar{K}^{g\bar{u}}(x) \ 
\langle \mbox{B}4u \,|  \mbox{B}4u \rangle \,, 
\\
&2.(y_{b},y_{a}) \ \ \ 
\mbox{P}(\mbox{R}_{2},x_{b}:\mbox{B}4u,y_{b},y_{a}) 
=\frac{\al}{2\pi} \,
\frac{1}{\mbox{C}_{\mbox{{\tiny F}}}} \,
P^{g\bar{u}}(x) \,
\ln \frac{\mu_{F}^{2}}{x\,s_{x_{b} y_{a}}}
\ \langle \mbox{B}4u \ | 
\mbox{T}_{y_{{\scriptscriptstyle b}}} \cdot 
\mbox{T}_{y_{{\scriptscriptstyle a}}}
| \ \mbox{B}4u \rangle, \\
&\hspace{20mm}
\mbox{K}(\mbox{R}_{2},x_{b}:\mbox{B}4u,y_{b},y_{a}) 
=-\frac{\al}{2\pi} \,
\frac{1}{\mbox{C}_{\mbox{{\tiny F}}}} \, 
\widetilde{K}^{g\bar{u}}(x) \ 
\langle \mbox{B}4u | 
\mbox{T}_{y_{{\scriptscriptstyle b}}} \cdot 
\mbox{T}_{y_{{\scriptscriptstyle a}}}
| \ \mbox{B}4u \rangle.
\end{align}
The outputs are written as
\begin{align}
\mbox{P}(\mbox{R}_{2}) 
&=\mbox{P}(\mbox{R}_{2},x_{b})
=\mbox{P}(\mbox{R}_{2},x_{b}:\mbox{B}4u,y_{b},y_{a})\,,
\label{dy2po}
\\
\mbox{K}(\mbox{K}_{2})
&=\mbox{K}(\mbox{R}_{2},x_{b})
=\mbox{K}(\mbox{R}_{2},x_{b}:\mbox{B}4u,y_{b},y_{0}) 
+\mbox{K}(\mbox{R}_{2},x_{b}:\mbox{B}4u,y_{b},y_{a})\,.
\label{dy2ko}
\end{align}
The P and K terms
P($\mbox{R}_{3}$) and K($\mbox{R}_{3}$)
are created in the same way as
P($\mbox{R}_{2}$) and K($\mbox{R}_{2}$).
\subsubsection*{Summary of  creation}
The P and K terms created from the inputs
of the real processes
$\{\mbox{R}_{1},\mbox{R}_{2},\mbox{R}_{3}\}$
are summarized in Table.\,\ref{s3_tab3}.
\begin{table}[t]
  \centering
\begin{align}
\mbox{P/K}\,&(\mbox{R}_{1}=u\bar{u} \to \mu^{-}\mu^{+}g)
\nonumber\\
&
\begin{array}{|c|c|c|c|c|c|} \hline
\mbox{Leg-}\,x_{a/b}
&{\tt Dip}\,j
& \mbox{B}j 
& S_{\mbox{{\tiny B}}_{j}}
& \mbox{Splitting} 
& (y_{a/b},y_{K}) 
\\[4pt] \hline
a
&{\tt Dip\,1}  
& \mbox{B}1=u\bar{u} \to \mu^{-}\mu^{+}
& S_{\mbox{{\tiny B}}_{1}}=1 
& (3)-0   
& 1.\,(a,0)
\\[4pt]  
&
&  
&   
& (3)-2 
& 2.\,(a,b)
\\[4pt]  \hline
b
&{\tt Dip\,1}  
& \mbox{B}1=u\bar{u} \to \mu^{-}\mu^{+}
& S_{\mbox{{\tiny B}}_{1}}=1 
& (3)-0 
& 3.\,(b,0)
\\[4pt]  
&
&  
&   
& (3)-2
& 4.\,(b,a)
\\[4pt]  \hline
\end{array} \nonumber
\end{align}
\begin{align}
\mbox{P/K}\,&(\mbox{R}_{2}=ug \to \mu^{-}\mu^{+}u)
\nonumber\\
&
\begin{array}{|c|c|c|c|c|c|} \hline
\mbox{Leg-}\,x_{a/b}
&{\tt Dip}\,j
& \mbox{B}j 
& S_{\mbox{{\tiny B}}_{j}}
& \mbox{Splitting} 
& (y_{a/b},y_{K}) 
\\[4pt] \hline
b
&{\tt Dip\,4}u  
& \mbox{B}4u=u\bar{u} \to \mu^{-}\mu^{+} 
& S_{\mbox{{\tiny B}}_{4u}}=1 
& (7)-0
& 1.\,(b,0)
\\[4pt]  
&
&  
&   
& (7)-2
& 2.\,(b,a)
\\[4pt]  \hline
\end{array} \nonumber
\end{align}
\begin{align}
\mbox{P/K}\,&(\mbox{R}_{3}=
\bar{u}g \to \mu^{-}\mu^{+}\bar{u})
\nonumber\\
&
\begin{array}{|c|c|c|c|c|c|} \hline
\mbox{Leg-}\,x_{a/b}
&{\tt Dip}\,j
& \mbox{B}j 
& S_{\mbox{{\tiny B}}_{j}}
& \mbox{Splitting} 
& (y_{a/b},y_{K}) 
\\[4pt] \hline
b
&{\tt Dip\,4}\bar{u}  
& \mbox{B}4\bar{u}=\bar{u}u \to \mu^{-}\mu^{+}
& S_{\mbox{{\tiny B}}_{4\bar{u}}}=1 
& (7)-0
& 1.\,(b,0)
\\[4pt]  
&
&  
&   
& (7)-2
& 2.\,(b,a)
\\[4pt]  \hline
\end{array} \nonumber
\end{align}
\caption{\small
Summary tables of the P and K term creation.
\label{s3_tab3}}
\end{table}
%
%
%
\subsection{NLO correction:\,
$\sigma_{\mbox{{\tiny NLO}}}$ \label{s3_5}}
In {\bfseries Step 5}, we obtain the NLO correction
$\sigma_{\mbox{{\tiny NLO}}}$ in Eq.\,(\ref{masternlo})
as
\begin{equation}
\sigma_{\mbox{{\tiny NLO}}} = \sum_{i=1}^{3} 
\sigma(\mbox{R}_{i})\,.
\end{equation}
The cross section $\sigma(\mbox{R}_{1})$ 
is concretely written as
\begin{align}
\sigma(\mbox{R}_{1}) 
&= \int dx_{1} \int dx_{2} \ 
f_{u}(x_{1}) 
f_{\bar{u}}(x_{2}) \ \times  \nonumber \\
& \ \ \biggl[
\bigl(\hat{\sigma}_{\mbox{{\tiny R}}}(\mbox{R}_{1}) - 
\hat{\sigma}_{\mbox{{\tiny D}}}(\mbox{R}_{1}) \bigr) + 
\bigl(\hat{\sigma}_{\mbox{{\tiny V}}}(\mbox{B}1(\mbox{R}_{1})) + 
\hat{\sigma}_{\mbox{{\tiny I}}}(\mbox{R}_{1}) \bigr) + 
\hat{\sigma}_{\mbox{{\tiny P}}}(\mbox{R}_{1}) + 
\hat{\sigma}_{\mbox{{\tiny K}}}(\mbox{R}_{1}) \biggr]\,,
\label{dyhcr1}
\end{align}
where the finite combinations of the partonic cross 
sections are written separately as
\begin{align}
\hat{\sigma}_{\mbox{{\tiny R}}}(\mbox{R}_{1}) - 
\hat{\sigma}_{\mbox{{\tiny D}}}(\mbox{R}_{1}) 
&=\frac{1}{S_{\mbox{{\tiny R}}_{1}}} 
\Phi(\mbox{R}_{1})_{4} \cdot \Bigl[ \ 
|\mbox{M}(\mbox{R}_{1})|_{4}^{2} \ - 
\ \frac{1}{n_{s}(u) n_{s}(\bar{u})} \mbox{D}(\mbox{R}_{1}) 
\ \Bigr]\,, 
\label{dypcr1r} \\
\hat{\sigma}_{\mbox{{\tiny V}}}(\mbox{B}1(\mbox{R}_{1})) + 
\hat{\sigma}_{\mbox{{\tiny I}}}(\mbox{R}_{1}) 
&= 
\frac{1}{S_{\mbox{{\tiny B1}} }} \ 
\Phi( \mbox{B}1  )_{d} \cdot 
\Bigl[ \ |\mbox{M}_{\mbox{{\tiny virt}}}(\mbox{B}1)|_{d}^{2} \ + \ 
\mbox{I}(\mbox{R}_{1}) \ \Bigr],
\label{dypcr1v} \\
\hat{\sigma}_{\mbox{{\tiny P}}}(\mbox{R}_{1}) 
+
\hat{\sigma}_{\mbox{{\tiny K}}}(\mbox{R}_{1}) 
&= \int_{0}^{1}dx\,  
\frac{1}{S_{\mbox{{\tiny B}}_{1} }}
\,\Phi_{a}(\mbox{R}_{1}:\mbox{B}_{1},x)_{4} \ \cdot \nonumber \\
&\biggl(\mbox{P}(\mbox{R}_{1},x_{a}:\mbox{B}_{1}) \ + \
\mbox{K}(\mbox{R}_{1},x_{a}:\mbox{B}_{1})
\biggr) \ + \
(a \leftrightarrow b)\,.
\label{dypcr1pk} 
\end{align} 
The subtraction terms $\mbox{D}(\mbox{R}_{1}),$ 
$\mbox{I}(\mbox{R}_{1}),$
$\mbox{P}(\mbox{R}_{1},x_{a/b}),$ and
$\mbox{K}(\mbox{R}_{1},x_{a/b})$ are
written in 
Eqs.\,(\ref{dy1do}),
(\ref{dy1io}), (\ref{dy1po}), and (\ref{dy1ko}),
respectively.
$\sigma(\mbox{R}_{2})$ is written as
\begin{align}
\sigma(\mbox{R}_{2}) 
&= \int dx_{1} \int dx_{2} \ 
f_{u}(x_{1}) 
f_{g}(x_{2}) 
\ \biggl[
\bigl(\hat{\sigma}_{\mbox{{\tiny R}}}(\mbox{R}_{2}) - 
\hat{\sigma}_{\mbox{{\tiny D}}}(\mbox{R}_{2}) \bigr) + 
\hat{\sigma}_{\mbox{{\tiny P}}}(\mbox{R}_{2}) + 
\hat{\sigma}_{\mbox{{\tiny K}}}(\mbox{R}_{2}) \biggr],  
\label{dyhcr2}
\end{align}
where the finite cross sections are written as
\begin{align}
\hat{\sigma}_{\mbox{{\tiny R}}}(\mbox{R}_{2}) - 
\hat{\sigma}_{\mbox{{\tiny D}}}(\mbox{R}_{2}) 
&=\frac{1}{S_{\mbox{{\tiny R}}_{2}}} 
\Phi(\mbox{R}_{2})_{4} \cdot \Bigl[ \ 
|\mbox{M}(\mbox{R}_{2})|_{4}^{2} \ - 
\ \frac{1}{n_{s}(u) n_{s}(g)} \mbox{D}(\mbox{R}_{2}) 
\ \Bigr]\,, 
\label{dypcr2r} \\
\hat{\sigma}_{\mbox{{\tiny P}}}(\mbox{R}_{2}) 
+
\hat{\sigma}_{\mbox{{\tiny K}}}(\mbox{R}_{2}) 
&= \int_{0}^{1}dx  
\frac{1}{S_{\mbox{{\tiny B}}_{4u} }}
\,\Phi_{b}(\mbox{R}_{2}:\mbox{B}_{4u},x)_{4}  
\biggl[\mbox{P}(\mbox{R}_{2},x_{b}:\mbox{B}_{4u}) + 
\mbox{K}(\mbox{R}_{2},x_{b}:\mbox{B}_{4u})
\biggr]\,.
\label{dypcr2pk}
\end{align} 
The subtraction terms $\mbox{D}(\mbox{R}_{2}),$ 
$\mbox{P}(\mbox{R}_{2},x_{b}),$ and
$\mbox{K}(\mbox{R}_{2},x_{b})$ are
written in 
Eqs.\,(\ref{dy2do}),
(\ref{dy2po}), and (\ref{dy2ko}), respectively.
$\sigma(\mbox{R}_{3})$ is written in
similar expressions to $\sigma(\mbox{R}_{2})$.
%
%
%
%
The contributions from the exchanged initial partons
must be added. For example, in $\sigma(\mbox{R}_{1})$,
the contribution from the process 
$\bar{u}u \to \mu^{-}\mu^{+}g$ is added with the 
multiplication of the exchanged PDFs,
$f_{\bar{u}}(x_{1})f_{u}(x_{2})$.
Furthermore, the contribution from the other quark
flavors must be added. 
The full expression is clarified in Sec.\,\ref{sec_5}.






\newpage
\section{Dijet\,: $pp \to 2\,jets + X$ \label{sec_4}}
In the present section, we apply the DSA to the dijet process.
Like the Drell--Yan process in
Sec.\,\ref{sec_3}, the five steps are executed
in Sec.\,\ref{s4_1},\, \ref{s4_2},\, \ref{s4_3},
\ref{s4_4}, and \ref{s4_5}, respectively.
\subsection{List of $\mbox{R}_{i}$ \label{s4_1}}
In {\bfseries Step 1} we make a list of the 
real processes $\{\mbox{R}_{i}\}$ as follows\,:
\begin{align}
\mbox{R}_{1u} 
&= u\bar{u} \to u\bar{u}g\,,
\ \ \ (\mbox{R}_{1d})
\nonumber\\
\mbox{R}_{2u} 
&= uu \to uug\,,
\ \ \ (\mbox{R}_{2\bar{u}},\mbox{R}_{2d},
\mbox{R}_{2\bar{d}})  
\nonumber\\
\mbox{R}_{3u} 
&= ug \to uu\bar{u}\,,
\ \ \ (\mbox{R}_{3\bar{u}},\mbox{R}_{3d},
\mbox{R}_{3\bar{d}})  
\nonumber\\
\mbox{R}_{4u} 
&= u\bar{u} \to d\bar{d}g\,,
\ \ \ (\mbox{R}_{4d}) 
\nonumber\\
\mbox{R}_{5ud} 
&= ud \to udg\,,
\ \ \ (\mbox{R}_{5\bar{u}\bar{d}}) 
\nonumber\\
\mbox{R}_{6u\bar{d}} 
&= u\bar{d} \to u\bar{d}g\,,
\ \ \ (\mbox{R}_{6\bar{u}d}) 
\nonumber\\
\mbox{R}_{7u} 
&= ug \to ud\bar{d}\,,
\ \ \  (\mbox{R}_{7\bar{u}},\mbox{R}_{7d},
\mbox{R}_{7\bar{d}}) 
\nonumber\\
\mbox{R}_{8u} 
&= u\bar{u} \to ggg\,,
\ \ \ (\mbox{R}_{8d}) 
\nonumber\\
\mbox{R}_{9u} 
&= ug \to ugg\,,
\ \ \  (\mbox{R}_{9\bar{u}},\mbox{R}_{9d},
\mbox{R}_{9\bar{d}}) 
\nonumber\\
\mbox{R}_{10u} 
&= gg \to u\bar{u}g\,,
\ \ \ (\mbox{R}_{10d}) 
\nonumber\\
\mbox{R}_{11} 
&= gg \to ggg\,. 
\label{listdi}
\end{align}
There are eleven independent processes so
$n_{\mbox{{\tiny real}}}=11$\,.
The number of final states is $(n+1)=3$.
We assume five massless quark flavors
$u,d,s,c,$ and $b$.
The contributing real processes 
can be exhausted 
by the independent processes
that are produced from the field contents
of only two quark flavors and a gluon.
We take the $u$ and $d$ quarks as the two flavors.
The real processes in the curly brackets in 
Eq.\,(\ref{listdi}) represent the processes
that are obtained by the replacements of the
quark flavors. For example, the process
$\mbox{R}_{1d}$ is concretely written as
$\mbox{R}_{1d}=d\bar{d} \to d\bar{d}g$,
which is obtained by the replacements
$u \to d$ and $\bar{u} \to \bar{d}$
in the process $\mbox{R}_{1u}=u\bar{u} \to u\bar{u}g$.
It is sometimes useful to categorize the partonic
processes by the crossing symmetry.
The processes $\mbox{R}_{1u},\mbox{R}_{2u}$, and 
$\mbox{R}_{3u}$ are categorized into the 
master process $0 \to uu\bar{u}\bar{u}g$.
The processes
$\mbox{R}_{4u},\mbox{R}_{5ud},\mbox{R}_{6u\bar{d}},$ 
and $\mbox{R}_{7u}$ 
are categorized into the process
$0 \to ud\bar{u}\bar{d}g$,
and $\mbox{R}_{8u},\mbox{R}_{9u},$ and $\mbox{R}_{10u}$
into $0 \to u\bar{u}ggg$.
The process $\mbox{R}_{11}$ is categorized
into $0 \to ggggg$.
%
%
%
\subsection{D term \label{s4_2}}
In {\bfseries Step 2} we create the dipole terms 
$\mbox{D}\,(\mbox{R}_{i})$ from the inputs
$\{\mbox{R}_{i}\}$ in Eq.\,(\ref{listdi}).
Summary tables of all the dipole terms created
are shown in Tables \ref{ap_B_1_tab1}--\ref{ap_B_1_tab11}
in Appendix \ref{ap_B_1}.
The details of the creation of 
$\mbox{D}\,(\mbox{R}_{1})$
have already been presented 
with the concrete expressions of some dipole terms
in Eqs.\,(\ref{dipexin})--(\ref{dr1last}).
In this section, we present only one dipole term in 
$\mbox{D}\,(\mbox{R}_{9u})$. 
Hereafter we drop the flavor index $u$,
leaving $\mbox{R}_{9}$, for simplicity.
The input process
$\mbox{R}_{9}$ in Eq.\,(\ref{listdi})
defines set $\{x\}$ as
\begin{align}
\{x\} &= \{x_{a},x_{b};x_{1},x_{2},x_{3}  \}\,,
\label{didipr9x}\\
\mbox{F}(\{x\}) &= \{ u,g; u,g, g \}\,,
\label{didipr9f} \\
\mbox{Momenta}: & \ \ \ \ \ \{ p_{a},p_{b};p_{1},p_{2},p_{3} \}\,. 
\label{didipr9m}
\end{align}
We select the dipole term 
$16.\,\mbox{D}\,(\mbox{R}_{9},{\tt dip}1,
(4)\mbox{-}1)_{b2,1}$
from the 27 dipole terms in 
Table \ref{ap_B_1_tab9}.
For the process $\mbox{R}_{9}$, the reduced Born process
B1 is fixed as
\begin{equation}
\mbox{B}1(\mbox{R}_{9})=ug \to ug\,,
\label{r9b1}
\end{equation}
which determines set $\{y\}$ as
\begin{align}
\{ y \} 
&= \{y_{a},y_{b};y_{1}, y_{2} \}\,, \label{diy} \\
\mbox{F}(\{ y \}) 
&= \{ u,g;u,g \}\,.  \label{diyF} 
\end{align}
For the dipole term $\mbox{D}_{b2,1}$,
set $\{\tilde{x}\}$ is defined as
\begin{align}
\{\tilde{x}\} 
&= \{x_{\widetilde{b2}},x_{a};x_{3},x_{\widetilde{1}} 
\}, \\
\mbox{F}(\{\tilde{x}\}) 
&= \{ g,u; g,u \}, \\
\mbox{P}(\{\tilde{x}\}) 
& =\{ \tilde{p}_{b2},p_{a};p_{3}, \tilde{p}_{1} \}\,, 
\end{align}
where the reduced momenta $\tilde{p}_{b2}$
and $\tilde{p}_{1}$ are defined in 
Eqs.\,(\ref{rmifem}) and (\ref{rmifsp}).
The field mapping is specified as
\begin{equation}
(y_{a},y_{b};y_{1}, y_{2})=
(x_{a},x_{\widetilde{b2}};x_{\widetilde{1}},x_{3})\,,
\end{equation}
which is abbreviated as 
$(a,\widetilde{b2};\widetilde{1},3)$
in Table \ref{ap_B_1_tab9}.
The field mapping determines the momenta of set 
$\{y\}$ as
\begin{equation}
\mbox{P}(\{ y \}) = \{p_{a}, \tilde{p}_{b2};\tilde{p}_{1}, p_{3} \}\,.
\end{equation}
The concrete expression of the dipole term is written in 
Eq.\,(\ref{dp141}) as
\begin{align}
\mbox{D}\,(\mbox{R}_{9},\,{\tt dip}1,\,
(4)\mbox{-}1)_{b2,1}
&= -\frac{1}{s_{b2}} \frac{1}{x_{21,b}}
\frac{1}{\mbox{C}_{\mbox{{\tiny A}}}} \ 
\langle \mbox{B}1 | 
\mbox{T}_{y_{b}} 
\cdot \mbox{T}_{y_{1}} \ 
\mbox{V}_{{\scriptscriptstyle b2,1}}^{y_{b}} | 
\ \mbox{B}1 \rangle,  
\end{align}
where
the dipole splitting function $\mbox{V}_{b2,1}$
and the Lorentz scalar $x_{21,b}$ are defined in 
Eqs.\,(\ref{ds141}) and (\ref{xaikif})
respectively.
The contribution to the cross section reads in Eq.\,(\ref{hatd}) as
\begin{equation}
\hat{\sigma}_{\mbox{{\tiny D}}}(\mbox{R}_{9}) 
= \frac{1}{S_{\mbox{{\tiny R}}_{9}}} \ \Phi(\mbox{R}_{9})_{4} 
\, \frac{1}{n_{s}(u) n_{s}(g)} \,
\biggl[
\mbox{D}\,(\mbox{R}_{9},\,{\tt dip}1,\,(4)\mbox{-}1)_{b2,1}
+ \cdots
\biggr]\,.
\end{equation}
It is noted that
the symmetric factor of the reduced Born process
$\mbox{B}1(\mbox{R}_{9})$ is $S_{\mbox{{\tiny B}}_{1}}=1$,
but the contribution of the dipole term is not divided 
by the factor.
The contribution of all the dipole terms involved in 
$\mbox{D}\,(\mbox{R}_{9})$ must be divided by 
the symmetric factor of the input real process 
$\mbox{R}_{9}$, $S_{\mbox{{\tiny R}}_{9}}=2$.

\subsection{I term \label{s4_3}}
In {\bfseries Step 3} we create the I terms
$\mbox{I}\,(\mbox{R}_{i})$ from the inputs
$\mbox{B}1\,(\mbox{R}_{i})$.
Summary tables of
all the I terms created are shown in
Tables \ref{ap_B_2_tab1}--\ref{ap_B_2_tab5}
in Appendix \ref{ap_B_2}.
Since the real processes
$\mbox{R}_{3}$ and $\mbox{R}_{7}$
do not have the reduced Born 
process B1, the I terms
I($\mbox{R}_{3}$) and I($\mbox{R}_{7}$)
do not exist.
The details of the creation of
I\,($\mbox{R}_{1}$)
have been explained in Sec.\,\ref{s2_3}.
Here we see I\,($\mbox{R}_{9}$) concretely.
The input to create I\,($\mbox{R}_{9}$) is the 
process $\mbox{B}1(\mbox{R}_{9})$
in Eq.\,(\ref{r9b1}) and the associated set
$\{y\}$ is defined in Eqs.\,(\ref{diy})
and (\ref{diyF}).
Twelve I terms are created and they are
listed in Table \ref{ap_B_2_tab3}.
Referring to the formula in Eq.\,(\ref{iform}),
the concrete expression of I\,($\mbox{R}_{9}$)
is written as
\begin{align}
\mbox{I}\,(\mbox{R}_{9})
&=- A_{d} \,
\biggl[\,
\frac{{\cal V}_{f}}{\mbox{C}_{\mbox{{\tiny F}}}} 
\,\bigl( \
[1,2]+[1,a]+[1,b]+[a,1]+[a,2]+[a,b] 
\bigr)
\nonumber\\
&\hspace{14mm}
+\frac{{\cal V}_{g}}{\mbox{C}_{\mbox{{\tiny A}}}} 
\,\bigl( \
[2,1]+[2,a]+[2,b]+[b,1]+[b,2]+[b,a] 
\bigr)
\biggr]\,,
\end{align}
where ${\cal V}_{f/g}$ are defined 
in Eqs.\,(\ref{usf1}) and (\ref{usf2}).
The contribution to the cross section reads
in Eq.\,(\ref{hati}) as
\begin{equation}
\hat{\sigma}_{\mbox{{\tiny I}}}(\mbox{R}_{9}) 
= \frac{1}{S_{\mbox{{\tiny B1}} }} \ \Phi(\mbox{B}1)_{d} 
\cdot \mbox{I}(\mbox{R}_{9})\,, 
\end{equation}
where the cross section is divided by 
the symmetric factor of the $\mbox{B}1(\mbox{R}_{9})$,
$S_{\mbox{{\tiny B}}_{1}}=1$.
%
%
%
\subsection{P and K terms \label{s4_4}}
In {\bfseries Step 4} 
we create the P and K terms
P($\mbox{R}_{i}$) and K($\mbox{R}_{i}$).
Summary tables of all the P and K terms created
are shown in Tables \ref{ap_B_3_tab1}--\ref{ap_B_3_tab11}
in Appendix \ref{ap_B_3}.
The details of the creation of 
$\mbox{P/K}\,(\mbox{R}_{1})$
are presented with some examples 
in Sec.\,\ref{s2_4}.
In this section, we show only one P term 
in $\mbox{P}\,(\mbox{R}_{9})$
and one K term in
$\mbox{K}\,(\mbox{R}_{9})$.
The input $\mbox{R}_{9}$ defines set $\{x\}$
in Eqs.\,(\ref{didipr9x}), (\ref{didipr9f}),
and (\ref{didipr9m}).
The possible reduced Born processes 
$\mbox{B}_{j}\,(\mbox{R}_{9})$ are
fixed during the creation of the dipole term
$\mbox{D}\,(\mbox{R}_{9})$, and are explicitly
shown in Table\,\ref{ap_B_1_tab9} as
\begin{align}
\mbox{B}1  &: \, \mbox{F}(\{ y\})= \{u,g; u,g \}\,,
 \\
\mbox{B}3u &: \, \mbox{F}(\{ y\})= \{g,g; g,g \}\,,
 \\
\mbox{B}4u &: \, \mbox{F}(\{ y\})= \{u,\bar{u}; g,g \}\,.
\end{align}
Here we show concrete expressions for the
P and K terms 
$10.\,\mbox{P/K}(\mbox{R}_{9},x_{b}:\mbox{B}1,y_{b},y_{1})$
in Table\,\ref{ap_B_3_tab9}.
The expressions are written
in Eqs.\,(\ref{pform}) and (\ref{kkform}) as
\begin{align}
\mbox{P}(\mbox{R}_{9},x_{b}:\mbox{B}1,y_{b},y_{1})
&=\frac{\al}{2\pi} \,
\frac{1}{\mbox{C}_{\mbox{{\tiny A}}}} \,
P^{gg}(x) \,
\ln \frac{\mu_{F}^{2}}{x\,s_{x_{b} y_{1}}}
\ \cdot \langle \mbox{B}1 \ | 
\mbox{T}_{y_{{\scriptscriptstyle b}}} \cdot 
\mbox{T}_{y_{{\scriptscriptstyle 1}}}
| \ \mbox{B}1 \rangle, \\
\mbox{K}(\mbox{R}_{9},x_{b}:\mbox{B}1,y_{b},y_{1})
&=\frac{\al}{2\pi} \,
\frac{\gamma_{u}}
{\mbox{C}_{\mbox{{\tiny F}}}} \, h(x)
\ \cdot \langle \mbox{B}1 | 
\mbox{T}_{y_{{\scriptscriptstyle b}}} \cdot 
\mbox{T}_{y_{1}}
| \ \mbox{B}1 \rangle\,.
\end{align}
The functions of the argument $x$, $P^{gg}(x)$ and $h(x)$,
are defined in 
Eqs.\,(\ref{alpgg}) and (\ref{hxd}).
The contribution to the cross section 
is written in Eq.\,(\ref{hatpk2}) as
\begin{align}
\hat{\sigma}_{\mbox{{\tiny P}}}(\mbox{R}_{9}) 
+
\hat{\sigma}_{\mbox{{\tiny K}}}(\mbox{R}_{9}) 
&= \int_{0}^{1}dx  
\biggl[
\frac{1}{S_{\mbox{{\tiny B}}_{1} }}
\,\Phi_{b}(\mbox{R}_{9}:\mbox{B}_{1},x)_{4} \ \cdot \nonumber \\
&\hspace{18mm}\bigl(
\mbox{P}(\mbox{R}_{9},x_{b}:\mbox{B}1,y_{b},y_{1})+
\mbox{K}(\mbox{R}_{9},x_{b}:\mbox{B}1,y_{b},y_{1})+
\cdots
\bigr) 
\nonumber\\
&\hspace{18mm}+ 
\frac{1}{S_{\mbox{{\tiny B}}_{4u} }}
\,\Phi_{b}(\mbox{R}_{9}:\mbox{B}_{4u},x)_{4} \ \cdot
\nonumber\\
&\hspace{18mm}\bigl(
\mbox{P}(\mbox{R}_{9},x_{b}:\mbox{B}_{4u})+
\mbox{K}(\mbox{R}_{9},x_{b}:\mbox{B}_{4u})+
\cdots
\bigr) 
+\cdots\,
\biggr]\,.
\end{align} 
The contributions of the P and K terms
$\mbox{P/K}(\mbox{R}_{9},x_{b}:\mbox{B}1,y_{b},y_{1})$
are divided by the symmetric factor of the
reduced Born process $\mbox{B}1$,
$S_{\mbox{{\tiny B}}_{1}}=1$. 
The contributions from the terms
$\mbox{P/K}(\mbox{R}_{9},x_{b}:\mbox{B}_{4u})$
are divided by the symmetric factor of the
reduced Born process $\mbox{B}_{4u}$,
$S_{\mbox{{\tiny B}}_{4u}}=2$.
In this way, the P and K terms are divided 
by the symmetric factor of the 
reduced Born processes,
$S_{\mbox{{\tiny B}}_{j}}$,
not by the symmetric factor of the
input real process, 
$S_{\mbox{{\tiny R}}_{i}}$.
In this sense, the symmetric
factor for the P and K terms is
in an inverse manner to that
for the dipole terms,
which is explained at the end of 
Sec.\,\ref{s4_2}.
%
%
%
\subsection{NLO correction:\,
$\sigma_{\mbox{{\tiny NLO}}}$ \label{s4_5}}
In {\bfseries Step 5} we obtain the NLO correction
$\sigma_{\mbox{{\tiny NLO}}}$ in Eq.\,(\ref{masternlo})
as
\begin{equation}
\sigma_{\mbox{{\tiny NLO}}} = \sum_{i=1}^{11} 
\sigma(\mbox{R}_{i})\,,
\end{equation}
where the summation over the different quark flavors
is suppressed.
The NLO cross sections
$\sigma(\mbox{R}_{i})$ 
are written in the formula in Eq.\,(\ref{master}) as
\begin{align}
\sigma(\mbox{R}_{i}) 
&= \int dx_{1} \int dx_{2} \ 
f_{\mbox{{\tiny F}}(x_{a})}(x_{1}) 
f_{\mbox{{\tiny F}}(x_{b})}(x_{2}) \ \times  \nonumber \\
& \ \ \biggl[
\bigl(\hat{\sigma}_{\mbox{{\tiny R}}}(\mbox{R}_{i}) - 
\hat{\sigma}_{\mbox{{\tiny D}}}(\mbox{R}_{i}) \bigr) + 
\bigl(\hat{\sigma}_{\mbox{{\tiny V}}}(\mbox{B}1(\mbox{R}_{i})) + 
\hat{\sigma}_{\mbox{{\tiny I}}}(\mbox{R}_{i}) \bigr) + 
\hat{\sigma}_{\mbox{{\tiny P}}}(\mbox{R}_{i}) + 
\hat{\sigma}_{\mbox{{\tiny K}}}(\mbox{R}_{i}) \biggr].
\end{align}
In the cases of the NLO cross sections
$\sigma(\mbox{R}_{3})$ and
$\sigma(\mbox{R}_{7})$,
the formula is simplified as
\begin{align}
\sigma(\mbox{R}_{i}) 
&= \int dx_{1} \int dx_{2} \ 
f_{\mbox{{\tiny F}}(x_{a})}(x_{1}) 
f_{\mbox{{\tiny F}}(x_{b})}(x_{2}) \,
\biggl[
\bigl(\hat{\sigma}_{\mbox{{\tiny R}}}(\mbox{R}_{i}) - 
\hat{\sigma}_{\mbox{{\tiny D}}}(\mbox{R}_{i}) \bigr) + 
\hat{\sigma}_{\mbox{{\tiny P}}}(\mbox{R}_{i}) + 
\hat{\sigma}_{\mbox{{\tiny K}}}(\mbox{R}_{i}) \biggr].
\end{align}
The formulae for the partonic cross sections 
are written in Eqs.\,(\ref{hatr})--(\ref{hatk}).
Like the Drell--Yan process in
Sec.\,\ref{s3_5},
the contribution from the exchanged PDFs must be
added. Furthermore, the contributions from
the remaining three quark flavors,
$s,c$, and $b$, 
in addition to the $u$ and $d$ flavors,
must be taken into account.







\section{Analytical check at Drell-Yan \label{sec_5}}
In the present section, we give an analytical check of 
the Drell--Yan process. In Sec.\,\ref{s5_1} we review
the well known analytical results obtained 
by the traditional
method. In Sec.\,\ref{s5_2} we obtain the analytical
results by the DSA. We will show that both sets of
results exactly coincide.
%
%
%
\subsection{Traditional method \label{s5_1}}
We review the well known results
that were obtained for the first time
in the pioneering works of Refs.
\cite{Altarelli:1978id,Altarelli:1979ub,Abad:1978nr,
Abad:1978ke,Humpert:1979qk,Humpert:1979hb,
Humpert:1980uv,KubarAndre:1978uy,Harada:1979bj}.
The method used in these works
became the traditional method
to calculate QCD NLO corrections in
hadron collider processes.
In the method, both the real and virtual
corrections are calculated in 
$d$ dimensions, i.e.,
the matrix elements, the PS integrations,
and the spin-average factors are all defined 
in $d$ dimensions.
In the method, not only the UV, soft, and
collinear divergences in the virtual correction
but also the soft and collinear divergences 
in the real correction are
regularized as poles
$1/\ep$ and $1/\ep^{2}$.

\begin{figure}[t]
\begin{center}
\includegraphics[width=7cm]{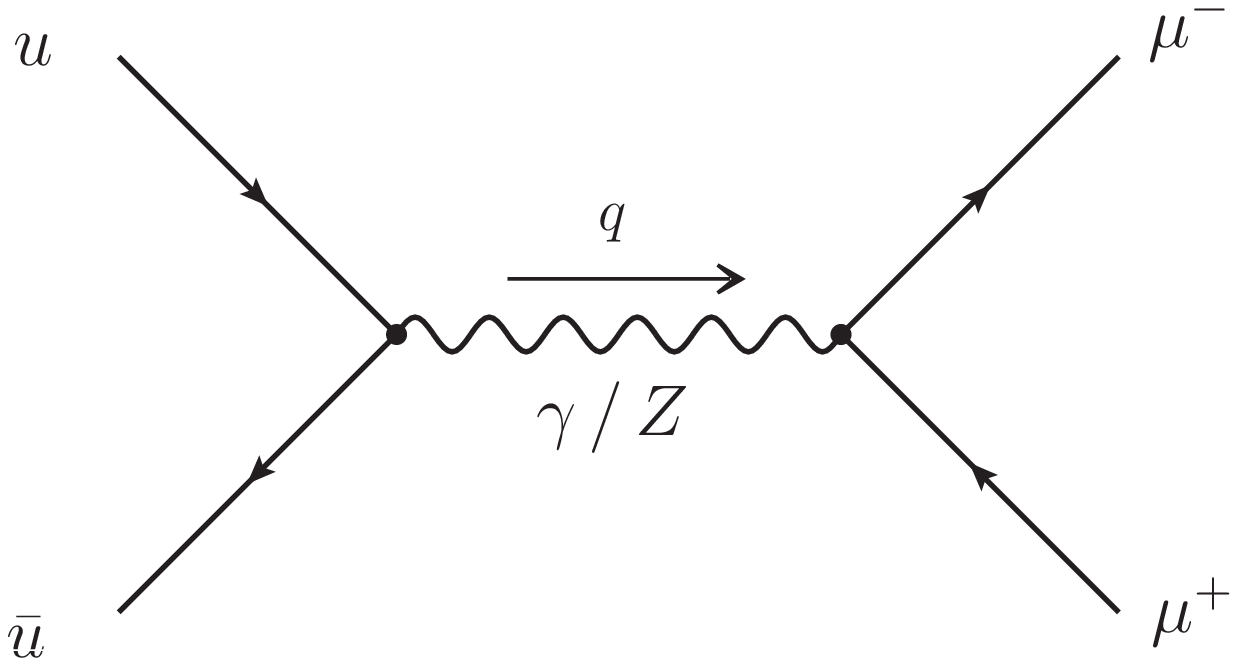}
\end{center} 
\caption{ Diagrams of the LO process
$\mbox{L}_{1} = u\bar{u} \to \mu^{-}\mu^{+}$.
\label{fig_s5_1} }
\end{figure}

We start with a review of the LO contribution.
The general formulae are given in 
Eqs.\,(\ref{masterlo}) and (\ref{mastlocr}).
In this method, we redefine the partonic
cross section in Eq.\,(\ref{mastlopa})\
in $d$ dimensions.
In the Drell--Yan process,
we assume one quark flavor, $u$,
and finally generalize to the five massless 
flavors.
At LO, only one independent process exists as
\begin{equation}
\mbox{L}_{1} = u\bar{u} \to \mu^{-}\mu^{+}\,.
\end{equation}
A Feynman diagram is shown in Fig.\,\ref{fig_s5_1}
\footnote{
All the Feynman diagrams in the 
present article are drawn using the {\tt JaxoDraw} 
package \cite{Binosi:2003yf}.
}.
The contribution is written as
\begin{align}
\sigma_{\mbox{{\tiny LO}}} 
&= \sigma(\mbox{L}_{1})
\nonumber\\
&= \int dx_{1} \int dx_{2} \, f_{u}(x_{1})\, 
f_{\bar{u}}(x_{2}) \ 
\hat{\sigma}(\mbox{L}_{1})\,,  
\end{align}
where the partonic cross section is defined 
in $d$ dimensions as
\begin{align}
\hat{\sigma} (\mbox{L}_{1}) 
&= \frac{1}{S_{\mbox{{\tiny L}}_{1}}} \ 
\Phi(\mbox{L}_{1})_{d} \cdot |\mbox{M}
(\mbox{L}_{1})|_{d}^{2}\,.
\end{align}
We fix the kinematical values
as follows.
The squared energy of two protons 
in the initial state is denoted as 
$S=(P_{a} + P_{b})^{2}=2P_{a} \cdot P_{b}$
with the momenta $P_{a/b}$\,.
The squared energy of two partons 
in the initial state is 
denoted as 
$\hat{s}=(p_{a}+p_{b})^{2}=2p_{a} \cdot p_{b}=x_{1}x_{2}S$, 
where the momenta are denoted as
$p_{a/b}=x_{1/2}P_{a/b}$.
The square of the muon-pair invariant
mass is denoted as
$\mbox{M}_{\mu^{+}\mu^{-}}^{2}=q^{2}=(p_{\mu^{+}}+ p_{\mu^{-}})^{2}$
with the momenta of the antimuon/muon
$p_{\mu^{+}/\mu^{-}}$. 
We write the total cross section as 
$\hat{\sigma}_{tot}(\mbox{L}_{1})=\hat{\sigma}_{\mbox{{\tiny LO}}}(\qs)$
and calculate it as
\begin{align}
\hat{\sigma}_{\mbox{{\tiny LO}}}(\qs)
= \ale^{2} \frac{2 \pi}{N_{c}} 
\frac{\Gamma(1/2)}{\Gamma(3/2 - \ep)}
\frac{(1 - \ep)^{2}}{3 - 2 \ep} 
\biggl(\frac{16 \pi \mu^4}{\qs} \biggr)^{\ep}
q^{2} \ \mbox{P}(\qs)\,,
\end{align}
where the color degree of freedom
of quark is denoted as $N_{c}=3$, and
the symbol $\mu$ represents a dimensionful free parameter 
introduced in the $d$-dimensional space-time.
The last factor $\mbox{P}(\qs)$ is defined as
\begin{align}
 \mbox{P}(\qs) 
&= \frac{1}{(q^{2})^{2}} Q_{u}^{\, 2} Q_{\mu}^{\, 2} 
 + \frac{2}{q^{2}(q^{2}-\mbox{M}_{z}^2)}Q_{u}Q_{\mu}
\frac{1}{(sc)^2}v_{u}v_{l} 
\nonumber \\
& + \frac{1}{(\qs -\mbox{M}_{z}^2)^2}
\frac{1}{(sc)^4}(v_{u}^2 + a_{u}^2)(v_{l}^2 + a_{l}^2)\,,
\end{align}
with the electric charges
$(Q_{u},Q_{\mu})=(2/3,-1)$ and
the constants from the Z--boson coupling,
$(v_{u},v_{l},a_{u},a_{\mu})=
(1/4 - 2 s^2/3,
-1/4 + s^2,
-1/4,1/4)$.
The square of the sin of the weak-mixing angle 
and the Z--boson mass
are denoted
as $s^{2}=\sin^{2}\theta_{W}$
and $\mbox{M}_{z}$,
respectively.
We omit the contribution from the 
decay of the on-shell Z boson 
for simplicity.
For reference, the total cross section 
of the process 
$\mbox{L}_{1d} = d\bar{d} \to \mu^{-}\mu^{+}$
is obtained by the replacements
$(Q_{u},v_{u},a_{u}) \to$
$(Q_{d},v_{d},a_{d})=
(-1/3,
-1/4 + s^2/3,
1/4)$\,.
We predict the distribution of
the squared muon-pair invariant
mass, $\mbox{M}_{\mu^{+}\mu^{-}}^{2}=q^{2}$,
because the observable may be the simplest and
most typical one in the Drell-Yan event.
The $\qs$-distribution is calculated as
\begin{equation}
\frac{d \hat{\sigma}(\mbox{L}_{1}) }{d \qs} 
= \hat{\sigma}_{\mbox{{\tiny LO}}}
(\qs) \, \delta(\qs - \hat{s})
 =\hat{\sigma}_{\mbox{{\tiny LO}}}
(\qs) \frac{1}{\qs} \, \delta(1-z)\,,
\label{partlo}
\end{equation}
where the variable $z$ is defined as $z=q^{2}/\hat{s}$.

The NLO corrections are written 
for general processes as
\begin{equation}
\sigma_{\mbox{{\tiny NLO}}} = \sum_{i=1}^{n_{\mbox{{\tiny real}}}} 
\sigma_{{\tiny \mbox{trad}}}(\mbox{R}_{i}), \label{masttrad}
\end{equation}
where the processes $\{ \mbox{R}_{i} \}$
are the real emission processes, which 
are the same as those listed in {\bfseries Step 1} in the
DSA. Each cross section
$\sigma_{{\tiny \mbox{trad}}}(\mbox{R}_{i})$
is written as
\begin{align}
\sigma_{{\tiny \mbox{trad}}}(\mbox{R}_{i}) 
= \int dx_{1} \int dx_{2} \ 
f_{\mbox{{\tiny F}}(x_{a})}(x_{1}) 
f_{\mbox{{\tiny F}}(x_{b})}(x_{2}) 
\,
\biggl[
\hat{\sigma}_{\mbox{{\tiny R}}}(\mbox{R}_{i}) + 
\hat{\sigma}_{\mbox{{\tiny V}}}(\mbox{B}1(\mbox{R}_{i})) + 
\hat{\sigma}_{\mbox{{\tiny C}}}(\mbox{R}_{i}) 
\biggr],  
\label{parttrad}
\end{align}
where the real correction
$\hat{\sigma}_{\mbox{{\tiny R}}}(\mbox{R}_{i})$
is defined in $d$ dimensions, in contrast
with the quantity in Eq.\,(\ref{hatr}).
The virtual correction
$\hat{\sigma}_{\mbox{{\tiny V}}}(\mbox{B}1(\mbox{R}_{i}))$
is the same as in Eq.\,(\ref{hatv}). 
The symbol $\hat{\sigma}_{\mbox{{\tiny C}}}(\mbox{R}_{i})$
represents the collinear subtraction term,
which is concretely shown later.

The Drell--Yan process has the three independent real
emission processes, listed in 
Eq.\,(\ref{listdy}) as
\begin{align}
\mbox{R}_{1} &= u\bar{u} \to \mu^{-}\mu^{+}g\,, \nonumber\\
\mbox{R}_{2} &= ug \to \mu^{-}\mu^{+}u\,,  \nonumber\\
\mbox{R}_{3} &= \bar{u}g \to \mu^{-}\mu^{+}\bar{u}\,.
\end{align}
Each cross section is written as
\begin{align}
\sigma_{{\tiny \mbox{trad}}}(\mbox{R}_{1}) 
&= \int dx_{1} \int dx_{2} \ 
f_{u}(x_{1}) 
f_{\bar{u}}(x_{2}) 
\,
\biggl[
\hat{\sigma}_{\mbox{{\tiny R}}}(\mbox{R}_{1}) + 
\hat{\sigma}_{\mbox{{\tiny V}}}(\mbox{B}1(\mbox{R}_{1})) + 
\hat{\sigma}_{\mbox{{\tiny C}}}(\mbox{R}_{1}) 
\biggr], 
\label{tradcr1}\\
\sigma_{{\tiny \mbox{trad}}}(\mbox{R}_{2}) 
&= \int dx_{1} \int dx_{2} \ 
f_{u}(x_{1}) 
f_{g}(x_{2}) 
\,
\biggl[
\hat{\sigma}_{\mbox{{\tiny R}}}(\mbox{R}_{2}) + 
\hat{\sigma}_{\mbox{{\tiny C}}}(\mbox{R}_{2}) 
\biggr]\,,
\label{tradcr2}
\end{align}
and the cross section
$\sigma_{{\tiny \mbox{trad}}}(\mbox{R}_{3})$
is obtained by the replacement $u \to \bar{u}$
in $\sigma_{{\tiny \mbox{trad}}}(\mbox{R}_{2})$.
First we show the partonic cross sections included
in the $\sigma_{{\tiny \mbox{trad}}}(\mbox{R}_{1})$
in Eq.\,(\ref{tradcr1}).
%
%
\begin{itemize}
\item $\hat{\sigma}_{\mbox{{\tiny R}}}(\mbox{R}_{1})$
\\
The real correction
$\hat{\sigma}_{\mbox{{\tiny R}}}(\mbox{R}_{i})$
is defined in $d$ dimensions as
\begin{equation}
\hat{\sigma}_{\mbox{{\tiny R}}}(\mbox{R}_{1}) 
= \frac{1}{S_{\mbox{{\tiny R}}_{1}}} \ \Phi(\mbox{R}_{1})_{d} 
\cdot |\mbox{M}(\mbox{R}_{1})|_{d}^{2}\,.
\end{equation}
A Feynman diagram of the real correction is shown
in Fig.\,\ref{fig_s5_2}.
\begin{figure}[t]
\begin{center}
\includegraphics[width=13cm]{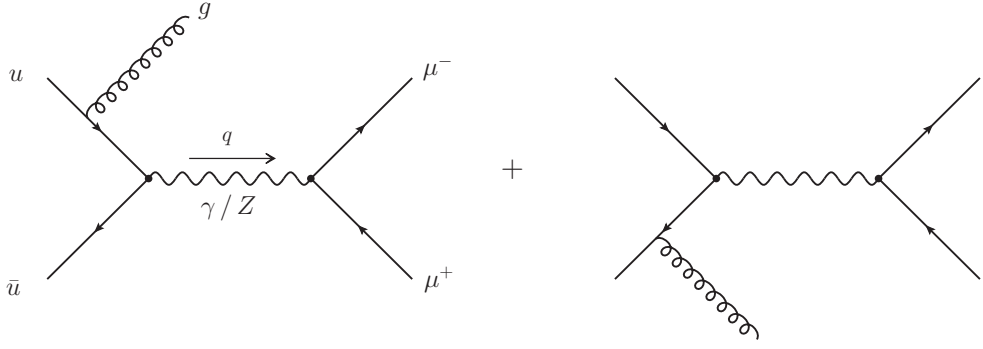}
\end{center} 
\caption{Diagrams of the real emission process 
$\mbox{R}_{1} = u\bar{u} \to \mu^{-}\mu^{+}g$\,.
\label{fig_s5_2} }
\end{figure}
The $\qs$-distribution is calculated as
\begin{eqnarray}
\frac{d 
\hat{\sigma}_{\mbox{{\tiny R}}}(\mbox{R}_{1}) 
}{d \qs} 
= C(q^2,\ep) \, a_{\mbox{{\tiny R}}}\,,
\label{r1r}
\end{eqnarray}
where the factor $C(q^2,\ep)$
is defined as
\begin{equation}
C(q^2,\ep) = 
\frac{\hat{\sigma}_{\mbox{{\tiny LO}}}(\qs)}{\qs}
\, \frac{\al}{\pi}\mbox{C}_{\mbox{{\tiny F}}}
\, \biggl(\frac{4\pi \mu^{2}}{\qs} \biggr)^{\ep}
\, \frac{\Gamma(1-\ep)}{\Gamma(1-2\ep)}\,.
\label{cdef}
\end{equation}
$a_{\mbox{{\tiny R}}}$ is written as 
\begin{equation}
a_{\mbox{{\tiny R}}}=
\delta(1-z)\frac{1}{\ep^{2}}+
z(1+z^2)\biggl[
-\frac{1}{\ep}\frac{1}{(1-z)_{\mbox{\tiny +}}} + 
2\biggl(\frac{\ln(1-z)}{1-z} \biggr)_{\mbox{\tiny +}}
-\frac{1}{1-z}\ln z \biggr] \, + \, O(\ep)\,.
\label{ar}
\end{equation}
This expression includes the so-called 
\textquoteleft +\textquoteright-distribution,
which is defined for
arbitrary functions $g(z)$ and $h(z)$ as
\begin{equation}
\int_{\tau}^{1} dz \ g(z) \ (h(z))_{+} = 
\int_{\tau}^{1} dz \ g(z) \ h(z) - \int_{0}^{1} dz \ g(z=1) \ h(z)\,,
\end{equation}
with a value $0 \le \tau < 1$.
\item $\hat{\sigma}_{\mbox{{\tiny V}}}(\mbox{B}1(\mbox{R}_{1}))$
\\
The virtual correction 
$\hat{\sigma}_{\mbox{{\tiny V}}}(\mbox{B}1(\mbox{R}_{1}))$
is defined in 
Eq.\,(\ref{hatv}) as
\begin{equation}
\hat{\sigma}_{\mbox{{\tiny V}}}(\mbox{B}1) 
= \frac{1}{S_{\mbox{{\tiny B1}} }} \ \Phi( \mbox{B}1  )_{d} 
\cdot |\mbox{M}_{\mbox{{\tiny virt}}}(\mbox{B}1)|_{d}^{2}\,.  
\end{equation}
A Feynman diagram of the virtual one-loop correction is 
shown in Fig.\,\ref{fig_s5_3}.
\begin{figure}[t]
\begin{center}
\includegraphics[width=8cm]{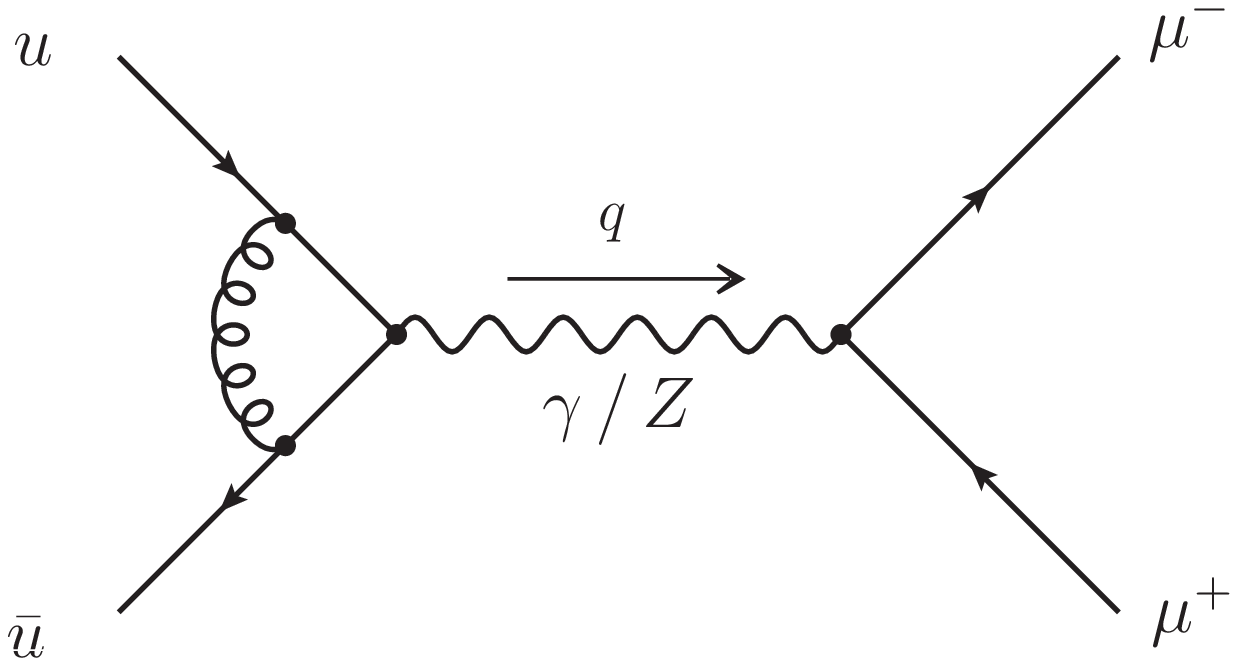}
\end{center} 
\caption{ A one-loop diagram of the process
$\mbox{B}1(\mbox{R}_{1})=u\bar{u} \to \mu^{-}\mu^{+}$.
\label{fig_s5_3} }
\end{figure}
The $\qs$-distribution is calculated as
\begin{eqnarray}
\frac{d 
\hat{\sigma}_{\mbox{{\tiny V}}}(\mbox{B}1) 
}{d \qs} 
= C(q^2,\ep) \, a_{\mbox{{\tiny V}}}\,,
\label{r1v}
\end{eqnarray}
where the factor $C(q^2,\ep)$
is the same as in Eq.\,(\ref{cdef})
and,
after the subtraction of the UV pole
$1/\ep_{{\tiny \mbox{UV}}}$ by the renormalization
program,
$a_{\mbox{{\tiny V}}}$ is obtained 
as 
\begin{equation}
a_{\mbox{{\tiny V}}} = \delta (1-z) 
\biggl[ -\frac{1}{\ep^2} - \frac{3}{2}\frac{1}{\ep} + 
\biggl(\frac{\pi^2}{3} -4 \biggr)   \biggr]\,.
\label{av}
\end{equation}
\item $\hat{\sigma}_{\mbox{{\tiny C}}}(\mbox{R}_{1})$
\\
The $\qs$-distribution of the
collinear subtraction term is written as
\begin{equation}
\frac{d
\,\hat{\sigma}_{\mbox{{\tiny C}}}\,(\mbox{R}_{1})}{d q^{2}}
=-\frac{\al}{\pi}\frac{1}{\Gamma(1-\ep)}
\int_{0}^{1} dx \ 
\biggl[-\frac{1}{\ep} 
\biggl(\frac{4\pi \mu^{2}}{\mu_{F}^{2}} \biggr)^{\ep} P^{\,uu}(x)
\biggr]
\frac{d \hat{\sigma}(\mbox{L}_{1}, x\hat{s})}{d \qs}\,,
\label{collst}
\end{equation}
where the $\qs$-distribution of the LO process
$\mbox{L}_{1}$ with the rescaled initial energy $x\hat{s}$
is defined as
\begin{equation}
\frac{d \hat{\sigma}(\mbox{L}_{1}, x \hat{s})}{d \qs} 
= \hat{\sigma}_{\mbox{{\tiny LO}}}
(\qs) \, \delta(\qs - x\hat{s})
 =\hat{\sigma}_{\mbox{{\tiny LO}}}
(\qs) \frac{z}{\qs} \, \delta(z-x)\,.
\label{rescalelo}
\end{equation}
The symbol $\mu_{F}$ represents the mass factorization
scale.
The {\em four-dimensional} Altarelli-Parisi splitting
function
$P^{\,uu}(w)=P^{\,ff}(w)$
is written in Eq.\,(\ref{alpff}).
There are two contributions to the collinear subtraction
term. One is the contribution of the gluon radiation
from the $u$-quark leg
in the initial state in the process 
$\mbox{R}_{1} = u\bar{u} \to \mu^{-}\mu^{+}g$.
The other is the contribution of the
radiation from the
$\bar{u}$-quark leg in $\mbox{R}_{1}$.
Both contributions have identical expressions
and the expression in Eq.\,(\ref{collst})
includes both contributions.
The $\qs$-distribution is calculated as
\begin{eqnarray}
\frac{d
\,\hat{\sigma}_{\mbox{{\tiny C}}}\,(\mbox{R}_{1})}{d q^{2}}
= C(q^2,\ep) \, a_{\mbox{{\tiny C}}}\,,
\label{r1c}
\end{eqnarray}
where the factor $C(q^2,\ep)$
is the same as in Eq.\,(\ref{cdef})
and $a_{\mbox{{\tiny C}}}$ is written as 
\begin{equation}
a_{\mbox{{\tiny C}}} 
= \frac{zP^{\, ff}(z)}{\mbox{C}_{\mbox{{\tiny F}}}} 
\biggl(\frac{1}{\ep} + 
\ln \frac{\qs}{\mu_{F}^{2}} 
\biggr) + O(\ep).
\end{equation}
\end{itemize}
%
%
The summation of the three contributions 
in Eqs.\,(\ref{r1r}),\,(\ref{r1v}),
and (\ref{r1c}), is free from divergences as follows\,:
\begin{align}
\frac{d
\hat{\sigma}_{{\tiny \mbox{trad}}}(\mbox{R}_{1}) 
}{d \qs}
&=
\frac{d 
\hat{\sigma}_{\mbox{{\tiny R}}}(\mbox{R}_{1}) 
}{d \qs}
+
\frac{d 
\hat{\sigma}_{\mbox{{\tiny V}}}(\mbox{B}1) 
}{d \qs} 
+
\frac{d
\,\hat{\sigma}_{\mbox{{\tiny C}}}\,(\mbox{R}_{1})}{d q^{2}}
\nonumber\\
&=C(q^2,\ep) \, 
(
a_{\mbox{{\tiny R}}}+
a_{\mbox{{\tiny V}}}+
a_{\mbox{{\tiny C}}}
)\,,
\label{partr1}
\end{align}
with the summation
\begin{align}
a_{\mbox{{\tiny V}}} + a_{\mbox{{\tiny R}}} + a_{\mbox{{\tiny C}}} 
&= \delta(1-z)\biggl(\frac{\pi^2}{3} -4  \biggr)
+
\frac{zP^{\,ff}(z)}{\mbox{C}_{\mbox{{\tiny F}}}}\ln \frac{\qs}{\mu_{F}^{2}} 
\nonumber\\
&\ \ + z(1+z^2) 
\biggl[ 2\biggl(\frac{\ln(1-z)}{1-z} \biggr)_{+}- \frac{1}{1-z} \ln z 
\biggr]\,.
\label{trar1res}
\end{align}
At this stage, with the finite results in 4 dimensions,
we can also return the common factor $C(q^2,\ep)$
to 4 dimensions as
\begin{equation}
C(q^2,\ep=0)=\frac{\hat{\sigma}_{\mbox{{\tiny LO}}}^{\,(4)}(\qs)}
{\qs}\, \frac{\al}{\pi}\mbox{C}_{\mbox{{\tiny F}}}\,,
\end{equation}
with the total cross section at LO in 4 dimensions
\begin{equation}
\hat{\sigma}_{\mbox{{\tiny LO}}}^{\,(4)}(\qs)
=\ale^{2} \, \frac{4 \pi}{3 \, N_{c}} 
\, q^{2} \ \mbox{P}(\qs)\,.
\end{equation}

Next we show the partonic cross sections in
$\sigma_{{\tiny \mbox{trad}}}(\mbox{R}_{2})$
in Eq.\,(\ref{tradcr2}).
%
%
\begin{itemize}
\item $\hat{\sigma}_{\mbox{{\tiny R}}}(\mbox{R}_{2})$
\\
The real correction
$\hat{\sigma}_{\mbox{{\tiny R}}}(\mbox{R}_{2})$
is defined as
\begin{equation}
\hat{\sigma}_{\mbox{{\tiny R}}}(\mbox{R}_{2}) 
= \frac{1}{S_{\mbox{{\tiny R}}_{2}}} \ \Phi(\mbox{R}_{2})_{d} 
\cdot |\mbox{M}(\mbox{R}_{2})|_{d}^{2}\,.
\end{equation}
Feynman diagrams of the real correction are shown
in Fig.\,\ref{fig_s5_4}.
\begin{figure}[t]
\begin{center}
\includegraphics[width=14cm]{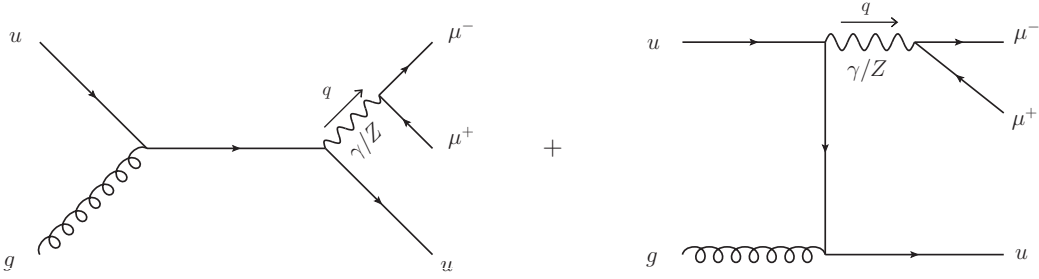}
\end{center} 
\caption{The diagrams of the real emission process
$\mbox{R}_{2} = ug \to \mu^{-}\mu^{+}u$\,.
\label{fig_s5_4} }
\end{figure}
The $\qs$-distribution is calculated as
\begin{eqnarray}
\frac{d 
\hat{\sigma}_{\mbox{{\tiny R}}}(\mbox{R}_{2}) 
}{d \qs} 
= C(q^2,\ep) \, a_{\mbox{{\tiny R}},ug}\,,
\label{r2r}
\end{eqnarray}
with
\begin{align}
a_{\mbox{{\tiny R}},ug} =  
\frac{z}{2\,\mbox{C}_{\mbox{{\tiny F}}}} \biggl[-\frac{1}{\ep} P^{\,gf}(z) 
+  \ln \frac{(1-z)^{2}}{z}P^{\,gf}(z)
+\frac{1}{4}(1+6z-7z^{2})\biggr] + O(\ep)\,,
\label{arug}
\end{align}
where the splitting function $P^{\,gf}(z)$
is written in Eq.\,(\ref{alpgf}).
\item $\hat{\sigma}_{\mbox{{\tiny C}}}(\mbox{R}_{2})$
\\
The $\qs$-distribution of the
collinear subtraction term is written as
\begin{equation}
\frac{d
\,\hat{\sigma}_{\mbox{{\tiny C}}}\,(\mbox{R}_{2})}{d q^{2}}
=-\frac{\al}{2 \pi}\frac{1}{\Gamma(1-\ep)}
\int_{0}^{1} dx \ 
\biggl[-\frac{1}{\ep} 
\biggl(\frac{4\pi \mu^{2}}{\mu_{F}^{2}} \biggr)^{\ep} P^{\,g\bar{u}}(x)
\biggr]
\frac{d \hat{\sigma}(\mbox{L}_{1}, x\hat{s})}{d \qs}\,, 
\end{equation}
where the $\qs$-distribution of the LO process
is the same as in Eq.\,(\ref{rescalelo})
and the splitting function
$P^{\,g\bar{u}}(z)=P^{\,gf}(z)$
is shown in Eq.\,(\ref{alpgf}).
The distribution is written in the form
\begin{align}
\frac{d
\,\hat{\sigma}_{\mbox{{\tiny C}}}\,(\mbox{R}_{2})}{d q^{2}}
&= C(q^2,\ep) \, a_{\mbox{{\tiny C}},\,ug}\,,
\label{r2c} \\
a_{\mbox{{\tiny C}},\,ug} 
&= \frac{z\,P^{\,gf}(z) }{\mbox{C}_{\mbox{{\tiny F}}}}
\biggl(
\frac{1}{\ep}  
+ \ln \frac{\qs}{\mu_{F}^{2}}
\biggr)
+ O(\ep)\,.
\end{align}
\end{itemize}
%
%
The summation of the two contributions 
in Eqs.\,(\ref{r2r})
and (\ref{r2c}) is free from divergences as follows\,:
\begin{align}
\frac{d
\hat{\sigma}_{{\tiny \mbox{trad}}}(\mbox{R}_{2}) 
}{d \qs}
&=
\frac{d 
\hat{\sigma}_{\mbox{{\tiny R}}}(\mbox{R}_{2}) 
}{d \qs}
+
\frac{d
\,\hat{\sigma}_{\mbox{{\tiny C}}}\,(\mbox{R}_{2})}{d q^{2}}
\nonumber \\
&=C(q^2,\ep=0) \, 
(
a_{\mbox{{\tiny R}},ug}+
a_{\mbox{{\tiny C}},ug}
)\,,
\label{partr2}
\end{align}
with
\begin{align}
a_{\mbox{{\tiny R}},ug}+a_{\mbox{{\tiny C}},ug}
=\frac{z}{2\,\mbox{C}_{\mbox{{\tiny F}}}}\biggl[
P^{\,gf}(z) \cdot \ln \frac{(1-z)^{2}\qs}{z \, \mu_{F}^{2}} 
+ \frac{1}{4}(1+6z-7z^{2}) \ \biggr]\,.
\label{arcug}
\end{align}

Taking account of the exchanged initial states
and all five massless flavors,
we obtain the prediction of the 
$\qs$-distribution at NLO accuracy as
\begin{equation}
\frac{d \,
\sigma_{\mbox{{\tiny prediction}}}
}{d q^{2}} =
\frac{d \,
\sigma(\mbox{L}_{1})
}{d q^{2}} +
\sum_{i=1}^{3}
\frac{d \,
\sigma_{{\tiny \mbox{trad}}}(\mbox{R}_{i}) 
}{d q^{2}}\,.
\end{equation}
The contributions from the distributions
$d \sigma(\mbox{L}_{1})/d \qs$
and
$d \sigma_{{\tiny \mbox{trad}}}(\mbox{R}_{1})/d \qs$
are written as
\begin{align}
\frac{d \,
\sigma(\mbox{L}_{1})
}{d q^{2}} +
\frac{d \,
\sigma_{{\tiny \mbox{trad}}}(\mbox{R}_{1}) 
}{d q^{2}}
&= \sum_{q=u,c,d,b,s} \int_{\tau}^{1} dx_{1}
\int_{\tau/x_{1}}^{1} dz \ \frac{\tau}{x_{1}z^2} 
\ H_{q\bar{q}}(x_{1},x_{2}) 
\nonumber \\
&\hspace{35mm}
\times 
\biggl[\,
\frac{d \,
\hat{\sigma}(\mbox{L}_{1})
}{d q^{2}} 
\, + \,
\frac{d \,
\hat{\sigma}_{{\tiny \mbox{trad}}}(\mbox{R}_{1}) 
}{d q^{2}}
\, \biggr]\,,
\end{align}
where the combination of PDFs is denoted as
$H_{q\bar{q}}(x_{1},x_{2})=f_{q}(x_{1})f_{\bar{q}}(x_{2}) 
+ f_{\bar{q}}(x_{1})f_{q}(x_{2})$.
The value $x_{2}$ is 
expressed as $x_{2}=\tau/(x_{1}z)$ with $\tau=\qs/S$.
The partonic distributions
$d \,\hat{\sigma}(\mbox{L}_{1})/d q^{2}$
and
$d \,\hat{\sigma}_{{\tiny \mbox{trad}}}(\mbox{R}_{1}) 
/d q^{2}$
are written in
Eqs.\,(\ref{partlo}) and (\ref{partr1}), 
respectively.
The contributions from the distributions
$d \sigma_{{\tiny \mbox{trad}}}(\mbox{R}_{2})/d \qs$
and
$d \sigma_{{\tiny \mbox{trad}}}(\mbox{R}_{3})/d \qs$
are written as
\begin{align}
\frac{d \,
\sigma_{{\tiny \mbox{trad}}}(\mbox{R}_{2}) 
}{d q^{2}}
+
\frac{d \,
\sigma_{{\tiny \mbox{trad}}}(\mbox{R}_{3}) 
}{d q^{2}}
&= \sum_{q=u,c,d,b,s} \int_{\tau}^{1} dx_{1}
\int_{\tau/x_{1}}^{1} dz \ \frac{\tau}{x_{1}z^2} 
\,\biggl[
\nonumber \\
&\hspace{15mm}
H_{qg}(x_{1},x_{2})
\frac{d \,
\hat{\sigma}_{{\tiny \mbox{trad}}}(\mbox{R}_{2}) 
}{d q^{2}}
\, + \,H_{\bar{q}g}(x_{1},x_{2})  
\frac{d \,
\hat{\sigma}_{{\tiny \mbox{trad}}}(\mbox{R}_{3}) 
}{d q^{2}}
\, \biggr]\,,
\end{align}
with the PDFs $H_{qg}(x_{1},x_{2})=f_{q}(x_{1}) f_{g}(x_{2}) 
+ f_{g}(x_{1}) f_{q}(x_{2})$.
The $q^{2}$-distribution of the partonic distribution
$d \,\hat{\sigma}_{{\tiny \mbox{trad}}}(\mbox{R}_{2}) 
/d q^{2}$
is written in Eq.\,(\ref{partr2}).
The expression of the partonic distribution
$d \,\hat{\sigma}_{{\tiny \mbox{trad}}}(\mbox{R}_{3})$ 
$/d q^{2}$
is identical to
$d \,\hat{\sigma}_{{\tiny \mbox{trad}}}(\mbox{R}_{2}) 
/d q^{2}$.
%
%
%
\subsection{DSA \label{s5_2}}
We use the dipole subtraction 
procedure through the DSA to obtain
analytical results of the $\qs$-distribution 
in the Drell--Yan process.
The DSA has already been applied 
to the Drell--Yan process 
in Sec.\,\ref{sec_3}.
The results are summarized 
in Sec.\,\ref{s3_5}.

We start with the explicit calculation of
$\sigma(\mbox{R}_{1})$ in Eq.\,(\ref{dyhcr1}).
The subtracted real cross section
$(\hat{\sigma}_{\mbox{{\tiny R}}}(\mbox{R}_{1}) 
-\hat{\sigma}_{\mbox{{\tiny D}}}(\mbox{R}_{1}))$
in Eq.\,(\ref{dypcr1r}) is defined in 4 dimensions.
However, the analytical integration of the PS 
in 4 dimensions does not seem easy.
Instead, we redefine the cross section 
in $d$ dimensions and regularize
the soft and collinear singularities 
as the poles of 
$1/\ep$ and $1/\ep^{2}$, 
which are produced 
by the $d$-dimensional PS integration.
The poles from
$\hat{\sigma}_{\mbox{{\tiny R}}}(\mbox{R}_{1})$
and
$\hat{\sigma}_{\mbox{{\tiny D}}}(\mbox{R}_{1})$
cancel each other, and a finite 
analytical expression in 4 dimensions is 
obtained. The distribution 
$d\hat{\sigma}_{\mbox{{\tiny R}}}(\mbox{R}_{1}) 
/d\qs$
is calculated in 
Eqs.\,(\ref{r1r}) and (\ref{ar}).
Then we proceed to the calculation of
$\hat{\sigma}_{\mbox{{\tiny D}}}(\mbox{R}_{1})$,
which is now defined in $d$ dimensions as
\begin{equation}
\hat{\sigma}_{\mbox{{\tiny D}}}(\mbox{R}_{1}) 
= \frac{1}{S_{\mbox{{\tiny R}}_{1}}} \ \Phi(\mbox{R}_{1})_{d} 
\, \frac{1}{n_{s}(u) n_{s}(\bar{u})} \, \mbox{D}(\mbox{R}_{1})\,.
\label{r1dipind}
\end{equation}
There are two dipole terms for the process $\mbox{R}_{1}$,
as shown in Eq.\,(\ref{dy1do}).
The first dipole term,
$\mbox{D}( {\tt dip}1,\,(3)\mbox{-}2)_{a3,b}$\,,
is written in Eq.\,(\ref{dyr1dip1}),
which is now defined in $d$ dimensions.
The dipole splitting function $\mbox{V}_{a3,b}$
in $d$ dimensions reads in Ref.\,\cite{Catani:1996vz} as
\begin{equation}
\mbox{V}_{a3,b} 
= 8 \pi \mu^{2 \ep} \al \mbox{C}_{\mbox{{\tiny F}}}
\biggl[ \frac{2}{1 - x_{3,ab}} - 1 - x_{3,ab}
- \ep(1 - x_{3,ab}) \biggr]\,,
\end{equation}
and the square of the reduced Born process
$\langle \mbox{B}1  \ | 
\mbox{T} \cdot \mbox{T}
 | \ \mbox{B}1 \rangle$
is also defined in $d$ dimensions.
The contribution to the partonic cross section 
of the dipole terms can be analytically 
integrated over the PS region including
the soft and collinear singularities.
The analytical integration of the dipole terms
for an arbitrary process is one core part
of the construction of the dipole 
subtraction procedure \cite{Catani:1996vz}.
The contribution of the first dipole term
to the cross section in Eq.\,(\ref{r1dipind}) 
is written as 
$\hat{\sigma}_{\mbox{{\tiny D}}}(\mbox{R}_{1} : \mbox{D}_{a3,b})$,
and
the integration formula is applied
to the cross section as
\begin{align}
\hat{\sigma}_{\mbox{{\tiny D}}}(\mbox{R}_{1} : \mbox{D}_{a3,b}) 
&= 
-\int_{0}^{1}dx\,  
\frac{1}{S_{\mbox{{\tiny R}}_{1}}}
\,\Phi_{a}(\mbox{R}_{1}:\mbox{B}_{1},x)_{d} 
\, \frac{1}{n_{s}(u) n_{s}(\bar{u})} 
\,\frac{\al}{2\pi}
\, \frac{1}{\Gamma(1-\ep)}\,
\nonumber\\
&\hspace{20mm} \cdot \biggl(\frac{4\pi \mu^{2}}{s_{ab}} \biggr)^{\ep}
{\widetilde {\cal V}}^{u,u}(x;\ep)
\frac{1}{\mbox{C}_{\mbox{{\tiny F}}}} 
\langle \mbox{B}1  \ | 
\mbox{T}_{y_{a}} \cdot \mbox{T}_{y_{b}}
 | \ \mbox{B}1 \rangle\,, 
\end{align}
where the Lorentz scalar is denoted
as $s_{ab}=2 p_{a} \cdot p_{b}$
and the function
${\widetilde {\cal V}}^{u,u}(x;\ep)$
reads in Ref.\,\cite{Catani:1996vz} as
\begin{align}
{\widetilde {\cal V}}^{u,u}(x;\ep)
&= - \frac{1}{\ep} P^{ff}(x) + 
\delta(1-x)
\biggl[
{\cal V}_{fg} (\ep) 
+
\mbox{C}_{\mbox{{\tiny F}}}
\Bigl(
\frac{\pi^2}{3} - 5
\Bigr)
\biggr]
\nonumber\\
& \ \ 
+\mbox{C}_{\mbox{{\tiny F}}} \biggl[
-\Bigl(\frac{4}{1-x} \ln \frac{1}{1-x} \Bigr)_{+} 
+1-x - 2\,(1+x)\ln \, (1-x)
\biggl]\,.
\end{align}
The correlation of the two color insertion operators
in the square of the Born process $\mbox{B}1$
is fully factorized as
$\langle \mbox{B}1  \ | 
\mbox{T}_{y_{a}} \cdot \mbox{T}_{y_{b}}
 | \ \mbox{B}1 \rangle
=- \mbox{C}_{\mbox{{\tiny F}}}
 \langle \mbox{B}1  | \mbox{B}1 \rangle$.
The $\qs$-distribution is written as
\begin{align}
\frac{d 
\hat{\sigma}_{\mbox{{\tiny D}}}(\mbox{R}_{1} : \mbox{D}_{a3,b}) 
}{d \qs}
&=\frac{\al}{2\pi}\frac{1}{\Gamma(1-\ep)}
\biggl(\frac{4\pi \mu^{2}}{s_{ab}} \biggr)^{\ep}
\int_{0}^{1} dx \ 
{\widetilde {\cal V}}^{u,u}(x;\ep)
\frac{d \hat{\sigma}(\mbox{L}_{1}, x\hat{s})}{d \qs} 
\nonumber\\
&=\frac{\al}{2\pi}\frac{1}{\Gamma(1-\ep)}
\biggl(\frac{4\pi \mu^{2}}{s_{ab}} \biggr)^{\ep}
{\widetilde {\cal V}}^{u,u}(z;\ep)
\frac{\hat{\sigma}_{\mbox{{\tiny LO}}}(\qs)}
{\qs} z\,,
\end{align}
where the distribution with the scaled initial 
energy, 
$d \hat{\sigma}(\mbox{L}_{1}, x\hat{s})/ d \qs$,
is defined in Eq.\,(\ref{rescalelo}).
The contribution from the second dipole
in Eq.\,(\ref{dyr1dip2})
is written identically.
Then the contribution of all dipole
terms in $\mbox{D}(\mbox{R}_{1})$
is written in the form
\begin{eqnarray}
\frac{d 
\hat{\sigma}_{\mbox{{\tiny D}}}(\mbox{R}_{1}) 
}{d \qs} 
= C(q^2,\ep) \, a_{\mbox{{\tiny D}}}\,,
\end{eqnarray}
where the common factor $C(q^2,\ep)$ is the same 
as in Eq.\,(\ref{cdef}),
and $a_{\mbox{{\tiny D}}}$
is written as
\begin{equation}
a_{\mbox{{\tiny D}}} = 
\frac{1}{\mbox{C}_{\mbox{{\tiny F}}}} 
\, \frac{\Gamma(1-2\ep)}{\Gamma(1-\ep)^{2}}\,
\, \biggl(\frac{\qs}{s_{ab}} \biggr)^{\ep}
{\widetilde {\cal V}}^{f,f}(z;\ep)
\,z\,,
\end{equation}
which is expanded by $1/\ep$ as
\begin{align}
a_{\mbox{{\tiny D}}}
&=
\delta(1-z)\frac{1}{\ep^{2}}+
z 	\biggl[
-\frac{1}{\ep}\frac{1+z^2}{(1-z)_{\mbox{\tiny +}}} 
-
\biggl(
\frac{4}{1-z}
\ln \frac{1}{1-z}
\biggr)_{\mbox{\tiny +}}
+1-z-z(1+z) \ln (1-z) 
\nonumber\\
&\hspace{30mm}-
\frac{1+z^{2}}{
(1-z)_{\mbox{\tiny +}}
}
\ln z
\biggr] 
\, + \, O(\ep)\,.
\end{align}
Recalling the expression of $a_{\mbox{{\tiny R}}}$
in Eq.\,(\ref{ar}),
the $\qs$-distribution of the subtracted
real correction is written as
\begin{equation}
\frac{d 
\hat{\sigma}_{\mbox{{\tiny R}}}(\mbox{R}_{1}) 
}{d \qs} 
-
\frac{d 
\hat{\sigma}_{\mbox{{\tiny D}}}(\mbox{R}_{1}) 
}{d \qs} 
=
C(q^2,\ep=0) \, 
(
a_{\mbox{{\tiny R}}}-a_{\mbox{{\tiny D}}}
)\,,
\end{equation}
with the difference
\begin{equation}
a_{\mbox{{\tiny R}}}-a_{\mbox{{\tiny D}}}=
-z(1-z)\,.
\label{armd}
\end{equation}
Next we calculate the subtracted
virtual correction 
$(\hat{\sigma}_{\mbox{{\tiny V}}}(\mbox{B}1(\mbox{R}_{1})) + 
\hat{\sigma}_{\mbox{{\tiny I}}}(\mbox{R}_{1}))$
in Eq.\,(\ref{dypcr1v}).
The concrete expression of 
the virtual correction
$\hat{\sigma}_{\mbox{{\tiny V}}}(\mbox{B}1(\mbox{R}_{1}))$
is written in 
Eqs.\,(\ref{r1v}) and (\ref{av}).
The expression of the I term
$\mbox{I}(\mbox{R}_{1})$
is given in Eq.\,(\ref{dy1io})
and the contribution to the 
partonic cross section is 
written as
\begin{align}
\hat{\sigma}_{\mbox{{\tiny I}}}(\mbox{R}_{1}) 
&= 
\frac{1}{S_{\mbox{{\tiny B1}} }} \ 
\Phi( \mbox{B}1  )_{d} 
\ \mbox{I}(\mbox{R}_{1}) 
\\
&=
\frac{1}{S_{\mbox{{\tiny B1}} }} \ 
\Phi( \mbox{B}1  )_{d} 
\ 2 A_{d} \, {\cal V}_{f}
\, s_{ab}^{-\ep}
\, \langle \mbox{B}1  \ | 
 \ \mbox{B}1 \rangle\,.
\end{align}
The $\qs$-distribution is calculated in the form
\begin{equation}
\frac{d 
\hat{\sigma}_{\mbox{{\tiny I}}}(\mbox{R}_{1}) 
}{d \qs} 
= C(q^2,\ep) \, a_{\mbox{{\tiny I}}}\,,
\end{equation}
with
\begin{align}
a_{\mbox{{\tiny I}}} 
&=
\delta(1-z) \,
\frac{1}{\mbox{C}_{\mbox{{\tiny F}}}} 
\, \frac{\Gamma(1-2\ep)}{\Gamma(1-\ep)^{2}}\,
\, {\cal V}_{f}
\nonumber \\
&=\delta (1-z) 
\biggl[ \frac{1}{\ep^2} + \frac{3}{2}\frac{1}{\ep} 
- \frac{\pi^2}{3} +5  
\biggr]\, + \, O(\ep)\,.
\end{align}
Then we obtain the $\qs$-distribution
of the subtracted virtual correction as
\begin{equation}
\frac{d 
\hat{\sigma}_{\mbox{{\tiny V}}}
(\mbox{B}1(\mbox{R}_{1}))
}{d \qs}
-
\frac{d 
\hat{\sigma}_{\mbox{{\tiny I}}}(\mbox{R}_{1}) 
}{d \qs} 
= C(q^2,\ep=0) \, 
(a_{\mbox{{\tiny V}}} - a_{\mbox{{\tiny I}}})\,,
\end{equation}
with the difference
\begin{equation}
a_{\mbox{{\tiny V}}} - a_{\mbox{{\tiny I}}} 
= \delta(1-z)\,.
\label{avmi}
\end{equation}
The cross section of the P and K terms 
is written in 
Eq.\,(\ref{dypcr1pk}).
The P and K terms are obtained
in Eqs.\,(\ref{dyr1pout}) and (\ref{dyr1kout}),
and 
the cross section is explicitly written down as
\begin{align}
\hat{\sigma}_{\mbox{{\tiny P}}}(\mbox{R}_{1}) 
+
\hat{\sigma}_{\mbox{{\tiny K}}}(\mbox{R}_{1}) 
&= \int_{0}^{1}dx\,  
\frac{1}{S_{\mbox{{\tiny B}}_{1} }}
\,\Phi_{a}(\mbox{R}_{1}:\mbox{B}_{1},x)_{4} \,
\frac{\al}{2\pi} 
\nonumber \\
&\hspace{15mm}\biggl[
-P^{uu}(x) \,
\ln \frac{\mu_{F}^{2}}{x\,s_{ab}}
+
\bar{K}^{uu}(x) 
+
\widetilde{K}^{uu}(x) 
\biggr]
\, \langle \mbox{B}1  \ | 
 \ \mbox{B}1 \rangle \,
\nonumber \\
&\hspace{5mm}
\ + \
(a \leftrightarrow b)\,.
\end{align}
The $\qs$-distribution is calculated as
\begin{align}
\frac{d 
\hat{\sigma}_{\mbox{{\tiny P}}}(\mbox{R}_{1}) 
}{d \qs}
+
\frac{d 
\hat{\sigma}_{\mbox{{\tiny K}}}(\mbox{R}_{1}) 
}{d \qs}
&= \int_{0}^{1}dx\,  
\frac{\al}{\pi} 
\biggl[
-P^{uu}(x) \,
\ln \frac{\mu_{F}^{2}}{x\,s_{ab}}
+
\bar{K}^{uu}(x) 
+
\widetilde{K}^{uu}(x) 
\biggr]
\frac{d \hat{\sigma}(\mbox{L}_{1}, x\hat{s})}{d \qs} 
\nonumber \\
&=
C(q^2,\ep=0) \ a_{\mbox{{\tiny PK}}}\,,
\end{align}
with the expression
\begin{align}
a_{\mbox{{\tiny PK}}}
&=
\frac{z}{\mbox{C}_{\mbox{{\tiny F}}}} 
\biggl[
-P^{ff}(z) \,
\ln \frac{\mu_{F}^{2}}{\qs}
+
\bar{K}^{ff}(z) 
+
\widetilde{K}^{ff}(z) 
\biggr] \\
&=
\delta(1-z)
\Bigl(
\frac{\pi^2}{3} - 5
\Bigr)
+
\frac{z P^{ff}(z)}{\mbox{C}_{\mbox{{\tiny F}}}} 
\, \ln \frac{\qs}{\mu_{F}^{2}}
\nonumber \\
&\hspace{8mm}+z(1+z^{2})
\biggl[
\biggl(
\frac{2}{1-z}
\ln (1-z)
\biggr)_{\mbox{\tiny +}}
-
\frac{1}{(1-z)}
\ln z
\biggr]\,.
\label{apk}
\end{align}
Then the $\qs$-distribution
$d \hat{\sigma}(\mbox{R}_{1}) /d \qs$
is written as
\begin{align}
\frac{d 
\hat{\sigma}(\mbox{R}_{1}) 
}{d \qs} 
&=
\frac{d}{d \qs} 
\Bigl(
\hat{\sigma}_{\mbox{{\tiny R}}}(\mbox{R}_{1}) 
-
\hat{\sigma}_{\mbox{{\tiny D}}}(\mbox{R}_{1}) 
+
\hat{\sigma}_{\mbox{{\tiny V}}}
(\mbox{B}1(\mbox{R}_{1}))
+
\hat{\sigma}_{\mbox{{\tiny I}}}(\mbox{R}_{1}) 
+
\hat{\sigma}_{\mbox{{\tiny P}}}(\mbox{R}_{1}) 
+
\hat{\sigma}_{\mbox{{\tiny K}}}(\mbox{R}_{1}) 
\Bigr)
\nonumber\\
&=
C(q^2,\ep=0) \,
\bigl[
(a_{\mbox{{\tiny R}}} - a_{\mbox{{\tiny D}}})
+
(a_{\mbox{{\tiny V}}} + a_{\mbox{{\tiny I}}})
+
a_{\mbox{{\tiny PK}}}
\bigr]\,,
\end{align}
where the finite quantities 
$(a_{\mbox{{\tiny R}}} - a_{\mbox{{\tiny D}}})$,
$(a_{\mbox{{\tiny V}}} + a_{\mbox{{\tiny I}}})$,
and $a_{\mbox{{\tiny PK}}}$
are written in
Eqs.\,(\ref{armd}), (\ref{avmi}), 
and (\ref{apk}).
The summation of the three contributions is
calculated as
\begin{align}
(a_{\mbox{{\tiny R}}} - a_{\mbox{{\tiny D}}})
+
(a_{\mbox{{\tiny V}}} + a_{\mbox{{\tiny I}}})
+
a_{\mbox{{\tiny PK}}}
&=
 \delta(1-z)\biggl(\frac{\pi^2}{3} -4  \biggr)
+
\frac{zP^{\,ff}(z)}{\mbox{C}_{\mbox{{\tiny F}}}}
\ln \frac{\qs}{\mu_{F}^{2}} 
\nonumber\\
&\ \ + z(1+z^2) 
\biggl[ 2\biggl(\frac{\ln(1-z)}{1-z} \biggr)_{+}- 
\frac{1}{1-z} \ln z 
\biggr]\,.
\end{align}
The results exactly coincide with the results obtained
by the traditional methods in
Eq.\,(\ref{trar1res}).

Next we calculate the $\qs$-distribution
of $\sigma(\mbox{R}_{2})$
in Eq.\,(\ref{dyhcr2}).
The subtracted real correction is 
written in 
Eq.\,(\ref{dypcr2r}).
Similarly to the case of $\sigma(\mbox{R}_{1})$,
we redefine the cross section in $d$ dimensions.
The distribution 
$d\hat{\sigma}_{\mbox{{\tiny R}}}(\mbox{R}_{2}) 
/d\qs$
is obtained in 
Eqs.\,(\ref{r2r}) and (\ref{arug}).
Then we proceed to the calculation of
$\hat{\sigma}_{\mbox{{\tiny D}}}(\mbox{R}_{2})$,
which is defined in $d$ dimensions as
\begin{equation}
\hat{\sigma}_{\mbox{{\tiny D}}}(\mbox{R}_{2}) 
=\frac{1}{S_{\mbox{{\tiny R}}_{2}}} 
\Phi(\mbox{R}_{2})_{d} 
\ \frac{1}{n_{s}(u) n_{s}(g)} \mbox{D}(\mbox{R}_{2})\,. 
\end{equation}
For the process $\mbox{R}_{2}$, only one dipole
exists in Eq.\,(\ref{dy2do}) and the 
expression is given in Eq.\,(\ref{dy2dip1}).
The cross section is analytically integrated
over the soft and collinear regions,
and the $\qs$-distribution is calculated as
\begin{align}
\frac{d 
\hat{\sigma}_{\mbox{{\tiny D}}}(\mbox{R}_{2}) 
}{d \qs}
&=\frac{\al}{2\pi}\frac{1}{\Gamma(1-\ep)}
\biggl(\frac{4\pi \mu^{2}}{s_{ab}} \biggr)^{\ep}
{\widetilde {\cal V}}^{\,g,f}(z;\ep)
\frac{\hat{\sigma}_{\mbox{{\tiny LO}}}(\qs)}
{\qs} z\,,
\end{align}
where the function
${\widetilde {\cal V}}^{\,g,f}(z;\ep)$
reads in Ref.\cite{Catani:1996vz} as
\begin{align}
{\widetilde {\cal V}}^{\,g,f}(z;\ep)
&= - \frac{1}{\ep} P^{\,gf}(z) 
+2\,P^{\,gf}(z) \ln (1-z)
+\mbox{T}_{\mbox{{\tiny R}}}
2 z (1-z)\,.
\end{align}
The result is written in the form
\begin{eqnarray}
\frac{d 
\hat{\sigma}_{\mbox{{\tiny D}}}(\mbox{R}_{2}) 
}{d \qs} 
= C(q^2,\ep) \, a_{\mbox{{\tiny D}},ug}\,,
\end{eqnarray}
with the factor
\begin{equation}
a_{\mbox{{\tiny D}},ug} = 
\frac{z}{2\,\mbox{C}_{\mbox{{\tiny F}}}} 
\biggl[
- \frac{1}{\ep} P^{\,gf}(z) 
+2\,P^{\,gf}(z) \ln (1-z)
- \,P^{\,gf}(z) \ln z
+\mbox{T}_{\mbox{{\tiny R}}}
2 z (1-z)
\biggr]\,.
\label{adug}
\end{equation}
The difference between
$a_{\mbox{{\tiny R}},ug}$
in Eq.\,(\ref{arug})
and
$a_{\mbox{{\tiny D}},ug}$
in Eq.\,(\ref{adug})
is calculated as
\begin{equation}
a_{\mbox{{\tiny R}},ug}
-
a_{\mbox{{\tiny D}},ug}
=
\frac{z}{8\,\mbox{C}_{\mbox{{\tiny F}}}} 
(1+2z-3z^{2})\,.
\end{equation}
The cross sections of the P and K terms
are written in Eq.\,(\ref{dypcr2pk}).
The P and K terms
$\mbox{P}(\mbox{R}_{2})$
and
$\mbox{K}(\mbox{R}_{2})$
are created  in 
Eqs.\,(\ref{dy2po}) and (\ref{dy2ko}).
The cross section is explicitly written as
\begin{align}
\hat{\sigma}_{\mbox{{\tiny P}}}(\mbox{R}_{2}) 
+
\hat{\sigma}_{\mbox{{\tiny K}}}(\mbox{R}_{2}) 
&= \int_{0}^{1}dx\,  
\frac{1}{S_{\mbox{{\tiny B}}_{4u} }}
\,\Phi_{a}(\mbox{R}_{2}:\mbox{B}_{4u},x)_{4} \,
\frac{\al}{2\pi} 
\nonumber \\
&\hspace{15mm}\biggl[
-P^{\,gf}(x) \,
\ln \frac{\mu_{F}^{2}}{x\,s_{ab}}
+
\bar{K}^{gf}(x) 
+
\widetilde{K}^{gf}(x) 
\biggr]
\, \langle \mbox{B}4u  \ | 
 \ \mbox{B}4u \rangle \,.
\end{align}
Then the $\qs$-distribution is 
calculated in the form
\begin{align}
\frac{d 
\hat{\sigma}_{\mbox{{\tiny P}}}(\mbox{R}_{2}) 
}{d \qs}
+
\frac{d 
\hat{\sigma}_{\mbox{{\tiny K}}}(\mbox{R}_{2}) 
}{d \qs}
&=
C(q^2,\ep=0) \ a_{\mbox{{\tiny PK}},\,ug}\,,
\end{align}
with the factor
\begin{equation}
a_{\mbox{{\tiny PK}},ug} = 
\frac{z}{2\,\mbox{C}_{\mbox{{\tiny F}}}} 
\biggl[
P^{\,gf}(z) \,
\ln \frac{(1-z)^{2} \qs}{z \mu_{F}^{2}}
+\mbox{T}_{\mbox{{\tiny R}}}
2 z (1-z)
\biggr]\,.
\end{equation}
Finally, we obtain the $\qs$-distribution
$d \hat{\sigma}(\mbox{R}_{2}) /d \qs$ 
as
\begin{align}
\frac{d 
\hat{\sigma}(\mbox{R}_{2}) 
}{d \qs} 
&=
\frac{d}{d \qs} 
\Bigl(
\hat{\sigma}_{\mbox{{\tiny R}}}(\mbox{R}_{2}) 
-
\hat{\sigma}_{\mbox{{\tiny D}}}(\mbox{R}_{2}) 
+
\hat{\sigma}_{\mbox{{\tiny P}}}(\mbox{R}_{2}) 
+
\hat{\sigma}_{\mbox{{\tiny K}}}(\mbox{R}_{2}) 
\Bigr)
\nonumber\\
&=
C(q^2,\ep=0) \,
\bigl[
(a_{\mbox{{\tiny R}},ug} - a_{\mbox{{\tiny D}},ug})
+
a_{\mbox{{\tiny PK}},ug}
\bigr]\,,
\end{align}
with the difference
\begin{align}
a_{\mbox{{\tiny R}},ug} - a_{\mbox{{\tiny D}},ug}
+
a_{\mbox{{\tiny PK}},ug}
&=
\frac{z}{2\,\mbox{C}_{\mbox{{\tiny F}}}} 
\biggl[
P^{\,gf}(z) \,
\ln \frac{(1-z)^{2} \qs}{z \mu_{F}^{2}}
+
\frac{1}{4}
\bigl(
1+6z-7z^{2}
\bigr)
\biggr]\,.
\end{align}
The results exactly coincide with the results in 
Eq.\,(\ref{arcug}).
The calculation of the distribution
$d \hat{\sigma}(\mbox{R}_{3}) /d \qs$ 
is completely analogous to
$d \hat{\sigma}(\mbox{R}_{2}) 
/d \qs$,
and the results are identical.
In this way, it is shown that the analytical
results by the DSA exactly agree with
the well known results by the traditional
methods as
\begin{equation}
\frac{d \,
\hat{\sigma}(\mbox{R}_{i}) 
}{d q^{2}}\
=
\frac{d \,
\hat{\sigma}_{{\tiny \mbox{trad}}}(\mbox{R}_{i}) 
}{d q^{2}}\
\ \mbox{for} \ \ i=1,2,\ \mbox{and} \ 3\,.
\end{equation}







\section{Summary \label{sec_6}}
We study the QCD NLO 
corrections in hadron collider processes.
In simple processes,
the analytical results are obtained by the
traditional method which may originate
in the pioneering works on the Drell--Yan process
\cite{Altarelli:1978id,Altarelli:1979ub,Abad:1978nr,
Abad:1978ke,Humpert:1979qk,Humpert:1979hb,
Humpert:1980uv,KubarAndre:1978uy,Harada:1979bj}.
The traditional method is reviewed in Sec.\,\ref{s5_1}.
In complex processes, namely, multiparton
leg processes like $pp \to n jets$,
it is almost impossible to obtain analytical
results for the NLO corrections.
The dipole subtraction procedure overcomes 
some difficulties
of the calculations and makes it possible to
obtain NLO corrections in the multiparton
leg processes. 
The price of employing the dipole subtraction 
mainly involves two things.
One is that many subtraction terms are created
and the expressions are not so simple.
The other one is that a large amount of the calculation 
is executed as numerical evaluation of the
Monte Carlo integration. As a consequence,
the person who has obtained the results of 
the NLO corrections
has the difficulty confirming whether 
the obtained results are true or false.
For the other person who does not do
the calculations him- or herself, 
the confirmation is
more difficult. In order to solve some of 
the difficulties, we need a practical algorithm
to use the dipole subtraction that allows
the clear presentation of the following items\,:
\begin{enumerate}
\item[-] Input, output, creation order,
and all formulae in the document,
\vspace{-3.5mm}
\item[-] Necessary information to specify each subtraction term,
\vspace{-3.5mm}
\item[-] Summary table of all created subtraction terms,
\vspace{-3.5mm}
\item[-] Associated proof algorithm.
\end{enumerate}
A clear definition of the practical algorithm in the
document produces the merit that we can precisely 
communicate about the subtraction terms through a
common language, and quickly compare those by two or 
more people, and also that, when the algorithm is 
implemented as a computer package, the users can 
understand the outputs properly. 
In this article, we have presented such an algorithm
that satisfies all the above requirements.
We call it the dipole splitting algorithm (DSA)
and define it in Sec.\,\ref{sec_2}.
The master formulae and all the steps
are defined in Sec.\,\ref{s2_1}
and the formulae for the subtraction terms
are collected in Appendix \ref{ap_A}.
The creation algorithm and concrete expressions
for the D, I, and P/K terms are
explained in Sec.\,\ref{s2_2}, \ref{s2_3}, and 
\ref{s2_4}, respectively.
The advantage of the DSA is clarified in
Sec.\,\ref{s2_5}.
We demonstrate the DSA in the Drell--Yan process
in Sec.\,\ref{sec_3} and in the dijet process
in Sec.\,\ref{sec_4}.
Summary tables for the Drell--Yan process
are shown in Sec.\,\ref{sec_3}.
Summary tables for the dijet process
are shown in Appendix \ref{ap_B}.
These tables can be a template for summary tables
specifying all the subtraction terms created
by the DSA.
Regarding the use of the summary tables,
we intend that the other person who does not
execute the DSA him- or herself can specify 
the subtraction terms created
and write down concrete expressions
for them simply
by reading the tables in a document.
As one particularly reliable confirmation of the DSA,
we have made an the analytical check against
the results by the traditional method
in the Drell--Yan process in 
Sec.\,\ref{sec_5}.
The associated algorithm proving
the consistency relation
is presented in Ref.\,\cite{Hasegawa:2014nna}.

We plan to study the following subjects in the future.
We will apply the DSA to some processes at the LHC 
and make predictions at the NLO accuracy.
The DSA can be applied to processes
at the $e^{-}e^{+}$ and $e^{-}p$ colliders
as well. The application is easy,
because the DSA becomes simpler 
than the hadron collider case.
The DSA is presented for an arbitrary process
including only massless quarks
in this article.
We will present the DSA for
processes including massive quarks.
The extension should be straightforward,
because the construction algorithm
of the subtraction terms 
for the massive quarks presented in 
Ref.\cite{Catani:2002hc} is
the same as in the case of the massless quarks
in Ref.\cite{Catani:1996vz}.
Regarding automation of the DSA
as a computer package, AutoDipole
\cite{Hasegawa:2009tx} is a good candidate
for its implementation, because
the creation algorithm of the dipole
and I terms implemented in AutoDipole
is essentially the same as the DSA
and only the creation algorithm
of the P and K terms is different
from the DSA. Thus, it is sufficient that
only some of the code 
to create the P and K terms is modified.
We hope that the DSA will help users to
obtain reliable predictions
at QCD NLO accuracy.

\hspace{5mm}
\subsection*{Acknowledgements}
We are grateful to V. Ravindran for helpful discussion.
The present article was completed during
a stay at the Harish-Chandra Research Institute.
We are grateful for the warm support
of the institute.


\clearpage
\appendix




\section{Formulae for DSA \label{ap_A}}
\subsection{Step\,2: D term  \label{ap_A_1}}
\begin{figure}[h!]
\begin{center}
\includegraphics[width=8.0cm]{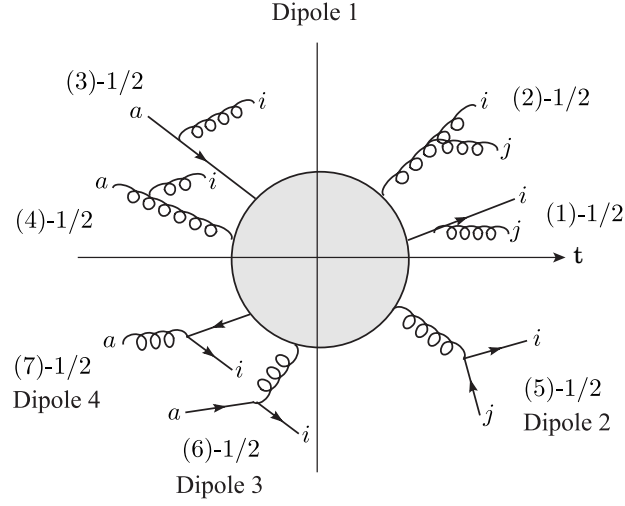}
\end{center} 
\caption{
The creation order of the dipole terms
is shown.
There are seven possible splittings
(1)--(7).
Splittings (1)--(4)
in the upper half are
diagonal splittings, which are grouped 
into category {\tt Dipole\,1}.
Splittings (5)--(7) 
in the lower half are
the nondiagonal splittings.
The indices $a$ and $i/j$ represent
the legs in the initial and final 
states, respectively.
\label{fig_Dterm}}
\end{figure}
\begin{align}
\hat{\sigma}_{\mbox{{\tiny D}}}(\mbox{R}_{i}) 
&= \frac{1}{S_{\mbox{{\tiny R}}_{i}}} \ \Phi(\mbox{R}_{i})_{4} 
\cdot \frac{1}{n_{s}(a) n_{s}(b)} \, \mbox{D}(\mbox{R}_{i})\,,
\\
\mbox{D}(\mbox{R}_{i},\,{\tt dip}j\ )_{IJ,K} 
&=
 -\frac{1}{s_{{\scriptscriptstyle IJ}}} 
\frac{1}{x_{{\scriptscriptstyle IJK}}}
\frac{1}{\mbox{T}_{\mbox{{\tiny F}}(y_{emi})}^{2}} 
\langle \mbox{B}j \ | 
\mbox{T}_{y_{emi}} \cdot 
\mbox{T}_{y_{spe}}
\ \mbox{V}_{{\scriptscriptstyle IJ,K}}^{y_{emi}}
 | \ \mbox{B}j \rangle\,. 
\end{align}

\leftline{\underline{ {\tt Dipole1} (1)-1 : $(ij, k) = (fg, k)$}}
\begin{align}
\mbox{D}_{ij,k} &= -\frac{1}{s_{ij}} 
\frac{1}{\mbox{C}_{\mbox{{\tiny F}}}} 
\mbox{V}_{ij,k} \
\langle \mbox{B}1  \ | 
\mbox{T} \cdot \mbox{T}
 | \ \mbox{B}1 \rangle, \label{dp111} \\
\mbox{V}_{ij,k} &= 8 \pi \al 
\mbox{C}_{\mbox{{\tiny F}}}
\biggl[ \frac{2}{1 - z_{i} (1 - y_{ij,k})} - 1 - z_{i}
\biggr]. \label{ds111}
\end{align}

\leftline{\underline{ {\tt Dipole1} (1)-2 : $(ij, a) = (fg, a)$}} 
\begin{align}
\mbox{D}_{ij,a} &= -\frac{1}{s_{ij}} 
\frac{1}{x_{ij,a}}
\frac{1}{\mbox{C}_{\mbox{{\tiny F}}}} 
\mbox{V}_{ij,a} \
\langle \mbox{B}1  \ | 
\mbox{T} \cdot \mbox{T}
 | \ \mbox{B}1 \rangle, \label{dp112} \\
\mbox{V}_{ij,a} &= 8 \pi \al 
\mbox{C}_{\mbox{{\tiny F}}}
\biggl[ \frac{2}{1 - z_{i} (1 - x_{ij,a})} - 1 - z_{i}
\biggr]. \label{ds112}
\end{align}

\leftline{\underline{ {\tt Dipole1} (2)-1 : $(ij, k) = (gg, k)$}} 
\begin{align}
\mbox{D}_{ij,k} &= -\frac{1}{s_{ij}} 
\frac{1}{\mbox{C}_{\mbox{{\tiny A}}}} 
\langle \mbox{B}1 | 
\mbox{T} \cdot 
\mbox{T}
\ \mbox{V}_{ij,k}
 | \ \mbox{B}1 \rangle, 
\label{dp121} \\
\mbox{V}_{ij,k}^{\mu \nu} 
&= 16 \pi \al \mbox{C}_{\mbox{{\tiny A}}}
\biggl[ 
-g^{\mu \nu} \biggl(
\frac{1}{1- z_i (1-y_{ij,k})}  
\nonumber \\
& +\frac{1}{1-  z_j (1-y_{ij,k})} - 2 
\biggr) + \frac{1}{p_ip_j} \,
( z_i p_i^{\mu} - 
z_j p_j^{\mu} )
\,(  z_i p_i^{\nu} - 
z_j p_j^{\nu} ) \,
\biggr]. \label{ds121}
\end{align}

\leftline{\underline{ {\tt Dipole1} (2)-2 : $(ij, a) = (gg, a)$}} 
\begin{align}
\mbox{D}_{ij,a} 
&= -\frac{1}{s_{ij}} 
\frac{1}{x_{ij,a}}
\frac{1}{\mbox{C}_{\mbox{{\tiny A}}}} 
\langle \mbox{B}1 | 
\mbox{T} \cdot 
\mbox{T}
\ \mbox{V}_{ij,a}
 | \ \mbox{B}1 \rangle, \label{dp122} \\
\mbox{V}_{ij,a}^{\mu \nu} 
&= 16\pi \al \mbox{C}_{\mbox{{\tiny F}}}
\biggl[ 
-g^{\mu \nu} \biggl(
\frac{1}{1- z_i+(1-x_{ij,a})} 
\nonumber \\
& + \frac{1}{1-  z_j+(1-x_{ij,a})} 
- 2 \biggr) +  
\frac{1}{p_ip_j}
( z_i p_i^{\mu} - 
z_j p_j^{\mu} )
( z_i p_i^{\nu} - 
z_j p_j^{\nu} ) \biggr].
\label{ds122}
\end{align}

\leftline{\underline{ {\tt Dipole1} (3)-1 : $(ai, k) = (fg, k)$}} 
\begin{align}
\mbox{D}_{ai,k}
&= -\frac{1}{s_{ai}} 
\frac{1}{x_{ik,a}}
\frac{1}{\mbox{C}_{\mbox{{\tiny F}}}} 
\mbox{V}_{ai,k} \
\langle \mbox{B}1  \ | 
\mbox{T} \cdot \mbox{T}
 | \ \mbox{B}1 \rangle,
\label{dp131} \\
\mbox{V}_{ai,k} 
&= 8 \pi \al \mbox{C}_{\mbox{{\tiny F}}}
\biggl[ \frac{2}{1 - x_{ik,a} + u_{i}} - 1 - x_{ik,a}
\biggr]\,.  
\label{ds131}
\end{align}

\leftline{\underline{ {\tt Dipole1} (3)-2 : $(ai, b) = (fg, b)$}} 
\begin{align}
\mbox{D}_{ai,b}
&= -\frac{1}{s_{ai}} 
\frac{1}{x_{i,ab}}
\frac{1}{\mbox{C}_{\mbox{{\tiny F}}}} 
\mbox{V}_{ai,b} \
\langle \mbox{B}1  \ | 
\mbox{T} \cdot \mbox{T}
 | \ \mbox{B}1 \rangle,
\label{dp132} \\
\mbox{V}_{ai,b} 
&= 8 \pi \al \mbox{C}_{\mbox{{\tiny F}}}
\biggl[ \frac{2}{1 - x_{i,ab}} - 1 - x_{i,ab}
\biggr]\,. 
\label{ds132}
\end{align}

\leftline{\underline{ {\tt Dipole1} (4)-1 : $(ai, k) = (gg, k)$}} 
\begin{align}
\mbox{D}_{ai,k} 
&= -\frac{1}{s_{ai}} 
\frac{1}{x_{ik,a}}
\frac{1}{\mbox{C}_{\mbox{{\tiny A}}}} 
\langle \mbox{B}1 | 
\mbox{T} \cdot 
\mbox{T}
\ \mbox{V}_{ai,k}
 | \ \mbox{B}1 \rangle,
\label{dp141} \\
\mbox{V}_{ai,k}^{\mu \nu} 
&= 16 \pi \al \mbox{C}_{\mbox{{\tiny A}}}
\biggl[ -g^{\mu \nu} \biggl(
\frac{1}{1-x_{ik,a} +u_{i}} -1
+ x_{ik,a}(1-x_{ik,a}) \biggr)  \nonumber \\
&+
\frac{1-x_{ik,a}}{x_{ik,a}} 
\frac{u_i(1-u_i)}{p_ip_k} 
\biggl( \frac{p_i^{\mu}}{u_i} - 
\frac{p_k^{\mu}}{1-u_i} \biggr)
\biggl( \frac{p_i^{\nu}}{u_i} - 
\frac{p_k^{\nu}}{1-u_i} \biggr)
\,\biggr].
\label{ds141}
\end{align}

\leftline{\underline{ {\tt Dipole1} (4)-2 : $(ai, b) = (gg, b)$}} 
\begin{align}
\mbox{D}_{ai,b} 
&= -\frac{1}{s_{ai}} 
\frac{1}{x_{i,ab}}
\frac{1}{\mbox{C}_{\mbox{{\tiny A}}}} 
\langle \mbox{B}1 | 
\mbox{T} \cdot 
\mbox{T}
\ \mbox{V}_{ai,b}
 | \ \mbox{B}1 \rangle, 
\label{dp142} \\
\mbox{V}_{ai,b}^{\mu \nu} 
&= 16\pi \al C_A
\biggl[ -g^{\mu \nu} \biggl(
\frac{x_{i,ab}}{1-x_{i,ab}} + x_{i,ab}(1-x_{i,ab}) \biggr)
\nonumber \\
&+ \frac{1-x_{i,ab}}{x_{i,ab}} 
\frac{p_a \cdot p_b}{p_i \cdot p_a \;p_i \cdot p_b} 
\biggl( p_i^{\mu} - \frac{p_ip_a}{p_bp_a} p_b^{\mu} \biggr)
\biggl( p_i^{\nu} - \frac{p_ip_a}{p_bp_a} p_b^{\nu} \biggr)
\biggr].
\label{ds142}
\end{align}

\leftline{\underline{ {\tt Dipole2} (5)-1 : $(ij,k) = (f\bar{f},k)$}} 
\begin{align}
\mbox{D}_{ij,k} &= -\frac{1}{s_{ij}} 
\frac{1}{\mbox{C}_{\mbox{{\tiny A}}}} 
\langle \mbox{B}2 | 
\mbox{T} \cdot 
\mbox{T}
\ \mbox{V}_{ij,k}
 | \ \mbox{B}2 \rangle,
\label{dp251} \\
\mbox{V}_{ij,k}^{\mu \nu} 
&= 8\pi  \al 
\mbox{T}_{\mbox{{\tiny R}}}
\biggl[ 
-g^{\mu \nu} - \frac{2}{p_ip_j} 
( z_i p_i^{\mu} - 
z_j p_j^{\mu} )
( z_i p_i^{\nu} - 
z_j p_j^{\nu} ) 
\biggr]\,.
\label{ds251}
\end{align}

\leftline{\underline{ {\tt Dipole2} (5)-2 : $(ij, b) = (f\bar{f}, b)$}} 
\begin{align}
\mbox{D}_{ij,a} 
&= -\frac{1}{s_{ij}} 
\frac{1}{x_{ij,a}}
\frac{1}{\mbox{C}_{\mbox{{\tiny A}}}} 
\langle \mbox{B}2 | 
\mbox{T} \cdot 
\mbox{T}
\ \mbox{V}_{ij,a}
 | \ \mbox{B}2 \rangle, 
\label{dp252} \\
\mbox{V}_{ij,a}^{\mu \nu} 
&= 8\pi \al \mbox{T}_{\mbox{{\tiny R}}}
\biggl[ -g^{\mu \nu} - \frac{2}{p_ip_j} \,
(  z_i p_i^{\mu} -  z_j p_j^{\mu} )
(  z_i p_i^{\nu} -  z_j p_j^{\nu} ) \biggr] \,.
\label{ds252}
\end{align}

\leftline{\underline{ {\tt Dipole3} (6)-1 : $(ai, k) = (ff, k)$}} 
\begin{align}
\mbox{D}_{ai,k} 
&= -\frac{1}{s_{ai}} 
\frac{1}{x_{ik,a}}
\frac{1}{\mbox{C}_{\mbox{{\tiny A}}}} 
\langle \mbox{B}3 | 
\mbox{T} \cdot 
\mbox{T}
\ \mbox{V}_{ai,k}
 | \ \mbox{B}3 \rangle, 
\label{dp361} \\
\mbox{V}_{ai,k}^{\mu \nu} 
&= 8\pi \al  \mbox{C}_{\mbox{{\tiny F}}}
\biggl[ -g^{\mu \nu}x_{ik,a} \nonumber \\
& + \frac{1-x_{ik,a}}{x_{ik,a}} 
\frac{2u_i(1-u_i)}{p_ip_k} \,
\biggr( \frac{p_i^{\mu}}{u_i} - 
\frac{p_k^{\mu}}{1-u_i} \biggl)
\biggr( \frac{p_i^{\nu}}{u_i} - 
\frac{p_k^{\nu}}{1-u_i} \biggl)
\, \biggr] \,.
\label{ds361}
\end{align}

\leftline{\underline{ {\tt Dipole3} (6)-2 : $(ai, b) = (ff, b)$}}
\begin{align}
\mbox{D}_{ai,b} 
&= -\frac{1}{s_{ai}} 
\frac{1}{x_{i,ab}}
\frac{1}{\mbox{C}_{\mbox{{\tiny A}}}} 
\langle \mbox{B}3 | 
\mbox{T} \cdot 
\mbox{T}
\ \mbox{V}_{ai,b}
 | \ \mbox{B}3 \rangle,
\label{dp362} \\
\mbox{V}_{ai,b}^{\mu \nu} 
&= 8\pi \al \mbox{C}_{\mbox{{\tiny F}}}
\biggl[ -g^{\mu \nu}x_{i,ab}  \nonumber \\
&+\frac{1-x_{i,ab}}{x_{i,ab}}
\frac{2p_a \cdot p_b}{p_i \cdot p_a \;p_i \cdot p_b} \,
\biggl( p_i^{\mu} - \frac{p_ip_a}{p_bp_a} p_b^{\mu} \biggr)
\biggl( p_i^{\nu} - \frac{p_ip_a}{p_bp_a} p_b^{\nu} \biggr)
\,\biggr] \,.
\label{ds362}
\end{align}

\leftline{\underline{ {\tt Dipole4} (7)-1 : $(ai, k) = (gf, k)$}} 
\begin{align}
\mbox{D}_{ai,k}
&= -\frac{1}{s_{ai}} 
\frac{1}{x_{ik,a}}
\frac{1}{\mbox{C}_{\mbox{{\tiny F}}}} 
\mbox{V}_{ai,k} \
\langle \mbox{B}4  \ | 
\mbox{T} \cdot \mbox{T}
 | \ \mbox{B}4 \rangle\,, 
\label{dp471} \\
\mbox{V}_{ai,k} 
&= 8 \pi \al \mbox{T}_{\mbox{{\tiny R}}}
\biggl[ 1 - 2 x_{ik,a}(1 - x_{ik,a}) 
\biggr]\,.
\label{ds471}
\end{align}

\leftline{\underline{ {\tt Dipole4} (7)-2 : $(ai, b) = (gf, b)$}} 
\begin{align}
\mbox{D}_{ai,b}
&= -\frac{1}{s_{ai}} 
\frac{1}{x_{i,ab}}
\frac{1}{\mbox{C}_{\mbox{{\tiny F}}}} 
\mbox{V}_{ai,b} \
\langle \mbox{B}4  \ | 
\mbox{T} \cdot \mbox{T}
 | \ \mbox{B}4 \rangle\,,
\label{dp472} \\
\mbox{V}_{ai,b} 
&= 8\pi  \al \mbox{T}_{\mbox{{\tiny R}}} \,
\bigl[ 1 -2 x_{i,ab}(1-x_{i,ab}) \bigr]\,. 
\label{ds472}
\end{align}

\subsubsection*{Reduced momenta and some functions\,:}
\leftline{\underline{$\mbox{D}_{ij,k}$: Final--Final dipole}} 
\begin{verbatim}
Dipole 1 (1)-1, (2)-1,  
Dipole 2 (5)-1:
\end{verbatim}
\begin{align}
\tilde{p}_{ij}^{\mu} 
&= p_{i}^{\mu} + p_{j}^{\mu} - 
\frac{y_{ij,k}}{1 - y_{ij,k}}p_{k}^{\mu}\,,  \label{rmffem} \\
\tilde{p}_{k}^{\mu} 
&= \frac{1}{1 - y_{ij,k}}p_{k}^{\mu}\,.  \label{rmffsp}
\end{align}
\begin{align}
x_{ijk}
&=1\,, \\
y_{ij,k} 
&= \frac{p_{i} \cdot p_{j}}{p_{i} \cdot p_{j} + 
p_{j} \cdot p_{k} + p_{k} \cdot p_{i}}\,,
\label{ffyijk} \\
z_{i} 
&= \frac{p_{i} \cdot p_{k}}{(p_{i} + p_{j}) 
\cdot p_{k}}\,, \label{ffzi} \\
z_{j} 
&= 1 - z_{i}\,. \label{ffzj} 
\end{align}

\leftline{\underline{$\mbox{D}_{ij,a}$: Final--Initial dipole}} 
\begin{verbatim}
Dipole 1 (1)-2, (2)-2,
Dipole 2 (5)-2:
\end{verbatim}
\begin{align}
\tilde{p}_{ij}^{\mu} 
&= p_{i}^{\mu} + p_{j}^{\mu} - 
(1 - x_{ij,b})p_{b}^{\mu}\,,  \label{rmfiem} \\
\tilde{p}_{a}^{\mu} 
&= x_{ij,a} \ p_{a}^{\mu}\,. \label{rmfisp}
\end{align}
\begin{align}
x_{ija}&=x_{ij,a}= \frac{p_{i} \cdot p_{a} + p_{j} \cdot 
p_{a} - p_{i} \cdot p_{j}}{(p_{i} + p_{j})\cdot 
p_{a}}\,, \\
z_{i} 
&= \frac{p_{i} \cdot p_{a}}{(p_{i} + p_{j}) 
\cdot p_{a}}\,, \label{fizi} \\
z_{j} 
&= 1 - z_{i}\,. \label{fizj} 
\end{align}

\leftline{\underline{$\mbox{D}_{ai,k}$: Initial--Final dipole}} 
\begin{verbatim}
Dipole 1 (3)-1, (4)-1, 
Dipole 3 (6)-1, 
Dipole 4 (7)-1:
\end{verbatim}
\begin{align}
\tilde{p}_{ai}^{\mu} 
&= x_{ik,a} \ p_{a}^{\mu}\,,  \label{rmifem} \\
\tilde{p}_{k}^{\mu} 
&= p_{i}^{\mu} + p_{k}^{\mu} - 
(1 - x_{ik,a})p_{a}^{\mu}\,.  \label{rmifsp}
\end{align}
\begin{align}
x_{aik}&=x_{ik,a} 
= \frac{p_{i} \cdot p_{a} + p_{k} \cdot p_{a} - 
p_{i} \cdot p_{k}}{(p_{i} + p_{k})\cdot p_{a}}\,, 
\label{xaikif}\\
u_{i} 
&= \frac{p_{i} \cdot p_{a}}{(p_{i} + p_{k}) 
\cdot p_{a}}\,. 
\end{align}

\leftline{\underline{$\mbox{D}_{ai,b}$: Initial--Initial dipole}} 
\begin{verbatim}
Dipole 1 (3)-2,  (4)-2, 
Dipole 3 (6)-2,  
Dipole 4 (7)-2:
\end{verbatim}
\begin{align}
\tilde{p}_{ai}^{\mu} 
&= x_{i,ab} \ p_{a}^{\mu}\,,   \label{rmiiem} \\
\tilde{k}_{j}^{\mu} 
&= k_{j}^{\mu} - 
\frac{2k_{j} \cdot (K + \tilde{K})}{(K + \tilde{K})^2}
(K + \tilde{K})^{\mu} + 
\frac{2k_{j} \cdot K}{K^2}\tilde{K}^{\mu}\,, \label{rmiisp} \\
x_{aib}&=x_{i,ab} 
= \frac{p_{a} \cdot p_{b} - p_{i} \cdot p_{a} - 
p_{i} \cdot p_{b}}{p_{a} \cdot p_{b}}\,, \label{iixiab} \\
K^{\mu} 
&= p_{a} + p_{b} - p_{i}\,,  \label{iikmu} \\
\tilde{K}^{\mu} 
&= \tilde{p}_{ai}^{\mu} + p_{b}^{\mu}\,.
\label{iiktilmu}
\end{align}
\clearpage
\subsection{Step\,3: I term \label{ap_A_2}}
\begin{figure}[h!]
\begin{center}
\includegraphics[width=8.0cm]{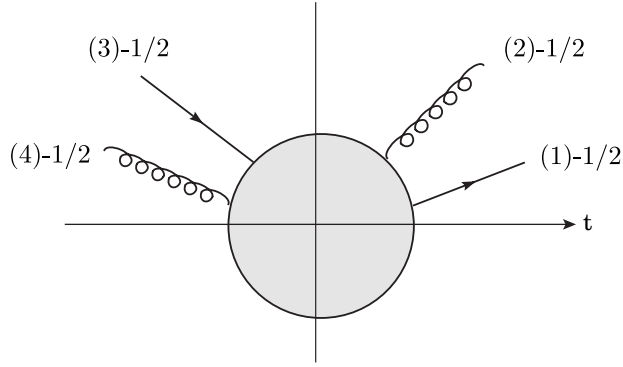}
\end{center} 
\caption{ 
The creation order of the I terms is shown.
\label{fig_Iterm}}
\end{figure}
\begin{align}
\hat{\sigma}_{\mbox{{\tiny I}}}(\mbox{R}_{i}) 
&= \frac{1}{S_{\mbox{{\tiny B1}} }} \ \Phi(\mbox{B}1)_{d} 
\cdot \mbox{I}(\mbox{R}_{i})\,, 
\\
\mbox{I}_{I,\,K}
&= - A_{d} \cdot
\frac{1}{\mbox{T}_{\mbox{{\tiny F}}(I)}^{2}} \,
{\cal V}_{\mbox{{\tiny F}}(I)}
\cdot s_{{\scriptscriptstyle IK}}^{-\ep} \,
\langle 
\mbox{T}_{I} \cdot 
\mbox{T}_{K}
\rangle\,.
\label{iform}
\end{align}
\begin{equation}
A_{d}=\frac{\al}{2\pi} \frac{(4\pi \mu^{2})^{\ep}}{\Gamma(1-\ep)}\,.
\end{equation}
Universal singular functions\,:
\begin{align}
{\cal V}_{f} &= {\cal V}_{fg} (\ep)\,, \label{usf1} \\
{\cal V}_{g} &= \half {\cal V}_{gg} (\ep) + 
N_{f}{\cal V}_{f\bar{f}} (\ep)\,, 
\label{usf2}
\end{align}
with
\begin{align}
{\cal V}_{fg} (\ep) 
&= \mbox{C}_{\mbox{{\tiny F}}} \biggl[\frac{1}{\ep^2} + \frac{3}{2\ep} +5 
-\frac{\pi^2}{2}   \biggr]\,, 
\label{isfg} \\
{\cal V}_{gg} (\ep) 
&= 2 \mbox{C}_{\mbox{{\tiny A}}} \biggl[\frac{1}{\ep^2} + \frac{11}{6\ep} +
\frac{50}{9} -\frac{\pi^2}{2} \biggr]\,, 
\label{isgg} \\
{\cal V}_{f\bar{f}} (\ep) 
&= \mbox{T}_{\mbox{{\tiny R}}} \biggl[- \frac{2}{3\ep} - \frac{16}{9} 
\biggr]\,.
\label{isff} 
\end{align}
\clearpage
\subsection{Step\,4: P and K terms \label{ap_A_3}}
\vspace{3mm}
\begin{figure}[h!]
\begin{center}
\includegraphics[width=7.5cm]{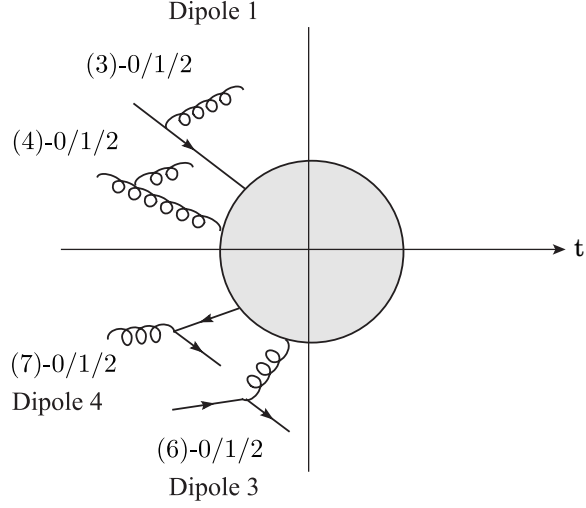}
\end{center} 
\caption{ 
The creation order of the P and K terms
is shown.
\label{fig_PKterm}}
\end{figure}
\begin{equation}
\hat{\sigma}_{\mbox{{\tiny P/K}}}(\mbox{R}_{i}) 
= \int_{0}^{1}dx \sum_{\mbox{{\tiny B}}_{j}} 
\frac{1}{S_{\mbox{{\tiny B}}_{j} }}
\Phi_{a}(\mbox{R}_{i}:\mbox{B}_{j},x)_{4} \cdot 
\mbox{P/K}(\mbox{R}_{i},x_{a}:\mbox{B}_{j},xp_{a}) \ + \ 
(a \leftrightarrow b)\,, 
\end{equation}
\subsubsection*{P term}
\begin{align}
\mbox{P}(\mbox{R}_{i},x_{a}:\mbox{B}j,y_{a},y_{K}) 
&= A_{4} \cdot
\frac{1}{\mbox{T}_{\mbox{{\tiny F}}(y_{a})}^{2}} \,
P^{
\mbox{{\tiny F}}(x_{a})\mbox{{\tiny F}}(y_{a})
}(x) \,
\ln \frac{\mu_{F}^{2}}{x\,s_{x_{a} y_{{\scriptscriptstyle K}}}}
\ \langle \mbox{B}j 
\ | 
\mbox{T}_{y_{{\scriptscriptstyle a}}} \cdot 
\mbox{T}_{y_{{\scriptscriptstyle K}}}
| \ \mbox{B}j \rangle\,.
\label{pform}
\end{align}
\begin{equation}
A_{4}=\frac{\al}{2\pi}.
\end{equation}


\leftline{Altarelli-Parisi splitting functions\,:}

\vspace{1mm}

\leftline{\underline{ {\tt Dipole1} (3) : $(a,i) = (f,g)$}} 
\begin{equation}
P^{ff}(x)
=\mbox{C}_{\mbox{{\tiny F}}} 
\biggl(\frac{1+x^2}{1-x}\biggr)_{+}
=\mbox{C}_{\mbox{{\tiny F}}} 
\biggl[
\frac{1+x^2}{(1-x)_{+}}
+
\frac{3}{2}\delta(1-x)
\biggr]
 \,.
\label{alpff} 
\end{equation}

\leftline{\underline{ {\tt Dipole1} (4) : $(a,i) = (g,g)$}} 
\begin{equation}
P^{gg}(x)
=2 \mbox{C}_{\mbox{{\tiny A}}}
\biggl[ \biggl(\frac{1}{1-x}\biggr)_{+}
+\frac{1-x}{x} -1 +x(1-x) \biggr]  
+ \delta(1-x)
\Bigl(\frac{11}{6}\mbox{C}_{\mbox{{\tiny A}}}
 - \frac{2}{3}N_{f}\mbox{T}_{\mbox{{\tiny R}}}
\Bigr) \,.
\label{alpgg} 
\end{equation}

\leftline{\underline{ {\tt Dipole3} (6) : $(a,i) = (f,f)$}} 
\begin{equation}
P^{fg}(x)
= \mbox{C}_{\mbox{{\tiny F}}}
\frac{1+(1-x)^2}{x}\,.
\label{alpfg} 
\end{equation}

\leftline{\underline{ {\tt Dipole4} (7) : $(a,i) = (g,f)$}} 
\begin{equation}
P^{gf}(x)
=\mbox{T}_{\mbox{{\tiny R}}}
[x^2 + (1-x)^{2}]\,.
\label{alpgf}
\end{equation}

\subsubsection*{K term}
\begin{align}
\mbox{K}(\mbox{R}_{i},x_{a}:\mbox{B}j,y_{a},y_{0}) 
&=A_{4} \cdot
\bar{K}^{
\mbox{{\tiny F}}(x_{a})\mbox{{\tiny F}}(y_{a})
}(x) \ \cdot \langle \mbox{B}j \,|  \mbox{B}j \rangle, 
\label{k0form}\\
\mbox{K}(\mbox{R}_{i},x_{a}:\mbox{B}1,y_{a},y_{k}) 
&=A_{4} \cdot
\frac{\gamma_{\mbox{{\tiny F}}(y_{k})}}
{\mbox{T}_{\mbox{{\tiny F}}(y_{k})}^{2}} \, h(x)
\ \cdot \langle \mbox{B}1 | 
\mbox{T}_{y_{{\scriptscriptstyle a}}} \cdot 
\mbox{T}_{y_{{\scriptscriptstyle k}}}
| \ \mbox{B}1 \rangle,
\label{kkform} \\
\mbox{K}(\mbox{R}_{i},x_{a}:\mbox{B}j,y_{a},y_{b}) 
&=- A_{4} \cdot
\frac{1}{\mbox{T}_{\mbox{{\tiny F}}(y_{a})}^{2}} \, 
\tilde{K}^{
\mbox{{\tiny F}}(x_{a})\mbox{{\tiny F}}(y_{a})
}(x) \ \cdot
\langle \mbox{B}j | 
\mbox{T}_{y_{{\scriptscriptstyle a}}} \cdot 
\mbox{T}_{y_{{\scriptscriptstyle b}}}
| \ \mbox{B}j \rangle.
\label{kbform}
\end{align}
\begin{align}
\gamma_{f} 
&= \frac{3}{2}\mbox{C}_{\mbox{{\tiny F}}} \,, 
\label{gamf}\\ 
\gamma_{g} 
&= \frac{11}{6}\mbox{C}_{\mbox{{\tiny A}}}
- \frac{2}{3}\mbox{T}_{\mbox{{\tiny R}}} N_{f}
\,.
\end{align}
\begin{equation}
h(x)=\biggl(\frac{1}{1-x} \biggr)_{+} 
+ \delta(1-x).
\label{hxd}
\end{equation}
\leftline{\underline{ {\tt Dipole1} (3) : $(a,i) = (f,g)$}} 
\begin{align}
\bar{\mbox{K}}^{ff}(x)
&=\mbox{C}_{\mbox{{\tiny F}}} \biggl[
\Bigl(\frac{2}{1-x} \ln \frac{1-x}{x} \Bigr)_{+} -(1+x)\ln \frac{1-x}{x}
+(1-x)  \nonumber \\
&  -\delta(1-x)(5-\pi^2)
\biggr]\,,
\label{apK13b} \\
\widetilde{\mbox{K}}^{ff}(x)
&=
P_{reg}^{ff}(x)\ln(1-x) +\mbox{C}_{\mbox{{\tiny F}}}
\biggl[
\Bigl(
\frac{2}{1-x}\ln(1-x) 
\Bigr)_{+} 
- \frac{\pi^2}{3}\delta(1-x)
\biggr]\,.
\end{align}

\leftline{\underline{ {\tt Dipole1} (4) : $(a,i) = (g,g)$}} 
\begin{align}
\bar{\mbox{K}}^{gg}(x)
&=
2 \mbox{C}_{\mbox{{\tiny A}}}
\biggl[\Bigl(\frac{1}{1-x} \ln \frac{1-x}{x} \Bigr)_{+} 
+\Bigl(
\frac{1-x}{x} -1 +x(1-x)
\Bigr)
\ln \frac{1-x}{x}
\biggr] \nonumber \\
&-\delta(1-x)
\biggl[
\Bigl(\frac{50}{9} - \pi^2 \Bigr)
\mbox{C}_{\mbox{{\tiny A}}}
- \frac{16}{9}\mbox{T}_{\mbox{{\tiny R}}}
N_{f}
\biggr]\,, \\
\widetilde{\mbox{K}}^{gg}(x)
&=
P_{reg}^{gg}(x)\ln(1-x) + \mbox{C}_{\mbox{{\tiny A}}}
\biggl[
\Bigl(
\frac{2}{1-x}\ln(1-x) 
\Bigr)_{+} 
- \frac{\pi^2}{3}\delta(1-x)
\biggr]\,.
\end{align}

\leftline{\underline{ {\tt Dipole3} (6) : $(a,i) = (f,f)$}} 
\begin{align}
\bar{\mbox{K}}^{fg}(x) 
&= P^{fg}(x) \ln \frac{1-x}{x} + 
\mbox{C}_{\mbox{{\tiny F}}} x\,,
\\
\widetilde{\mbox{K}}^{fg}(x) 
&=P^{fg}(x)\ln(1-x)\,.
\label{apKtil36}
\end{align}

\leftline{\underline{ {\tt Dipole4} (7) : $(a,i) = (g,f)$}} 
\begin{align}
\bar{\mbox{K}}^{gf}(x) &= P^{gf}(x) \ln \frac{1-x}{x} + 
\mbox{T}_{\mbox{{\tiny R}}} 2x(1-x)\,, \\
\widetilde{\mbox{K}}^{gf}(x)&=P^{gf}(x)\ln(1-x)\,.
\end{align}

Regular part of the Altarelli-Parisi splitting function\,:
\begin{align}
P^{ff}_{reg}(x)
&= -\mbox{C}_{\mbox{{\tiny F}}}
(1+x)\,,  \\
P^{gg}_{reg}(x)
&= 2 \mbox{C}_{\mbox{{\tiny A}}}
\biggl[\frac{1-x}{x} -1 + x(1-x) \biggr]\,.
\end{align}

The above expressions for the K terms
are in the  $\overline{\mbox{MS}}$ 
subtraction scheme.
The other factorization scheme is specified by the term 
$K_{FS}^{ab}(x)$, which is exactly defined 
in Eq.\,(6.6) of Ref.\cite{Catani:1996vz}.
The scheme-dependent term is introduced in the dipole subtraction
procedure at the K term.
The inclusion is realized by the replacement of
$\bar{K}^{
\mbox{{\tiny F}}(x_{a})\mbox{{\tiny F}}(y_{a})
}(x) 
$ in Eq.\,(\ref{k0form}) as
\begin{equation}
\bar{K}^{
\mbox{{\tiny F}}(x_{a})\mbox{{\tiny F}}(y_{a})
}(x) 
\ \to \ 
\bar{K}^{
\mbox{{\tiny F}}(x_{a})\mbox{{\tiny F}}(y_{a})
}(x) 
-
K_{FS}^{
\mbox{{\tiny F}}(x_{a})\mbox{{\tiny F}}(y_{a})
}
(x)\,.
\end{equation}

\clearpage




\section{Summary tables for dijet process \label{ap_B}}
\subsection{D term  \label{ap_B_1}}
%
%
\begin{table}[h!]
  \centering
\begin{align}
\hat{\sigma}_{\mbox{{\tiny D}}}(\mbox{R}_{i}) 
&= \frac{1}{S_{\mbox{{\tiny R}}_{i}}} \ \Phi(\mbox{R}_{i})_{4} 
\cdot \frac{1}{n_{s}(a) n_{s}(b)} \, \mbox{D}(\mbox{R}_{i})\,,
\nonumber \\
\mbox{D}(\,\mbox{R}_{i},\,{\tt dip}j\ )_{IJ,K} 
&=
 -\frac{1}{s_{{\scriptscriptstyle IJ}}} 
\frac{1}{x_{{\scriptscriptstyle IJK}}}
\frac{1}{\mbox{T}_{\mbox{{\tiny F}}(y_{emi})}^{2}} 
\langle \mbox{B}j \ | 
\mbox{T}_{y_{emi}} \cdot 
\mbox{T}_{y_{spe}}
\ \mbox{V}_{{\scriptscriptstyle IJ,K}}^{y_{emi}}
 | \ \mbox{B}j \rangle\,. 
\nonumber
\end{align}
\begin{align}
&\mbox{D}\,
(\mbox{R}_{1u} = u\bar{u} \to u\bar{u}g): 
\ \ S_{\mbox{{\tiny R}}_{1}}=1
\nonumber\\
&
\begin{array}{|c|c|c|c|c|} \hline
{\tt Dip}\,j
& \mbox{B}j 
& \mbox{Splitting} 
& (x_{I}x_{J},x_{K}) 
& (y_{a},y_{b}:y_{1},y_{2}) 
\\[4pt] \hline
& & & & \\[-12pt]
{\tt Dip\,1}  
& \mbox{B}1=u\bar{u} \to u\bar{u}
& (1)-1   
& 1.\,(13,2)
& (a,b\,;\widetilde{13},\widetilde{2} ) 
\\[4pt]  
&  
&   
& 2.\,(23,1)
& (a,b\,;\widetilde{1},\widetilde{23}) 
\\[4pt]  
&  
& (1)-2  
& 3.\,(13,a)
& (\widetilde{a},b\,;\widetilde{13},2) 
\\[4pt]  
&  
&   
& 4.\,(13,b)
& (a,\widetilde{b}\,;\widetilde{13},2) 
\\[4pt]  
&  
&  
& 5.\,(23,a)
& (\widetilde{a},b\,;1,\widetilde{23}) 
\\[4pt]  
&  
&   
& 6.\,(23,b)
& (a,\widetilde{b}\,;1,\widetilde{23}) 
\\[4pt]  
&  
& (3)-1  
& 7.\,(a3,1)
& (\widetilde{a3},b\,;\widetilde{1},2) 
\\[4pt]  
&  
&   
& 8.\,(a3,2)
& (\widetilde{a3},b\,;1,\widetilde{2}) 
\\[4pt]  
&  
&   
& 9.\,(b3,1)
& (a,\widetilde{b3}\,;\widetilde{1},2) 
\\[4pt]  
&  
&   
& 10.\,(b3,2)
& (a,\widetilde{b3}\,;1,\widetilde{2})
\\[4pt]  
&  
& (3)-2  
& 11.\,(a3,b)
& (\widetilde{a3},\widetilde{b}\,;1,2) 
\\[4pt]  
&  
&   
& 12.\,(b3,a)
& (\widetilde{a},\widetilde{b3}\,;1,2) 
\\[4pt]  \hline
& & & & \\[-12pt]
{\tt Dip\,2u}  
& \mbox{B}2u=u\bar{u} \to gg
& (5)-1  
& 13.\,(12,3)
& (a,b\,;\widetilde{12},\widetilde{3}) 
\\[4pt]  
&  
& (5)-2    
& 14.\,(12,a)
& (\widetilde{a},b\,;\widetilde{12},3) 
\\[4pt]  
&  
&   
& 15.\,(12,b)
& (a,\widetilde{b}\,;\widetilde{12},3) 
\\[4pt]  \hline
& & & & \\[-12pt]
{\tt Dip\,3u}  
&  \mbox{B}3u=g\bar{u} \to \bar{u}g
& (6)-1    
& 16.\,(a1,2)
& (\widetilde{a1},b\,;\widetilde{2},3) 
\\[4pt]  
&  
&   
& 17.\,(a1,3)
& (\widetilde{a1},b\,;2,\widetilde{3}) 
\\[4pt]  
&  
& (6)-2  
& 18.\,(a1,b)
& (\widetilde{a1},\widetilde{b}\,;2,3) 
\\[4pt]  \hline
& & & & \\[-12pt]
{\tt Dip\,3\bar{u}}  
& \mbox{B}3\bar{u}=ug \to ug
& (6)-1   
& 19.\,(b2,1)
& (a,\widetilde{b2}\,;\widetilde{1},3)  
\\[4pt]  
&  
&   
& 20.\,(b2,3)
& (a,\widetilde{b2}\,;1,\widetilde{3})  
\\[4pt]  
&  
& (6)-2   
& 21.\,(b2,a)
& (\widetilde{a},\widetilde{b2}\,;1,3)  
\\[4pt] \hline
\end{array} \nonumber
\end{align}
\caption{\small
Summary table of
$\mbox{D}\,(\mbox{R}_{1u})$
\label{ap_B_1_tab1}}
\end{table}
%
%
\begin{table}[h!]
  \centering
\begin{align}
&\mbox{D}\,
(\mbox{R}_{2u} = uu \to uug\,): 
\ \ S_{\mbox{{\tiny R}}_{2}}=2
\nonumber\\
&
\begin{array}{|c|c|c|c|c|} \hline
{\tt Dip}\,j
& \mbox{B}j 
& \mbox{Splitting} 
& (x_{I}x_{J},x_{K}) 
& (y_{a},y_{b}:y_{1},y_{2}) 
\\[4pt] \hline
& & & & \\[-12pt]
{\tt Dip\,1}  
& \mbox{B}1=uu \to uu
& (1)-1   
& 1.\,(13,2)
& (a,b\,;\widetilde{13},\widetilde{2} ) 
\\[4pt]  
&  
&   
& 2.\,(23,1)
& (a,b\,;\widetilde{1},\widetilde{23}) 
\\[4pt]  
&  
& (1)-2  
& 3.\,(13,a)
& (\widetilde{a},b\,;\widetilde{13},2) 
\\[4pt]  
&  
&   
& 4.\,(13,b)
& (a,\widetilde{b}\,;\widetilde{13},2) 
\\[4pt]  
&  
&  
& 5.\,(23,a)
& (\widetilde{a},b\,;1,\widetilde{23}) 
\\[4pt]  
&  
&   
& 6.\,(23,b)
& (a,\widetilde{b}\,;1,\widetilde{23}) 
\\[4pt]  
&  
& (3)-1  
& 7.\,(a3,1)
& (\widetilde{a3},b\,;\widetilde{1},2) 
\\[4pt]  
&  
&   
& 8.\,(a3,2)
& (\widetilde{a3},b\,;1,\widetilde{2}) 
\\[4pt]  
&  
&   
& 9.\,(b3,1)
& (a,\widetilde{b3}\,;\widetilde{1},2) 
\\[4pt]  
&  
&   
& 10.\,(b3,2)
& (a,\widetilde{b3}\,;1,\widetilde{2})
\\[4pt]  
&  
& (3)-2  
& 11.\,(a3,b)
& (\widetilde{a3},\widetilde{b}\,;1,2) 
\\[4pt]  
&  
&   
& 12.\,(b3,a)
& (\widetilde{a},\widetilde{b3}\,;1,2) 
\\[4pt]  \hline
& & & & \\[-12pt]
{\tt Dip\,3u}  
&  \mbox{B}3u=gu \to ug
& (6)-1    
& 13.\,(a1,2)
& (\widetilde{a1},b\,;\widetilde{2},3) 
\\[4pt]  
&  
&   
& 14.\,(a1,3)
& (\widetilde{a1},b\,;2,\widetilde{3}) 
\\[4pt]  
&  
&   
& 15.\,(a2,1)
& (\widetilde{a2},b\,;\widetilde{1},3) 
\\[4pt]  
&  
&   
& 16.\,(a2,3)
& (\widetilde{a2},b\,;1,\widetilde{3}) 
\\[4pt]  
&  
&   
& 17.\,(b1,2)
& (\widetilde{b1},a\,;\widetilde{2},3) 
\\[4pt]  
&  
&   
& 18.\,(b1,3)
& (\widetilde{b1},a\,;2,\widetilde{3}) 
\\[4pt]  
&  
&   
& 19.\,(b2,1)
& (\widetilde{b2},a\,;\widetilde{1},3) 
\\[4pt]  
&  
&   
& 20.\,(b2,3)
& (\widetilde{b2},a\,;1,\widetilde{3}) 
\\[4pt]  
&  
& (6)-2  
& 21.\,(a1,b)
& (\widetilde{a1},\widetilde{b}\,;2,3) 
\\[4pt]  
&  
& 
& 22.\,(a2,b)
& (\widetilde{a2},\widetilde{b}\,;1,3) 
\\[4pt]  
&  
& 
& 23.\,(b1,a)
& (\widetilde{b1},\widetilde{a}\,;2,3) 
\\[4pt]  
&  
& 
& 24.\,(b2,a)
& (\widetilde{b2},\widetilde{a}\,;1,3) 
\\[4pt]  \hline
\end{array} \nonumber
\end{align}
\caption{\small
Summary table of $\mbox{D}\,(\mbox{R}_{2u})$
\label{ap_B_1_tab2}}
\end{table}
%
%
\begin{table}[h!]
  \centering
\begin{align}
&\mbox{D}\,
(\mbox{R}_{3u}= ug \to uu\bar{u}): 
\ \ S_{\mbox{{\tiny R}}_{3u}}=2
\nonumber\\
&
\begin{array}{|c|c|c|c|c|} \hline
{\tt Dip}\,j
& \mbox{B}j 
& \mbox{Splitting} 
& (x_{I}x_{J},x_{K}) 
& (y_{a},y_{b}:y_{1},y_{2}) 
\\[4pt] \hline
& & & & \\[-12pt]
{\tt Dip\,2u}  
& \mbox{B}2u=ug \to gu
& (5)-1  
& 1.\,(13,2)
& (a,b\,;\widetilde{13},\widetilde{2}) 
\\[4pt] 
& 
&   
& 2.\,(23,1)
& (a,b\,;\widetilde{23},\widetilde{1}) 
\\[4pt] 
&  
& (5)-2    
& 3.\,(13,a)
& (\widetilde{a},b\,;\widetilde{13},2) 
\\[4pt]  
&  
&   
& 4.\,(13,b)
& (a,\widetilde{b}\,;\widetilde{13},2) 
\\[4pt]  
&  
&   
& 5.\,(23,a)
& (\widetilde{a},b\,;\widetilde{23},1) 
\\[4pt] 
&  
&   
& 6.\,(23,b)
& (a,\widetilde{b}\,;\widetilde{23},1) 
\\[4pt]  \hline
& & & & \\[-12pt]
{\tt Dip\,3u}  
&  \mbox{B}3u=gg \to u\bar{u}
& (6)-1    
& 7.\,(a1,2)
& (\widetilde{a1},b\,;\widetilde{2},3) 
\\[4pt]  
&  
&   
& 8.\,(a1,3)
& (\widetilde{a1},b\,;2,\widetilde{3}) 
\\[4pt] 
&  
&   
& 9.\,(a2,1)
& (\widetilde{a2},b\,;\widetilde{1},3) 
\\[4pt]  
&  
&   
& 10.\,(a2,3)
& (\widetilde{a2},b\,;1,\widetilde{3}) 
\\[4pt]  
&  
& (6)-2  
& 11.\,(a1,b)
& (\widetilde{a1},\widetilde{b}\,;2,3) 
\\[4pt]
&  
&  
& 12.\,(a2,b)
& (\widetilde{a2},\widetilde{b}\,;1,3) 
\\[4pt]  \hline
& & & & \\[-12pt]
{\tt Dip\,4u}  
& \mbox{B}4u=u\bar{u} \to u\bar{u}
& (7)-1   
& 13.\,(b1,2)
& (a,\widetilde{b1}\,;\widetilde{2},3)  
\\[4pt]  
&  
&   
& 14.\,(b1,3)
& (a,\widetilde{b1}\,;2,\widetilde{3})  
\\[4pt]  
&  
&   
& 15.\,(b2,1)
& (a,\widetilde{b2}\,;\widetilde{1},3)  
\\[4pt]  
&  
&   
& 16.\,(b2,3)
& (a,\widetilde{b2}\,;1,\widetilde{3})  
\\[4pt]  
&  
& (7)-2   
& 17.\,(b1,a)
& (\widetilde{a},\widetilde{b1}\,;2,3)  
\\[4pt] 
&  
&   
& 18.\,(b2,a)
& (\widetilde{a},\widetilde{b2}\,;1,3)  
\\[4pt] \hline
& & & & \\[-12pt]
{\tt Dip\,4\bar{u}}  
& \mbox{B}4\bar{u}=uu \to uu
& (7)-1   
& 19.\,(b3,1)
& (a,\widetilde{b3}\,;\widetilde{1},2)  
\\[4pt]  
&  
&   
& 20.\,(b3,2)
& (a,\widetilde{b3}\,;1,\widetilde{2})  
\\[4pt]  
&  
& (7)-2   
& 21.\,(b3,a)
& (\widetilde{a},\widetilde{b3}\,;1,2)  
\\[4pt] \hline
\end{array} \nonumber
\end{align}
\caption{\small
Summary table of $\mbox{D}\,(\mbox{R}_{3u})$
\label{ap_B_1_tab3}}
\end{table}
%
%
\begin{table}[h!]
  \centering
\begin{align}
&\mbox{D}\,
(\mbox{R}_{4u} = u\bar{u} \to d\bar{d}g): 
\ \ S_{\mbox{{\tiny R}}_{4u}}=1
\nonumber\\
&
\begin{array}{|c|c|c|c|c|} \hline
{\tt Dip}\,j
& \mbox{B}j 
& \mbox{Splitting} 
& (x_{I}x_{J},x_{K}) 
& (y_{a},y_{b}:y_{1},y_{2}) 
\\[4pt] \hline
& & & & \\[-12pt]
{\tt Dip\,1}  
& \mbox{B}1=u\bar{u} \to d\bar{d}
& (1)-1   
& 1.\,(13,2)
& (a,b\,;\widetilde{13},\widetilde{2} ) 
\\[4pt]  
&  
&   
& 2.\,(23,1)
& (a,b\,;\widetilde{1},\widetilde{23}) 
\\[4pt]  
&  
& (1)-2  
& 3.\,(13,a)
& (\widetilde{a},b\,;\widetilde{13},2) 
\\[4pt]  
&  
&   
& 4.\,(13,b)
& (a,\widetilde{b}\,;\widetilde{13},2) 
\\[4pt]  
&  
&  
& 5.\,(23,a)
& (\widetilde{a},b\,;1,\widetilde{23}) 
\\[4pt]  
&  
&   
& 6.\,(23,b)
& (a,\widetilde{b}\,;1,\widetilde{23}) 
\\[4pt]  
&  
& (3)-1  
& 7.\,(a3,1)
& (\widetilde{a3},b\,;\widetilde{1},2) 
\\[4pt]  
&  
&   
& 8.\,(a3,2)
& (\widetilde{a3},b\,;1,\widetilde{2}) 
\\[4pt]  
&  
&   
& 9.\,(b3,1)
& (a,\widetilde{b3}\,;\widetilde{1},2) 
\\[4pt]  
&  
&   
& 10.\,(b3,2)
& (a,\widetilde{b3}\,;1,\widetilde{2})
\\[4pt]  
&  
& (3)-2  
& 11.\,(a3,b)
& (\widetilde{a3},\widetilde{b}\,;1,2) 
\\[4pt]  
&  
&   
& 12.\,(b3,a)
& (\widetilde{a},\widetilde{b3}\,;1,2) 
\\[4pt]  \hline
& & & & \\[-12pt]
{\tt Dip\,2d}  
& \mbox{B}2d=u\bar{u} \to gg
& (5)-1  
& 13.\,(12,3)
& (a,b\,;\widetilde{12},\widetilde{3}) 
\\[4pt]  
&  
& (5)-2    
& 14.\,(12,a)
& (\widetilde{a},b\,;\widetilde{12},3) 
\\[4pt]  
&  
&   
& 15.\,(12,b)
& (a,\widetilde{b}\,;\widetilde{12},3) 
\\[4pt]  \hline
\end{array} \nonumber
\end{align}
\caption{\small
Summary table of $\mbox{D}\,(\mbox{R}_{4u})$
\label{ap_B_1_tab4}}
\end{table}
%
%
\begin{table}[h!]
  \centering
\begin{align}
&\mbox{D}\,
(\mbox{R}_{5ud} = ud \to udg): 
\ \ S_{\mbox{{\tiny R}}_{5ud}}=1
\nonumber\\
&
\begin{array}{|c|c|c|c|c|} \hline
{\tt Dip}\,j
& \mbox{B}j 
& \mbox{Splitting} 
& (x_{I}x_{J},x_{K}) 
& (y_{a},y_{b}:y_{1},y_{2}) 
\\[4pt] \hline
& & & & \\[-12pt]
{\tt Dip\,1}  
& \mbox{B}1=ud \to ud
& (1)-1   
& 1.\,(13,2)
& (a,b\,;\widetilde{13},\widetilde{2} ) 
\\[4pt]  
&  
&   
& 2.\,(23,1)
& (a,b\,;\widetilde{1},\widetilde{23}) 
\\[4pt]  
&  
& (1)-2  
& 3.\,(13,a)
& (\widetilde{a},b\,;\widetilde{13},2) 
\\[4pt]  
&  
&   
& 4.\,(13,b)
& (a,\widetilde{b}\,;\widetilde{13},2) 
\\[4pt]  
&  
&  
& 5.\,(23,a)
& (\widetilde{a},b\,;1,\widetilde{23}) 
\\[4pt]  
&  
&   
& 6.\,(23,b)
& (a,\widetilde{b}\,;1,\widetilde{23}) 
\\[4pt]  
&  
& (3)-1  
& 7.\,(a3,1)
& (\widetilde{a3},b\,;\widetilde{1},2) 
\\[4pt]  
&  
&   
& 8.\,(a3,2)
& (\widetilde{a3},b\,;1,\widetilde{2}) 
\\[4pt]  
&  
&   
& 9.\,(b3,1)
& (a,\widetilde{b3}\,;\widetilde{1},2) 
\\[4pt]  
&  
&   
& 10.\,(b3,2)
& (a,\widetilde{b3}\,;1,\widetilde{2})
\\[4pt]  
&  
& (3)-2  
& 11.\,(a3,b)
& (\widetilde{a3},\widetilde{b}\,;1,2) 
\\[4pt]  
&  
&   
& 12.\,(b3,a)
& (\widetilde{a},\widetilde{b3}\,;1,2) 
\\[4pt]  \hline
& & & & \\[-12pt]
{\tt Dip\,3u}  
&  \mbox{B}3u=gd \to dg
& (6)-1    
& 13.\,(a1,2)
& (\widetilde{a1},b\,;\widetilde{2},3) 
\\[4pt]  
&  
&   
& 14.\,(a1,3)
& (\widetilde{a1},b\,;2,\widetilde{3}) 
\\[4pt]  
&  
& (6)-2  
& 15.\,(a1,b)
& (\widetilde{a1},\widetilde{b}\,;2,3) 
\\[4pt]  \hline
& & & & \\[-12pt]
{\tt Dip\,3d}  
& \mbox{B}3d=ug \to ug
& (6)-1   
& 16.\,(b2,1)
& (a,\widetilde{b2}\,;\widetilde{1},3)  
\\[4pt]  
&  
&   
& 17.\,(b2,3)
& (a,\widetilde{b2}\,;1,\widetilde{3})  
\\[4pt]  
&  
& (6)-2   
& 18.\,(b2,a)
& (\widetilde{a},\widetilde{b2}\,;1,3)  
\\[4pt] \hline
\end{array} \nonumber
\end{align}
\caption{\small
Summary table of $\mbox{D}\,(\mbox{R}_{5ud})$
\label{ap_B_1_tab5}}
\end{table}
%
%
\begin{table}[h!]
  \centering
\begin{align}
&\mbox{D}\,
(\mbox{R}_{6u\bar{d}}=u\bar{d} \to u\bar{d}g): 
\ \ S_{\mbox{{\tiny R}}_{6u\bar{d}}}=1
\nonumber\\
&
\begin{array}{|c|c|c|c|c|} \hline
{\tt Dip}\,j
& \mbox{B}j 
& \mbox{Splitting} 
& (x_{I}x_{J},x_{K}) 
& (y_{a},y_{b}:y_{1},y_{2}) 
\\[4pt] \hline
& & & & \\[-12pt]
{\tt Dip\,1}  
& \mbox{B}1=u\bar{d} \to u\bar{d}
& (1)-1   
& 1.\,(13,2)
& (a,b\,;\widetilde{13},\widetilde{2} ) 
\\[4pt]  
&  
&   
& 2.\,(23,1)
& (a,b\,;\widetilde{1},\widetilde{23}) 
\\[4pt]  
&  
& (1)-2  
& 3.\,(13,a)
& (\widetilde{a},b\,;\widetilde{13},2) 
\\[4pt]  
&  
&   
& 4.\,(13,b)
& (a,\widetilde{b}\,;\widetilde{13},2) 
\\[4pt]  
&  
&  
& 5.\,(23,a)
& (\widetilde{a},b\,;1,\widetilde{23}) 
\\[4pt]  
&  
&   
& 6.\,(23,b)
& (a,\widetilde{b}\,;1,\widetilde{23}) 
\\[4pt]  
&  
& (3)-1  
& 7.\,(a3,1)
& (\widetilde{a3},b\,;\widetilde{1},2) 
\\[4pt]  
&  
&   
& 8.\,(a3,2)
& (\widetilde{a3},b\,;1,\widetilde{2}) 
\\[4pt]  
&  
&   
& 9.\,(b3,1)
& (a,\widetilde{b3}\,;\widetilde{1},2) 
\\[4pt]  
&  
&   
& 10.\,(b3,2)
& (a,\widetilde{b3}\,;1,\widetilde{2})
\\[4pt]  
&  
& (3)-2  
& 11.\,(a3,b)
& (\widetilde{a3},\widetilde{b}\,;1,2) 
\\[4pt]  
&  
&   
& 12.\,(b3,a)
& (\widetilde{a},\widetilde{b3}\,;1,2) 
\\[4pt]  \hline
& & & & \\[-12pt]
{\tt Dip\,3u}  
&  \mbox{B}3u=g\bar{d} \to \bar{d}g
& (6)-1    
& 13.\,(a1,2)
& (\widetilde{a1},b\,;\widetilde{2},3) 
\\[4pt]  
&  
&   
& 14.\,(a1,3)
& (\widetilde{a1},b\,;2,\widetilde{3}) 
\\[4pt]  
&  
& (6)-2  
& 15.\,(a1,b)
& (\widetilde{a1},\widetilde{b}\,;2,3) 
\\[4pt]  \hline
& & & & \\[-12pt]
{\tt Dip\,3\bar{d}}  
& \mbox{B}3\bar{d}=ug \to ug
& (6)-1   
& 16.\,(b2,1)
& (a,\widetilde{b2}\,;\widetilde{1},3)  
\\[4pt]  
&  
&   
& 17.\,(b2,3)
& (a,\widetilde{b2}\,;1,\widetilde{3})  
\\[4pt]  
&  
& (6)-2   
& 18.\,(b2,a)
& (\widetilde{a},\widetilde{b2}\,;1,3)  
\\[4pt] \hline
\end{array} \nonumber
\end{align}
\caption{\small
Summary table of $\mbox{D}\,(\mbox{R}_{6u\bar{d}})$
\label{ap_B_1_tab6}}
\end{table}
%
%
\begin{table}[h!]
  \centering
\begin{align}
&\mbox{D}\,
(\mbox{R}_{7u} = ug \to ud\bar{d}): 
\ \ S_{\mbox{{\tiny R}}_{7u}}=1
\nonumber\\
&
\begin{array}{|c|c|c|c|c|} \hline
{\tt Dip}\,j
& \mbox{B}j 
& \mbox{Splitting} 
& (x_{I}x_{J},x_{K}) 
& (y_{a},y_{b}:y_{1},y_{2}) 
\\[4pt] \hline
& & & & \\[-12pt]
{\tt Dip\,2u}  
& \mbox{B}2u=ug \to ug
& (5)-1  
& 1.\,(23,1)
& (a,b\,;\widetilde{1},\widetilde{23}) 
\\[4pt] 
&  
& (5)-2    
& 2.\,(23,a)
& (\widetilde{a},b\,;1,\widetilde{23}) 
\\[4pt]  
&  
&   
& 3.\,(23,b)
& (a,\widetilde{b}\,;1,\widetilde{23}) 
\\[4pt]  \hline
& & & & \\[-12pt]
{\tt Dip\,3u}  
&  \mbox{B}3u=gg \to d\bar{d}
& (6)-1    
& 4.\,(a1,2)
& (\widetilde{a1},b\,;\widetilde{2},3) 
\\[4pt]  
&  
&   
& 5.\,(a1,3)
& (\widetilde{a1},b\,;2,\widetilde{3}) 
\\[4pt] 
&  
& (6)-2  
& 6.\,(a1,b)
& (\widetilde{a1},\widetilde{b}\,;2,3) 
\\[4pt] \hline
& & & & \\[-12pt]
{\tt Dip\,4u}  
& \mbox{B}4u=u\bar{u} \to d\bar{d}
& (7)-1   
& 7.\,(b1,2)
& (a,\widetilde{b1}\,;\widetilde{2},3)  
\\[4pt]  
&  
&   
& 8.\,(b1,3)
& (a,\widetilde{b1}\,;2,\widetilde{3})  
\\[4pt] 
&  
& (7)-2 
& 9.\,(b1,a)
& (\widetilde{a},\widetilde{b1}\,;2,3)  
\\[4pt] \hline
& & & & \\[-12pt]
{\tt Dip\,4d}  
& \mbox{B}4d=u\bar{d} \to u\bar{d}
& (7)-1   
& 10.\,(b2,1)
& (a,\widetilde{b2}\,;\widetilde{1},3)  
\\[4pt]  
&  
&   
& 11.\,(b2,3)
& (a,\widetilde{b2}\,;1,\widetilde{3})  
\\[4pt]  
&  
& (7)-2   
& 12.\,(b2,a)
& (\widetilde{a},\widetilde{b2}\,;1,3)  
\\[4pt] \hline
& & & & \\[-12pt]
{\tt Dip\,4\bar{d}}  
& \mbox{B}4\bar{d}=ud \to ud
& (7)-1   
& 13.\,(b3,1)
& (a,\widetilde{b3}\,;\widetilde{1},2)  
\\[4pt]  
&  
&   
& 14.\,(b3,2)
& (a,\widetilde{b3}\,;1,\widetilde{2})  
\\[4pt]  
&  
& (7)-2   
& 15.\,(b3,a)
& (\widetilde{a},\widetilde{b3}\,;1,2)  
\\[4pt] \hline
\end{array} \nonumber
\end{align}
\caption{\small
Summary table of: 
$\mbox{D}\,(\mbox{R}_{7u})$
\label{ap_B_1_tab7}}
\end{table}
%
%
\begin{table}[h!]
  \centering
\begin{align}
&\mbox{D}\,
(\mbox{R}_{8u} = u\bar{u} \to ggg): 
\ \ S_{\mbox{{\tiny R}}_{8u}}=6
\nonumber\\
&
\begin{array}{|c|c|c|c|c|} \hline
{\tt Dip}\,j
& \mbox{B}j 
& \mbox{Splitting} 
& (x_{I}x_{J},x_{K}) 
& (y_{a},y_{b}:y_{1},y_{2}) 
\\[4pt] \hline
& & & & \\[-12pt]
{\tt Dip\,1}  
& \mbox{B}1=u\bar{u} \to gg
& (2)-1   
& 1.\,(12,3)
& (a,b\,;\widetilde{12},\widetilde{3} ) 
\\[4pt]  
&  
&   
& 2.\,(13,2)
& (a,b\,;\widetilde{13},\widetilde{2}) 
\\[4pt]  
&  
&   
& 3.\,(23,1)
& (a,b\,;\widetilde{23},\widetilde{1}) 
\\[4pt]  
&  
& (2)-2  
& 4.\,(12,a)
& (\widetilde{a},b\,;\widetilde{12},3) 
\\[4pt]  
&  
&   
& 5.\,(12,b)
& (a,\widetilde{b}\,;\widetilde{12},3) 
\\[4pt]  
&  
&  
& 6.\,(13,a)
& (\widetilde{a},b\,;\widetilde{13},2)
\\[4pt]  
&  
&   
& 7.\,(13,b)
& (a,\widetilde{b}\,;\widetilde{13},2) 
\\[4pt] 
&  
&   
& 8.\,(23,a)
& (\widetilde{a},b\,;\widetilde{23},1)
\\[4pt] 
&  
&   
& 9.\,(23,b)
& (a,\widetilde{b}\,;\widetilde{23},1) 
\\[4pt] 
&  
& (3)-1  
& 10.\,(a1,2)
& (\widetilde{a1},b\,;\widetilde{2},3) 
\\[4pt]  
&  
&   
& 11.\,(a1,3)
& (\widetilde{a1},b\,;2,\widetilde{3}) 
\\[4pt]  
&  
&   
& 12.\,(a2,1)
& (\widetilde{a2},b\,;\widetilde{1},3)  
\\[4pt]  
&  
&   
& 13.\,(a2,3)
& (\widetilde{a2},b\,;1,\widetilde{3}) 
\\[4pt]  
&  
&   
& 14.\,(a3,1)
& (\widetilde{a3},b\,;\widetilde{1},2)
\\[4pt]  
&  
&   
& 15.\,(a3,2)
& (\widetilde{a3},b\,;1,\widetilde{2})
\\[4pt]   
&  
&   
& 16.\,(b1,2)
& (a,\widetilde{b1}\,;\widetilde{2},3)
\\[4pt]  
&  
&   
& 17.\,(b1,3)
& (a,\widetilde{b1}\,;2,\widetilde{3})
\\[4pt]  
&  
&   
& 18.\,(b2,1)
& (a,\widetilde{b2}\,;\widetilde{1},3)
\\[4pt]  
&  
&   
& 19.\,(b2,3)
& (a,\widetilde{b2}\,;1,\widetilde{3})
\\[4pt]  
&  
&   
& 20.\,(b3,1)
& (a,\widetilde{b3}\,;\widetilde{1},2)
\\[4pt]   
&  
&   
& 21.\,(b3,2)
& (a,\widetilde{b3}\,;1,\widetilde{2})
\\[4pt]  
&  
& (3)-2  
& 22.\,(a1,b)
& (\widetilde{a1},\widetilde{b}\,;2,3) 
\\[4pt]  
&  
& 
& 23.\,(a2,b)
& (\widetilde{a2},\widetilde{b}\,;1,3) 
\\[4pt] 
&  
& 
& 24.\,(a3,b)
& (\widetilde{a3},\widetilde{b}\,;1,2) 
\\[4pt] 
&  
& 
& 25.\,(b1,a)
& (\widetilde{a},\widetilde{b1}\,;2,3) 
\\[4pt] 
&  
& 
& 26.\,(b2,a)
& (\widetilde{a},\widetilde{b2}\,;1,3) 
\\[4pt] 
&  
&   
& 27.\,(b3,a)
& (\widetilde{a},\widetilde{b3}\,;1,2) 
\\[4pt]  \hline
\end{array} \nonumber
\end{align}
\caption{\small
Summary table of 
$\mbox{D}\,(\mbox{R}_{8u})$
\label{ap_B_1_tab8}}
\end{table}
%
%
\begin{table}[h!]
  \centering
\begin{align}
&\mbox{D}\,
(\mbox{R}_{9u} = ug \to ugg): 
\ \ S_{\mbox{{\tiny R}}_{9u}}=2
\nonumber\\
&
\begin{array}{|c|c|c|c|c|} \hline
{\tt Dip}\,j
& \mbox{B}j 
& \mbox{Splitting} 
& (x_{I}x_{J},x_{K}) 
& (y_{a},y_{b}:y_{1},y_{2}) 
\\[4pt] \hline
& & & & \\[-12pt]
{\tt Dip\,1}  
& \mbox{B}1=ug \to ug
& (1)-1   
& 1.\,(12,3)
& (a,b\,;\widetilde{12},\widetilde{3} ) 
\\[4pt]  
&  
&   
& 2.\,(13,2)
& (a,b\,;\widetilde{13},\widetilde{2}) 
\\[4pt]  
&  
& (1)-2  
& 3.\,(12,a)
& (\widetilde{a},b\,;\widetilde{12},3) 
\\[4pt]  
&  
&   
& 4.\,(12,b)
& (a,\widetilde{b}\,;\widetilde{12},3) 
\\[4pt]  
&  
&  
& 5.\,(13,a)
& (\widetilde{a},b\,;\widetilde{13},2) 
\\[4pt]  
&  
&   
& 6.\,(13,b)
& (a,\widetilde{b}\,;\widetilde{13},2) 
\\[4pt]  
&  
& (2)-1  
& 7.\,(23,1)
& (a,b\,;\widetilde{1},\widetilde{23}) 
\\[4pt]  
&  
& (2)-2
& 8.\,(23,a)
& (\widetilde{a},b\,;1,\widetilde{23}) 
\\[4pt]  
&  
&   
& 9.\,(23,b)
& (a,\widetilde{b}\,;1,\widetilde{23}) 
\\[4pt]  
&  
& (3)-1  
& 10.\,(a2,1)
& (\widetilde{a2},b\,;\widetilde{1},3)
\\[4pt]  
&  
&
& 11.\,(a2,3)
& (\widetilde{a2},b\,;1,\widetilde{3}) 
\\[4pt]  
&  
&   
& 12.\,(a3,1)
& (\widetilde{a3},b\,;\widetilde{1},2) 
\\[4pt]
&  
& 
& 13.\,(a3,2)
& (\widetilde{a3},b\,;1,\widetilde{2}) 
\\[4pt]
&  
& (3)-2
& 14.\,(a2,b)
& (\widetilde{a2},\widetilde{b}\,;1,3) 
\\[4pt]
&  
&   
& 15.\,(a3,b)
& (\widetilde{a3},\widetilde{b}\,;1,2) 
\\[4pt]
&  
& (4)-1  
& 16.\,(b2,1)
& (a,\widetilde{b2}\,;\widetilde{1},3)
\\[4pt]  
&  
&
& 17.\,(b2,3)
& (a,\widetilde{b2}\,;1,\widetilde{3}) 
\\[4pt]  
&  
&   
& 18.\,(b3,1)
& (a,\widetilde{b3}\,;\widetilde{1},2) 
\\[4pt]
&  
& 
& 19.\,(b3,2)
& (a,\widetilde{b3}\,;1,\widetilde{2}) 
\\[4pt]
&  
& (4)-2
& 20.\,(b2,a)
& (\widetilde{a},\widetilde{b2}\,;1,3) 
\\[4pt]
&  
&   
& 21.\,(b3,a)
& (\widetilde{a},\widetilde{b3}\,;1,2) 
\\[4pt]  \hline
& & & & \\[-12pt]
{\tt Dip\,3u}  
&  \mbox{B}3u=gg \to gg
& (6)-1    
& 22.\,(a1,2)
& (\widetilde{a1},b\,;\widetilde{2},3) 
\\[4pt]  
&  
&   
& 23.\,(a1,3)
& (\widetilde{a1},b\,;2,\widetilde{3}) 
\\[4pt]  
&  
& (6)-2  
& 24.\,(a1,b)
& (\widetilde{a1},\widetilde{b}\,;2,3) 
\\[4pt]  \hline
& & & & \\[-12pt]
{\tt Dip\,4u}  
& \mbox{B}4u=u\bar{u} \to gg
& (7)-1   
& 25.\,(b1,2)
& (a,\widetilde{b1}\,;\widetilde{2},3)  
\\[4pt]  
&  
&   
& 26.\,(b1,3)
& (a,\widetilde{b1}\,;2,\widetilde{3})  
\\[4pt]  
&  
& (7)-2   
& 27.\,(b1,a)
& (\widetilde{a},\widetilde{b1}\,;2,3)  
\\[4pt] \hline
\end{array} \nonumber
\end{align}
\caption{\small
Summary table of 
$\mbox{D}\,(\mbox{R}_{9u})$
\label{ap_B_1_tab9}}
\end{table}
%
%
\begin{table}[h!]
  \centering
\begin{align}
&\mbox{D}\,
(\mbox{R}_{10u}=gg \to u\bar{u}g): 
\ \ S_{\mbox{{\tiny R}}_{10u}}=1
\nonumber\\
&
\begin{array}{|c|c|c|c|c|} \hline
{\tt Dip}\,j
& \mbox{B}j 
& \mbox{Splitting} 
& (x_{I}x_{J},x_{K}) 
& (y_{a},y_{b}:y_{1},y_{2}) 
\\[4pt] \hline
& & & & \\[-12pt]
{\tt Dip\,1}  
& \mbox{B}1=gg \to u\bar{u}
& (1)-1   
& 1.\,(13,2)
& (a,b\,;\widetilde{13},\widetilde{2} ) 
\\[4pt]  
&  
&   
& 2.\,(23,1)
& (a,b\,;\widetilde{1},\widetilde{23}) 
\\[4pt]  
&  
& (1)-2  
& 3.\,(13,a)
& (\widetilde{a},b\,;\widetilde{13},2) 
\\[4pt]  
&  
&   
& 4.\,(13,b)
& (a,\widetilde{b}\,;\widetilde{13},2) 
\\[4pt]  
&  
&  
& 5.\,(23,a)
& (\widetilde{a},b\,;1,\widetilde{23}) 
\\[4pt]  
&  
&   
& 6.\,(23,b)
& (a,\widetilde{b}\,;1,\widetilde{23}) 
\\[4pt]  
&  
& (4)-1  
& 7.\,(a3,1)
& (\widetilde{a3},b\,;\widetilde{1},2) 
\\[4pt]  
&  
&   
& 8.\,(a3,2)
& (\widetilde{a3},b\,;1,\widetilde{2}) 
\\[4pt]  
&  
&   
& 9.\,(b3,1)
& (a,\widetilde{b3}\,;\widetilde{1},2) 
\\[4pt]  
&  
&   
& 10.\,(b3,2)
& (a,\widetilde{b3}\,;1,\widetilde{2})
\\[4pt]  
&  
& (4)-2  
& 11.\,(a3,b)
& (\widetilde{a3},\widetilde{b}\,;1,2) 
\\[4pt]  
&  
&   
& 12.\,(b3,a)
& (\widetilde{a},\widetilde{b3}\,;1,2) 
\\[4pt]  \hline
& & & & \\[-12pt]
{\tt Dip\,2u}  
& \mbox{B}2u=gg \to gg
& (5)-1  
& 13.\,(12,3)
& (a,b\,;\widetilde{12},\widetilde{3}) 
\\[4pt]  
&  
& (5)-2    
& 14.\,(12,a)
& (\widetilde{a},b\,;\widetilde{12},3) 
\\[4pt]  
&  
&   
& 15.\,(12,b)
& (a,\widetilde{b}\,;\widetilde{12},3) 
\\[4pt]  \hline
& & & & \\[-12pt]
{\tt Dip\,4u}  
&  \mbox{B}4u=\bar{u}g \to \bar{u}g
& (7)-1    
& 16.\,(a1,2)
& (\widetilde{a1},b\,;\widetilde{2},3) 
\\[4pt]  
&  
&   
& 17.\,(a1,3)
& (\widetilde{a1},b\,;2,\widetilde{3}) 
\\[4pt] 
&  
&   
& 18.\,(b1,2)
& (\widetilde{b1},a\,;\widetilde{2},3) 
\\[4pt] 
&  
&   
& 19.\,(b1,3)
& (\widetilde{b1},a\,;2,\widetilde{3}) 
\\[4pt] 
&  
& (7)-2  
& 20.\,(a1,b)
& (\widetilde{a1},\widetilde{b}\,;2,3) 
\\[4pt] 
&  
&  
& 21.\,(b1,a)
& (\widetilde{b1},\widetilde{a}\,;2,3) 
\\[4pt]  \hline
& & & & \\[-12pt]
{\tt Dip\,4\bar{u}}  
& \mbox{B}4\bar{u}=ug \to ug
& (7)-1   
& 22.\,(a2,1)
& (\widetilde{a2},b\,;\widetilde{1},3)  
\\[4pt]  
&  
&   
& 23.\,(a2,3)
& (\widetilde{a2},b\,;1,\widetilde{3})  
\\[4pt] 
&  
&   
& 24.\,(b2,1)
& (\widetilde{b2},a\,;\widetilde{1},3)  
\\[4pt] 
&  
&   
& 25.\,(b2,3)
& (\widetilde{b2},a\,;1,\widetilde{3})  
\\[4pt] 
&  
& (7)-2   
& 26.\,(a2,b)
& (\widetilde{a2},\widetilde{b}\,;1,3)  
\\[4pt]
&  
&   
& 27.\,(b2,a)
& (\widetilde{b2},\widetilde{a}\,;1,3)  
\\[4pt] \hline
\end{array} \nonumber
\end{align}
\caption{\small
Summary table of
$\mbox{D}\,(\mbox{R}_{10u})$
\label{ap_B_1_tab10}}
\end{table}
%
%
\begin{table}[h!]
  \centering
\begin{align}
&\mbox{D}\,
(\mbox{R}_{11}=gg \to ggg): 
\ \ S_{\mbox{{\tiny R}}_{11}}=6
\nonumber\\
&
\begin{array}{|c|c|c|c|c|} \hline
{\tt Dip}\,j
& \mbox{B}j 
& \mbox{Splitting} 
& (x_{I}x_{J},x_{K}) 
& (y_{a},y_{b}:y_{1},y_{2}) 
\\[4pt] \hline
& & & & \\[-12pt]
{\tt Dip\,1}  
& \mbox{B}1=gg \to gg
& (2)-1   
& 1.\,(12,3)
& (a,b\,;\widetilde{12},\widetilde{3}) 
\\[4pt]  
&  
&   
& 2.\,(13,2)
& (a,b\,;\widetilde{13},\widetilde{2}) 
\\[4pt]  
&  
&   
& 3.\,(23,1)
& (a,b\,;\widetilde{23},\widetilde{1}) 
\\[4pt]  
&  
& (2)-2  
& 4.\,(12,a)
& (\widetilde{a},b\,;\widetilde{12},3) 
\\[4pt]  
&  
&   
& 5.\,(12,b)
& (a,\widetilde{b}\,;\widetilde{12},3) 
\\[4pt]  
&  
&  
& 6.\,(13,a)
& (\widetilde{a},b\,;\widetilde{13},2) 
\\[4pt]  
&  
&   
& 7.\,(13,b)
& (a,\widetilde{b}\,;\widetilde{13},2) 
\\[4pt]  
&  
&   
& 8.\,(23,a)
& (\widetilde{a},b\,;\widetilde{23},1) 
\\[4pt]  
&  
&   
& 9.\,(23,b)
& (a,\widetilde{b}\,;\widetilde{23},1) 
\\[4pt]  
&  
& (4)-1  
& 10.\,(a1,2)
& (\widetilde{a1},b\,;\widetilde{2},3) 
\\[4pt]  
&  
&   
& 11.\,(a1,3)
& (\widetilde{a1},b\,;2,\widetilde{3}) 
\\[4pt]  
&  
&   
& 12.\,(a2,1)
& (\widetilde{a2},b\,;\widetilde{1},3) 
\\[4pt]  
&  
&   
& 13.\,(a2,3)
& (\widetilde{a2},b\,;1,\widetilde{3})
\\[4pt]  
&  
&   
& 14.\,(a3,1)
& (\widetilde{a3},b\,;\widetilde{1},2)
\\[4pt]  
&  
&   
& 15.\,(a3,2)
& (\widetilde{a3},b\,;1,\widetilde{2})
\\[4pt]  
&  
&   
& 16.\,(b1,2)
& (a,\widetilde{b1}\,;\widetilde{2},3)
\\[4pt]  
&  
&   
& 17.\,(b1,3)
& (a,\widetilde{b1}\,;2,\widetilde{3})
\\[4pt]  
&  
&   
& 18.\,(b2,1)
& (a,\widetilde{b2}\,;\widetilde{1},3)
\\[4pt]  
&  
&   
& 19.\,(b2,3)
& (a,\widetilde{b2}\,;1,\widetilde{3})
\\[4pt]  
&  
&   
& 20.\,(b3,1)
& (a,\widetilde{b3}\,;\widetilde{1},2)
\\[4pt]  
&  
&   
& 21.\,(b3,2)
& (a,\widetilde{b3}\,;1,\widetilde{2})
\\[4pt]  
&  
& (4)-2  
& 22.\,(a1,b)
& (\widetilde{a1},\widetilde{b}\,;2,3) 
\\[4pt]  
&  
& 
& 23.\,(a2,b)
& (\widetilde{a2},\widetilde{b}\,;1,3) 
\\[4pt]  
&  
& 
& 24.\,(a3,b)
& (\widetilde{a3},\widetilde{b}\,;1,2) 
\\[4pt]  
&  
& 
& 25.\,(b1,a)
& (\widetilde{a},\widetilde{b1}\,;2,3) 
\\[4pt]  
&  
& 
& 26.\,(b2,a)
& (\widetilde{a},\widetilde{b2}\,;1,3) 
\\[4pt]  
&  
&   
& 27.\,(b3,a)
& (\widetilde{a},\widetilde{b3}\,;1,2) 
\\[4pt]  \hline
\end{array} \nonumber
\end{align}
\caption{\small
Summary table of $\mbox{D}\,(\mbox{R}_{11})$
\label{ap_B_1_tab11}}
\end{table}

\clearpage
\subsection{I term \label{ap_B_2}}
\begin{equation}
\hat{\sigma}_{\mbox{{\tiny I}}}(\mbox{R}_{i}) 
= \frac{1}{S_{\mbox{{\tiny B1}} }} \ \Phi(\mbox{B}1)_{d} 
\cdot \mbox{I}(\mbox{R}_{i})\,, 
\end{equation}
\begin{equation}
\mbox{I}_{I,\,K}
= - A_{d} \cdot
\frac{1}{\mbox{T}_{\mbox{{\tiny F}}(I)}^{2}}\,
{\cal V}_{\mbox{{\tiny F}}(I)}
\cdot [ I,K ]\,,
\end{equation}
with the definition,
\begin{equation}
A_{d} = \frac{\al}{2\pi} \frac{(4\pi \mu^{2})^{\ep}}{\Gamma(1-\ep)}\,,
\ \ \mbox{and} \ \ 
[ I,K ]
=s_{{\scriptscriptstyle IK}}^{-\ep} \,
\langle 
\mbox{T}_{I} \cdot 
\mbox{T}_{K}
\rangle\,.
\end{equation}
%
%
\begin{table}[h!]
  \centering
\begin{align}
&\mbox{I}\,(\mbox{R}_{1u}): \ 
\mbox{B}1=u\bar{u} \to u\bar{u}\,,
\ \ S_{\mbox{{\tiny B}}_{1}}=1
\nonumber\\
&
\begin{array}{|c|c|c|} \hline
& & \\[-12pt]
 \mbox{Leg-}\,y_{I} 
& 
{\cal V}_{\mbox{{\tiny F}}(I)}/\mbox{T}_{\mbox{{\tiny F}}(I)}^{2}
& (y_{I},y_{K}) 
\\[2pt] \hline
(1)-1   
& {\cal V}_{f}/\mbox{C}_{\mbox{{\tiny F}}}
& 1.\,(1,2)
\\[4pt]  
& 
& 2.\,(2,1)
\\[4pt]  
(1)-2
& 
& 3.\,(1,a)
\\[4pt]  
& 
& 4.\,(1,b)
\\[4pt]  
& 
& 5.\,(2,a)
\\[4pt]  
& 
& 6.\,(2,b)
\\[4pt]  
(3)-1
& {\cal V}_{f}/\mbox{C}_{\mbox{{\tiny F}}}
& 7.\,(a,1)
\\[4pt]  
& 
& 8.\,(a,2)
\\[4pt]  
& 
& 9.\,(b,1)
\\[4pt]  
& 
& 10.\,(b,2)
\\[4pt]  
(3)-2
& 
& 11.\,(a,b)
\\[4pt]  
& 
& 12.\,(b,a)
\\[4pt]  \hline
\end{array} \nonumber
\end{align}
\caption{\small
Summary table of
$\mbox{I}\,(\mbox{R}_{1u})$.
The I terms,
$\mbox{I}\,(\mbox{R}_{2u/4u/5ud/6u})$,
are created from the inputs,
$\mbox{B}1(\mbox{R}_{2u})=uu \to uu$,
$\mbox{B}1(\mbox{R}_{4u})=u\bar{u} \to d\bar{d}$,
$\mbox{B}1(\mbox{R}_{5ud})= ud \to ud$,
and
$\mbox{B}1(\mbox{R}_{6u\bar{d}})= u\bar{d} \to u\bar{d}$,
respectively.
The creations of the I terms are completely
analogous to the creation $\mbox{I}\,(\mbox{R}_{1u})$.
The summary tables are identical except for 
the differences about the field species, 
$\mbox{F}(y_{I})=u,d,\bar{u}$,
or $\bar{d}$, and the symmetric factor,
$S_{\mbox{{\tiny B}}_{2u}}=2$.
\label{ap_B_2_tab1}}
\end{table}
%
%
\begin{table}[h!]
  \centering
\begin{align}
&\mbox{I}\,(\mbox{R}_{8u}): \ 
\mbox{B}1=u\bar{u} \to gg\,,
\ \ S_{\mbox{{\tiny B}}_{1}}=2
\nonumber\\
&
\begin{array}{|c|c|c|} \hline
& & \\[-12pt]
 \mbox{Leg-}\,y_{I} 
& 
{\cal V}_{\mbox{{\tiny F}}(I)}/\mbox{T}_{\mbox{{\tiny F}}(I)}^{2}
& (y_{I},y_{K}) 
\\[2pt] \hline
(2)-1   
& {\cal V}_{g}/\mbox{C}_{\mbox{{\tiny A}}}
& 1.\,(1,2)
\\[4pt]  
& 
& 2.\,(2,1)
\\[4pt]  
(2)-2
& 
& 3.\,(1,a)
\\[4pt]  
& 
& 4.\,(1,b)
\\[4pt]  
& 
& 5.\,(2,a)
\\[4pt]  
& 
& 6.\,(2,b)
\\[4pt]  
(3)-1
& {\cal V}_{f}/\mbox{C}_{\mbox{{\tiny F}}}
& 7.\,(a,1)
\\[4pt]  
& 
& 8.\,(a,2)
\\[4pt]  
& 
& 9.\,(b,1)
\\[4pt]  
& 
& 10.\,(b,2)
\\[4pt]  
(3)-2
& 
& 11.\,(a,b)
\\[4pt]  
& 
& 12.\,(b,a)
\\[4pt]  \hline
\end{array} \nonumber
\end{align}
\caption{\small
Summary table of 
$\mbox{I}\,(\mbox{R}_{8u})$.
\label{ap_B_2_tab2}}
\end{table}
%
%
\begin{table}[h!]
  \centering
\begin{align}
&\mbox{I}\,(\mbox{R}_{9u}): \ 
\mbox{B}1=ug \to ug \,,
\ \ S_{\mbox{{\tiny B}}_{1}}=1
\nonumber\\
&
\begin{array}{|c|c|c|} \hline
& & \\[-12pt]
 \mbox{Leg-}\,y_{I} 
& 
{\cal V}_{\mbox{{\tiny F}}(I)}/\mbox{T}_{\mbox{{\tiny F}}(I)}^{2}
& (y_{I},y_{K}) 
\\[2pt] \hline
(1)-1   
& {\cal V}_{f}/\mbox{C}_{\mbox{{\tiny F}}}
& 1.\,(1,2)
\\[4pt]  
(1)-2
& 
& 2.\,(1,a)
\\[4pt]  
& 
& 3.\,(1,b)
\\[4pt]  
(2)-1   
& {\cal V}_{g}/\mbox{C}_{\mbox{{\tiny A}}}
& 4.\,(2,1)
\\[4pt]  
(2)-2    
& 
& 5.\,(2,a)
\\[4pt]  
& 
& 6.\,(2,b)
\\[4pt]  
(3)-1
& {\cal V}_{f}/\mbox{C}_{\mbox{{\tiny F}}}
& 7.\,(a,1)
\\[4pt]  
& 
& 8.\,(a,2)
\\[4pt]  
(3)-2   
& 
& 9.\,(a,b)
\\[4pt]  
(4)-1    
& {\cal V}_{g}/\mbox{C}_{\mbox{{\tiny A}}}
& 10.\,(b,1)
\\[4pt]  
& 
& 11.\,(b,2)
\\[4pt]  
(4)-2   
& 
& 12.\,(b,a)
\\[4pt]  \hline
\end{array} \nonumber
\end{align}
\caption{\small
Summary table of 
$\mbox{I}\,(\mbox{R}_{9u})$
\label{ap_B_2_tab3}}
\end{table}
%
%
\begin{table}[h!]
  \centering
\begin{align}
&\mbox{I}\,(\mbox{R}_{10u}): \ 
\mbox{B}1=gg \to u\bar{u}\,,
\ \ S_{\mbox{{\tiny B}}_{1}}=1
\nonumber\\
&
\begin{array}{|c|c|c|} \hline
& & \\[-12pt]
 \mbox{Leg-}\,y_{I} 
& 
{\cal V}_{\mbox{{\tiny F}}(I)}/\mbox{T}_{\mbox{{\tiny F}}(I)}^{2}
& (y_{I},y_{K}) 
\\[2pt] \hline
(1)-1   
& {\cal V}_{f}/\mbox{C}_{\mbox{{\tiny F}}}
& 1.\,(1,2)
\\[4pt]  
& 
& 2.\,(2,1)
\\[4pt]  
(1)-2
& 
& 3.\,(1,a)
\\[4pt]  
& 
& 4.\,(1,b)
\\[4pt]  
& 
& 5.\,(2,a)
\\[4pt]  
& 
& 6.\,(2,b)
\\[4pt]  
(4)-1
& {\cal V}_{g}/\mbox{C}_{\mbox{{\tiny A}}}
& 7.\,(a,1)
\\[4pt]  
& 
& 8.\,(a,2)
\\[4pt]  
& 
& 9.\,(b,1)
\\[4pt]  
& 
& 10.\,(b,2)
\\[4pt]  
(4)-2
& 
& 11.\,(a,b)
\\[4pt]  
& 
& 12.\,(b,a)
\\[4pt]  \hline
\end{array} \nonumber
\end{align}
\caption{\small
Summary table of 
$\mbox{I}\,(\mbox{R}_{10u})$
\label{ap_B_2_tab4}}
\end{table}
%
%
\begin{table}[h!]
  \centering
\begin{align}
&\mbox{I}\,(\mbox{R}_{11}): \ 
\mbox{B}1=gg \to gg\,,
\ \ S_{\mbox{{\tiny B}}_{1}}=2
\nonumber\\
&
\begin{array}{|c|c|c|} \hline
& & \\[-12pt]
 \mbox{Leg-}\,y_{I} 
&
{\cal V}_{\mbox{{\tiny F}}(I)}/\mbox{T}_{\mbox{{\tiny F}}(I)}^{2}
& (y_{I},y_{K}) 
\\[2pt] \hline
(2)-1   
& {\cal V}_{g}/\mbox{C}_{\mbox{{\tiny A}}}
& 1.\,(1,2)
\\[4pt]  
& 
& 2.\,(2,1)
\\[4pt]  
(2)-2
& 
& 3.\,(1,a)
\\[4pt]  
& 
& 4.\,(1,b)
\\[4pt]  
& 
& 5.\,(2,a)
\\[4pt]  
& 
& 6.\,(2,b)
\\[4pt]  
(4)-1
& 
& 7.\,(a,1)
\\[4pt]  
& 
& 8.\,(a,2)
\\[4pt]  
& 
& 9.\,(b,1)
\\[4pt]  
& 
& 10.\,(b,2)
\\[4pt]  
(4)-2
& 
& 11.\,(a,b)
\\[4pt]  
& 
& 12.\,(b,a)
\\[4pt]  \hline
\end{array} \nonumber
\end{align}
\caption{\small
Summary table of 
$\mbox{I}\,(\mbox{R}_{11})$
\label{ap_B_2_tab5}}
\end{table}

\clearpage
\subsection{P and K terms \label{ap_B_3}}
\begin{equation}
\hat{\sigma}_{\mbox{{\tiny P/K}}}(\mbox{R}_{i}) 
= \int_{0}^{1}dx \sum_{\mbox{{\tiny B}}_{j}} 
\frac{1}{S_{\mbox{{\tiny B}}_{j} }}
\Phi_{a}(\mbox{R}_{i}:\mbox{B}_{j},x)_{4} \cdot 
\mbox{P/K}(\mbox{R}_{i}, {\tt dip}j
,x_{a}) \ + \ 
(a \leftrightarrow b)\,. 
\nonumber
\end{equation}
\begin{align}
\mbox{P}(\mbox{R}_{i},
{\tt dip}j,
\,x_{a/b},y_{emi},y_{spe}
)
&=A_{4} \cdot
\frac{1}{\mbox{T}_{\mbox{{\tiny F}}(y_{emi})}^{2}} \,
P^{
\mbox{{\tiny F}}(x_{a/b})\mbox{{\tiny F}}(y_{emi})
}(x) \,
\ln \frac{\mu_{F}^{2}}{x\,s_{x_{a/b} y_{spe}}}
\times
\nonumber \\
&\hspace{40mm}
\ \langle \mbox{B}j 
\ | 
\mbox{T}_{y_{{\scriptscriptstyle emi}}} \cdot 
\mbox{T}_{y_{{\scriptscriptstyle spe}}}
| \ \mbox{B}j \rangle\,.
\end{align}
\begin{align}
&\mbox{K}(\mbox{R}_{i},{\tt dip}1/3/4\,,
(3)/(4)/(6)/(7)\mbox{-}0,
x_{a/b},y_{emi},y_{0}) 
= A_{4} \cdot
\overline{K}^{
\mbox{{\tiny F}}(x_{a/b})\mbox{{\tiny F}}(y_{emi})
}(x) \ \cdot \langle \mbox{B}j \rangle, 
\\
&\mbox{K}(\mbox{R}_{i},{\tt dip}1, (3)/(4)\mbox{-}1, x_{a/b},y_{emi},y_{spe}) 
= A_{4} \cdot
\frac{\gamma_{\mbox{{\tiny F}}(y_{spe})}}
{\mbox{T}_{\mbox{{\tiny F}}(y_{spe})}^{2}} \, h(x)
\ \cdot \langle \mbox{B}1 | 
\mbox{T}_{y_{emi}} \cdot 
\mbox{T}_{y_{spe}}
| \ \mbox{B}1 \rangle,
\\
&\mbox{K}(\mbox{R}_{i},{\tt dip}1/3/4\,,
(3)/(4)/(6)/(7)\mbox{-}2
,x_{a/b},y_{emi},y_{spe}) 
=- A_{4} \cdot
\frac{1}{\mbox{T}_{\mbox{{\tiny F}}(y_{emi})}^{2}} \, 
\tilde{K}^{
\mbox{{\tiny F}}(x_{a/b})\mbox{{\tiny F}}(y_{emi})
}(x) \ \times
\nonumber \\
& \hspace{100mm} \langle \mbox{B}j | 
\mbox{T}_{y_{emi}} \cdot 
\mbox{T}_{y_{spe}}
| \ \mbox{B}j \rangle\,.
\end{align}
%
%
\begin{table}[h!]
  \centering
\begin{align}
&\mbox{P/K}\,(\mbox{R}_{1u} = u\bar{u} \to u\bar{u}g)
\nonumber\\
&
\begin{array}{|c|c|c|c|c|c|c|} \hline
& & & & & & \\[-12pt]
\mbox{Leg}
&{\tt Dip}\,j
& \mbox{B}j 
& S_{\mbox{{\tiny B}}_{j}}
& \mbox{Splitting}
& P^{
\mbox{{\tiny F}}(x_{a/b})\mbox{{\tiny F}}(y_{emi})
}/\mbox{T}_{\mbox{{\tiny F}}(y_{emi})}^{2}
& (y_{emi},y_{spe}) 
\\[4pt] \hline
a
&{\tt Dip\,1}  
& \mbox{B}1=u\bar{u} \to u\bar{u}
& 1 
& (3)-0   
&
& 1.\,(a,0)
\\[4pt]  
&
&  
&   
& (3)-1
& P^{ff}/\mbox{C}_{\mbox{{\tiny F}}}
& 2.\,(a,1)
\\[4pt]  
&
&  
&   
&
& 
& 3.\,(a,2)
\\[4pt]  
&
&  
&   
& (3)-2
&
& 4.\,(a,b)
\\[4pt]  \cline{2-7}
&{\tt Dip\,3u}
&\mbox{B}3u=g\bar{u} \to \bar{u}g
& 1    
& (6)-0
&
& 5.\,(a,0)
\\[4pt] 
&
&  
&   
& (6)-1
& P^{fg}/\mbox{C}_{\mbox{{\tiny A}}}
& 6.\,(a,1)
\\[4pt] 
&
&  
&   
& 
&
& 7.\,(a,2)
\\[4pt] 
&
&  
&   
& (6)-2
&
& 8.\,(a,b)
\\[4pt]  \hline
b
&{\tt Dip\,1}  
& \mbox{B}1=u\bar{u} \to u\bar{u}
& 1 
& (3)-0   
&
& 9.\,(b,0)
\\[4pt]  
&
&  
&   
& (3)-1
& P^{ff}/\mbox{C}_{\mbox{{\tiny F}}}
& 10.\,(b,1)
\\[4pt]  
&
&  
&   
& 
&
& 11.\,(b,2)
\\[4pt]  
&
&  
&   
& (3)-2
&
& 12.\,(b,a)
\\[4pt]  \cline{2-7}
&{\tt Dip\,3\bar{u}}
&\mbox{B}3\bar{u}=ug \to ug
& 1    
& (6)-0
&
& 13.\,(b,0)
\\[4pt] 
&
&  
&   
& (6)-1
& P^{fg}/\mbox{C}_{\mbox{{\tiny A}}}
& 14.\,(b,1)
\\[4pt] 
&
&  
&   
& 
&
& 15.\,(b,2)
\\[4pt] 
&
&  
&   
& (6)-2
&
& 16.\,(b,a)
\\[4pt]  \hline
\end{array} \nonumber
\end{align}
\caption{\small
Summary table of 
$\mbox{P/K}\,(\mbox{R}_{1u})$
\label{ap_B_3_tab1}}
\end{table}
%
%
\begin{table}[h!]
  \centering
\begin{align}
&\mbox{P/K}\,(\mbox{R}_{2u} = uu \to uug)
\nonumber\\
&
\begin{array}{|c|c|c|c|c|c|c|} \hline
& & & & & & \\[-12pt]
\mbox{Leg}
&{\tt Dip}\,j
& \mbox{B}j 
& S_{\mbox{{\tiny B}}_{j}}
& \mbox{Splitting}
& P^{
\mbox{{\tiny F}}(x_{a/b})\mbox{{\tiny F}}(y_{emi})
}/\mbox{T}_{\mbox{{\tiny F}}(y_{emi})}^{2}
& (y_{emi},y_{spe})
\\[4pt] \hline
a
&{\tt Dip\,1}  
& \mbox{B}1=uu \to uu
& 2 
& (3)-0   
&
& 1.\,(a,0)
\\[4pt]  
&
&  
&   
& (3)-1
& P^{ff}/\mbox{C}_{\mbox{{\tiny F}}}
& 2.\,(a,1)
\\[4pt]  
&
&  
&   
&
&
& 3.\,(a,2)
\\[4pt]  
&
&  
&   
& (3)-2
&
& 4.\,(a,b)
\\[4pt]  \cline{2-7}
&{\tt Dip\,3u}
&\mbox{B}3u=gu \to ug
& 1    
& (6)-0
&
& 5.\,(a,0)
\\[4pt] 
&
&  
&   
& (6)-1
& P^{fg}/\mbox{C}_{\mbox{{\tiny A}}}
& 6.\,(a,1)
\\[4pt] 
&
&  
&   
&
& 
& 7.\,(a,2)
\\[4pt] 
&
&  
&   
& (6)-2
&
& 8.\,(a,b)
\\[4pt]  \hline
b
&{\tt Dip\,1}  
& \mbox{B}1=u\bar{u} \to u\bar{u}
& 1 
& (3)-0   
&
& 9.\,(b,0)
\\[4pt]  
&
&  
&   
& (3)-1
& P^{ff}/\mbox{C}_{\mbox{{\tiny F}}}
& 10.\,(b,1)
\\[4pt]  
&
&  
&   
& 
&
& 11.\,(b,2)
\\[4pt]  
&
&  
&   
& (3)-2
&
& 12.\,(b,a)
\\[4pt]  \cline{2-7}
&{\tt Dip\,3u}
&\mbox{B}3u=gu \to ug
& 1    
& (6)-0
&
& 13.\,(a,0)
\\[4pt] 
&
&  
&   
& (6)-1
& P^{fg}/\mbox{C}_{\mbox{{\tiny A}}}
& 14.\,(a,1)
\\[4pt] 
&
&  
&   
& 
&
& 15.\,(a,2)
\\[4pt] 
&
&  
&   
& (6)-2
&
& 16.\,(a,b)
\\[4pt]  \hline
\end{array} \nonumber
\end{align}
\caption{\small
Summary table of 
$\mbox{P/K}\,(\mbox{R}_{2u})$
\label{ap_B_3_tab2}}
\end{table}
%
%
\begin{table}[h!]
  \centering
\begin{align}
&\mbox{P/K}\,(\mbox{R}_{3u}= ug \to uu\bar{u})
\nonumber\\
&
\begin{array}{|c|c|c|c|c|c|c|} \hline
& & & & & & \\[-12pt]
\mbox{Leg}
&{\tt Dip}\,j
& \mbox{B}j 
& S_{\mbox{{\tiny B}}_{j}}
& \mbox{Splitting} 
& P^{
\mbox{{\tiny F}}(x_{a/b})\mbox{{\tiny F}}(y_{emi})
}/\mbox{T}_{\mbox{{\tiny F}}(y_{emi})}^{2}
& (y_{emi},y_{spe})
\\[4pt] \hline
a
&{\tt Dip\,3u}
&\mbox{B}3u=gg \to u\bar{u}
& 1    
& (6)-0
&
& 1.\,(a,0)
\\[4pt] 
&
&  
&   
& (6)-1
& P^{fg}/\mbox{C}_{\mbox{{\tiny A}}}
& 2.\,(a,1)
\\[4pt] 
&
&  
&   
&
& 
& 3.\,(a,2)
\\[4pt] 
&
&  
&   
& (6)-2
&
& 4.\,(a,b)
\\[4pt]  \hline
b
&{\tt Dip\,4u}
&\mbox{B}4u=u\bar{u} \to u\bar{u}
& 1    
& (7)-0
&
& 5.\,(b,0)
\\[4pt] 
&
&  
&   
& (7)-1
& P^{gf}/\mbox{C}_{\mbox{{\tiny F}}}
& 6.\,(b,1)
\\[4pt] 
&
&  
&   
&
& 
& 7.\,(b,2)
\\[4pt] 
&
&  
&   
& (7)-2
&
& 8.\,(b,a)
\\[4pt]  \cline{2-7}
&{\tt Dip\,4\bar{u}}
&\mbox{B}4\bar{u}=uu \to uu
& 2    
& (7)-0
&
& 9.\,(b,0)
\\[4pt] 
&
&  
&   
& (7)-1
& P^{gf}/\mbox{C}_{\mbox{{\tiny F}}}
& 10.\,(b,1)
\\[4pt] 
&
&  
&   
&
& 
& 11.\,(b,2)
\\[4pt] 
&
&  
&   
& (7)-2
&
& 12.\,(b,a)
\\[4pt]  \hline
\end{array} \nonumber
\end{align}
\caption{\small
Summary table of 
$\mbox{P/K}\,(\mbox{R}_{3u})$
\label{ap_B_3_tab3}}
\end{table}
%
%
\begin{table}[h!]
  \centering
\begin{align}
&\mbox{P/K}\,(\mbox{R}_{4u} = u\bar{u} \to d\bar{d}g)
\nonumber\\
&
\begin{array}{|c|c|c|c|c|c|c|} \hline
& & & & & & \\[-12pt]
\mbox{Leg}
&{\tt Dip}\,j
& \mbox{B}j 
& S_{\mbox{{\tiny B}}_{j}}
& \mbox{Splitting} 
& P^{
\mbox{{\tiny F}}(x_{a/b})\mbox{{\tiny F}}(y_{emi})
}/\mbox{T}_{\mbox{{\tiny F}}(y_{emi})}^{2}
& (y_{emi},y_{spe})
\\[4pt] \hline
& & & & & & \\[-12pt]
a
&{\tt Dip\,1}  
& \mbox{B}1=u\bar{u} \to d\bar{d}
& 1 
& (3)-0   
&
& 1.\,(a,0)
\\[4pt]  
&
&  
&   
& (3)-1
& P^{ff}/\mbox{C}_{\mbox{{\tiny F}}}
& 2.\,(a,1)
\\[4pt]  
&
&  
&   
& 
&
& 3.\,(a,2)
\\[4pt]  
&
&  
&   
& (3)-2
&
& 4.\,(a,b)
\\[4pt]  \hline
& & & & & & \\[-12pt]
b
&{\tt Dip\,1}  
& \mbox{B}1=u\bar{u} \to d\bar{d}
& 1 
& (3)-0   
&
& 5.\,(b,0)
\\[4pt]  
&
&  
&   
& (3)-1
& P^{ff}/\mbox{C}_{\mbox{{\tiny F}}}
& 6.\,(b,1)
\\[4pt]  
&
&  
&   
& 
&
& 7.\,(b,2)
\\[4pt]  
&
&  
&   
& (3)-2
&
& 8.\,(b,a)
\\[4pt]  \hline
\end{array} \nonumber
\end{align}
\caption{\small
Summary table of 
$\mbox{P/K}\,(\mbox{R}_{4u})$
\label{ap_B_3_tab4}}
\end{table}
%
%
\begin{table}[h!]
  \centering
\begin{align}
&\mbox{P/K}\,(\mbox{R}_{5ud} = ud \to udg)
\nonumber\\
&
\begin{array}{|c|c|c|c|c|c|c|} \hline
& & & & & & \\[-12pt]
\mbox{Leg}
&{\tt Dip}\,j
& \mbox{B}j 
& S_{\mbox{{\tiny B}}_{j}}
& \mbox{Splitting} 
& P^{
\mbox{{\tiny F}}(x_{a/b})\mbox{{\tiny F}}(y_{emi})
}/\mbox{T}_{\mbox{{\tiny F}}(y_{emi})}^{2}
& (y_{emi},y_{spe})
\\[4pt] \hline
& & & & & & \\[-12pt]
a
&{\tt Dip\,1}  
&\mbox{B}1=ud \to ud
& 1 
& (3)-0   
&
& 1.\,(a,0)
\\[4pt]  
&
&  
&   
& (3)-1
& P^{ff}/\mbox{C}_{\mbox{{\tiny F}}}
& 2.\,(a,1)
\\[4pt]  
&
&  
&   
&
& 
& 3.\,(a,2)
\\[4pt]  
&
&  
&   
& (3)-2
&
& 4.\,(a,b)
\\[4pt]  \cline{2-7}
& & & & \\[-12pt]
&{\tt Dip\,3u}
&\mbox{B}3u=gd \to dg
& 1    
& (6)-0
&
& 5.\,(a,0)
\\[4pt] 
&
&  
&   
& (6)-1
& P^{fg}/\mbox{C}_{\mbox{{\tiny A}}}
& 6.\,(a,1)
\\[4pt] 
&
&  
&   
& 
&
& 7.\,(a,2)
\\[4pt] 
&
&  
&   
& (6)-2
&
& 8.\,(a,b)
\\[4pt]  \hline
& & & & \\[-12pt]
b
&{\tt Dip\,1}  
&\mbox{B}1=ud \to ud
& 1 
& (3)-0   
&
& 9.\,(b,0)
\\[4pt]  
&
&  
&   
& (3)-1
& P^{ff}/\mbox{C}_{\mbox{{\tiny F}}}
& 10.\,(b,1)
\\[4pt]  
&
&  
&   
& 
&
& 11.\,(b,2)
\\[4pt]  
&
&  
&   
& (3)-2
&
& 12.\,(b,a)
\\[4pt]  \cline{2-7}
& & & & & &\\[-12pt]
&{\tt Dip\,3d}
&\mbox{B}3d=ug \to ug
& 1    
& (6)-0
&
& 13.\,(b,0)
\\[4pt] 
&
&  
&   
& (6)-1
& P^{fg}/\mbox{C}_{\mbox{{\tiny A}}}
& 14.\,(b,1)
\\[4pt] 
&
&  
&   
& 
&
& 15.\,(b,2)
\\[4pt] 
&
&  
&   
& (6)-2
&
& 16.\,(b,a)
\\[4pt]  \hline
\end{array} \nonumber
\end{align}
\caption{\small
Summary table of 
$\mbox{P/K}\,(\mbox{R}_{5ud})$
\label{ap_B_3_tab5}}
\end{table}
%
%
\begin{table}[h!]
  \centering
\begin{align}
&\mbox{P/K}\,(\mbox{R}_{6u\bar{d}}=u\bar{d} \to u\bar{d}g)
\nonumber\\
&
\begin{array}{|c|c|c|c|c|c|c|} \hline
& & & & & & \\[-12pt]
\mbox{Leg}
&{\tt Dip}\,j
& \mbox{B}j 
& S_{\mbox{{\tiny B}}_{j}}
& \mbox{Splitting}
& P^{
\mbox{{\tiny F}}(x_{a/b})\mbox{{\tiny F}}(y_{emi})
}/\mbox{T}_{\mbox{{\tiny F}}(y_{emi})}^{2}
& (y_{emi},y_{spe})
\\[4pt] \hline
& & & & & & \\[-12pt]
a
&{\tt Dip\,1}  
&\mbox{B}1=u\bar{d} \to u\bar{d}
& 1 
& (3)-0 
&
& 1.\,(a,0)
\\[4pt]  
&
&  
&   
& (3)-1
& P^{ff}/\mbox{C}_{\mbox{{\tiny F}}}
& 2.\,(a,1)
\\[4pt]  
&
&  
&   
& 
&
& 3.\,(a,2)
\\[4pt]  
&
&  
&   
& (3)-2
&
& 4.\,(a,b)
\\[4pt]  \cline{2-7}
& & & & & & \\[-12pt]
&{\tt Dip\,3u}
&\mbox{B}3u=g\bar{d} \to \bar{d}g
& 1    
& (6)-0
&
& 5.\,(a,0)
\\[4pt] 
&
&  
&   
& (6)-1
& P^{fg}/\mbox{C}_{\mbox{{\tiny A}}}
& 6.\,(a,1)
\\[4pt] 
&
&  
&   
& 
&
& 7.\,(a,2)
\\[4pt] 
&
&  
&   
& (6)-2
&
& 8.\,(a,b)
\\[4pt]  \hline
& & & & & & \\[-12pt]
b
&{\tt Dip\,1}  
&\mbox{B}1=u\bar{d} \to u\bar{d}
& 1 
& (3)-0   
&
& 9.\,(b,0)
\\[4pt]  
&
&  
&   
& (3)-1
& P^{ff}/\mbox{C}_{\mbox{{\tiny F}}}
& 10.\,(b,1)
\\[4pt]  
&
&  
&   
& 
&
& 11.\,(b,2)
\\[4pt]  
&
&  
&   
& (3)-2
&
& 12.\,(b,a)
\\[4pt]  \cline{2-7}
& & & & & &\\[-12pt]
&{\tt Dip\,3\bar{d}}
&\mbox{B}3\bar{d}=ug \to ug
& 1    
& (6)-0
&
& 13.\,(b,0)
\\[4pt] 
&
&  
&   
& (6)-1
& P^{fg}/\mbox{C}_{\mbox{{\tiny A}}}
& 14.\,(b,1)
\\[4pt] 
&
&  
&   
& 
&
& 15.\,(b,2)
\\[4pt] 
&
&  
&   
& (6)-2
&
& 16.\,(b,a)
\\[4pt]  \hline
\end{array} \nonumber
\end{align}
\caption{\small
Summary table of 
$\mbox{P/K}\,(\mbox{R}_{6u\bar{d}})$
\label{ap_B_3_tab6}}
\end{table}
%
%
\begin{table}[h!]
  \centering
\begin{align}
&\mbox{P/K}\,(\mbox{R}_{7u} = ug \to ud\bar{d})
\nonumber\\
&
\begin{array}{|c|c|c|c|c|c|c|} \hline
& & & & & & \\[-12pt]
\mbox{Leg}
&{\tt Dip}\,j
& \mbox{B}j 
& S_{\mbox{{\tiny B}}_{j}}
& \mbox{Splitting} 
& P^{
\mbox{{\tiny F}}(x_{a/b})\mbox{{\tiny F}}(y_{emi})
}/\mbox{T}_{\mbox{{\tiny F}}(y_{emi})}^{2}
& (y_{emi},y_{spe})
\\[4pt] \hline
& & & & & & \\[-12pt]
a
&{\tt Dip\,3u}
&\mbox{B}3u=gg \to d\bar{d}
& 1    
& (6)-0
&
& 1.\,(a,0)
\\[4pt] 
&
&  
&   
& (6)-1
& P^{fg}/\mbox{C}_{\mbox{{\tiny A}}}
& 2.\,(a,1)
\\[4pt] 
&
&  
&   
& 
&
& 3.\,(a,2)
\\[4pt] 
&
&  
&   
& (6)-2
&
& 4.\,(a,b)
\\[4pt]  \hline
& & & & & & \\[-12pt]
b
&{\tt Dip\,4u}
&\mbox{B}4u=u\bar{u} \to d\bar{d}
& 1    
& (7)-0
&
& 5.\,(b,0)
\\[4pt] 
&
&  
&   
& (7)-1
& P^{gf}/\mbox{C}_{\mbox{{\tiny F}}}
& 6.\,(b,1)
\\[4pt] 
&
&  
&   
& 
&
& 7.\,(b,2)
\\[4pt] 
&
&  
&   
& (7)-2
&
& 8.\,(b,a)
\\[4pt]  \cline{2-7}
& & & & & & \\[-12pt]
&{\tt Dip\,4d}
&\mbox{B}4d=u\bar{d} \to u\bar{d}
& 1    
& (7)-0
&
& 9.\,(b,0)
\\[4pt] 
&
&  
&   
& (7)-1
& P^{gf}/\mbox{C}_{\mbox{{\tiny F}}}
& 10.\,(b,1)
\\[4pt] 
&
&  
&   
& 
&
& 11.\,(b,2)
\\[4pt] 
&
&  
&   
& (7)-2
&
& 12.\,(b,a)
\\[4pt]  \cline{2-7}
& & & & & & \\[-12pt]
&{\tt Dip\,4\bar{d}}
&\mbox{B}4\bar{d}=ud \to ud
& 1    
& (7)-0
&
& 13.\,(b,0)
\\[4pt] 
&
&  
&   
& (7)-1
& P^{gf}/\mbox{C}_{\mbox{{\tiny F}}}
& 14.\,(b,1)
\\[4pt] 
&
&  
&   
& 
&
& 15.\,(b,2)
\\[4pt] 
&
&  
&   
& (7)-2
&
& 16.\,(b,a)
\\[4pt]  \hline
\end{array} \nonumber
\end{align}
\caption{\small
Summary table of 
$\mbox{P/K}\,(\mbox{R}_{7u})$
\label{ap_B_3_tab7}}
\end{table}
%
%
\begin{table}[h!]
  \centering
\begin{align}
&\mbox{P/K}\,(\mbox{R}_{8u} = u\bar{u} \to ggg)
\nonumber\\
&
\begin{array}{|c|c|c|c|c|c|c|} \hline
& & & & & & \\[-12pt]
\mbox{Leg}
&{\tt Dip}\,j
& \mbox{B}j 
& S_{\mbox{{\tiny B}}_{j}}
& \mbox{Splitting}
& P^{
\mbox{{\tiny F}}(x_{a/b})\mbox{{\tiny F}}(y_{emi})
}/\mbox{T}_{\mbox{{\tiny F}}(y_{emi})}^{2}
& (y_{emi},y_{spe})
\\[4pt] \hline
a
&{\tt Dip\,1}  
& \mbox{B}1=u\bar{u} \to gg
& 2 
& (3)-0   
&
& 1.\,(a,0)
\\[4pt]  
&
&  
&   
& (3)-1
& P^{ff}/\mbox{C}_{\mbox{{\tiny F}}}
& 2.\,(a,1)
\\[4pt]  
&
&  
&   
& 
&
& 3.\,(a,2)
\\[4pt]  
&
&  
&   
& (3)-2
&
& 4.\,(a,b)
\\[4pt]  \hline
b
&{\tt Dip\,1}  
& \mbox{B}1=u\bar{u} \to gg
& 2
& (3)-0   
&
& 5.\,(b,0)
\\[4pt]  
&
&  
&   
& (3)-1
& P^{ff}/\mbox{C}_{\mbox{{\tiny F}}}
& 6.\,(b,1)
\\[4pt]  
&
&  
&   
& 
&
& 7.\,(b,2)
\\[4pt]  
&
&  
&   
& (3)-2
&
& 8.\,(b,a)
\\[4pt]  \hline
\end{array} \nonumber
\end{align}
\caption{\small
Summary table of 
$\mbox{P/K}\,(\mbox{R}_{8u})$
\label{ap_B_3_tab8}}
\end{table}
%
%
\begin{table}[h!]
  \centering
\begin{align}
&\mbox{P/K}\,(\mbox{R}_{9u} = ug \to ugg)
\nonumber\\
&
\begin{array}{|c|c|c|c|c|c|c|} \hline
& & & & & & \\[-12pt]
\mbox{Leg}
&{\tt Dip}\,j
& \mbox{B}j 
& S_{\mbox{{\tiny B}}_{j}}
& \mbox{Splitting} 
& P^{
\mbox{{\tiny F}}(x_{a/b})\mbox{{\tiny F}}(y_{emi})
}/\mbox{T}_{\mbox{{\tiny F}}(y_{emi})}^{2}
& (y_{emi},y_{spe})
\\[4pt] \hline
a
&{\tt Dip\,1}  
&\mbox{B}1=ug \to ug
& 1 
& (3)-0   
&
& 1.\,(a,0)
\\[4pt]  
&
&  
&   
& (3)-1
& P^{ff}/\mbox{C}_{\mbox{{\tiny F}}}
& 2.\,(a,1)
\\[4pt]  
&
&  
&   
& 
&
& 3.\,(a,2)
\\[4pt]  
&
&  
&   
& (3)-2
&
& 4.\,(a,b)
\\[4pt]  \cline{2-7}
&{\tt Dip\,3u}
&\mbox{B}3u=gg \to gg
& 2   
& (6)-0
&
& 5.\,(a,0)
\\[4pt] 
&
&  
&   
& (6)-1
& P^{fg}/\mbox{C}_{\mbox{{\tiny A}}}
& 6.\,(a,1)
\\[4pt] 
&
&  
&   
& 
&
& 7.\,(a,2)
\\[4pt] 
&
&  
&   
& (6)-2
&
& 8.\,(a,b)
\\[4pt]  \hline
b
&{\tt Dip\,1}  
&\mbox{B}1=ug \to ug
& 1 
& (4)-0   
&
& 9.\,(b,0)
\\[4pt]  
&
&  
&   
& (4)-1
& P^{gg}/\mbox{C}_{\mbox{{\tiny A}}}
& 10.\,(b,1)
\\[4pt]  
&
&  
&   
&
& 
& 11.\,(b,2)
\\[4pt]  
&
&  
&   
& (4)-2
&
& 12.\,(b,a)
\\[4pt]  \cline{2-7}
&{\tt Dip\,4u}
&\mbox{B}4u=u\bar{u} \to gg
& 2
& (7)-0
&
& 13.\,(b,0)
\\[4pt] 
&
&  
&   
& (7)-1
& P^{gf}/\mbox{C}_{\mbox{{\tiny F}}}
& 14.\,(b,1)
\\[4pt] 
&
&  
&   
& 
&
& 15.\,(b,2)
\\[4pt] 
&
&  
&   
& (7)-2
&
& 16.\,(b,a)
\\[4pt]  \hline
\end{array} \nonumber
\end{align}
\caption{\small
Summary table of 
$\mbox{P/K}\,(\mbox{R}_{9u})$
\label{ap_B_3_tab9}}
\end{table}
%
%
\begin{table}[h!]
  \centering
\begin{align}
&\mbox{P/K}\,(\mbox{R}_{10u}=gg \to u\bar{u}g)
\nonumber\\
&
\begin{array}{|c|c|c|c|c|c|c|} \hline
& & & & & & \\[-12pt]
\mbox{Leg}
&{\tt Dip}\,j
& \mbox{B}j 
& S_{\mbox{{\tiny B}}_{j}}
& \mbox{Splitting} 
& P^{
\mbox{{\tiny F}}(x_{a/b})\mbox{{\tiny F}}(y_{emi})
}/\mbox{T}_{\mbox{{\tiny F}}(y_{emi})}^{2}
& (y_{emi},y_{spe})
\\[4pt] \hline
a
&{\tt Dip\,1}  
&\mbox{B}1=gg \to u\bar{u}
& 1 
& (4)-0   
&
& 1.\,(a,0)
\\[4pt]  
&
&  
&   
& (4)-1
& P^{gg}/\mbox{C}_{\mbox{{\tiny A}}}
& 2.\,(a,1)
\\[4pt]  
&
&  
&   
& 
&
& 3.\,(a,2)
\\[4pt]  
&
&  
&   
& (4)-2
&
& 4.\,(a,b)
\\[4pt]  \cline{2-7}
&{\tt Dip\,4u}
&\mbox{B}4u=\bar{u}g \to \bar{u}g
& 1    
& (7)-0
&
& 5.\,(a,0)
\\[4pt] 
&
&  
&   
& (7)-1
& P^{gf}/\mbox{C}_{\mbox{{\tiny F}}}
& 6.\,(a,1)
\\[4pt] 
&
&  
&   
& 
&
& 7.\,(a,2)
\\[4pt] 
&
&  
&   
& (7)-2
&
& 8.\,(a,b)
\\[4pt]  \cline{2-7}
&{\tt Dip\,4\bar{u}}
&\mbox{B}4\bar{u}=ug \to ug
& 1    
& (7)-0
&
& 9.\,(a,0)
\\[4pt] 
&
&  
&   
& (7)-1
& P^{gf}/\mbox{C}_{\mbox{{\tiny F}}}
& 10.\,(a,1)
\\[4pt] 
&
&  
&   
&
& 
& 11.\,(a,2)
\\[4pt] 
&
&  
&   
& (7)-2
&
& 12.\,(a,b)
\\[4pt]  \hline
b
&{\tt Dip\,1}  
&\mbox{B}1=gg \to u\bar{u}
& 1 
& (4)-0   
&
& 13.\,(b,0)
\\[4pt]  
&
&  
&   
& (4)-1
& P^{gg}/\mbox{C}_{\mbox{{\tiny A}}}
& 14.\,(b,1)
\\[4pt]  
&
&  
&   
&
& 
& 15.\,(b,2)
\\[4pt]  
&
&  
&   
& (4)-2
&
& 16.\,(b,a)
\\[4pt]  \cline{2-7}
&{\tt Dip\,4u}
&\mbox{B}4u=\bar{u}g \to \bar{u}g
& 1    
& (7)-0
&
& 17.\,(a,0)
\\[4pt] 
&
&  
&   
& (7)-1
& P^{gf}/\mbox{C}_{\mbox{{\tiny F}}}
& 18.\,(a,1)
\\[4pt] 
&
&  
&   
&
& 
& 19.\,(a,2)
\\[4pt] 
&
&  
&   
& (7)-2
&
& 20.\,(a,b)
\\[4pt]  \cline{2-7}
&{\tt Dip\,4\bar{u}}
&\mbox{B}4\bar{u}=ug \to ug
& 1    
& (7)-0
&
& 21.\,(a,0)
\\[4pt] 
&
&  
&   
& (7)-1
& P^{gf}/\mbox{C}_{\mbox{{\tiny F}}}
& 22.\,(a,1)
\\[4pt] 
&
&  
&   
&
& 
& 23.\,(a,2)
\\[4pt] 
&
&  
&   
& (7)-2
&
& 24.\,(a,b)
\\[4pt]  \hline
\end{array} \nonumber
\end{align}
\caption{\small
Summary table of 
$\mbox{P/K}\,(\mbox{R}_{10u})$
\label{ap_B_3_tab10}}
\end{table}
%
%
\begin{table}[h!]
  \centering
\begin{align}
&\mbox{P/K}\,(\mbox{R}_{11}=gg \to ggg)
\nonumber\\
&
\begin{array}{|c|c|c|c|c|c|c|} \hline
& & & & & & \\[-12pt]
\mbox{Leg}
&{\tt Dip}\,j
& \mbox{B}j 
& S_{\mbox{{\tiny B}}_{j}}
& \mbox{Splitting} 
& P^{
\mbox{{\tiny F}}(x_{a/b})\mbox{{\tiny F}}(y_{emi})
}/\mbox{T}_{\mbox{{\tiny F}}(y_{emi})}^{2}
& (y_{emi},y_{spe})
\\[4pt] \hline
a
&{\tt Dip\,1}  
&\mbox{B}1=gg \to gg
& 2 
& (4)-0   
&
& 1.\,(a,0)
\\[4pt]  
&
&  
&   
& (4)-1
& P^{gg}/\mbox{C}_{\mbox{{\tiny A}}}
& 2.\,(a,1)
\\[4pt]  
&
&  
&   
&
& 
& 3.\,(a,2)
\\[4pt]  
&
&  
&   
& (4)-2
&
& 4.\,(a,b)
\\[4pt]  \hline
b
&{\tt Dip\,1}  
&\mbox{B}1=gg \to gg
& 2 
& (4)-0   
&
& 5.\,(b,0)
\\[4pt]  
&
&  
&   
& (4)-1
& P^{gg}/\mbox{C}_{\mbox{{\tiny A}}}
& 6.\,(b,1)
\\[4pt]  
&
&  
&   
&
& 
& 7.\,(b,2)
\\[4pt]  
&
&  
&   
& (4)-2
&
& 8.\,(b,a)
\\[4pt]  \hline
\end{array} \nonumber
\end{align}
\caption{\small
Summary table of 
$\mbox{P/K}\,(\mbox{R}_{11})$
\label{ap_B_3_tab11}}
\end{table}


\clearpage
{\footnotesize


}

\end{document}